\UseRawInputEncoding
\documentclass[sn-mathphys,pdflatex]{sn-jnl}

\usepackage{amssymb}
\usepackage{amsmath}
\usepackage{bm}
\usepackage[utf8]{inputenc}
\usepackage{longtable}
\usepackage{lineno}

\jyear{2021}%

\theoremstyle{thmstyleone}%
\theoremstyle{thmstyletwo}%

\theoremstyle{thmstylethree}%

\begin{document}

\title[Nuclear electromagnetic moments]{Nuclear electromagnetic moments by spin-precession methods}

\author*[1]{\fnm{Georgi} \sur{Georgiev}}\email{georgi.georgiev@ijclab.in2p3.fr}

\author*[2]{\fnm{Dimiter L.} \sur{Balabanski}}\email{dimiter.balabanski@eli-np.ro}
\equalcont{These authors contributed equally to this work.}

\author*[3]{\fnm{Andrew E.} \sur{Stuchbery}}\email{andrew.stuchbery@anu.edu.au}
\equalcont{These authors contributed equally to this work.}

\author*[4]{\fnm{Hideki} \sur{Ueno}}\email{ueno@riken.jp}
\equalcont{These authors contributed equally to this work.}

\affil[1]{\orgdiv{IJCLab}, \orgname{CNRS/IN2P3, Universit\'{e} Paris-Saclay, UMR9012}, \orgaddress{\street{15 rue Georges Clemenceau}, \city{Orsay}, \postcode{91405},\country{France}}}

\affil[2]{\orgdiv{Extreme Light Infrastructure - Nuclear Physics (ELI-NP)}, \orgname{Horia Hulubei National Institute for R$\&$D in Physics and Nuclear Engineering (IFIN-HH)}, \orgaddress{\city{Bucharest-Magurele}, \postcode{077125}, \country{Romania}}}

\affil[3]{\orgdiv{Department of Nuclear Physics and Accelerator Applications, RSPhys}, \orgname{Australian National University}, \orgaddress{\city{Canberra}, \postcode{2601}, \state{ACT}, \country{Australia}}}

\affil[4]{\orgname{RIKEN Nishina Center for Accelerator-Based Science}, \orgaddress{\street{2-1 Hirosawa}, \city{Wako}, \postcode{351-0198}, \state{Saitama}, \country{Japan}}}

\abstract{Nuclear moment studies carried out with spin-precession methods at and after the turn of the millennium are critically assessed. A period of about 30 years is covered, during which  much of} the focus of nuclear structure research shifted from high-spin physics to studies of neutron-rich exotic nuclei. The formalism for the extraction of nuclear moments is described. The $\beta$-nuclear magnetic resonance/nuclear quadrupole resonance ($\beta$-NMR/NQR), the time-dependent perturbed angular distribution (TDPAD), the transient field, the recoil-in-vacuum (RIV), and the tilted-foils methods for measurements of nuclear magnetic dipole and electric quadrupole moments are described in detail, as well as the requirements for their application in studies of exotic nuclei. The impact of nuclear-moment measurements on the understanding of key topics of nuclear structure research is discussed.  {Key results on short-lived states, mainly from transient-field measurements, are reviewed. Included are comparisons with large-basis shell model calculations, discussions on the nature of weakly-collective nuclei, insights into emerging collectivity away from closed shells, and electromagnetic properties of odd-$A$ rotors.} In the field of high-spin physics, research related to high-spin yrast and $\mathrm{K}$ isomers, superdeformation, magnetic, anti-magnetic, and chiral rotation is covered. In neutron-rich exotic nuclei, studies related to the $\mathrm{N=20}$, $\mathrm{N=28}$ and $\mathrm{N=40}$ ``islands of inversion'', the structure of nuclei around $^{68-78}$Ni and $^{132}$Sn, and in the $A \sim 100$ mass region are discussed.




\maketitle
\tableofcontents

\section{Introduction}
\label{intro}
Nuclear moments are basic characteristics of a given nuclear state. Magnetic dipole moments provide information about the nuclear wave function, and electric quadrupole moments are a measure of the deformation of the state of interest. For this reason, measurements of nuclear moments and their relation to different nuclear models have long  been central
to understanding the structure of nuclear excitations.

The principle of the various techniques that are used to measure nuclear moments is to force the nucleus to interact with some external magnetic and/or electric fields and measure the outcome of this interaction. Hence, the experimental observables are the results of the interactions, not the nuclear moments themselves. For an optimum measurement, the magnitude of the interaction should be chosen such that it has a considerable impact on the nucleus. It is clear that for the extraction of the nuclear moments, it is necessary to know the parameters of the applied external fields. In many cases, nuclei interact with atomic or lattice hyperfine fields. Thus, nuclear moment studies reside at the interface of nuclear, atomic, molecular, solid-state and laser physics. For basic texts related to nuclear moments, see Refs.~\cite {cast90,reck74,blin53,rams53,kopf58}.
The 1983 volume on \textit{Hyperfine Interactions of Radioactive Nuclei} \cite{chri83}, although becoming dated, provides a useful overview with an emphasis on the exploitation and application of hyperfine interactions.

Over the decades, the progress in the field has been reviewed frequently, often from the point of view of the different experimental techniques. Review articles related to the techniques addressed in this paper can be found in Ref.~\cite{spei02,benc07} on the transient field measurements, Ref.~\cite{neye03} on the use of spin-oriented radioactive beams, and on recoil in vacuum in Refs.~\cite{gold82,gold85}
and Ref.~\cite{berr82} for the tilted-foil technique.

\subsection{Scope and outline of the paper}
\label{intro_scope}
The present paper aims to review the recent advances and experimental results in nuclear moment studies  {by spin-precession techniques}, with a focus on nuclear structure, exotic nuclei and exotic nuclear excitations in the period at and after the turn of the millennium. The concept and basic definitions for nuclear dipole and electric quadrupole moments are introduced, and their relation to different nuclear structure models is discussed.

Within the period covered by this review, a major transformation of the focus of nuclear structure research took place. Nuclear structure studies at high angular momenta, which were dominating the field starting in the 1970s, were gradually replaced by studies in neutron-rich nuclei. This shift is related to the technological developments in the field of accelerator technologies. In the 1970s, heavy-ion accelerators were installed in all major laboratories worldwide, which provided the foundation of the era of high-spin physics. The development of accelerators able to deliver heavy-ion beams with sufficient intensities and beam energies went hand-in-hand with the detector developments, a game-changer being the construction of $4\pi$ multi-detector $\gamma$-ray arrays. As a result, the method of coincidence $\gamma\gamma$ spectroscopy was applied in these studies, which reached a sensitivity level for weak excitations at high angular momenta of~$\sim 10^{-6}$. Such an improvement of the sensitivity of the nuclear physics experiment can be compared with the replacement of the magnifying glass with the optical microscope. Fusion-evaporation reactions were the workhorse in these studies, which resulted in extremely detailed nuclear structure studies in neutron-deficient nuclei. An important feature of these reactions is that a significant amount of spin alignment of the reaction residues is created, of the order of 40\%,  which enables nuclear moment measurements. \\
The next step in the development of the methodology of nuclear structure experiments was related to the production of beams of radioactive isotopes (RIBs) using projectile fragmentation reactions or the isotope separation on-line (ISOL) technique. This resulted in intense studies of the structure of neutron-rich nuclei. Numerous innovative techniques for alignment or polarization of spin ensembles were developed, which made it possible to investigate nuclear moments far away from stability.

Several widely used experimental methods, such as the $\beta$-nuclear magnetic resonance/nuclear quadrupole resonance ($\beta$-NMR/NQR) technique for ground-state studies, the time-dependent perturbed angular distribution technique (TDPAD) technique for isomeric studies, and the transient field (TF) and the recoil in vacuum (RIV) techniques for investigations of short-lived excited states, are discussed in detail. Research with the tilted-foil method is also summarized. The basic formulae for the extraction of nuclear electromagnetic moments, when using the experimental techniques of interest, are provided. The review covers measurements of nuclear moments of ground states in exotic, neutron-rich nuclei, which are produced in experiments with intermediate-energy or relativistic beams, studies of nuclear moments at high angular momenta, and studies of nuclear moments of short-lived states.  {The aim of the present paper is to review the experimental techniques for nuclear moment studies, particularly based on spin-rotation techniques.} Some of the methods described here were didactically introduced in Ref.~\cite{neug06}. Laser-based methods for nuclear moment measurements, such as collinear laser spectroscopy, resonance ionization spectroscopy, and in-source laser spectroscopy,  {have been widely used over the last decades and have provided a plethora of new experimental results. The laser spectroscopy techniques for exotic nuclei (ground states and long-lived isomeric states) have been recently reviewed, see {\it e.g.} Refs.~\cite{otte89,neug06,yang23,kosz24,kosz24a}. Those reviews can be considered complementary to the content of the present work. Where relevant, the physics results obtained by laser methods will be discussed in the present review, but the laser experimental methods go beyond the scope of the present paper.}

Next to the experimental methods and the requirements for their application, we discuss key nuclear structure topics, which were addressed by nuclear-moment measurements in the last decades, such as the structure of the 'islands of inversion' in neutron-rich nuclei at $N=20$, $N=28$ and $N=40$, studies around $^{132}$Sn, the onset of deformation at $A \approx 100$, high-$K$ isomers, magnetic, anti-magnetic and chiral rotation, quasiparticle alignment, superdeformation, multi-quasiparticle excitations, {\it etc}.

\subsection{Basic definitions of nuclear moments}
\label{intro_basic_concepts}
The idea that the matter around us is composed of indivisible units or building blocks was proposed in ancient Greece more than two millennia ago. The word atom is derived from the Greek word $\alpha\tau$o$\mu$o$\varsigma$, which means indivisible or uncuttable. With the advances of physics in the nineteenth century, it was discovered that the entities named atoms are not elementary particles but much more complex structures. In the center of the atom, which has a radius of about an Angstrom, 1\: \AA~=~$10^{-10}$~m, is the atomic nucleus having a radius which is five orders of magnitude smaller $\sim 10^{-15}$~m. The nucleus is a composite system, too. It contains $ Z$ positively charged protons and $N$ electrically neutral neutrons, together commonly called nucleons. Neither the protons nor the neutrons are elementary particles; however, for the purposes of the present paper, we will neglect the underlying quark structure because the excitation energies that are necessary to excite a nucleon are about three orders of magnitude higher than the energy scale of the nuclear excitations discussed in this paper.

Although the structure of nuclei is dominated by the strong interaction, which is about two orders of magnitude stronger than the electromagnetic one, a large amount of the experimental nuclear structure information is obtained through electromagnetic probes. The electric charge of the nuclei and the currents, resulting from the movement of the nucleons, are predominantly determined by the protons. They are influenced by the neutrons as well through the proton-neutron interaction.

 Any object that produces a magnetic field has a magnetic moment. The magnetic moment depends on the magnetic field strength and its orientation.
Among the examples of objects that possess a magnetic moment are current loops, permanent bar magnets, and composite quantum objects such as molecules, nuclei, {\it etc}. When placed in an external magnetic field $\boldsymbol{B}$, such an object experiences a torque, $\boldsymbol{\tau}$, which tends to align its moment with the field direction. The magnetic dipole moment, $\boldsymbol{\mu}$, is a measure of the strength of this torque. The torque is defined as the vector product of the magnetic field and the magnetic moment $\boldsymbol{\tau}$ = $\boldsymbol{\mu} \times \boldsymbol{B}$. The potential energy, related to the orientation of the magnetic dipole immersed in the magnetic field, is the scalar product of these two vectors $E = -\boldsymbol{\mu} \cdot \boldsymbol{B}$.

A magnetic dipole moment can be created by a loop current, $\boldsymbol{\mu} = i\boldsymbol{A}$, where $i$ is the current and $\boldsymbol{A}$ is the area of the loop with direction perpendicular to the plane of the loop. This concept can be generalized to the classical case of a particle with a mass $m$ and a charge $q$, orbiting with a velocity $v$ on a radius $r$.  The magnetic moment is:
\begin{equation}\label{mu_classical}
  \boldsymbol{\mu} = \frac{q\boldsymbol{r}\times\boldsymbol{v}}{2} = \frac{q\bm{\ell}}{2m}
\end{equation}
where $\boldsymbol{\ell} = m\boldsymbol{r}\times\boldsymbol{v}$ is the orbital angular momentum of the particle.

The electric charge and the current densities of the nuclei can be described using a multipole expansion in terms of spherical harmonics. This involves multipole moments of different orders, such as monopole, dipole, quadrupole, octupole, and higher orders. Due to the symmetries of the nuclear structure and their respective operators, some of these moments can be identical to zero for some or for all nuclear states. Here, we will discuss the magnetic dipole and the electric quadrupole moments. It is worth mentioning that the electric monopole moment corresponds to the nuclear charge. The magnitude of the interaction between the nuclear moment and the external field generally decreases with the increasing multipole order. As a result, there are few higher-order multipole moment measurements.
On the theoretical side, higher-order multipole moments, up to magnetic triakontadupole moments ($M5$), have been calculated \cite{brow80}. However, from the comparison between the higher-order theoretical moments and the existing experimental values for \textit{sd}-shell nuclei, simple conclusions cannot be drawn. It has been suggested that this could be due to either unrealistic assignments of the experimental uncertainties in specific cases or to possible fundamental defects of the theoretical approach \cite{brow80}. However, several measurements of magnetic octupole moments,  $\Omega_3 = -\langle II \vert \hat{M}_3^0 \vert II \rangle$, where $\hat{M}_3^0$ is the magnetic octupole moment operator, were carried out in stable nuclei. For compilations see Ref.~\cite{degr21,bofo24}.

The magnetic dipole moment of a nucleus arises from both the intrinsic spins and the orbital motions of the protons and neutrons. Any nuclear state with a total angular momentum (spin) larger than zero, $I \geq 1/2$, has a non-zero magnetic moment. The concept of nuclear angular momentum and the nuclear magnetic moment was introduced by Pauli in 1924 \cite{paul24} in order to explain the experimentally observed hyperfine structure in some optical spectra. Already the first experimental observations of nuclear electromagnetic moments, dating back to the 1930's, had an important impact on our understanding of nuclei and their structure. For example, the first classification of the nuclear magnetic moments by Schmidt in 1937 \cite{schm37} demonstrated the dependence of the magnetic dipole moments on the orbital angular momentum of the valence nucleon. These observations provided experimental evidence for the importance of the shell structure in atomic nuclei.  {This was nicely illustrated with some specific examples in a contribution to the recent Handbook of Nuclear Physics \cite{degr20a}.}

Atomic nuclei are quantum objects, and their properties are defined in terms of quantum mechanics. Replacing the  orbital angular momentum of Eq.~(\ref{mu_classical}) by $\bm{\ell} \hbar$, where  $\hbar$ is the reduced Planck constant, gives
\begin{equation}\label{mu_qm}
  \boldsymbol{\mu} = \frac{q \hbar \bm{\ell}}{2m}.
\end{equation}
Based on this expression, nuclear magnetic moments are given in units of nuclear magnetons,
\begin{equation}\label{eq:muN}
\mu_N = \frac{e \hbar} {2m_p},
\end{equation}
where $e$ is the elementary charge and $m_p$ is the proton mass. The experimental value is $\mu_N = 5.050 783 7461(15) \cdot 10^{-27} [J T^{-1}]$.

It is worth noting that nuclear magnetism is considerably weaker than atomic magnetism because the basic unit of atomic magnetism is the Bohr magneton, which replaces the proton mass in Eq.~(\ref{eq:muN}) by the electron mass; thus, the proton-to-electron mass ratio, $m_p/m_e = 1836.152 673 43(11)$, defines the ratio between atomic and nuclear magnetism.

The electric quadrupole moment is a measure of the deviation of the nuclear charge distribution from sphericity. Therefore, it provides information on the deformation and shape of nuclei. These quantities are related to collective nuclear properties. Although a macroscopic object can possess an electric quadrupole moment, it is a property that is more often related to the properties of microscopic systems, such as molecules, atoms, and nuclei. Due to the properties of the quadrupole moment operator, see Eq.~(\ref{eq:q_moment_spec}) below, no electric quadrupole moment can be defined for nuclear states with spin $I\lt 1$.

\subsection{Nucleon magnetic moments}
\label{intro_nucleon}
Within the Dirac theory, the magnetic moment associated with the intrinsic spin of an elementary spin-$1/2$ particle is
\begin{equation}\label{eqn:mu_nucleon}
  \boldsymbol{\mu} = \frac{q\hbar}{2m} \boldsymbol{\sigma},
\end{equation}
where $\bm{\sigma }$ is the Pauli spin matrix.
Inserting the electron charge and mass gives an expression identical to the Bohr magneton. Thus, the intrinsic magnetic moment of the electron, in Bohr magnetons, should be identical to unity, {\it i.e.}, by the definition of the $g$ factor, $g = 2$. This is very close to the experimental value. The deviations of the order of 0.2$\%$ are closely scrutinized in terms of quantum electrodynamic corrections, which can be calculated to extremely high precision but are not within the scope of the present paper.

By using the same arguments, the $g$ factors of the proton and the neutron would be $g^{p,\rm{free}}_s=2$ and $g^{n,\rm{free}}_s=0$, respectively, if the proton and the neutron were structureless particles. However, they have a complex quantum chromodynamic structure, and experimentally, the intrinsic (spin) $g$~factors of the free proton and neutron are $g^{p,\rm{free}}_s = +5.5856946893(16)$ and $g^{n,\rm{free}}_s = -3.82608552(90)$. These large intrinsic $g$~factors, which exceed the Dirac values and have opposite signs for protons and neutrons, provide a sensitive tool for probing the structure of nuclear states. The nuclear magnetism in general, and the nuclear magnetic dipole moments in particular, depend on the origin of the nuclear angular momentum, {\it e.g.}, protons {\it vs} neutrons and intrinsic spin {\it vs} orbital motion.

\subsection{Nuclear magnetic moments}
\label{intro_magnetic_moments}
All nuclear states with angular momentum $I \geq 1/2$ have a magnetic moment. Eq.~(\ref{mu_qm}) can be generalized  considering the fact that nuclear magnetism is not only due to the orbital movement of the nucleons, but also to their spin

\begin{equation}\label{eqn:mu_nuclear}
  \boldsymbol{\mu} = g\boldsymbol{I}\mu_N,
\end{equation}

\noindent where $\boldsymbol{I}$ is the vector of the total angular momentum of a given nuclear state and $g$ is its gyromagnetic factor, a dimensionless quantity that characterizes the relation between the magnetic moment and the angular momentum of the system. As mentioned earlier, it is appropriate to specify the nuclear magnetic moments in nuclear magnetons. Therefore, one can write for the expectation value of the magnetic moment $\mu = gI$, the $\mu_N$ unit being understood. As will be noted in the next paragraph, spin precession measurements directly measure the $g$~factor. Also, in comparisons of theory and experiment, it is often the $g$~factor (sometimes called the gyromagnetic factor) that gives the more transparent nuclear structure information.

In some works, a distinction can be found between the $g$ factor, $g$, and the gyromagnetic ratio, $\gamma_R$. These two quantities are closely related, originating from the spin precession (Larmor) frequency, $\omega_\textrm{L}$, of a magnetic moment in an external magnetic field with strength $B$

\begin{equation}\label{eq:gammaR}
 \omega _\textrm{L} = -\frac{g\mu_N}{\hbar}B = -\gamma_R B,
\end{equation}
 thus $\gamma_R = g\mu_N/\hbar$. The $g$ factor is dimensionless, while the gyromagnetic ratio has the SI units of radian per second per tesla $[s^{-1} T^{-1}]$.
 It is worth noting that in many nuclear structure studies, the terms $g$~factor and gyromagnetic ratio are used interchangeably, the usage usually being clear from the context.

\subsection{Nuclear quadrupole moments}
\label{sect:intro_quadrupole}
The nuclear electric quadrupole moment is a measure for the deviation from sphericity of the nuclear charge distribution. The electric quadrupole moment operator in Cartesian coordinates is:
\begin{equation}\label{eq:q_moment_Cart}
  \hat{Q} = \sum_{i=1}^{A}e_i(3z_i^2 - r_i^2)
\end{equation}
where $\emph{e}_i$  {is the charge of the \emph{$i_{th}$} nucleon, and $z_i$ and $r_i$ are its position coordinates, $r_i$ being the distance from the origin  \cite{cast90}}. However, it is more common to express the quadrupole moment in spherical coordinates, and hence in terms of spherical harmonics, giving
\begin{equation}\label{eq:q_moment_spherical}
  \hat{Q} = \sqrt{\frac{16\pi}{5}}\sum_{i=1}^{A}e_ir_i^2Y^0_2(\theta_i, \phi_i).
\end{equation}

The nuclear moments of a state with spin $I$ are defined as the expectation value of the relevant multipole operator evaluated in the substate with $m=I$. Thus, for the quadrupole moment,
\begin{equation}\label{eq:q_moment_spec}
  Q_s(I) = \langle II \vert \hat{Q} \vert II \rangle = \sqrt{\frac{I(2I-1)}{(2I+1)(I+1)(2I+3)}} \langle I \vert\vert \hat{Q} \vert\vert I \rangle,
\end{equation}
where $Q_s$ is the spectroscopic quadrupole moment of the nucleus, a quantity that can be determined experimentally, and $ \langle I \vert\vert \hat{Q} \vert\vert I \rangle$
is the reduced matrix element. From this expression, it becomes evident that the spectroscopic quadrupole moments of nuclear states with spins $I = 0$ or $1/2$ are zero, although they may have an intrinsic structure that deviates from sphericity, {\it i.e.} the spherical symmetry is broken, and they are deformed, but in these cases the deformation cannot be observed via a quadrupole moment.

The relation between the spectroscopic quadrupole moment, $Q_\textrm{s}$, and the intrinsic quadrupole moment in the rest frame of a deformed nucleus, $Q_0$, is model dependent. An assumption, referred to as the strong-coupling limit, is that the nuclear state is axially deformed and that the projection of the nuclear spin on the symmetry axis, $K$, is well defined. In this case,
\begin{equation}\label{eq:q_s-q_0}
  Q_\textrm{s} = \frac{3K^2 - I(I+1)}{(I+1)(2I+3)}Q_0.
\end{equation}
This model is often used, but is certainly not always valid.

The intrinsic quadrupole moment $Q_0$ for an odd-$A$ nuclide has two components, a single-particle $Q_{sp}$ and a core $Q_c$ part. The single-particle quadrupole moment of a nucleon on a $j$ orbit is
\begin{equation}\label{eq:q_sp}
  Q_{sp}^j = -e_j\frac{2j-1}{2j+2}\langle r_j^2\rangle,
\end{equation}
where the $\langle r_j^2 \rangle$ is the mean square radius of the orbit of the nucleon, and $e_j = e_\pi, e_\nu$ is its charge. The negative sign in Eq.~(\ref{eq:q_sp}) is because a particle moving outside of a closed shell has a negative moment corresponding to oblate deformation, while a hole would create a positive, {\it i.e.}, prolate deformation, $e_j^{\mathrm{hole}} = -e_j^{\mathrm{particle}}$. The single-particle moment $Q_{sp}^j$ can be calculated using bare charges of $e_{\pi}= 1$ and $e_{\nu} = 0$ since the core polarization effect can be taken care of by the core quadrupole moment. The single-particle component of the quadrupole moment is zero for even-even nuclei, for which only the core quadrupole moment remains.

The core quadrupole moment can be evaluated using the hydrodynamical model. There are formulae in the literature that give the relation between the core quadrupole moment, $Q_c$, which is equal to $Q_0$ for even nuclei, and the various nuclear deformation parameters. L\"obner et al. \cite{lobn70} give a useful tabulation of the relationships. Here we give expressions for $Q_0$ in terms of the Woods-Saxon deformation parameter $\beta$ and note the difference between assuming a sharp versus a diffuse nuclear surface.

In the case of a sharp nuclear surface, the core quadrupole moment can be expressed as
\begin{equation}\label{eq:q_moment_core_sharp}
  Q_c = \frac{3}{\sqrt{5\pi}}eZR^2\left\{\beta + \frac{2}{7}\sqrt{\frac{5}{\pi}}\beta^2 \right\},
\end{equation}
where the nuclear radius is taken as $R = 1.2 \cdot A^{1/3}$ fm. If the diffuseness of the nuclear surface $a$ is also taken into account, then an additional term is introduced~\cite{saga88}
\begin{equation}\label{eq:q_moment_core}
  Q_c = \frac{3}{\sqrt{5\pi}}eZR^2\left\{\left[1+\pi^2\left(\frac{a}{R}\right)^2\right]\beta + \frac{2}{7}\sqrt{\frac{5}{\pi}}\beta^2 \right\},
\end{equation}
where the surface thickness may be taken as $a = 0.54$~fm. Clearly, the diffuseness is more significant for low-mass nuclei than high-mass nuclei.  For example, the additional term in Eq.~(\ref{eq:q_moment_core}) brings in a correction of less than 10\% for the mid-mass ($A \sim 100$) nuclei, while for very light nuclei ($A \sim 10$) the correction is as high as 40\%.

\subsection{Experimental techniques}
\label{sec:techniques}
 {
The experimental determination of nuclear electromagnetic moments relies on their interaction with an electromagnetic field. For example, in the classical approximation, the nucleus can be treated as a spinning charged body. The rotation of this charge creates a current loop, which in turn generates a magnetic dipole moment. The interaction of the nuclear magnetic moment with a magnetic field results in a torque, causing the nuclear spin to rotate around the magnetic field direction. The main goal of a nuclear moment measurement technique is therefore to determine the frequency of the rotation of the nuclear spin, relative to the strength of a known applied electromagnetic field.
}

 {
In the quantum-mechanical approach, the frequency of the nuclear spin rotation can be regarded as an additional energy, removing the degeneracy of the $m$ substates, which are associated with the orientation of the nucleus. This has indeed been observed as the hyperfine splitting of the atomic spectra, caused by the electromagnetic fields created by the electron cloud at the nuclear site.
}

\begin{figure}[!ht]
\centering
\includegraphics[width=0.95\columnwidth]{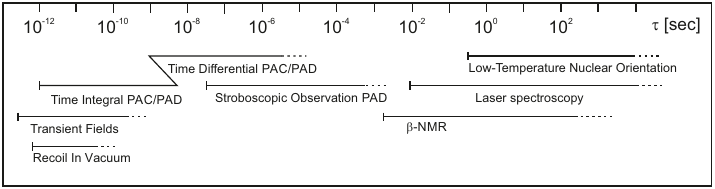}
\caption{Summary of the techniques for nuclear moment studies, considered in the present review. The figure is taken with modifications from Ref.~\cite{reck74}}
\label{fig:Recknagel}
\end{figure}

 {
In the present paper, we will focus on techniques that can be considered as a rotation of the nuclear spin ensemble.
This rotation of the nuclear spins can be observed through the variation of the angular distribution of the emitted radiation ($\beta$- or $\gamma$-rays) within the lifetime of the state of interest, which requires a proper matching between the strength of the interacting electromagnetic field and the nuclear lifetime. The use of an appropriate technique allows covering time ranges from several seconds (as in NMR) down to picoseconds (e.g., for Transient Fields or Recoil in Vacuum). A schematic overview of the techniques as a function of the nuclear lifetime is presented in Fig. \ref{fig:Recknagel} and serves as a roadmap for the discussion that follows.
}

 {
As a general guidance, nuclear moments of picosecond-to-nanosecond lifetimes can be measured using the (time-integral) Recoil In Vacuum, the Transient Field, and the Time-Integral PAD/PAC techniques. In these cases, the rotation of the nuclear spin ensemble is of the order of a few degrees (some ten milliradians), and electromagnetic (hyperfine) fields of the order of a few kilotesla are typically employed. More details can be found in Sect. \ref{sect:short-lived-meth}.
}

 {
The nanosecond-to-microsecond time range can be addressed using the Time-Differential Perturbed Angular Distribution (TDPAD) or the Time-Differential Perturbed Angular Correlation (TDPAC) techniques. Using magnetic fields of the order of a few hundred militesla up to a few tesla allows the observation of a few full rotations of the nuclear spin ensembles within the nuclear lifetime. The TDPAD/TDPAC techniques are discussed in Sect.\ref{sec:isomers}.
}

 {
Nuclear lifetimes in the tens of microseconds up to a few milliseconds represent a particularly challenging regime for nuclear moment studies. The Stroboscopic Observation Perturbed Angular Distribution (SOPAD) technique \cite{chri70} can partially cover this time range, though a detailed treatment of this method lies beyond the scope of the present paper.
}

 {
Nuclear states with lifetimes in the millisecond to a few-second time range are usually decaying through $\beta$ decay, and their nuclear moments can be measured using the Nuclear Magnetic Resonance (NMR) and the Nuclear Quadrupole Resonance (NQR) techniques, which are presented in Sect. \ref{sec:ground}. For longer-lived states up to stable nuclei, laser spectroscopy and Low-Temperature Nuclear Orientation (LTNO)\cite{post86} are well-established tools, with laser spectroscopy covering lifetimes from the millisecond range upward \cite{otte89,neug06,yang23,kosz24,kosz24a}. The present paper focuses on the techniques spanning the picosecond-to-second lifetime range, with results from laser spectroscopy and nuclear orientation included where relevant for comparison with the nuclear structure and regions of interest discussed here.
}

\section{Nuclear moments and nuclear structure }
\label{sec:1}

\subsection{Nuclear moment of a compound nuclear state from shell model perspective: Additivity relation and Schmidt values}
\label{sec:sm}
The magnetic moment, created by a nucleon moving in an orbit with a total angular momentum $\bm{j} = \bm{\ell} + \bm{s}$,  can be considered as having two components, one due to the loop current created by the moving charge (for protons), and the other due to the intrinsic spin $\bm{s}$.
The magnetic dipole moment operator, $\hat{\mu}^{(i)}$ for an individual nucleon therefore has the form
\begin{equation}\label{eq:M1}
  \hat{\mu}^{(i)} = g_{\ell}^{(i)}\hat{\ell} + g_s^{(i)}\hat{s},
\end{equation}
where the hats indicate operators and the nominal values for the orbital $g$ factors of the protons and the neutrons are respectively $g_{\ell}^{p,\rm{free}} = 1$ and $g_{\ell}^{n,\rm{free}} = 0$. The spin $g$ factors are, in principle, those of the free nucleons as indicated above.
The magnetic moment operator for the entire nucleus, $\hat{\mu}$, is simply the sum over the single-particle operators
\begin{equation}\label{eq:M1_sum}
  \hat{\mu} = \sum_{i = 1}^{A}\hat{\mu}^{(i)}.
\end{equation}
The magnetic moment of the nucleus can be obtained by evaluating the expectation value of the $z$ component of Eq.~(\ref{eq:M1_sum}) for the nuclear substate for which $m=I$.
\begin{equation}\label{eq:mu_substate}
  \mu = gI = \langle II \vert \hat{\mu}_z \vert II \rangle.
\end{equation}
As the nuclear Hamiltonian is rotationally invariant, and the operators can be written as spherical tensors, applying the Wigner-Eckart theorem gives
\begin{equation}\label{eq:mu_WE}
  \mu = gI = \sqrt{\frac{I}{(I+1)(2I+1)}}\langle I \vert\vert \hat{\mu} \vert\vert I \rangle,
\end{equation}
where $\langle I \vert\vert \hat{\mu} \vert\vert I \rangle$ is the reduced matrix element. It is worth noting that the expression of Eq.~(\ref{eq:mu_WE}) differs from the definition of Castel and Towner \cite{cast90} by a factor of $(2I+1)^{-1/2}$ due to a difference in the definition of the reduced matrix element. Here we are following the notation of Bohr and Mottelson \cite{bohr75}. Note that depending on how the $\hat{M1}$ operator is defined (or evaluated in particular computer codes), an additional factor of $\sqrt{4\pi/3}$ may also be required in Eq.~(\ref{eq:mu_WE}).

Using these definitions, the magnetic moment (or the $g$ factor) of a composite state can be evaluated. Consider a state of a total angular momentum $I$, composed of two components with angular momenta $I_1$ and $I_2$, which can be denoted as $(I_1 \otimes I_2)_I$ or $\vert I_1, I_2; I \rangle$. The total $\hat{M1}$ operator is $\hat{\mu} = \hat{\mu}_1 + \hat{\mu}_2$, where $\hat{\mu}_1$ acts on the $I_1$ part of the wave function and $\hat{\mu}_2$ acts on the $I_2$ part. The result can also be presented in terms of $g$ factors, $g = \mu/I$, $g_1 = \mu_1/I_1$ and $g_2 = \mu_2/I_2$:
\begin{equation}\label{eq:additivity}
  g = \frac{1}{2}(g_1 + g_2) + \frac{1}{2}(g_1 - g_2)\frac{I_1(I_1+1) - I_2(I_2+1)}{I(I+1)}.
\end{equation}
This expression is known as the \textit{additivity relation} or the Land\'e formula. It has two important features to be noted:
\begin{itemize}
    \item{if there are two identical nucleons in the same orbit, the $g$ factor of the compound state is $g(I) = g(j)$. Thus, the $g$ factor of all configurations of type $j^n$ for $n$ identical nucleons in the same orbit are equal to the single-nucleon $g$ factor, independent of $n$ and $I$.  }
    \item{ the $g$ factor of an individual nucleon in a shell-model orbit $j = \ell \pm 1/2$ can be calculated by replacing $I = j$, $I_1 = \ell$ and $I_2 = 1/2$} in Eq.~(\ref{eq:additivity}). With some algebra, the result is:
        \begin{equation}\label{eq:sm_gl_gs}
          g\left(j = \ell \pm \frac{1}{2}\right) = g_{\ell} \pm \frac{g_s - g_{\ell}}{2\ell + 1}.
        \end{equation}
\end{itemize}
When this expression is evaluated using free-nucleon $g$ factors, it gives the so-called Schmidt values. Indeed, although the work of Schmidt \cite{schm37} came about a decade before the seminal work of Goeppert Mayer \cite{maye48}, it gives a very clear indication of the validity of the shell model concept in the structure of the atomic nuclei because it identifies the $\ell$ and $j$ values of single-particle orbits adjacent to doubly closed shells.

\subsection{The effective \texorpdfstring{$\hat{M}1$}{} operator}
\label{sec:eff_M1}
\begin{figure}
    \centering
    \includegraphics[width=0.75\linewidth]{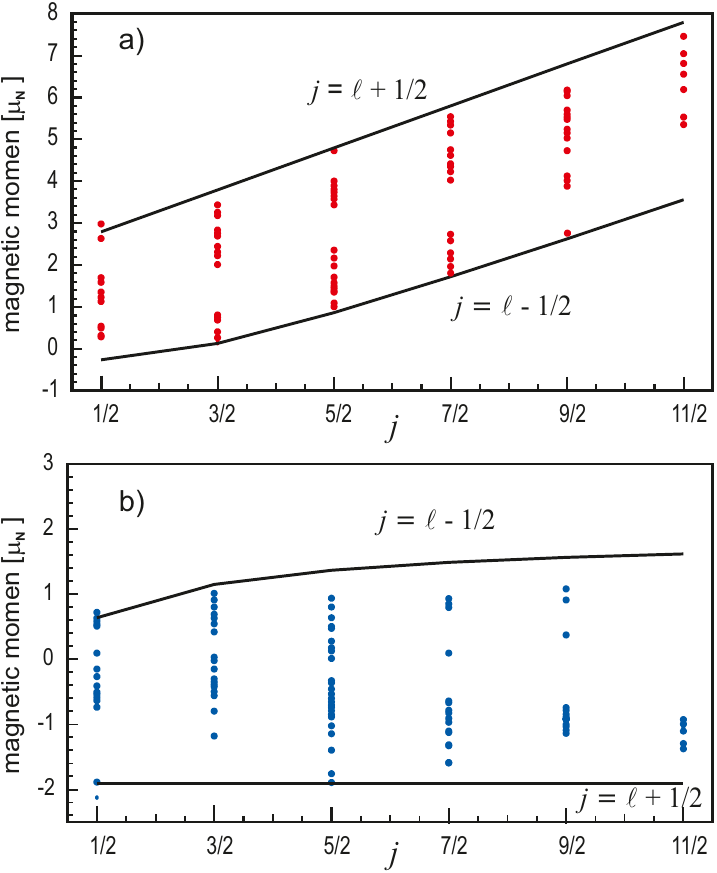}
    \caption{Comparison between the experimental magnetic dipole moments and the Schmidt lines for (a) protons and (b) neutrons.}
    \label{fig:schmidt_pn}
\end{figure}

When comparing the experimental magnetic moments of odd-mass nuclei near closed shells, it was observed that they all lie close to, but not exactly on, the Schmidt lines. The general observation is that the experimental magnetic moments lie between the Schmidt lines for $j = \ell + 1/2$ and $j = \ell -1/2$, see Fig.~\ref{fig:schmidt_pn}. The assumption that the odd nucleon occupies a pure spherical shell model orbit is clearly not a valid one in most cases. Note that deformed nuclear states are excluded from the discussion in this section.

Even if we limit the comparison to odd-mass nuclei adjacent to doubly magic nuclei, the data still show a significant deviation from the Schmidt lines. There are two main reasons for this disparity. The first one is the configuration mixing of the nuclear wave function, also called \textit{core polarization}, which is due to the core not being strictly inert. The second is due to the meson exchange currents, \textit{i.e.}, additional currents originating from the mesons that carry the nucleon-nucleon interactions.

These two effects can modify both the orbital and spin $g$ factors from their free-nucleon, {\it i.e.}, ``bare", values.
To evaluate the effects of the core polarization and the meson exchange on nuclear magnetic moments, it is worth noting that the ${\hat{M1}}$ operator, due to its properties, can change neither the principal quantum number \textit{n} nor the orbital angular momentum $\ell$. Therefore, there are only three independent non-vanishing single-particle matrix elements for a given \textit{n} and $\ell$, namely $\langle j = \ell + 1/2 \vert \vert \hat{\mu} \vert \vert j = \ell + 1/2 \rangle$, $\langle j = \ell - 1/2 \vert \vert \hat{\mu} \vert \vert j = \ell - 1/2 \rangle$ and $\langle j = \ell + 1/2 \vert \vert \hat{\mu} \vert \vert j = \ell - 1/2 \rangle$. A completely general effective $\hat{M1}$ operator can be written in the form
\begin{equation}\label{eq:mu_eff}
  \hat{\mu} = (g_{\ell}^\textrm{free} + \delta g_{\ell})\hat{\ell} + (g_s^\textrm{free} + \delta g_s)\hat{s} + g_p [\hat{Y_2},\hat{s}]_1,
\end{equation}
where $g_{\ell}^\textrm{free}$ and $g_s^\textrm{free}$ are the free-nucleon orbital and spin $g$ factors, respectively, $\delta g_{\ell}$ and $\delta g_s$ are the corrections to them, and $g_p$ is the tensor-term $g$ factor. In this general expression, there is an additional term, compared to the bare $\hat{M1}$ operator, $[\hat{Y_2},\hat{s}]_1$, consisting of a spherical harmonic coupled to a spin operator to give a spherical tensor of rank~1.

In heavier nuclei, due to the spin-orbit term of the interaction, it appears quite often that the $j_1 = \ell + 1/2$ orbit is filled, while its spin-orbit partner $j_2 = \ell - 1/2$ is empty. Particle-hole excitations of the form $(j_1^{-1}j_2^1)^{(1^+)}$ couple strongly to the $\hat{M1}$ operator and give rise to the main contribution of the core-polarization corrections. Even small contributions of these core-excited $1^+$ states can give significant corrections to the magnetic moment since the off-diagonal matrix element of $\hat{M1}$ is large, compared to other contributions. These effects were first discussed by Arima and Horie \cite{arim54a}. Extensive calculations on core polarization and meson exchange contributions to the effective $\hat{M1}$ operator for nuclei with a single nucleon outside a doubly magic core have since been reported by Towner \cite{town87} and Arima~{\it et al.}~\cite{arim87}.

An alternative approach is to treat $\delta g_{\ell}$, $\delta g_s$ and $g_p$ as parameters in a fit to extensive experimental nuclear data. Such an approach has been adopted for the $sd$ shell by Richter, Mkhize and Brown \cite{rich08}, updating earlier work by Brown and Wildenthal \cite{brow87}. A similar type of fit for the $^{208}$Pb region has been performed by Maier et al. \cite{maie72}.

The deviations of the orbital part of the nuclear magnetism, $\delta g_{\ell}$, originate predominantly from meson exchange currents (see Sec. \ref{Sec:MEC}), whereas anomalous spin $g$ factors, $g_s$, and the tensor component, $g_p$, are mainly due to first-order core polarization. The tensor component is relatively small compared to the $\delta g_s$ term, and it can often be omitted.

It is generally accepted that the corrections to the magnetic moment operator, $\delta g_s$, $\delta g_{\ell}$ and $g_p$, vary from orbit to orbit and from nucleus to nucleus. Calculating them accurately for cases with more than a single valence nucleon becomes difficult. In practice, it is common to adopt 'standard' values of $\delta g_s$ and $\delta g_{\ell}$ through a range of nuclei. It is also common to ignore the tensor term $g_p$ and to set the orbital correction $\delta g_{\ell} = 0$, although there are regions in the nuclear chart, {\it e.g.}, around $^{132}$Sn and $^{208}$Pb, where the magnetic moments of higher-spin orbitals are better reproduced by using $\delta g_{\ell} \sim 0.1$.
It is very common to choose $\delta g_s$ such that the effective spin $g$ factor is quenched to about $0.7\cdot g_s^{free}$ both for protons and neutrons. This quenching of $g_s$ moves the single-nucleon moment estimates to the region  ``between'' the Schmidt limits for $j = \ell \pm 1/2$, in better agreement with the experiment. The 0.7 quenching factor is often invoked in analogy to the quenching of the Gamow-Teller strength in $\beta$ decay. However, this poses the question about the reason for such an analogy and the origin of effective values in physics in general. For example, some recent works on $\beta$-decay studies, see, {\it e.g.}, Ref.~\cite{gysb19}, have shown that the discrepancy between experimental and theoretical $\beta$-decay strength around $^{100}$Sn can be explained without the necessity to employ a quenching of the Gamow-Teller strength. Similar conclusions have also appeared recently in nuclear moment studies \cite{vern22}. These new cases suggest that the adoption of the ``standard'' spin quenching factor, which has been taken for granted for the past several decades, should be carefully reexamined.

\subsection{Core polarization and meson exchange contributions to the \textit{M}1 operator}

\subsubsection{First order core polarization theory}

In a simple single-particle shell model, the ground state of an
odd-mass nucleus is described by the outermost unpaired valence
nucleons, with the other protons and neutrons, acting as a core, are grouped
in pairs to form the core angular momentum and parity
$I_\textrm{c}^{\pi}= 0^+$.  The magnetic moment in such a nucleus
is given by the Schmidt formulae, Eq.~(\ref{eq:sm_gl_gs}), which can be re-written in the form
\begin{eqnarray}
  \mu =
  \begin{cases}
    g_{\ell}\ell + \dfrac{1}{2}g_s
        & (j = \ell + 1/2)\\
    \dfrac{j}{j+1}\left( g_{\ell}({\ell}+1)
                                  -\dfrac{1}{2}g_s \right)
        & (j = \ell - 1/2)
  \end{cases}
\label{Eq:MEC_muSchmidt}
\end{eqnarray}
in units of $\mu_N$.
As already discussed, most of the measured magnetic moments deviate from the Schmidt values.  This is caused by two main mechanisms, {\it i.e.}, certain correlations in the nuclear wave function and higher-order effects such as meson exchange currents (MECs). The correlations in the nuclear wave functions were initially investigated based on configuration mixing within first-order perturbation theory.

As already noted, the deviation of the moments from the Schmidt values, with very few exceptions, places the values between the two Schmidt lines corresponding to each of $j=\ell \pm 1/2$.
This systematic tendency of magnetic moments is explained by the first-order effect of the $M1$ core polarization~\cite{arim54,arim54a,noya58,blin53}.  It considers a dominant configuration of the outermost single particle that gives the Schmidt value and a nucleon configuration corresponding to an excitation of the core to the spin-orbit partner orbit, which contributes to the magnetic moment as an off-diagonal matrix element of the ${\hat{M1}}$ operator.

More specifically, the dominant single-particle state configuration
\begin{equation}\label{eq:wf_sp_conf}
\psi_0 = \lvert I=j \rangle = \lvert j \otimes [(j_1)^{2n}]^{I_\textrm{c}=0}\rangle^{I=j}
\end{equation}
gives the Schmidt value, and the configuration with which it mixes
\begin{equation}\label{eq:wf_miced_conf}
\psi_{M1} = \lvert j
\otimes [(j_1)^{2n-1}(j_2)]^{I_\textrm{c}=1}\rangle^{I=j}
\end{equation}
contributes to the shift of the magnetic moment, where $I$ is the spin of the nucleus, $j_1$ is the total angular momentum of the single-particle orbit in which $2n$ nucleons from the core are placed, $j_2$ is the spin of its $LS$-partner orbit (if $j_1=\ell\pm 1/2$ then $j_2=\ell\mp 1/2$) and $I_\textrm{c}$ denotes the total spin of the core. Notably, the mixing of $\psi_{M1}$ with respect to $\psi_0$ is usually small, but the off-diagonal matrix element ${\langle}\psi_0{\lvert}\hat{\mu}{\lvert}\psi_{M1}{\rangle}$, where $\hat{\mu}$ is the magnetic moment operator, affects at the first order.  Therefore, even a small probability of the $M1$ core polarization component induces a large correction to the magnetic moment~\cite{mori76}.  However, in nuclei where the $LS$ partner orbit is occupied by another particle, blocking of the $M1$ core polarization occurs~\cite{naga73}.  This phenomenon has been experimentally confirmed, for example, in the systematic $g$-factor measurement of the nuclei comprising a ${}^{208}$Pb core plus $n$-protons. The main configuration of such states can be written as $\lvert ({}^{208}\textrm{Pb})^{0^+}{\otimes}({\pi}{h_\textrm{9/2}})^{n}\rangle^{I}$, where the quenching becomes smaller with increasing $n$, indicating that the $({\pi}h_\textrm{11/2}^{\ -1}{\pi}h_\textrm{9/2})^{I=1^+}$-type $M1$ core polarization is blocked~\cite{yama74}. Note that these isomers are also impacted by coupling to the octupole vibration of the $^{208}$Pb core, but this blocking effect remains evident \cite{stuc93}.

In addition to the magnitude of the contribution from the $M1$ core polarization discussed above, its sign is extremely important.  The deviations in the first-order core polarization theory have a common quenching factor of $-[1+(-)^{1/2+\ell-j}(j+1/2)](g_s-g_{\ell})$ whose sign always moves the moments between the Schmidt values, thus explaining the systematic properties of the quenching~\cite{noya58}.

Most of the experimental magnetic moments fall between the two Schmidt lines. However, there are several exceptions, including ${}^{3}$H, ${}^{3}$He, ${}^{13}$C, ${}^{15}$N, and ${}^{17}$N.  For example, for $^{17}$N, a 33\% ``outward'' deviation from the Schmidt value has been observed, the absolute value of the deviation being $\lvert \delta\mu \rvert$ $= 0.088(2)~\mu_N$~\cite{ueno96}. The last three nuclei listed here have a $p_{1/2}$-valence nucleon.

The quenching factor disappears in the approximation using $\delta$-function interactions~\cite{noya58}.  Therefore, the admixture of the $M1$ core polarization configuration is considered to be extremely hindered. Furthermore, excitations of the core nucleons, identical to the $p_{1/2}$-valence particle, are strictly forbidden when the angular momentum is considered, {\it i.e.}, the $M1$-type core polarization of the $p_{3/2}$ proton is not possible because it uniquely results in a total angular momentum $I=3/2$, which is in disharmony with the ground-state spin~1/2, without considering the configuration in which a pair of core particles are coupled to form $I^{\pi}=2^{+}$.  For this reason, the first-order effect is strongly hindered, and quenching is not observed, again suggesting the necessity of reconsidering the quenching using this mechanism.

\subsubsection{Effects of higher-order configuration mixing and mesonic-exchange currents}
\label{Sec:MEC}

The first-order $M1$ core polarization theory explains the systematics of the discrepancy between the experimental and Schmidt values of the magnetic moments in many cases. However, a deviation still persists in nuclei having an $LS$-closed shell $\pm$~one nucleon, where the first-order effects described above do not occur. Pioneering studies to explain this discrepancy were conducted by considering meson exchange current (MEC) effects~\cite{vill47,miya51}.  This assumption was later confirmed experimentally by a measurement of the magnetic moment of the $I^\pi = 11^{-}$ excited state in $^{210}$Po~\cite{yama70} and by the anomalous $g_{\ell}$ factors obtained from the analysis of the experimental $\mu$-moment values in a wide mass region~\cite{naga71}.  Subsequent theoretical studies were based on precise calculations that included higher-order corrections.

The effects of the second-order configuration mixing~\cite{ichi65,shim74,town83}, MECs~\cite{vill47,miya51}, and their cross terms~\cite{arim87} have been considered in detail, see Refs.~\cite{hyug80,town83,arim87}.
Hyuga et~al.~\cite{hyug80} introduced a rough criterion for the influence of the exchange currents, which is useful for an intuitive understanding. It was suggested that an $f^2$ contribution arises from the first-order configuration mixing and one-pion exchange currents, whereas an $f^4$ contribution arises from the second-order configuration mixing, two-pion exchange current, and interference between the one-pion exchange current and the first-order configuration mixing, where $f$ is the pion--nucleon coupling constant.

Higher-order corrections are introduced by additional correction terms to the isovector and isoscalar components of the one-body $\hat{M1}$~operator. The generalized magnetic moment operator given by Eq.~(\ref{eq:mu_eff}) can be rewritten as~\cite{town83}

\begin{eqnarray}
  \hat{\mu}^{(t)} = \bar{g}_{\ell}^{(t)}\hat{\ell} + \bar{g}_{s}^{(t)}\hat{s}
  + g_{p}^{(t)} [\hat{Y_2}, \hat{s}]_1,
\label{Eq:MEC_mu}
\end{eqnarray}
where $\bar{g}_{\ell}^{(t)}$, $\bar{g}_{s}^{(t)}$, $g_{p}^{(t)}$,  are respectively the effective orbital, spin, and tensor $g$~factors, now including the correction terms ${\delta}g_{\ell}$, ${\delta}g_s$, and ${\delta}g_p =g_p$.
The superscript ($t$) simply denotes whether the nucleon is a proton or a neutron, {\it i.e.}, $t = p$ or $n$ for the proton or neutron, respectively. The isoscalar and isovector combinations with $i=0,1$, are defined for any of the $g$~factors in Eq.~(\ref{Eq:MEC_mu}) as
\begin{eqnarray}
  \begin{aligned}
  g^{(0)} &= \frac{1}{2}
         \left( g^\textrm{(p)} + g^\textrm{(n)} \right),\\
  g^{(1)} &= \frac{1}{2}
         \left( g^\textrm{(p)} - g^\textrm{(n)} \right).
  \end{aligned}
\label{Eq:MEC_gISIV}
\end{eqnarray}
The corrections from the higher-order configuration mixing and MECs to the isovector and isoscalar $g$~factors are reflected in ${\delta}g_{\ell}^{(i)}$, ${\delta}g_{s}^{(i)}$, and ${\delta}g_{p}^{(i)}$ ($i$ $=0,\,1$) as a correction of the one-body operator.  These isoscalar and isovector corrections are defined in terms of those for the proton and neutron by the combinations given in Eq.~(\ref{Eq:MEC_gISIV}).  Note that the isoscalar and isovector $g$~factors determined from the $g$~factors of the free nucleons are
$g_{\ell}^{(0)}=0.5$, $g_{\ell}^{(1)}=0.5$,
$g_{s}^{(0)}$ $=$ $0.880$, $g_{s}^{(1)}$ $=$ $4.706$, and
$g_{p}^{(0)}$ $=$ $g_{p}^{(1)}$ $=$ $0$. In this case  ${\delta}g_{\ell}^{(i)}$, ${\delta}g_{s}^{(i)}$, and
${\delta}g_{p}^{(i)}$ are all zero, for both $i=0$ and $i=1$.

The effective $\hat{M1}$ operator, with correction terms ${\delta}g_{\ell}$,
${\delta}g_{s}$, and ${\delta}g_{p}$ to the orbital, spin and tensor $g$ factors, has been investigated theoretically~\cite{hyug80,town83,arim87}. The effects of second-order core polarization, isobar currents, and MECs were evaluated. The two-pion exchange process, in which the $\rho$-meson exchange process and related processes induced by the $\Delta$~admixture were considered, as well as the three-pion exchange process involving the $\omega$-meson exchange process, were studied in addition to the one-pion exchange process.  To avoid the ambiguity caused by the complexity of the nuclear wavefunction as much as possible, the magnetic moments of nuclei with an $LS$-closed core $\pm$~one nucleon were evaluated.  Although there were some differences in the magnitude of the values in these calculations for each single particle orbit in the \textit{sd}~shell, the signs of the isovector and isoscalar ${\delta}g_\ell$, ${\delta}g_s$, and ${\delta}g_p$ values as well as the behavior of high-order configuration mixing, the MECs involving the $\Delta$~isobar, and other MECs were the same.  The following conclusions emerged from these studies:
\begin{itemize}
    \item for the isovector moment, all three higher-order corrections have a certain magnitude, however, complex offsetting occurs among all contributions, and the calculated ${\delta}g$ values result from a delicate balance; and
    \item for the isoscalar moment, ${\delta}g_p^{(0)}$ is negligible because the second-order configuration mixing, which is caused by the tensor forces, dominates since $\Delta$-hole mixing does not apply in the first order, and MECs, which involve isoscalar $\sigma$ and $\omega$ mesons, do not occur.
\end{itemize}
In both cases, the ${\delta}g_{\ell}$ values turn out to be small due to the cancellation of the second-order configuration mixing and MEC contributions, but isoscalar ${\delta}g_{\ell}^{(0)}$ is slightly enhanced by the MEC.

A different approach was used in Ref.~\cite{brow87}, which broadened the observations beyond the limited data set adjacent to closed shells. The available experimental data for nuclei with $A$ $=$ 17--39 were analyzed. As a next step, empirical ${\hat{M1}}$ operators were determined using wave functions obtained from 0$\hbar\omega$ shell model calculations with the USD interaction~\cite{wild84} for all available experimental magnetic moments, and numerous non-diagonal matrix elements from $M1$ transitions. The results showed that the MEC has a significant effect on the isovector ${\delta}g_{s}$ term of the obtained effective $\hat{M1}$ operator, and that each ${\delta}g$ correction
term is in good agreement with the theoretical calculations of higher-order effects and MECs. The shell model calculations obtained using the $\hat{M1}$ operator accurately
reproduce the experimental data of the magnetic moment and $M1$ $\gamma$~decay.  The resulting empirical $\hat{M1}$ operator was later updated along with the interaction, which was revised by including new datasets accumulated at radioactive ion beam facilities~\cite{rich08}. The differences in the ${\delta}g$ values obtained in different theoretical calculations~\cite{town83,arim87,brow87} were discussed in Ref.~\cite{brow87}.

With the improvement of computational performance, direct simulation of quantum many-body effects has become possible.  The quantum Monte Carlo calculation method has been developed as an efficient method for computing multidimensional integrals in various forms in quantum many-body problems. Thus, first-principles \textit{ab~initio} calculations that describe nuclei using raw nuclear forces were employed for studies of MECs. The effects of two-body MECs on magnetic moments and $M1$ transitions were investigated based on the Green's function quantum Monte Carlo (GFMC) approach~\cite{pudl97,wiri00} with chiral effective field theory interactions~\cite{epel09,mach11}, in which the Argonne~v18 two-nucleon~\cite{wiri95} and Illinois-2 three-nucleon potentials~\cite{piep01,piep08} were adopted. A first study reported on nuclei with $A$ $=$ 2--7~\cite{marc08}, which later was  extended to $A$ $=$ 8--9~\cite{past13}. In these calculations, magnetic moments were calculated to include two-body MEC contributions. Marcucci {\it et~al.} \cite{marc08} calculated the magnetic moments in the impulse approximation. The one-body $\hat{M1}$ operator did not demonstrate any significant change using MECs for $T=0$ nuclei. However, the contribution of the two-body operator to the isovector component was large, reaching 15--17\% of the $M1$ matrix elements.  This suggests that the agreement between the calculated and experimental magnetic moments and transition rates, especially in the isovector channel, is within 1.5\% of the difference between the calculated and experimental amplitudes. Further investigation is needed related to the isoscalar channels.

For the $A$ $=$ 8--9 nuclei~\cite{past13}, an electromagnetic current operator was also
adopted that considered a two-body MEC in addition to the standard one-body convection and spin magnetization terms, where the standard nuclear physics approach, {\it i.e}., a model describing phenomenologically exchange currents, and the chiral effective field theory were discussed by comparing the magnetic moments and $M1$ transition probabilities.  The results for MEC corrections for isoscalar and isovector magnetic moments in both formulations were in qualitative agreement, demonstrating the importance of MEC correction, although the calculation based on the chiral effective field theory utilizing the electromagnetic current operator was closer to the experimental data.  In addition, this study addressed the issue of anomalies in the $^{9}$Li-$^{9}$C isoscalar spin expectation value, which will be discussed in Sect.~\ref{sec:9Li9C}.


 {The impact of two-body currents on magnetic dipole moments remains of interest. Recent work \cite{miya24} has investigated the magnetic moments of medium and heavy odd-mass nuclei near doubly magic nuclei using a shell-model based approach. The so-called valence-space in-medium similarity renormalization group  (VS-IMSRG) was used
with chiral effective field theory interactions and currents to determine many-body correlations. It was found that two-body currents generally improved agreement between theory and experiment, and that their influence increased with mass.}

Contemporary DFT can  {also} describe magnetic moments of odd-$A$ nuclei without reverting to effective charges and effective $g$~factors. It is found that time-odd fields are responsible for the quenching of the spin $g$~factor and that coupling to even-spin excitations of the core can be associated with shape polarization
\cite{sass22,vern22,gray23,bonn23,wibo25,wibo25a,doba26a,doba26b,doba26}.

\subsection{Collective models: Even-even and odd-\textit{A} nuclei }
\label{sec:collective}

\subsubsection{Even-even nuclei}
The low-excitation structure of atomic nuclei is strongly affected by pairing correlations, {\it i.e.}, like nucleons tend to pair in time-reversed orbits. One consequence is that all even-even nuclei have 0$^+$ ground states. When considering collective excitations built on the ground state, it is therefore reasonable to assume that the intrinsic spins of the nucleons largely cancel and that the angular momentum is carried by the orbital motion of the nucleons. In this approximation, $I=L_p + L_n$, where $L_p$ and $L_n$ are the orbital angular momenta of the protons and neutrons, respectively. Since $g_{\ell}^{p,\rm{free}}=1$ for protons and $g_{\ell}^{n,\rm{free}}=0$ for neutrons, in this approximation, $\mu=L_p$. Thus
\begin{equation}\label{eq:mucollectivsL}
\mu = g I = g(L_p+L_n)=L_p
\end{equation}
or
\begin{equation}
\label{eq:gcollectiveL}
g  = \frac{L_p}{L_p+L_n}.
\end{equation}

The concept that the collective $g$~factor represents the fraction of the total angular momentum carried by the protons is common to many collective models \cite{bohr75,EG87}.
For example, if all of the nucleons take part in the collective motion and the intrinsic spin contributions cancel, then
\begin{equation}\label{eq:g_collective}
g  \approx \frac{Z}{A}.
\end{equation}
This approximation to the collective $g$~factor applies to both rotational and vibrational states in the Bohr collective model. Typically, experimental values for collective $g$ factors are 70\% to 80\% of $Z/A$. Reasons for this reduction will be discussed below.

An alternative interpretation of Eq.~(\ref{eq:gcollectiveL}), which arises in the proton-neutron Interacting Boson Model (IBM-2), is to include only valence nucleons in the evaluation of the $g$~factor \cite{samb81,samb84}. Thus to a first approximation in IBM-2
\begin{equation}
\label{eq:gIBM2a}
g  = \frac{N_{\pi}}{N_{\pi}+N_{\nu}},
\end{equation}
where $N_{\pi}$ and $N_{\nu}$ are the numbers of proton and neutron bosons, respectively. Experimentally, collective $g$~factors are often smaller than predicted by this formula and vary more slowly with the number of valence protons. A more general form is therefore usually used:
\begin{equation}
\label{eq:gIBM2}
g  = g_{\pi} \frac{N_{\pi}}{N_{\pi}+N_{\nu}} + g_{\nu} \frac{N_{\nu}}{N_{\pi}+N_{\nu}} ,
\end{equation}
where typically the proton-boson $g$~factor $g_{\pi} < 1 $ and the neutron-boson $g$~factor $g_{\nu}>0$.

If the nucleus is considered as a rotating spheroid, with protons and neutrons rotating at a common rotational frequency $\omega$, Eq.~(\ref{eq:gcollectiveL}) can be rewritten in terms of moments of inertia ${\cal J}$. Since $L= {\cal J} \omega$,
\begin{equation}
\label{eq:gcollectiveMoI}
g  =  \frac{{\cal J}_p}{{\cal J}_p + {\cal J}_n} = \frac{{\cal J}_p}{{\cal J}},
\end{equation}
where ${\cal J}_p$ (${\cal J}_n$) is the moment of inertia of the protons (neutrons) and ${\cal J}$ is the total moment of inertia of the nucleus. This formulation points toward more microscopic cranking models~\cite{beng79a,ragn95,frau18} of $g$~factors in collective nuclei and provides a general motivation for the observation that $g <Z/A$ in collective nuclei. In particular, the reduction of nuclear moments of inertia from rigid body values has long been attributed to the effect of pairing correlations. The reduction of $g$ from $Z/A$ can be attributed to the difference in pairing strengths for protons and neutrons, which leads to a relatively larger contribution from ${\cal J}_n$ in the denominator of Eq.~(\ref{eq:gcollectiveMoI}), {\it i.e.}, the neutron fluid is more rigid than the proton fluid and consequently $g < Z/A$.

Microscopic models of collective $g$~factors can be formulated in terms of the cranking model for deformed nuclei and in terms of the random phase approximation for vibrational excitations~\cite{ring04}.

\subsubsection{Odd-A nuclei: General remarks}

The $g$~factors of odd-$A$ nuclei are determined largely by the unpaired (odd) nucleon. Indeed, the extreme single-particle model attributes the nuclear moment solely to the odd nucleon. Aside from the core polarization and MEC effects discussed above, departures of the magnetic moments of odd-$A$ nuclei from the Schmidt limits in semi-magic and open-shell nuclei can be understood largely in terms of coupling between the odd nucleon and excitations of the even-even core. There are two extremes: relatively weak coupling of the odd nucleon to the core excitation in near-spherical nuclei, and strong coupling of the odd nucleon to a deformed core. The first case is associated with the weak-coupling core-excitation model~\cite{desh61} and its extension to the particle-vibration model~\cite{bohr75,bohr52,bohr53}. The second case is described by the Nilsson model~\cite{nils55,ragn95}, and its extension to the particle-plus-rotor model~\cite{bohr75,ragn95}.

\subsubsection{Odd-A nuclei: Weak-coupling and particle-vibration models}
The particle-core weak-coupling model was introduced by de-Shalit \cite{desh61}. In this model, a single nucleon in an orbit of angular momentum $j$ is coupled to the 0$^+$ ground state and the first $2^+$ excitation of an even-even core. The odd-mass nuclide has a ground-state angular momentum of $j$ and a ``multiplet" of states near the excitation energy of the core 2$^+$ state with angular momentum $I$, where $ \vert j-2 \vert \leq I \leq \vert j+2 \vert$.

The ground state has the $g$~factor of the odd nucleon. The $g$ factors of the excited states in the weak-coupling multiplet are then given by the additivity formula, see Eq.~(\ref{eq:additivity}), where the two angular momentum components are those of the core excitation and the odd nucleon. This model does not specify the structure of the core excitation. If $j > 1/2$, then there can be mixing between the nominal ground-state configuration and the member of the multiplet with $I=j$, but within the model, there is no prescription to specify the strength of such mixing.

The particle-vibrational model developed by Bohr and  {Mottelson}~\cite{bohr75,bohr52,bohr53} can be considered as an extension of the weak-coupling model~\cite{desh61} in which the core excitations are associated with the vibrational model, more than one single-particle orbit may couple to the core states, and particle-core interactions are included. The weak-coupling model wave functions can serve as the basis states for a particle-vibration model calculation. There is extensive literature on the evaluation of observables, including magnetic moments \cite{chou54,chou67,heyd67,heyd68,chou69,heyd69,cast71,vand71,sen72,atal72}.
 {This model is discussed in relation to measured magnetic moments in section~\ref{sec:oddA-spherical} below.}

\subsubsection{Odd-A deformed nuclei: Nilsson and particle-plus-rotor model}\label{subsub:nilsson}
The Nilsson model~\cite{nils55,ragn95} describes deformed odd-$A$ nuclei. It treats them as a single unpaired nucleon strongly coupled to a spheroidal even-even core. In the deformed field, the angular momentum of the single nucleon, $j$, is no longer conserved. However, the magnitude of its projection on the symmetry axis of the spheroidal nucleus, denoted $\Omega$, is conserved. Even so, $\Omega$ is not directly measurable. It can be determined by invoking the particle-plus-rotor model~\cite{bohr75,ragn95}, {\it i.e.}, the band-head of a rotational band normally coincides with $I=\Omega$. The projection of the total angular momentum of the particle-plus-rotor is denoted $K$. In many cases, especially for the ground states of odd-$A$ nuclei, $\Omega = K$, which is assumed in the following discussion.

The $g$ factors of the states in a rotational band with pure $K$, {\it i.e.} in the strong coupling limit of the particle-rotor model where Coriolis mixing is ignored, are given by
\begin{equation} \label{eq:gnils}
g = g_{\rm R} + \frac{K^2}{I(I+1)}(g_K - g_{\rm R})
[1 + \delta_{K,\frac{1}{2}}(2I+1)(-)^{(I+\frac{1}{2})} b_0],
\end{equation}
where $g_{\rm R}$ is the $g$~factor of the rotational core,  $g_K$ is the intrinsic $g$ factor associated with the single nucleon, and $b_0$ is the magnetic decoupling parameter, which gives rise to a signature-dependent staggering effect for bands with $K=1/2$ (see e.g. Ref.~\cite{bohr75}, Vol. II, p.~45 and p.~57). The expression for $M1$ transitions within a band (with no Coriolis mixing) is given by
\begin{equation}\label{eq::bm1_strong_coupling}
B({\rm M}1;I \rightarrow I-1) = \frac{3}{4 \pi} \mu_{\rm N}^2
(g_K - g_{\rm R})^2 K^2
\langle I K 1 0 \vert I-1 K \rangle ^2
 [1 + \delta_{K,\frac{1}{2}}(-)^{(I+\frac{1}{2})} b_0]^2
\end{equation}
The parameters $g_K$ and $b_0$ can be evaluated from Nilsson model wave functions. The rotational $g$~factor may be taken from a neighboring even-even nucleus, if known, however, it is often estimated in heavy nuclei as $g_R \sim 0.7 \cdot Z/A$.

The Nilsson model treats the nucleus in an intrinsic or body-fixed frame of reference, {\it i.e.}, the core does not rotate. As noted above, the angular momentum of the odd nucleon ${\bm j}$ is not conserved, yet the total angular momentum of the nucleus in the laboratory frame, $\bm I$, must be conserved. The particle-rotor model is essentially a statement of angular momentum conservation that connects the intrinsic and laboratory frames in the particle-plus-core model by allowing core rotation, ${\bm R}$, around an axis perpendicular to the symmetry axis. Hence, angular momentum conservation is satisfied by the vector addition ${\bm I} = {\bm R} + {\bm j}$, or ${\bm R} = {\bm I} -{\bm j}$. The rotational Hamiltonian is $\hat{\mathcal{H}}_\textrm{R}= \frac{\hbar^2}{2 {\cal J}} {\bm R}^2$. Replacing ${\bm R}$ by  ${\bm I} -{\bm j}$ and evaluating the resulting expression in terms of angular momentum operators gives rise to a Coriolis term of the Hamiltonian of the form $(\hbar^2/2{\cal J} )(\hat{I}_+ \hat{j}_- + \hat{I}_- \hat{j}_+)$,  which connects states that differ in $K$ by one unit of angular momentum. Thus, Coriolis interactions mix intrinsic states with $\Delta K = \pm1$. As a first approximation, the particle can be considered as strongly coupled to the core, and the Coriolis interactions neglected. In this limit, the model predicts a series of rotational bands, each with a pure $K$ value. These bands are usually labeled by their $K$ value and the asymptotic quantum numbers of the intrinsic Nilsson state.

It is common practice to deduce the $M1$ properties of rotational bands in odd-$A$ nuclei from the $\gamma$-ray branching ratios by employing the strong-coupling (pure-$K$, no Coriolis mixing) limit of the particle-rotor model. The relevant expressions are summarized here.

The mixing ratios for in-band  $I \rightarrow I-1$ transitions can be determined from the $\gamma$-ray branching ratios of the $I \rightarrow I-1$ and $I \rightarrow I-2$ transitions:
\begin{equation}\label{eq::pureK-delta}
\frac{1}{\delta ^2} =
\frac{I_\gamma(I \rightarrow I-1)}{I_\gamma(I \rightarrow I-2)}
\frac{\vert \langle IK20 \vert I-2K\rangle \vert^2}{ \vert \langle IK20 \vert I-1K\rangle \vert ^2}
\frac{E_\gamma((I\rightarrow I-2)^5}{E_\gamma((I\rightarrow I-1)^5} -1.
\end{equation}
The $(g_K-g_R)$ values can be determined from
\begin{equation}\label{eq:gKgRQ0}
\frac{(g_K-g_R)}{Q_0} = 0.93 \frac{E_\gamma(I\rightarrow I-1)}
{\delta\sqrt{I^2-1}} ,
\end{equation}
where the intrinsic quadrupole moment $Q_0$ is in barns. The $B$(M1)/$B$(E2) ratios in $\mu_N^2/e^2$b$^2$ are given by
\begin{equation}\label{eq::bm1be2}
\frac{B(M1;I \rightarrow I-1)}{B(E2;I \rightarrow I-2)} =
0.6969\frac{I_\gamma(I \rightarrow I-1)}{I_\gamma(I \rightarrow I-2)}
\frac{E_\gamma(I \rightarrow I-2)^5}
{E_\gamma(I \rightarrow I-1)^3 (1+\delta^2)} ,
\end{equation}
assuming either experimental mixing ratios or values from Eq.~(\ref{eq::pureK-delta}). In the above expressions, $I_\gamma$ is the intensity of the designated $\gamma$-ray transition and $E_\gamma$ is the transition energy in MeV.

Thus, in-band $\gamma$-ray transition intensities and energies can be used to predict $g$~factors and $B(M1)$ values, given reasonable estimates of $Q_0$ and $g_R$.

 {
The application of these concepts to experimental $g$~factors of excited states in rotational bands is discussed in section~\ref{sec:oddA-rotors}.
}

\subsubsection{ {Other models of odd-\texorpdfstring{$A$}{} nuclei}}
\label{sec:other-oddA}

 {
\paragraph{Other phenomenological models}
}
 {
The above-mentioned models of odd-$A$ nuclei can be considered as benchmarks, which have been used to classify a huge amount of nuclear structure data. However, there are numerous other theoretical approaches. In section \ref{sec:IBFM}, some highlights of the application of the Interacting Boson-Fermion Model (IBFM) will be reviewed. A novel collective model of odd-$A$ nuclei, which invokes effective field theory concepts to determine theoretical uncertainties, will be discussed in section \ref{sec:EFT-oddA}. The collective part of the wavefunction in these models is treated phenomenologically.
}

 {
\paragraph{Microscopic models of odd-$A$ nuclei: Density Functional Theory}
}
 {
There has been great progress recently in the application of Density Functional Theory (DFT) to calculate nuclear moments. This approach gives a microscopic model of the nuclear moments - magnetic dipole, electric quadrupole and magnetic octupole moments have been calculated \cite{sass22,vern22,gray23,bonn23,wibo25,wibo25a,doba26a,doba26b,doba26}. It appears to be successful for both spherical and deformed nuclei and has the great advantage that the use of effective charges and effective $g$~factors can be avoided. Further remarks on the application of the latest DFT calculations to rotational nuclei will be made in section~\ref{sec:oddA-rotors}.
For a more focused review on the application of DFT to electromagnetic and exotic nuclear moments, see Dobaczewski et al. \cite{doba26}.
}\\

\section{Basic concepts of spin-precession nuclear moment studies }
\label{sec:concepts}

In this chapter, we discuss the interaction of atomic nuclei with external and hyperfine fields. The basic concept is that in the process of interaction, the nuclear spin precesses, and this precession is observed in experiments.

\subsection{Fundamentals of methodology. Interactions of nuclei with external fields}
\label{sec:fundamentals}

\subsubsection{Interaction of the magnetic moment with external magnetic field}
\label{sec:ext-mag-field}
The Hamiltonian of interaction between the nuclear magnetic dipole moment, $\boldsymbol{\mu}$ and an external magnetic field, $\boldsymbol{B}$, known as the Zeeman interaction, is the scalar product of the two vectors
\begin{equation}
\hat{\mathcal{H}}_\mu = - \boldsymbol{\mu} \cdot \boldsymbol{B} = \hbar \omega_L I,
\label{eq:Hamiltonian-magnetic}
\end{equation}
where $I$ is the nuclear spin and $\omega_L$ is the Larmor precession:
\begin{equation}
    \label{eq:Larmor}
    \omega_L = -\frac{g\mu_N}{\hbar} B,
\end{equation}
where $g$ is the nuclear $g$ factor, $\mu_N$ is the nuclear magneton and $B$ is the strength of the external field. The negative sign in Eqs.~(\ref{eq:Hamiltonian-magnetic}) and (\ref{eq:Larmor}) reflects the fact that, for a positive $g$ factor, the direction of the Larmor precession vector is opposite to the magnetic field direction.

\begin{figure}[!ht]
  \begin{center}
    \includegraphics[width=0.95\linewidth]{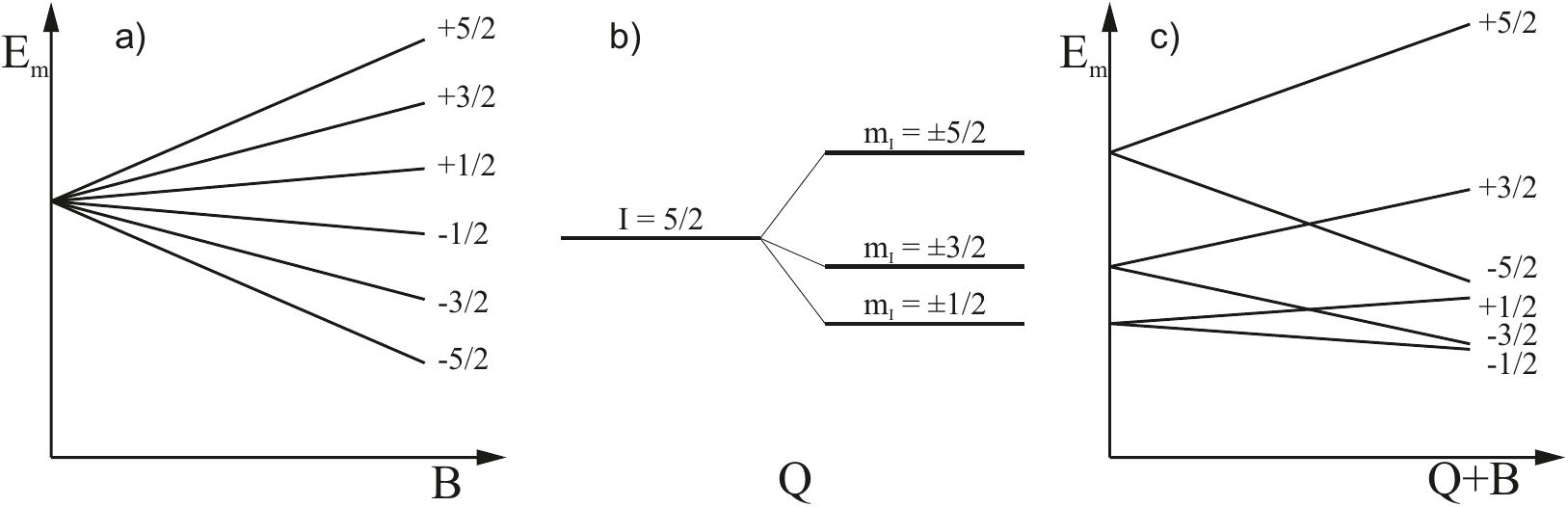}
  \end{center}
\caption{Energy splitting of the magnetic substates of a nucleus with a spin of $I=5/2$ (a) immersed in a magnetic field, (b) interacting with an electric field gradient, and  (c) subjected to combined magnetic dipole and electric quadrupole interactions. }
\label{fig:zeeman_q_b}
\end{figure}

In a semi-classical picture, the interaction between the nuclear magnetic moment and the external magnetic field can be presented as a rotation of the nuclear spin around the direction of the magnetic field.
In a quantum-mechanical representation, this results in a removal of the energy degeneracy between magnetic substates with different $m$
\begin{equation}
    \label{eq:E_zeeman}
    E_m = mgB\mu_N,
\end{equation}
which is known as the $\emph{Zeeman effect}$. The energy splitting of a state with spin $I=5/2$ is presented as an example in Fig. \ref{fig:zeeman_q_b}a.
Further details about the interaction of the nuclear spin with an external magnetic field will be given in Sect. \ref{sec:methods}.

\subsubsection{Interaction of the electric quadrupole moment with an electric field gradient. }
\label{sec:EFG}

For most nuclei, spherical symmetry is broken in their ground states, and they have a non-zero intrinsic quadrupole moment. However, for nuclei with spin $I=0$ or $I=1/2$, the spectroscopic quadrupole is trivially zero. The following discussion considers nuclear states with $I > 1/2$ and non-zero spectroscopic quadrupole moments. When placed in an inhomogeneous electric field, such nuclei interact with it. The quadrupole part of the interaction Hamiltonian is the inner product of two second-rank tensors, {\it i.e.}, the tensor of the electric field gradient (EFG), $V_{ij}$ and the tensor of the nuclear electric quadrupole moment, $Q_{ij}$.

The EFG tensor is defined as the second derivative in Cartesian coordinates of the Coulomb potential, $U$, at the position of the nucleus ($r=0)$:
\begin{eqnarray}
V_{ij} =
     \left.
     \left( \frac{\partial^2U}{\partial x_i \partial x_j} \right) \right \rvert_{r=0}.
\label{eq:V_tensor}
\end{eqnarray}
The axes $x_i = x, y, z$ are the principal axes of the EFG tensor, which in this representation is diagonal:
\begin{eqnarray}
V_{ij} =
    \left( \begin{array}{ccc} V_\textrm{xx} & 0 & 0 \\
    0 & V_\textrm{yy} & 0 \\
    0 & 0 & V_\textrm{zz}
    \end{array} \right).
\label{eq:V_tensor_diagonal}
\end{eqnarray}
The coordinate system, is chosen such that $\mid V_\textrm{zz} \mid \geq \mid V_\textrm{yy} \mid \geq \mid V_\textrm{xx} \mid$. The Laplace equation also holds, {\it i.e.}, $V_\textrm{xx} + V_\textrm{yy} + V_\textrm{zz}=0$. Thus, two independent parameters can be introduced, the largest component of the EFG, $V_\textrm{zz} = eq$, and its asymmetry parameter,
\begin{eqnarray}
\eta =
   \left \lvert \frac{V_\textrm{xx} - V_\textrm{yy}}{V_\textrm{zz}} \right \rvert,
\label{eq::eta}
\end{eqnarray}
which within this convention results in $0 \leq \eta \leq 1$, and $\eta=0$ means axial symmetry around the $z$ axis.

The interaction Hamiltonian of the nuclear quadrupole moment with the EFG is
\begin{eqnarray}
\hat{\mathcal{H}}_Q =
    \frac{1}{6h}\sum_{ij}\hat{Q}_{ij}V{ij} ,
\label{eq:Q_hamilt_general}
\end{eqnarray}
where $h$ is the Planck constant and the electric quadrupole moment operator, $\hat{Q}_{ij}$, is a symmetric, traceless second-rank tensor operator.

The degeneracy of the nuclear ground state with a spin $I$ is $2I + 1$. $\hat{\mathcal{H}}_Q$ may be treated as a small perturbation that at least partially removes the degeneracy of the nuclear ground state in the presence of an EFG. To calculate the energies of the nuclear quadrupole energy levels, it is necessary to evaluate the matrix elements $\langle I m \vert \mathcal{H}_Q \vert I m^\prime \rangle$, where $m$ and $m^\prime$ are the magnetic quantum numbers. The problem is reduced to the calculation of the matrix elements $\langle I m \vert \hat{Q}_{ij} \vert Im^\prime \rangle$ of the operator of the nuclear quadrupole moment. The energy splitting of the $m$ sub-states is shown schematically in the middle part of Fig.~\ref{fig:zeeman_q_b} and will be discussed in more detail in Sect.~\ref{beta-NMR method}.

The quadrupole moment operator is expressed by the spin operators $\hat{I}_i$ as
\begin{eqnarray}
\hat{Q}_{ij} =
    \frac{eQ_\textrm{s}}{h}\cdot \frac{1}{6I(2I-1)}
    \left(\frac{3}{2}(\hat{I}_i\hat{I}_j + \hat{I}_j\hat{I}_j) - \delta_{ij}\hat{I}^2
    \right),
\label{eq:Q_tensor}
\end{eqnarray}
where $Q_s$ is the spectroscopic quadrupole moment (see section \ref{sect:intro_quadrupole}). The components of the operator in Cartesian coordinates in the frame of the nucleus are:
\begin{eqnarray}
& \hat{Q}_{2,0} = & \frac{eQ_\textrm{s}}{h}\cdot \frac{1}{2I(2I-1)}(3\hat{I}_\textrm{z}^2 - \hat{I}^2) \nonumber \\
& \hat{Q}_{2,\pm 1} = & \frac{eQ_\textrm{s}}{h}\cdot \frac{\pm \sqrt{6}}{4I(2I-1)}(\hat{I}_\textrm{z}\hat{I}_{\pm} + \hat{I}_{\pm}\hat{I}_\textrm{z}) \nonumber \\
& \hat{Q}_{2,\pm 2} = & \frac{eQ_\textrm{s}}{h}\cdot \frac{\pm \sqrt{6}}{4I(2I-1)}\hat{I}_{\pm}^2
\label{eq:Q_tensor_comp}
\end{eqnarray}
where $\hat{I}_{\textrm{z}}$ is the projection of the spin operator on the $z$-axis, and $\hat{I}_{\pm} = \hat{I}_{\textrm{x}} \pm i\hat{I}_{\textrm{y}}$ are the spin ladder operators.

Finally, by choosing the $z$-axis as the quantization axis, the Hamiltonian of the quadrupole interaction is given by
\begin{eqnarray}
\hat{\mathcal{H}}_Q =
    \frac{e^2 qQ_{s}}{h}\cdot \frac{1}{4I(2I-1)}
    \left(
       3(\hat{I}_\textrm{z}^2 - \hat{I}^2 )
       + \frac{1}{2}{\eta}({\hat{I}_+}^2 + {\hat{I}_-}^2)
    \right),
\label{eq:Hamiltonian-Q+EFQ}
\end{eqnarray}
A generalized presentation of the nuclear magnetic and quadrupole interaction will be discussed further in relation to the formalism of the nuclear magnetic and quadrupole resonances, see Sect.~\ref{beta-NMR method}. Here we discuss the general properties of the nuclear quadrupole interactions.

Note that if the point group of the site of the nucleus in a crystal lattice is cubic, then due to the symmetry, all components are zero. In case of tetragonal or hexagonal crystals $\eta = 0$, but $V_{zz} \neq 0$. Hereafter, we consider an axially symmetric EFG tensor with $\eta=0$ for simplicity. In this case, the Hamiltonian of Eq.~(\ref{eq:Hamiltonian-Q+EFQ}) becomes
\begin{eqnarray}
\hat{\mathcal{H}}_Q =
    \frac{3}{4} \cdot
    \frac{e^2qQ_\textrm{s}}{h}\cdot \frac{1}{I(2I-1)} (\hat{I}_\textrm{z}^2 - \hat{I}^2).
\label{eq:Hamiltonian-etazeero}
\end{eqnarray}

The quantity
\begin{eqnarray}
\nu_\textrm{Q} = \frac{e^2qQ_\textrm{s}}{h},
\label{eq:quadrupole constatnt}
\end{eqnarray}
which is called nuclear quadrupole coupling constant (NQCC), or nuclear quadrupole frequency, is what is observed experimentally. In different papers, the reported quadrupole frequency may correspond to  $eqQ_\textrm{s}/h$, $eqQ_\textrm{s}$, $V_\textrm{zz}Q_{\textrm{s}}$, or the hyperfine quadrupole coupling constant $B=eQ_\textrm{s}V_\textrm{zz}$, all in frequency units.

It should be noted as well that $\nu_Q$ is most frequently used in NQR experiments. In TDPAD and/or TDPAC experiments, the nuclear quadrupole frequency is more often used in the form $\omega_Q$
\begin{eqnarray}
 \omega_Q = \frac{e^2qQ_\textrm{s}}{4I(2I-1)\hbar}.
\label{eq:quadrupole_frequency}
\end{eqnarray}

A compilation of the measured electric field gradients in metals can be found in Ref.~\cite{vian87}, and a compilation of the measured electric quadrupole moments can be found in Ref.~\cite{ston05a,ston14,pyyk18}. Recommended values of nuclear electric quadrupole moments can be found in Ref.~\cite{ston16,ston21}.

The EFGs of non-cubic metals are temperature dependent, and it is commonly accepted that they follow the ``$T^{3/2}$ law"
\begin{eqnarray}
V_\textrm{zz}(T) = V_\textrm{zz}(0) \cdot (1 - bT^{3/2}),
\label{eq:T3/2}
\end{eqnarray}
where $V_\textrm{zz}(0)$ is the EFG at $T = 0$~K and $b$ is the temperature dependence factor~\cite{kauf79,vian87}. However, recent studies of EFGs in Zn and Cd pointed out deviations from the $T^{3/2}$ dependence; see~\cite{haas24} and references therein.

 {\paragraph{Challenges of EFG calibration}}
\label{EFG_calibration}
The long-standing problem of calibration of the electric fields at the point of the nucleus still awaits its solution. On one hand, it is necessary to know precisely at least one nuclear moment, a {\it ``gold standard"}, which can be further used for a quantitative interpretation of nuclear probe studies in matter. On the other hand, it is necessary to know the EFG to deduce the quadrupole moment. Nowadays, Density Functional Theory (DFT) calculations of the electron density in the relevant crystal or molecular hosts are used to predict the EFGs. For a general text related to DFT, see, {\it e.g.}, Ref.~\cite{shol22}. DFT appears to be a good tool, with sufficient predictive power, to point the experiment on a new compound in the right direction. Ideally, DFT, by construction, provides the exact charge density and therefore exact EFGs. In practice, of course, various approximations are used. Initially, DFT was applied to EFGs in its all-electron formulation. One of the first full-potential, all-electron codes to implement this calculation was the WIEN2k package~\cite{blah85} implementing the full-potential linearized augmented-plane-wave (FLAPW) method, which was first applied to calculating EFGs and since then has been used for several classes of materials. It is essential for obtaining reliable results to have no additional approximations for the charge density and potential shape. On the other hand, pseudo-potential methods do not have any restrictions on the shape of the potential, being just plane wave expansions, and the calculations are very fast. The problem with such methods is that at the position of the nucleus they use pseudo-wave functions and pseudo-density, rather than the actual electronic charge. This problem was resolved by implementing the projector augmented waves (PAW) method~\cite{bloc94}, which allows extraction of the true electronic density from PAW pseudo-potential calculations. The formalism was found to be consistent with all-electron calculations~\cite{petr98}. Later, PAW pseudo-potentials were implemented in the software package VASP (Vienna ab-initio simulation package) using the formalism of Ref.~\cite{petr98}.

For example, for the study of the electric quadrupole moment of the $K=35/2$ isomer in $^{179}$W, Tl was used as a host material~\cite{bala01}. Tl has a hexagonal structure (hcp) for temperatures below 503~K, and a cubic (bcc) lattice for temperatures above it. Its lattice is close to the ideal crystal, and the EFGs of different atoms sitting at substitutional sites are known to be small~\cite{vian87}. The experiment was carried out at $T = 473(1)$~K. In the hcp phase, the EFG of Tl is strongly temperature dependent and decreases with temperature~\cite{scha82}. The temperature dependence of non-cubic metals follows the $T^{3/2}$ dependence, see Eq.~(\ref{eq:T3/2}). First, the system Tl in Tl (Tl\underline{Tl}) was studied. An experimental value of $V_\textrm{zz}(\textrm{Tl}\textrm{\underline{Tl}}) = 1.7(3) \cdot 10^{21}$~V/m$^2$ at 293~K was measured for this system, and the temperature dependence factor $b = 7.0(11)\cdot 10^{-5}$~K$^{-3/2}$ was also known~\cite{vian87}. These results were reproduced with band-structure calculations based on DFT using the full-potential linearized augmented plane wave (FLAPW) method~\cite{blah85}. The next step was to calculate the EFG of a W impurity in hcp Tl. A 54-atom $3 \times 3 \times 3$ super-cell lattice space was used in the calculations, allowing full structural relaxation of three cells of neighboring atoms. Spin-orbit interactions were included in the calculation because of the heavy nuclei involved. The EFG was derived from the self-consistent charge density without further approximations~\cite{schw92}, and a value of $V_\textrm{zz} (\textrm{W}\textrm{\underline{Tl}} ) = 2.54 \cdot 10^{21}$~V/m$^2$ at 0~K was obtained. A dedicated experiment was performed to measure the temperature dependence factor $b$. A value of $b = 7.6\binom{+0.2}{-0.4} \cdot 10^{-5}$ K$^{3/2}$ was derived. This results in a value for the EFG of $V_\textrm{zz}(\textrm{W}\textrm{\underline{Tl}}) = 0.55\binom{+0.12}{-0.08} \cdot 10^{21}$~V/m$^2$ at 473~K, which was used to derive the electric quadrupole moment from the measured quadrupole coupling constant. A theoretical uncertainty of 10\% for the FLAPW calculation is taken into consideration in the final value.

In short, DFT calculations for the EFG estimates are widely accepted to be an established tool. However, in recent studies, Haas {\it et al.} demonstrated that they should be taken with caution~\cite{haas17,haas21}. The accurate knowledge of nuclear quadrupole moments is of importance for our understanding of fundamental nuclear properties. As discussed, they are deduced from the measured NQCCs. Several different techniques are used for electric field gradient measurements, see Ref.~\cite{vian87} and references therein. A commonly used tool in materials physics for investigating magnetic fields and EFGs in condensed matter is the nuclear physics technique of perturbed $\gamma\gamma$ angular correlations (PAC), see Sect.~\ref{sec:isomers}. For example, in Ref.~\cite{haas17}, the quadrupole moments of the $I^\pi = 5/2^+$ isomer in $^{111}$Cd and the $I^\pi = 5/2^-$ isomer in $^{67}$Zn, measured with the PAC technique in metals and in molecules, were compared to DFT calculations for metals and solids, molecular calculations for free molecules, and atomic calculations for corrections due to the atomic $^3P_1$ states. The obtained values for the quadrupole moments differ from the accepted values~\cite{ston16}.
\begin{figure}[h]
  \begin{center}
    \includegraphics[width=0.95\linewidth]{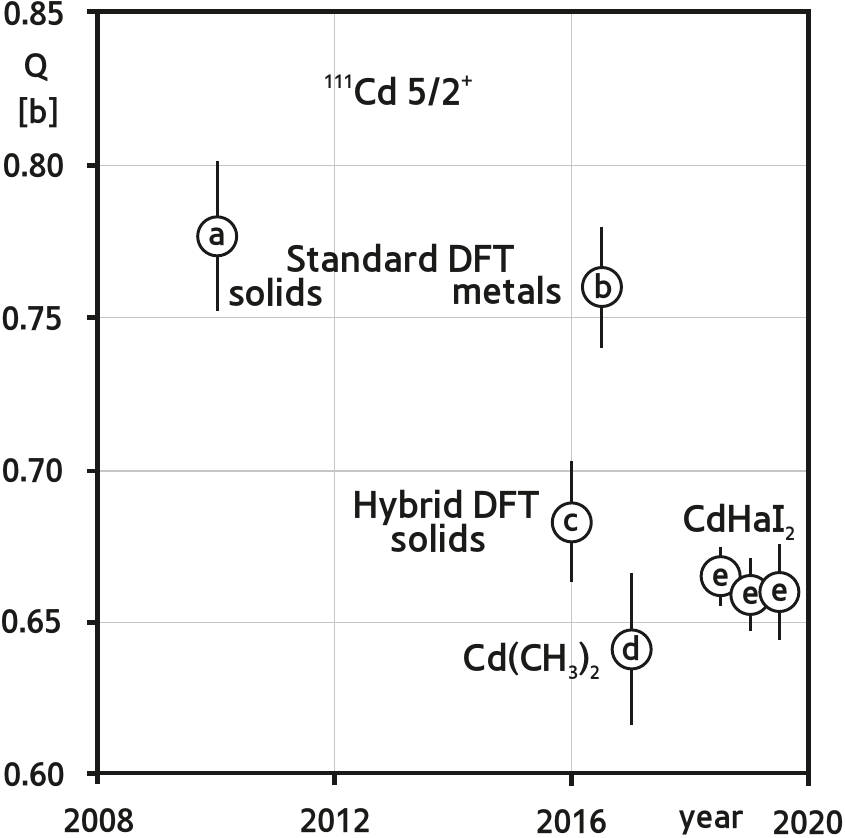}
  \end{center}
\caption{Resulting nuclear quadrupole moment for the $5/2^+$ isomer in $^{111}$Cd from recent EFG calculations using various methods: (a) Ref.~\cite{haas10}, (b) Ref.~\cite{erri16},
(c) Ref.~\cite{haas16}, (d) Ref.~\cite{haas17}, (e) Ref.~\cite{haas21}. The figure is taken from Ref.~\cite{haas21}.}
\label{fig:Q_value_111Cd}
\end{figure}

 {\paragraph{EFG calibrations using free molecules}}
In recent studies, aiming at improved precision, the technique has been applied to free molecules in a gas phase~\cite{haas21}. The basic idea of this approach is related to the fact that in a linear molecule, the principal axis of the EFG is, by symmetry, along the molecular axis, and the rotational angular momentum $\bm{J}$ is perpendicular to it for all states. The nuclear spin $\bm{I}$ being much smaller than $\bm{J}$ for practically all thermally populated rotational states, this leads to a total angular momentum $\bm{F} = \bm{J} + \bm{I}$ which is essentially aligned with $\bm{J}$. The resulting perturbation function is then expected to be very similar for all $J$ states, resembling the well-known one for a randomly oriented axially symmetric EFG. Hg and Cd halides were used in a proof-of-principle experiment~\cite{haas21}. The experimental results from PAC measurements in metals and in free molecules were compared to theoretical calculations. Furthermore, it has been demonstrated that the adopted value of the quadrupole moment $Q$ of the $5/2^+$ isomer in $^{111}$Cd has changed over the years, depending on the estimation of the EFG based on different DFT calculations, see Fig.~\ref{fig:Q_value_111Cd}. Finally, the deduced value for the quadrupole moment of the $5/2^+$ isomer in $^{111}$Cd from these studies is $Q_{\textrm{new}} = +0.664(7)$, compared to the 2016 accepted value $Q_{\textrm{acc}}(2016) = +0.74(7)$~\cite{ston16}. In a follow-up table of recommended values, it was readjusted to $Q_{\textrm{acc}}(2021) = +0.64(3)$~\cite{ston21}; however, the new result \cite{haas21} appears to have just missed the cut-off date for consideration in this evaluation. Thus, in summary, the value of the electric quadrupole moment of the $5/2^+$ isomer in $^{111}$Cd has been somewhat adjusted during the years, and the most recent developments increase the precision by an order of magnitude.

In conclusion, the experimental observable in different quadrupole-moment measurement techniques is the NQCC, and not the nuclear quadrupole moment itself. Therefore, it is necessary to consider with caution all deduced experimental values for the nuclear quadrupole moments based on hyperfine techniques, including the recommended values~\cite{ston16,ston21}, because they depend on knowledge of the EFG strength. \\

\subsection{Hyperfine magnetic fields overview}

The experimental methods of magnetic moment measurements indicated in Fig.~\ref{fig:Recknagel} that reach down to the picosecond lifetime regime require that the nucleus be subjected to intense magnetic fields with strengths that range from tens to thousands of tesla. Such fields can be produced at the nucleus by hyperfine interactions, which originate from the atomic electrons surrounding it. There are three types of hyperfine fields that are frequently exploited for magnetic moment measurements on short-lived excited states. These are: (i) the static hyperfine field at impurities at rest in a ferromagnetic host, (ii) the transient hyperfine field, which acts on ions moving through a polarized ferromagnetic host with a velocity of the order of a few percent of the speed of light, and (iii) the hyperfine fields acting on the nuclei of free ions moving through vacuum.

An overview of the static hyperfine field and aspects of its application to moment measurements will be given next in section \ref{sect:SF}. An overview of experimental methods applicable to short-lived states is given in section \ref{sect:short-lived-meth}. The transient-field method and the physics of the transient field are discussed in section \ref{sect:TF-method}, while free-ion hyperfine fields and recoil in vacuum methods are discussed in section \ref{sect:RIV-method}.

\subsubsection{Static hyperfine magnetic field} \label{sect:SF}

The static hyperfine magnetic field is present at the nuclei of impurity atoms within a ferromagnetic host. The impurity is usually considered to be on a substitutional site within the host matrix, although, as discussed further below, the impurity site may be neither substitutional nor unique.

The static hyperfine fields have been measured for most elements as impurities within the ferromagnetic $3d$ transition metals iron, cobalt and nickel. The ferromagnetic rare-earth hosts are less well studied. Gadolinium, for which the magnetism originates entirely from the intrinsic spin of the $4f$ electrons, is best studied. For the other rare-earth ferromagnetic metals, the magnetism has both orbital and spin contributions. The data and trends in iron, cobalt, nickel, and gadolinium hosts, which depend upon the outer electron configuration of the impurity, were summarized by Rao \cite{rao85}. Updates to the tabulation of hyperfine fields for impurities in these same four hosts have been compiled by Krane \cite{kran83} and Rao \cite{rao97}. Krane selects a single ``optimal" value for each element, whereas Rao lists all available measurements, i.e., on different isotopes, employing different sample preparation techniques, at various temperatures, and using alternative measurement methods.

 {\paragraph{Quality of hyperfine field data}}

While there are clear systematic trends in the hyperfine field data with the atomic number of the impurity, and the physical processes underlying these trends have been identified, precise calculations of the fields from first principles remain a challenge. Campbell \cite{camp69} gives a useful introduction to the origins of the static hyperfine field. He notes that there are three main groups of impurities: the rare earths (filling the $4f$ shell), the transition elements (filling $3d$, $4d$, $5d$), and non-magnetic elements (with partially filled $s$-$p$ shells).

Considering the $3d$ transition metal hosts, there are broadly three mechanisms that contribute to the observed hyperfine fields, which may be either in the same or the opposite direction to the polarization of the host. These contributions originate from so-called conduction electron polarization, core polarization, and fields due to a volume misfit between the impurity and the host.

{Conduction electron polarization} refers to the polarization of the conduction electrons in a ferromagnetic medium.  This polarization can give rise to an effective magnetic field at the nucleus of a non-magnetic impurity atom that is in the opposite direction to the overall polarization of the host. Such fields, in the opposite direction to the polarization of the host, are designated negative. By the same mechanism, the hyperfine fields experienced by muons in a ferromagnetic host are in the opposite direction to the polarization of the host \cite{estr82}. It is found that the hyperfine fields for the first part of the (nonmagnetic) $s$-$p$ series of elements are negative, but they become positive about halfway through the series. As noted by Campbell, the conduction electron polarization mechanism, as first proposed by Daniel and Friedel \cite{dani63}, is able to explain this trend. However, the conduction electron polarization alone is not sufficient to explain the hyperfine fields; this led Shirley, Rosenblum and Matthias \cite{shir68} to discuss additional concepts, including that known as { \em core polarization}.

Impurities that are themselves magnetic (e.g., transition metals and rare earths) have a {\em local moment} that can be enhanced by polarization of the impurity's core through the interactions of its outer electrons with the polarization of the host. In other words, unpaired vacancies of the magnetic impurity can be polarized by interactions with their outer polarized electrons, which in turn interact with the polarized electrons of the host lattice. This mechanism produces a negative hyperfine field at the nucleus. It is interesting to note that the hyperfine fields for Fe in iron, cobalt, and nickel are all about $-30$ tesla, decreasing only by 20\% as the magnetization of the host decreases by a factor of 3 from iron to nickel hosts.

Rare-earth ions behave differently to the $d$-series elements because the $4f$ shell, which gives rise to their magnetism, is more localized, and because rare-earth ions have both spin and orbital contributions to their magnetism. de Oliveira et al. \cite{deol98} have examined the spin and orbital contributions to the hyperfine fields of the rare earth series in iron and nickel hosts and are able to explain the observation that the total magnetic moment of the rare-earth impurity is parallel to the host magnetization for the first half of the rare-earth series and that the reverse occurs for the second half.

Along with the conduction electron polarization and core polarization contributions, a third contribution to the static hyperfine field can be associated with volume misfits between the implanted impurity and the host lattice. This contribution can oppose the conduction electron polarization effect \cite{stea79}.

For applications to $g$-factor measurements, the static hyperfine field should be calibrated through a measurement on a known moment under similar conditions, such as the method of introducing impurities into the host and the host temperature during the measurement. In practice, the microscopic origin of the static field is of secondary importance; a more pressing concern is whether the impurity atoms are occupying a unique substitutional site in the host lattice. For time-dependent measurements (e.g. TDPAD or TDPAC) that observe the nuclear precession through several periods, the departure of the $R(t)$ function (see Eq.~(\ref{eq:Rt_90}), Sect. \ref{sec:isomers}) from a pure cosine function signals more than one field at the nucleus and hence more than one site. Sometimes a two-site model is assumed, with a good site and a null site. In this case, the amplitude of the $R(t)$ function is reduced.

If the static field is used to measure short-lived states by integral precession measurements, great care must be taken to ensure that the average hyperfine field experienced by the impurities is well understood; an integral measurement cannot identify whether there is a single hyperfine field or several.

There is an additional concern if the impurity is introduced into the ferromagnetic host by implantation. In this case, the violence of the implantation process leads to a local heating of the host lattice around the final site of the implanted ion, which can quench the hyperfine field for several picoseconds \cite{stuc99}. Thus, as a generalization, the static field following recoil-implantation of an excited nucleus of interest onto a ferromagnetic host is applicable for $g$-factor measurements on states with lifetimes longer than about 100 ps, provided that the average hyperfine field is known.

In recent work, the static hyperfine fields of impurities recoil-implanted into gadolinium hosts have been employed with LaBr$_3$ detectors and the TDPAD method to measure $g$~factors of isomeric excited states in the Sn isotopes with lifetimes as short as $\tau \sim 3$~ns \cite{gray20,gray23}. In that work, Gray et al. \cite{gray23} characterized the hyperfine field of Sn in their gadolinium host through TDPAD measurements on the 11/2$^-$ ($\tau = 124(3)$~ns) isomer in $^{113}$Sn, which had a known $g$~factor (measured by TDPAD with an external magnetic field). They were then able to measure the $g$~factors of the 11/2$^-$ states in $^{111}$Sn ($\tau = 14.4(7)$~ns) and $^{109}$Sn ($\tau = 2.9(3)$~ns). To our knowledge, the case in $^{109}$Sn sets a record for the application of the TDPAD method to the shortest-lived level.

Despite its limitations and the need for attention to detail, the static hyperfine field remains a powerful tool for magnetic moment measurements.

 {\paragraph{Demagnetization correction}}

In some cases, for sufficiently long-lived states and small hyperfine fields, the frequency measurement by the TDPAD method can reach sufficiently high precision that the demagnetization of the host in the external polarizing field must be considered as a correction to the sum of the external and hyperfine field strengths.

In brief, the demagnetizing field is the magnetic field generated by the internal magnetization of a magnetic material. It can be viewed as a reduction of the applied polarizing field. For applications to the static hyperfine field, the effective field at the nucleus is
\begin{equation}\label{eq:Heff}
    H_{\rm eff}=H_{\rm static} - H_{\rm demag} + B(1+K_{\rm S}),
\end{equation}
where $H_{\rm static}$ is the hyperfine field strength, $B$ is the applied field, $K_{\rm S}$ is the Knight shift \cite{knig49}, and $H_{\rm demag}=d_{\rm f} M$ is the demagnetization field, with $d_{\rm f}$ the demagnetization factor and $M$ the foil magnetization. The evaluation of $d_{\rm f}$ for a sample of arbitrary shape is not trivial, but $d_{\rm f}$ is small and simple expressions can be given for thin rectangular foils magnetized in the plane of the foil. Thus, for a rectangular cuboid of sides $a$, $b$, and $c$ with the external field along the $c$ direction \cite{proz18},
\begin{equation}\label{eq:df}
d_{\rm f} \approx\frac{4ab}{4ab+3c(a+b)}.
\end{equation}
The demagnetizing field for thin foils polarized in the plane of the foil is typically of the order of a few mT, which is usually small compared to the strength of the static hyperfine field, which can be tens or even hundreds of T.

The Knight shift is the additional field induced at the nucleus by the conduction electrons in a metal host, and is not unique to ferromagnetic materials \cite{knig49,benn68}. It is typically on the order of a few percent, and small compared to hyperfine fields of tens of tesla or more. \\

\subsubsection{Hyperfine fields and low-temperature nuclear orientation}
\label{sec:low-temperature}

An experimental technique that makes broad use of static hyperfine fields is Low-Temperature Nuclear Orientation (LT/ON) \cite{post86}. Although the technique goes beyond the scope of the present paper, we will refer here to some of the recent results both for nuclear moments and hyperfine-field studies. For example, the hyperfine field of Se in Fe has been studied in Ref. \cite{ston01}, thereby obtaining a higher precision value for the magnetic moment of $^{75}\mathrm{Se}$. A study of the hyperfine fields of Sr and Y in Fe, as well as of Sr in Ni, has been reported in Ref. \cite{nish04}.

Among the nuclear moment measurements using the LT/ON technique since the year 2000 there have been studies in the $^{48}\mathrm{Ca}$ region \cite{ohts07, ohts12}, in the neutron-deficient mass 70 region \cite{golo05,seve05}, in the mass 100 \cite{ohya01, golo10} and 150 \cite{tani01} regions, and in heavy \cite{ohts04} and very heavy nuclei \cite{seve09}.

\subsection{Spin orientation concepts}
\label{sec:orientation}

All experimental methods for nuclear moment measurements considered in the present review are based on the observation of the modification of the angular distribution/correlation of the radiation emitted during the decay of the state of interest. The probability of emitting a particle, {\it e.g.}, $\alpha$, $\beta$, $\gamma$, etc., in a certain direction ${\bm k}$ depends on the direction of the spin ${\bm I}$ of the decaying nuclear state and thus on the angle between  ${\bm k}$ and ${\bm I}$.
Naturally, for nuclear states with zero spin, the angular distribution of the emitted particles is uniform. This is also true in the general case of a randomly oriented spin ensemble for which an isotropic radiation of the emitted particles is also observed. For a spin-oriented ensemble, the intensity and the polarization properties of the emitted radiation depend on the emission direction ${\bm k}$ and the polarization ${\bm P}$ (linear or circular). If the polarization of the radiation is not observed, the \textit{directional (angular) distribution} function $W({\bm k})$ is measured. If the detection system is sensitive to the polarization of the emitted quanta ${\bm P}$, then the \textit{polarization distribution} $W(\bm {k,P})$ can be measured for the emitted $\beta$-particles or $\gamma$-rays, for example. Angular correlation measurements can provide information on the properties of the nuclear states involved and on the angular momenta carried away by the emitted radiation. For example, $\alpha$-$ \gamma$ or $\gamma$-$\gamma$ correlations can yield the spins of the nuclear levels but not their parities. The relative parities can be determined if the linear polarization of the $\gamma$-rays is observed or by measuring the angular correlation between conversion electrons. The angular correlations of $\beta$-$\gamma$ cascades can provide information not only on the nuclear spins and parities but also on the matrix elements of the $\beta$-decay. Measurement of the $\textit{circular}$ polarization of $\gamma$-rays is tedious and inefficient. There are very few cases in which this has been done and in the following discussion, we will not consider such measurements.

\begin{figure}[h!!]
    \centering
    \includegraphics[width=0.6\linewidth]{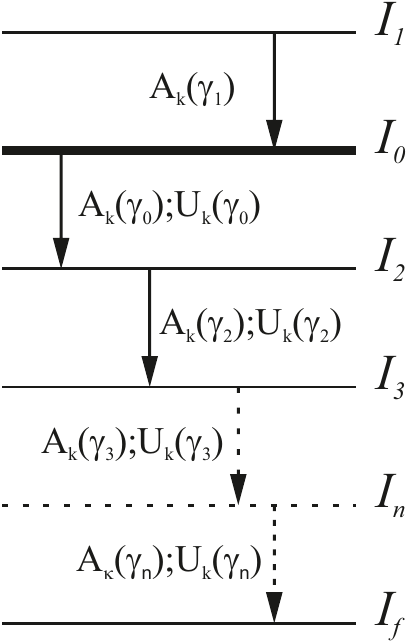}
    \caption{Schematic drawing of the states involved in angular distribution and/or angular correlation measurements. The state of interest ($\mathrm{I_0}$) is the one for which spin orientation is to be observed. Here $\mathrm{A_k}$ are the angular distribution coefficients (Eq.~(\ref{eq:angular_distribution_coeff})) and  $\mathrm{U_k}$ are the de-orientation coefficients (Eq.~(\ref{eq:de-orientation})). }
    \label{fig::scheme_dist_corr}
\end{figure}

There are two general ways by which a non-isotropic distribution of the emitted radiation can be observed. Naturally, this can take place following the decay of nuclei with oriented spins (typically aligned by a reaction mechanism), which is referred to as $\textit{angular distribution}$ (see Fig. \ref{fig::scheme_dist_corr}). Another option is to select a sub-ensemble of the decaying nuclei for which the spins have a defined orientation in space by detecting the direction of the radiation that populates the state of interest. This approach is referred to as $\textit{angular correlation}$.

The spin-oriented ensemble of nuclei ${\bm I}$ can be described quantum mechanically using the density matrices $\rho(I)$ and their multipole expansions $\rho^k_q$, which are statistical tensors. Here we will consider an axially symmetrical spin ensemble, which is often the case obtained in different types of nuclear reactions. The statistical tensors can then be expressed as (see e.g. \cite{stef75}):
\begin{equation}\label{eq:rho_axial}
  \rho^{k}_q = \sum_m (-1)^{I+m} \sqrt{2k+1} \begin{pmatrix}
        I & I & k \\
        -m & m & q \\
        \end{pmatrix} p(m)
\end{equation}
where $p(m)$ is the population probability of each $m$ substate and $k$ is the tensor rank, not to be confused with the particle emission direction ${\bm k}$ discussed above. The 3J symbol requires that $q=0$, thus, in the case of axial symmetry of the spin ensemble, all components with $q \ne 0$ vanish.

Axially symmetric ensembles can be classified as \emph{aligned} or \emph{polarized}, depending on the transformation properties under reversal of the quantization axis $z$. Axially symmetric ensembles that are invariant under the reversal of the quantization axis (e.g., $e_z \rightarrow -e_z$) are \emph{aligned}. Due to the symmetry properties of this transformation, only statistical tensors \emph{of even rank $k$} are non-zero. The diagonal elements of the corresponding density matrix  are symmetric between positive and negative $m$-values, which, expressed in terms of population probability for the different $m$ substates, can be written in the form:
\begin{equation}\label{eq:pm_alignment}
  p(m) = p(-m) \qquad\textrm{for all} \; m.
\end{equation}

A polarized ensemble is not symmetric with respect to a reversal of the $z$ axis. Such an ensemble is described by statistical tensors of both \emph{odd and even rank $k$}. At least one odd-rank $k$ statistical tensor must be non-zero for the ensemble to be polarized. In terms of population probability, this can be written in the form:
\begin{equation}\label{eq:pm_polarization}
  p(m_i) \ne  p(m_j) \qquad\textrm{for } \; i \ne j.
\end{equation}

For axially symmetric spin ensembles, all $q \ne 0$ components of the statistical tensor vanish, thus, one can define a series of orientation parameters $B_{k}(I)$:

\begin{equation}\label{eq:B_orientation}
    \begin{aligned}
  B_{k}(I) &= \sqrt{2I+1}\rho_0^{k}(I) = \sum_{m}(-1)^{I+m} \sqrt{(2k+1)(2I+1)}\begin{pmatrix}
     I & I & k \\
    -m & m & 0 \\
        \end{pmatrix} p(m)\\
        &= \frac{2k+1}{\sqrt{I(I+1)(2I+3)(2I-1)}}\sum_m{\alpha_k(m)p_m}
    \end{aligned}
\end{equation}
where $\alpha_k(m)= 3m^2-I(I+1)$ is an attenuation coefficient \cite{mori76} that accounts for an incomplete spin orientation.

In the case of broken axial symmetry of the interaction that has created the spin orientation, the non-diagonal elements of the density matrix $\rho(I)$ do not vanish. Thus the statistical tensors $\rho_q^{k}(I)$ with $q \ne 0$ must be considered:
\begin{equation}\label{eq:B_k^q}
  B_{k}^q(I) = \sqrt{2I+1}\rho_q^{k}(I)
\end{equation}
where $\rho_q^k$ is defined as in Eq.~(\ref{eq:rho_axial}).

Note that in the literature there are various definitions of the statistical tensor denoted $\rho_q^{k}$ that differ in the inclusion or not of the $\sqrt{2k+1}$ or  $\sqrt{2I+1}$ factors. However, $B_k^q(I)$ seems to be defined consistently.

The spin alignment $\cal{A}$ for an axially symmetric spin ensemble can be defined using the orientation parameters $B_k(I)$
\begin{equation}
    \label{eq:alignment}
   {\cal{A}} = \frac{\sqrt{I(I+1)(2I+3)(2I-1)}}{\sqrt{5}\vert \alpha_2(\mathrm{max})\vert}B_2,
\end{equation}
where we have limited ourselves only to the $k=2$ term. Note that in this definition, the alignment $\cal A$ is normalized $(-1\leq A \leq 1)$. $\cal A$ has positive values for prolate alignment and negative ones in the oblate case. The maximum values $\alpha_2(\mathrm{max})$ are different for the two types of alignment. For a full prolate alignment, all nuclear spins are aligned along the quantization axis, thus $m=I$, leading to $\alpha_2(\mathrm{max})_{pr} = I(2I-1)$. For a full oblate alignment, only the minimum $m$ values are populated. Thus:
\begin{equation}
    \label{eq:alpha2max}
    \begin{aligned}
        \vert \alpha_2(\mathrm{max}) \vert_{ob} &= \alpha_2 (m=0) &=& I(I+1) & \textrm{for integer spins}\\
        \vert \alpha_2(\mathrm{max}) \vert_{ob} &= \alpha_2 (m=1/2) &=& I(I+1) -3/4 & \textrm{for half-integer spins}.\\
    \end{aligned}
\end{equation}
Within this definition, the spin alignment can be readily defined using Eqs.~(\ref{eq:alignment}) and (\ref{eq:alpha2max}) since the orientation parameter $B_2$ can be extracted from the amplitude of the $R(t)$ function, see, \textit{e.g.}, Eq.~(\ref{eq:Rt_2}).
(There can be other definitions of alignment, see Ref.~\cite{stuc03}, for which the full alignment can exceed 100\%.)

In general, the angular distribution of the radiation emitted from an oriented state of spin $I_i$ can be described as~\cite{kran86}
\begin{equation}\label{eq:w_general}
  W(\theta,\phi) = \sqrt{4\pi}\sqrt{2I+1}\sum_{k,q}\frac{\rho^{k *}_{q}(I_i)A_{k}Y_{k}^{q}(\theta,\phi)}{\sqrt{2k+1}},
\end{equation}
where $Y_{k q}$ are the spherical harmonics and $A_{k}$ are the angular distribution coefficients that are defined in the general form using the ordinary $F$-coefficients $F_{k}(LL^{\prime} I I_i^{\prime})$ \cite{yama67},
\begin{equation}\label{eq:angular_distribution_coeff}
  A_{k} = \frac{F_{k}(LLI_fI_i)+2\delta F_{k}(LL^{\prime}I_fI_i)+\delta^2F_{k}(L^{\prime}L^{\prime}I_fI_i)}{1+\delta^2},
\end{equation}
where $k$ is the tensor rank, $L$ and $L^{\prime}$ are the angular momentum (multipolarity) contributions to the transition, $I_i$ and $I_f$ are the spins of, respectively, the initial and final state, and $\delta$ is the mixing ratio of the transition. The mixing ratio is defined as the ratio of the partial intensities of the different multipolarities, $\delta^2 = \frac{T_\gamma(L^{\prime})}{T_\gamma(L)}$.  Usually $L^{\prime}=L+1$. The sign of $\delta$ depends on the relative phase of the two reduced matrix elements and is of high importance since it can change the shape of the angular distribution or correlation considerably.

It should be noted that Eq.~(\ref{eq:angular_distribution_coeff}) is valid for the case of a mixing of two multipolarities. In some specific cases, the selection rules may operate so that three multipolarities could contribute to the transition. The expression for this general case can be found in \cite{kran86}.

The ordinary $F$-coefficients are defined by:
\begin{equation}\label{eq:f_ordinary}
 \begin{aligned}
 F_{k}(LL^{\prime}I_1I_0) & = (-1)^{I_1+I_0-1}\sqrt{(2k+1)(2L+1)(2L^{\prime}+1)(2I_0+1)}\\
 & \qquad\times\left( \begin{array}{ccc}
                L & L^{\prime} & k \\
                1 & -1 & 0
                \end{array}
  \right)\left\{\begin{array}{ccc}
                  L & L^{\prime} & k \\
                  I_0 & I_0 & I_1
                \end{array}
   \right\},
  \end{aligned}
\end{equation}
and are valid in the case when only one of the states is oriented, {\it i.e.}, its orientation is observed. In the more general case, when the orientation of both states is observed, the  generalized $F$-coefficients are to be used:
\begin{equation}\label{eq:f_generalized}
 \begin{aligned}
 F_{k}^{k_1 k_0}(LL^{\prime}I_1I_0) & = \sqrt{(2I_0+1)(2I_1+1)(2L+1)(2L^{\prime}+1)(2k_0+1)(2k_1+1)(2k+1)}\\
  &\times(-1)^{L^{\prime}+k_1+k_0+1}
  \left( \begin{array}{ccc}
                L & L^{\prime} & k \\
                1 & -1 & 0
                \end{array}\right)
  \left\{\begin{array}{ccc}
                I_1 & L & I_0 \\
                I_1 & L^{\prime} & I_0 \\
                k_1 & k & k_0
                \end{array}\right\}.
  \end{aligned}
\end{equation}
They are characteristic if two oriented states $I_0$ and $I_1$, of tensor rank $k_0$ and $k_1$, respectively, which are connected by a radiation tensor of rank $k$. If one of the oriented states $I_0$ or $I_1$ is random, $e.g.$ either $k_0$ or $k_1$ is zero, the generalized $F$-coefficients are reduced to the ordinary $F$-coefficients of Eq.~(\ref{eq:f_ordinary}). More details about the ordinary and the generalized $F$-coefficients, including their symmetry properties, can be found in e.g. \cite{stef75}. It should be noted that the angular momentum of the oriented state, $I_0$, is always written last in the $F$-coefficient. In a TDPAD experiment, the oriented state $I_0$ is the isomeric state of interest. Attention should be paid to the order of the arguments in the
$F$-coefficients in the case of angular correlation measurements, {\it e.g.} TDPAC, i.e., when calculating the $F$ coefficients for the transitions feeding the isomeric state of interest and its de-excitation.

In the case of an axial symmetry of the oriented state, Eq.~(\ref{eq:w_general}) can be simplified by reducing the statistical tensors $\rho^{k}_q$, Eq.~(\ref{eq:rho_axial}), to the orientation parameters $B_{k}$, see Eq.~(\ref{eq:B_orientation}), and reducing the spherical harmonics to Legendre polynomials:
\begin{equation}\label{eq:w_distribution_axial}
  W(\theta) = \sum_{k} A_{k}B_{k}P_{k}(\cos \theta).
\end{equation}
When the angular distribution of $\gamma$-rays is observed and their circular polarization is not detected, the sum in Eq.~(\ref{eq:w_distribution_axial}) runs only over even $k$ values.

So far, we have considered only the angular distribution of the emitted radiation in the absence of external interactions. The essence of all of the experimental methods discussed in this review is in the modification of the angular correlation or distribution as a function of time due to the presence of external electromagnetic fields. In the case of an external magnetic field, Eq.~(\ref{eq:w_distribution_axial}) must be written in the form
\begin{equation}
    \label{eq:W_theta_t}
    W(\theta,t,B) = \sum_{k} A_{k}B_{k}G_k(t,B)P_{k}(\cos \theta),
\end{equation}
where the $G_k(t,B)$ term stands for the interaction of the nuclear spin ensemble due to the magnetic field. The generalization of the perturbed angular correlation/distribution will be discussed further in the following, depending on whether the nuclei are subjected to magnetic or electric interactions.

In the case when the spin-orientation of the state of interest is not obtained by the reaction mechanism or by any other posterior method, {\it e.g}., optical pumping or low-temperature nuclear orientation, a non-isotropic angular distribution can be obtained for the successive particle emission. Most often, this is done for states populated in $\gamma$-$\gamma$ cascades, although, in some specific cases, one could observe conversion electron -- $\gamma$ or $\beta$-$\gamma$ correlations. In the $\gamma$-$\gamma$ case, the spin orientation of the state of interest is obtained by detecting the populating $\gamma$-ray at a specific angle with respect to the $\gamma$ depopulating it, which modifies Eq.~(\ref{eq:w_distribution_axial}) to:
\begin{equation}\label{eq:w_correlation_gamma}
  W(\theta) = \sum_{k} A_{k}(\gamma_1)A_{k}(\gamma_2)P_{k}(\cos \theta),
\end{equation}
where the coefficient $A_{k}(\gamma_1)$ depends on the spins of the initial and the intermediate nuclear states, while the $A_{k}(\gamma_2)$ is determined by the characteristics of the decay radiation $\gamma_2$ (see Fig. \ref{fig::scheme_dist_corr}). The angle $\theta$ is that between the directions of emission of $\gamma_1$ and  $\gamma_2$. The $A_{k}(\gamma)$ coefficients can be calculated again using Eq.~(\ref{eq:angular_distribution_coeff}) with special attention to be taken that for the ordinary $F$-coefficients, Eq.~(\ref{eq:f_ordinary}), the angular momentum of the studied state should always be written last. If the observed decay out transition from the state of interest is not the immediate transition (e.g. $\gamma_0$ in Fig. \ref{fig::scheme_dist_corr}) but a subsequent transition ($\gamma_2$, $\gamma_3$ ... $\gamma_n$), then the de-orientation of the preceding transitions should be taken into account by incorporation of the applicable ${U_{k}}$ coefficients.

For a pure transition of a multipole order ${L}$, emitted between states ${I_1}$ and ${I_2}$,  the de-orientation coefficients are:
\begin{equation}\label{eq:de-orientation}
 U_{k}(I_1I_2L)  = (-1)^{I_1 + I_2 + L +k}\sqrt{(2I_1+1)(2I_2+1)}\\
      \left\{\begin{array}{ccc}
                I_1 & I_1 & k \\
                I_2 & I_2 & L
                \end{array}\right\}.
\end{equation}
In the case of a mixed-multipolarity transition $(L,L^{\prime})$ this becomes:
\begin{equation}\label{eq:de-orientation_mixed}
 \begin{aligned}
U_{k}(I_1I_2LL^{\prime})  &= \frac{(-1)^{I_1 + I_2 + L +k}}{1+\delta^2}\sqrt{(2I_1+1)(2I_2+1)}
      \left\{\begin{array}{ccc}
                I_1 & I_1 & k \\
                I_2 & I_2 & L
                \end{array}\right\}\\
    &+ \delta^2(-1)^{I_1 + I_2 + L^{\prime} +k}\sqrt{(2I_1+1)(2I_2+1)}
      \left\{\begin{array}{ccc}
                I_1 & I_1 & k \\
                I_2 & I_2 & L^{\prime}
                \end{array}\right\}.
\end{aligned}
\end{equation}
If the path from the oriented state $(I_0)$ to the observed one $(I_i)$ goes through a cascade of many states $(I_1,I_2, ... I_n)$, then the de-orientation coefficient is the product of the successive de-orientation coefficients of each of the preceding transitions:
\begin{equation}
\label{eq:de-orientation_total}
U_{k}(I_0 ... I_i) = U_{k}(I_0I_1)U_{k}(I_1I_2)...U_{k}(I_nI_i),
\end{equation}
It is worth mentioning that for a stretched ${E2}$ cascade the angular distribution coefficient, which is a product of $A_{k}U_{k}$ is a constant through the cascade, i.e. while $A_{k}$ is increasing with the decreasing spin of the states, the $U_{k}$, always smaller than unity, keeps the product constant.

Finally, it is useful to note the relationships between the Wigner $D$ matrix, $D^k_{q0}$, spherical harmonics, $Y_{k}^{q}$, and associated Legendre polynomials, $P_k^q$. With the phase convention of Condon and Shortley,
\begin{equation}
\label{eq:DYP}
D^{k*}_{q0}(\phi,\theta,0)=\sqrt{\frac{4 \pi}{2k+1}}Y_{k}^{q}(\theta,\phi)=(-1)^q\sqrt{\frac{(k-q)!}{(k+q)!}} P_k^q(\cos\theta) e^{iq\phi}.
\end{equation}

Further details about the angular distribution coefficients $A_{k}$, the $F$-coefficients and the de-orientation coefficients $U_{k}$ can be found in \cite{kran86,stef75}. Also, a general formalism for perturbed angular correlations from reaction-aligned states followed by $\gamma\gamma$ coincidences has been described in \cite{stuc02a,robi02a}.

\subsubsection{Spin-orientation methods}
\label{subsubsec:spin_orientations}
Obtaining a spin-oriented ensemble of nuclei is an essential ingredient for most of the techniques considered in the present review. Spin-oriented ensembles, obtained in fusion-evaporation reactions, have been used widely for many decades and will not be included in detail here. The general procedure of parametrizing the $m$-substate population following heavy-ion reactions in terms of a Gaussian distribution is described by Yamazaki \cite{yama67}; see also \cite{mori76}. More specifically, $p(m)$ in Eq.~(\ref{eq:rho_axial}) is replaced by a Gaussian distribution centered on $m=0$. Typically, the standard deviation is $\sigma \approx 0.35I$.

We will focus here on nuclear spin orientation following projectile fragmentation reactions.
At the end of the section, some remarks will be made on other spin-orientation processes, namely, through particle detection following Coulomb excitation, and in the process of fission.

 {\paragraph{Spin orientation in projectile fragmentation}}

\begin{figure}
    \centering
    \includegraphics[width=0.8\linewidth]{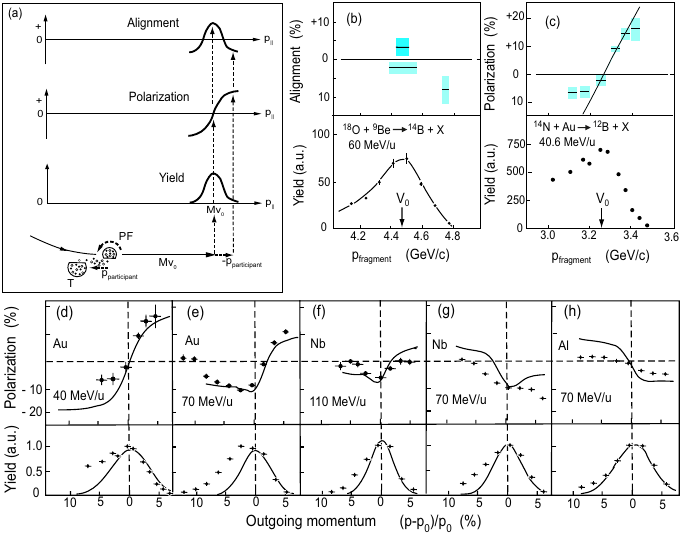}
    \caption{Schematic representation of the participant-spectator model a). Results from the first measurements of spin alignment \cite{asah91} and spin polarization \cite{asah90} in projectile fragmentation are presented, respectively, in parts b) and c). Parts d) to h) represent studies of the spin orientation as a function of the target mass and beam energy \cite{asah92,okun94}. Here, $p$ is the momentum of the outgoing particle and $p_0$ is the central value of the momentum distribution. The figure is taken from Ref.~ \cite{asah09}.
    }
    \label{fig:participant-spectator-model}
\end{figure}

The first observation of spin polarization at intermediate energies in projectile fragmentation reactions was reported by Asahi et al. \cite{asah90}. A sizable spin polarization, up to 20\%, was observed for $^{12}\mathrm{B}$ fragments produced in a $^{14}\mathrm{N} + ^{197}\mathrm{Au}$ reaction at about 40 MeV/u. The reaction products were selected at an angle of $\theta = 5^{\circ}$ off the beam axis, within a solid angle of 1.0 msr. The momentum selection was limited to $\Delta p/p = \pm 0.5\%$. A strong momentum dependence of the obtained polarization was observed (see Fig. \ref{fig:participant-spectator-model} c), with the maximum values detected at the wing of the distribution and a change of the sign of the polarization at the center of the momentum distribution.

The experimental results were explained in a simple kinematical model of the reaction. It is considered that in a peripheral collision, a cluster of nucleons (``participant'') from the projectile is removed from the overlapping volume when it interacts with the target nucleus. The remaining part of the projectile is assumed to behave as a ``spectator'', which is further detected as a fragment, and its momentum is reflected by the motion of the removed cluster. The name of the model was coined as the ``participant - spectator model''. It reproduced very well the experimental observations with the main features that the maximum polarization is observed at the (higher momentum) wing of the longitudinal momentum distribution, and a change of the sign of the polarization is observed in the center. A schematic drawing of the momentum dependence of the spin orientation and the yield in projectile fragmentation is presented in Fig. \ref{fig:participant-spectator-model}.

A follow-up study \cite{asah91} investigated the nuclear spin alignment under similar conditions, namely, $^{14}\mathrm{B}$ fragments, obtained following the reaction of 60 MeV/u $^{18}\mathrm{O}$ beam on a $^9\mathrm{Be}$ target. The spin alignment was observed through the detection of the angular distribution of the $\gamma$-rays in the daughter nucleus $^{14}\mathrm{C}$. Small positive alignment was observed in the center of the momentum distribution for a narrow selection ($\Delta p/p = 2\%$). Opening the momentum selection to $\Delta p/p = 4\%$ brought in still small but positive alignment. A selection at the higher wing of the momentum distribution (+6\%) with a $\Delta p/p = 1.1\%$ revealed a strong negative alignment of ${\cal A} = -8.0 \pm 3.7\%$. According to the participant-spectator model, further discussed in Ref.~\cite{asah91}, the positive alignment occurs at the center of the momentum distribution $p = p_0$. Its value decreases to negative values in the tails regions (wings). For a more quantitative description of the nuclear spin alignment, it should be considered that after the reaction, the fragments might be left in an excited state, thus de-exciting to the state of interest through $\gamma$-ray and particle emission, which might lead to a substantial reduction of the experimentally observed spin orientation.

Further systematic studies of the spin polarization in projectile fragmentation reactions have been reported in Refs. \cite{asah92,okun94}. It has been observed that the polarization obtained using heavy targets, e.g., $^{197}\mathrm{Au}$, is negative at the low-momentum wing and positive at the higher-momentum. This reverses completely when a light-mass target, e.g., $^{27}\mathrm{Al}$ is used. This behavior has been explained in terms of a change of the projectile trajectory from \emph{near-side} to \emph{far-side}. At intermediate masses, see Fig. \ref{fig:participant-spectator-model} f and g, the polarization does not follow a simple increase or a decrease function of the momentum selection anymore, and a maximum is observed at the center of the yield peak. A significant energy dependence of the spin polarization, as a function of the beam energy, has been observed as well.

A study of the spin alignment at higher beam energy (500 MeV/u) has been performed at GSI using the TDPAD technique on an isomeric state with a previously known $g$ factor, $I^\pi = 19/2^- \mathrm{\;in\;} ^{43}\mathrm{Sc}$~\cite{schm94}. A strong positive spin alignment ($\sim 35 \%$) was observed at the center of the momentum distribution, whereas a smaller alignment ($\sim 15\%$) was observed at the wing of the momentum distribution. While the spin-alignment at the wing of the momentum distribution is well reproduced by the participant-spectator model, the experimental alignment at the center of the momentum distribution is twice smaller than the value predicted by the model.

\begin{figure}
    \centering
    \includegraphics[width=0.8\linewidth]{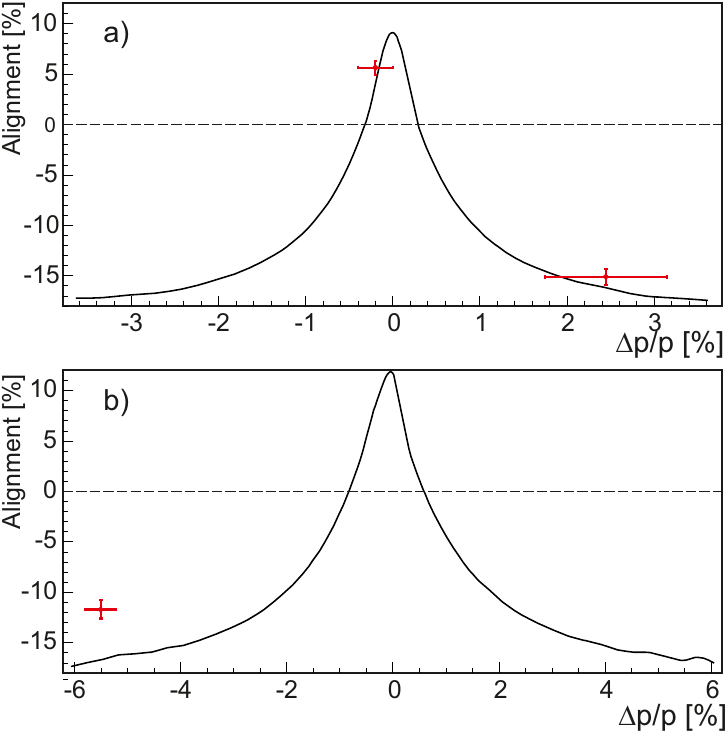}
    \caption{Nuclear spin alignment observed for $^{61}\mathrm{Fe}$ a) and $^{54}\mathrm{Fe}$ b), following the fragmentation of 55 MeV/u $^{64}\mathrm{Ni}$ beam on $^9\mathrm{Be}$ target. The solid lines represent the predictions of the participant-spectator model.}
    \label{fig:fe_alignment}
\end{figure}

Following the large spin alignment observed for $^{43}\mathrm{Sc}$,  further studies were initiated at GANIL. The first experiment, which measured isomeric states in the vicinity of $^{68}\mathrm{Ni}$ \cite{geor02}, encountered some difficulties with sub-optimal experimental conditions. After improvement of the setup, the second study \cite{mate04} observed significant spin alignment following the fragmentation of a $\sim 55$ MeV/u $^{64}\mathrm{Ni}$ beam on a $^9\mathrm{Be}$ target, see Fig. \ref{fig:fe_alignment}. It should be noted that the agreement between the experimentally observed alignment and the calculations of the participant-spectator model is very well in agreement for the $^{61}\mathrm{Fe}$ case (only 3-nucleon removal from the beam), while some 30\% deviation is observed in the case of $^{54}\mathrm{Fe}$ for a 10-nucleon removal reaction. It was suggested that the removal of a large number of nucleons from the projectile would significantly reduce the experimentally observed spin alignment, compared to the participant-spectator model. It should be noted that special care should be taken about the target thickness when aiming for a large amount of spin orientation in projectile fragmentation reactions. The use of a thicker production target will result in an increase of the yield of the isotope of interest. However, if the momentum spread of the beam, induced by a thicker target, is considerably larger than the Goldhaber distribution \cite{gold74}, the momentum of the fragments produced at the entrance of the target differs substantially from those produced at the exit. This mixes the actual momentum selection of the beam and leads to an attenuation of the observed spin orientation. The effect is most pronounced at the center of the momentum distribution, where a rapid change of the amplitude and the sign of the spin orientation is observed.

A study performed at the RIBF facility at RIKEN made it possible to quantify the influence of the number of removed nucleons on the degree of spin alignment. One of its key features was the use of two-step projectile fragmentation reactions \cite{ichi12}. A secondary beam, in the close vicinity of the isotope to be studied, is produced during the first step of the reaction. The state of interest is obtained as a tertiary beam by the second step of the reaction, following the removal of very few (usually a single) nucleons. This allows for maintaining the maximum possible spin orientation. Another key feature of the two-step reaction method is the use of the momentum dispersion matching technique. This allows the use of a very thick primary target, avoiding the drawback of the momentum mixing discussed above.

$^{32m}\mathrm{Al}\; (T_{1/2} = 187 \: ns, \; E_x = 957 \: keV)$ was used as a test case for the proof of principle for the two-step fragmentation approach \cite{ichi12}. The secondary beam of $^{33}\mathrm{Al}$ was produced following the fragmentation of a 345 MeV/u $^{48}\mathrm{Ca}$ beam on a 10 mm thick $^9\mathrm{Be}$ target. It was further subjected to a second-step reaction on an Al wedge-shaped degrader in order to populate the state of interest. The degree of spin alignment, obtained in this two-step approach, was derived as ${\cal A} = 8(1)\%$. This is about a factor of 4 smaller than what can be estimated by the participant-spectator model ($\sim 30\%$), which was attributed to the indirect population of the isomeric state of interest through the de-excitation of higher-lying states in the projectile fragmentation reaction. This indicates that the spin alignment in projectile fragmentation reactions depends strongly on both the reaction mechanism and the nuclear structure. The advantage of the two-step approach was confirmed in the same study by performing a single-step fragmentation of a 345 MeV/u $^{48}\mathrm{Ca}$ beam on a 4 mm thick $^9\mathrm{Be}$ target and populating the $^{32m}\mathrm{Al}$ state of interest directly. The degree of spin alignment obtained in this direct approach was determined as ${\cal A} < 0.8 \%$, clearly demonstrating the superiority of the two-step projectile fragmentation mechanism.

The peculiarity of this method is that it requires a two-stage fragment separator, which, until recently, could be performed only at Big-RIPS at RIKEN. With the development of new fragmentation facilities in the USA, Germany, Korea, and China, the two-step approach might become more widely applicable.

 {\paragraph{Spin orientation in Coulomb excitation}}

Turning now to the case of Coulomb excitation, considerable spin orientation can be achieved when the ground state of the excited nucleus has low spin and scattered ions are detected at specific angles. An advantage of this method is that the spin orientation can be calculated accurately based on the semiclassical theory of Coulomb excitation.
Considering the case of excitation and detection of backscattered beam ions, the excitation process is largely limited to $\Delta m=0$ (where the beam axis is the $z$ axis). Starting from $I=0$, this ensures that the excited states are predominantly populated in the $m=0$ substate, i.e., the spins are largely in the plane perpendicular to the beam. Many of the transient-field and recoil in vacuum measurements discussed in sections \ref{sect:TF-method} and \ref{sect:RIV-method} below make use of this feature.

When the ground state has high angular momentum $I_{\rm gs}$, however, there is little spin orientation. The  $\Delta m=0$ excitation process will populate all magnetic substates between $m=-I_{\rm gs}$ and $m=+I_{\rm gs}$, resulting in very little spin orientation. For this reason, moment measurements based on Coulomb excitation have been performed only for $I_{\rm gs} \leq 3/2$.

 {\paragraph{Spin orientation in fission}}
Finally, we note that spontaneous fission results in fragments at relatively high angular momentum with considerable spin orientation. Thus, there can be an anisotropic angular correlation between the fission fragment direction and $\gamma$-ray emission. The phenomenon was reported by Wilhemy et al. \cite{wilh72} in 1972 and remains a subject of investigation \cite{chal24}.

Wolf and Cheifetz \cite{wolf76} used this spin orientation from fission to measure the $g$~factor of the 6$^+_1$ state in neutron-rich $^{134}$Te. There appear to have been few subsequent measurements of this type. Rather, $g$ factors of fission fragments in the $A \sim 100$ region have been measured via integral perturbed $\gamma\gamma$ angular correlation measurements with a $^{252}$Cf source sandwiched between two magnetized iron foils and placed in Gammasphere \cite{pate02}. A variation of the method by Goodin et al. \cite{good08,good09} observed attenuations of the $\gamma\gamma$ angular correlations in a non-magnetized iron host.

Exploiting the spin orientation of fission fragments for moment measurements may provide opportunities for future research. Wolf and Cheifetz used a simple arrangement with a single $\gamma$-ray detector and three fission fragment detectors. In principle, enhanced selectivity and sensitivity could be achieved with a contemporary fission fragment detector coupled with a large $\gamma$-ray detector array. Depending on the lifetime of the state of interest, one could consider the time-dependent perturbed angular correlation technique with either an external field or an internal (static-hyperfine) field, or integral methods including the static-field, transient-field, and recoil in vacuum techniques.

\subsubsection{Tilted-foil technique}
\label{subsubsec:tiltedfoils}
An alternative method for obtaining a spin-oriented ensemble of short-lived radioactive isotopes is the tilted-foil technique (TFT). The electron spins are polarized by passing an ion beam through a foil which is tilted at an angle less than 90$^{\circ}$ with respect to the beam axis. The polarization is attained at the exit surface of the foil and is further transferred to the nuclear spin during the flight in vacuum via the hyperfine interaction, provided that the nuclei of interest have non-zero spin. The polarization process can be understood as due to the transfer of momentum from the surface electrons of the foil to the electrons of the exiting ion. The polarization transfer from the electron to the nuclear spins during the flight of the ions in vacuum is discussed in Sect.~\ref{sect:RIV-concepts}.

The possibility for obtaining spin orientation of an atomic beam by the beam-tilted-foil interaction and transferring it to the nuclear spin was suggested in Ref.~\cite{fano73,elli73}. The atomic spin orientation was confirmed soon after~\cite{berr74}. The nuclear polarization produced by the TFT was observed through a measurement of the anisotropy of the emitted $\beta$ radiation of $^{12}$B~\cite{noji83}, or of the $\gamma$ rays emitted from high-spin isomers~\cite{dafn83,hass84}. An enhancement of the spin polarization was achieved by using a multi-tilted-foil array~\cite{gold85a}. Different aspects of the TFT, namely, (i) production of atomic polarization in beam-tilted-foil interaction, including a description of the models describing the process, (ii) polarization transfer between an atom and a nucleus, and (iii) enhancement of the nuclear polarization by beam interaction with a multi-tilted-foil stack, are discussed in Ref.~\cite{momo02}.

 {Initially, the TFT was applied to short-lived excited states (from tens of picoseconds to microseconds), populated in fusion-evaporation or inelastic-scattering reactions at heavy-ion accelerators \cite{dafn82,hass84,dafn85}. In that respect, it could be considered as a method for producing nuclear spin polarization, which can be further combined with other techniques for nuclear moment studies. For example, the TFT has been used to determine the signs of the quadrupole moments of high-spin isomers in the Gd isotopes \cite{dafn85}. By polarizing the spin-aligned products of a heavy-ion-induced reaction, the perturbed angular distribution resulting from implantation into a noncubic crystal becomes sensitive to the sign of the quadrupole moment.
}

 {
Later on, a few studies were performed at radioactive beam facilities using ions with total energies of the order of 200 - 500 keV. Two examples can be mentioned, namely, the magnetic moment studies of the ground states of $^{17}$Ne~\cite{baby04} and $^{23}$Mg~\cite{lind97}, which were performed at the HV platform at ISOLDE.
}

 {
The next step was taken about a decade later when post-accelerated RIBs were used. For example, studies of $^8$Li~\cite{hira12} and $^{123}\mathrm{In}$ \cite{hira11} were performed at the TRIAC facility in Japan. The work on $^8$Li consists of a systematic investigation of the nuclear ground-state polarization as a function of the number of foils and the incoming beam energy, varied between $\sim$ 140 keV and $\sim$ 240 keV. A maximum observed polarization of $7\%$ was reported at the lowest measured beam energy. After the passage of the Li ions through the stack of foils, they should appear with three different charge states. It could be expected that the polarization will be induced predominantly, if not exclusively, by only one of those. If the optimum charge state could be selected after the tilted foils, a factor of 2 or 3 increase in the observed polarization might be expected. The authors of Ref. \cite{hira12} attempted to identify which of the different charge states could be responsible for the obtained polarization. They clearly state that the use of an electromagnetic separator is necessary to distinguish between the possibilities. \\
Lower nuclear polarization, $\sim 1\%$, was observed for the higher-$Z$ case of $^{123}\mathrm{In}$ \cite{hira11} at a beam energy of $\sim$~300~keV/u, which naturally raises the question whether the optimum beam energy for obtaining maximum spin polarization is $Z$-dependent.
}

\begin{figure}[h!!]
    \centering
    \includegraphics[width=0.8\linewidth]{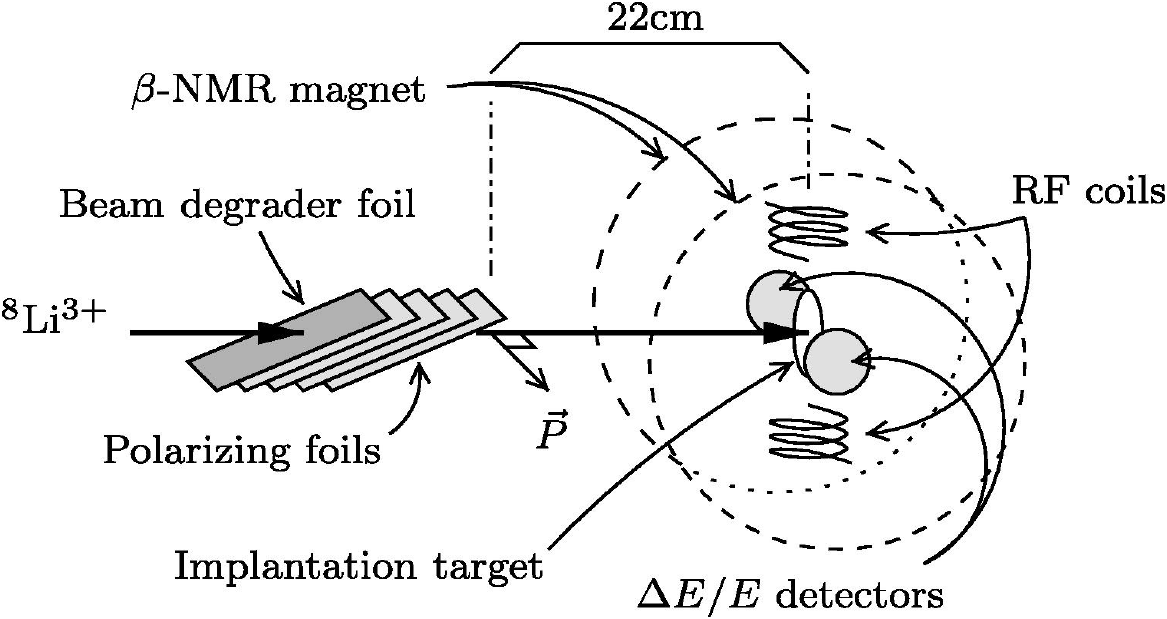}
    \caption{Schematic depiction of the TFT experimental setup at REX-ISOLDE. The foils were mounted in a holder that was rotatable around the beam axis to control the polarization vector, $\vec{P}$. Two scintillator $\beta$-particle $\Delta$E-E telescopes, the RF coils used to probe the nuclear spin polarization, and a platinum implantation host are positioned between the poles of the NMR magnet. The figure is taken from Ref.~\cite{torn13}.}
    \label{fig::TFT}
\end{figure}

 {The TFT on $^{8}\mathrm{Li}$ was performed at REX-ISOLDE as well~\cite {torn12}. A schematic drawing of the experimental setup is shown in Fig.~\ref{fig::TFT}. A stack of ten diamond-like carbon polarizing foils, each with a thickness of 4~$\mu$g/cm$^2$, was used. Polarization of $3.6 \pm 0.3$\% was measured, consistent with the results from other groups~\cite{momo02,hira12}. It is worth noting that the comparison of the REX-ISOLDE experimental results with those from TRIAC indicates that higher spin polarization is obtained at lower beam energies~\cite{torn12} for this case.
}

 {
Further studies are needed for a better understanding of the polarization process in the TFT technique. Future work should address at least the following points: i) a possible $Z$-dependence of the optimum beam energy; and ii) charge states and electron configurations, responsible for obtaining the atomic polarization and its transfer to the nuclear spin ensemble. Provided those two parameters are well understood and under control, the tilted-foils polarization technique could be used for producing spin-polarized, post-accelerated RIBs in the energy range of a few MeV/u or higher. For example, the polarizing foils could be installed at the appropriate energy stages of the accelerator before post-acceleration. This could open the possibility of nuclear moments and reaction studies, using, e.g., Coulomb excitation, transfer reactions, etc.
}

 {
It has to be noted that, in addition to the TFT, other polarization methods, for example, optical pumping, need to be considered as well. To our present understanding, the advantage of the TFT is that it is not chemically dependent, can create sizable polarization for higher-spin species, and the required apparatus is relatively simple.
}

\section{Spin-precession methods for nuclear moment studies}
\label{sec:methods}

\subsection{ {Ground states nuclear moment studies by \texorpdfstring{$\beta$}{}-NMR/NQR.}}
\label{sec:ground}
 {In the present paper, we will consider the $\beta$-NMR/NQR techniques for nuclear moment studies for ground-state and long-lived (millisecond) isomeric states in radioactive nuclei. These two techniques apply to radioactive ($\beta$-decaying) nuclei and can cover species with lifetimes typical for $\beta$ decay, namely from a few milliseconds usually up to a few seconds. An initially spin-polarized ensemble of nuclei, a starting point for those techniques, can be obtained either by the reaction mechanism by which the nuclei of interest are populated, which is the main focus of the present paper, or by any other approach, posterior to their production, e.g., by optical \cite{neug06} or low-temperature \cite{post86} techniques. The experimental accuracy, usually obtained by the $\beta$-NMR technique, is on the sub-per-mill level and can get down to the part-per-million scale \cite{hard20}. The technique is based on the modification of the initial spin polarization either by destroying it, as in the standard NMR, or by inverting its direction, as in the Adiabatic Fast Passage (AFP) method. The advantage of the latter approach is that it increases the experimental signal twice, bringing in a significant increase in the sensitivity for the scarcely produced radioactive species. The specificities of the AFP will be discussed in the following subsections.
}

\subsubsection{The \texorpdfstring{$\beta-$}{}NMR technique }
\label{beta-NMR method}
\paragraph{Characteristics}
In the conventional nuclear magnetic resonance (NMR) method~\cite{bloc46a,bloc46b,purc46}, the change in magnetization caused by NMR is observed as an electrical signal resulting from the induced electromotive force generated in a tuned circuit.  The magnetization, which is the source of the signal, is generated by the spin polarization, which is due to the difference in the distribution probability over the magnetic substates. It follows the Boltzmann distribution in an external magnetic field.  Since small spin polarization is produced, the intensity of the NMR signals is usually low. For example, in a magnetic resonance imaging (MRI) system using proton spins as probes, the spin polarization is only $\sim$0.003\% with a typical static magnetic field $B$ $=$ 1.5~T.  Nevertheless, the small contribution to the magnetization of the proton spins, due to the presence of a large amount of water, {\it i.e.}, protons in the human body, the individual signals are summed to a detectable size.

In case the method is applied to short-lived radioactive isotopes (RIs), the magnetization of the sample is so tiny compared to that of stable nuclei that the NMR signals cannot be detected at all. The $\beta$-ray NMR ($\beta$-NMR)~\cite{sugi66} enables NMR measurements for such short-lived RIs. In this case, instead of an electrical signal, the NMR signal is observed through a change in the asymmetry in the $\beta$-ray angular distribution, which is an extremely sensitive indicator, taking advantage of the asymmetric emission of $\beta$ rays with respect to the nuclear spins.  Furthermore, nuclear reactions and laser optical pumping, among others, can be used to obtain highly spin-polarized RIs compared to the usual method, which relies on the Boltzmann distribution of the population of the magnetic substates. The $\beta$-NMR is an RI-specific but extremely sensitive NMR technique. In Ref.~\cite{wich01}, the sensitivity of different methods based on the number of nuclear probes required to detect defects in semiconductors was compared, and it was demonstrated that the $\beta$-NMR method is ten orders of magnitude more sensitive than the conventional NMR techniques. The following steps need to be undertaken within a $\beta$-NMR/NQR experiment: (i) production of spin-polarized $\beta$-decaying nuclei, (ii) implantation of the nuclei of interest in a host material to preserve the orientation, (iii) detection of the polarization in terms of $\beta$-asymmetry, and (iv) resonant destruction or inversion of the spin polarization.

The $\beta$-NMR technique was developed by Sugimoto~et~al. at Osaka University in 1966~\cite{sugi66}, where the ground-state magnetic moment of $^{17}$F was measured.  This successful experiment opened the avenue for many nuclear-moment measurements of RIs with low production yield in light nuclei. It is now used worldwide in combination with various methods to produce spin polarization and has been applied not only to nuclear structure research through nuclear-moment measurements, but also to materials science studies.

\paragraph{Principle}

\begin{figure}[htbp]
  \begin{center}
    \includegraphics[width=0.8\linewidth]{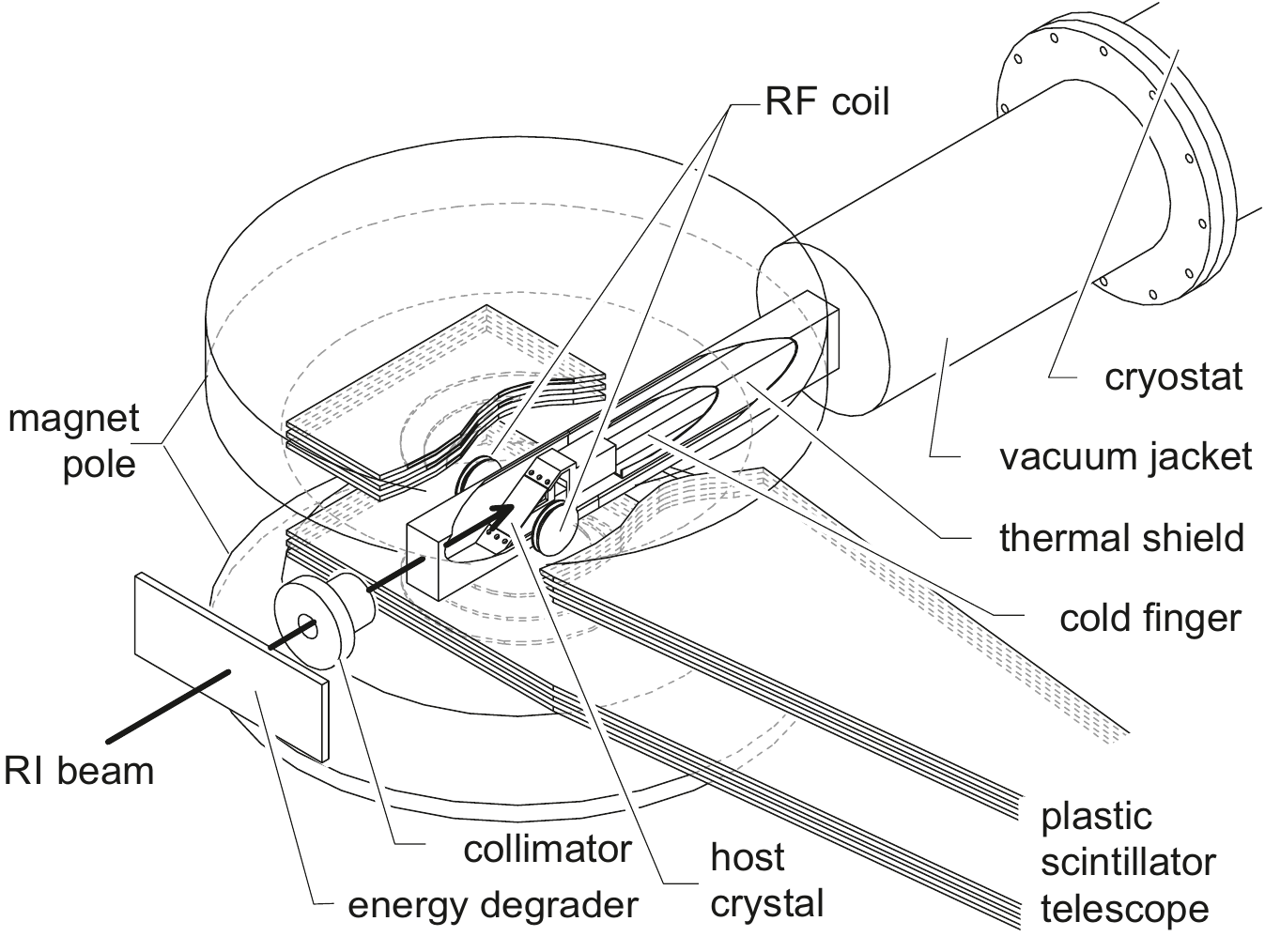}
  \end{center}
\caption{A typical $\beta$-NMR apparatus. Spin-polarized RI beams
  are implanted into a host crystal. The crystal is mounted on the
  cold finger tip of a cryostat to preserve the crystal at the
  appropriate temperature in terms of spin relaxation time. A static
  magnetic field $B$ is applied parallel to the spin polarization
  of the RI beam. The NMR RF coils are placed such that the oscillating field, $B_1$, and the static field, $B$, are perpendicular. The $\beta$~rays emitted from the
  implanted RIs are measured with plastic scintillator telescopes
  located above and below the crystal. The figure is taken from Ref.~\cite{naga09}.
}
\label{FIG:bNMR_Appratus}
\end{figure}

A typical $\beta$-NMR apparatus is shown in Fig.~\ref{FIG:bNMR_Appratus}.  As mentioned above, the NMR signal is detected from $\beta$~rays emitted from spin-polarized RIs. The angular distribution function $W(\theta)$ for the  $\beta$ rays emitted from nuclei with the spin polarization $P$ is given by
\begin{eqnarray}
W(\theta)=1+\frac{v}{c}A_{\beta}P\cos \theta ,
\label{Eq:bNMR_Wtheta}
\end{eqnarray}
where $\theta$ denotes the angle between the direction of the $\beta$~emission and the axis of the nuclear polarization, $v$ and $c$ are the velocities of the $\beta$~particles and light, respectively, and $A_{\beta}$ is the $\beta$-decay asymmetry parameter. This is the explicit form of Eq.~(\ref{eq:w_distribution_axial}) for the case $\beta$-decay with the angular distribution coefficient $A_1 = \frac{v}{c}A_{\beta}$, the spin orientation parameter $B_1 = P$ and the first order Legendre polynomial $P_1(\cos \theta) = \cos \theta$.

For simplicity, we consider an approximation that $v/c \simeq 1$. This is a good approximation when only a high-energy portion of the $\beta$~spectrum is included in the analysis.  When a static magnetic field $B$ is applied vertically in the laboratory frame and $\beta$-ray detectors are placed above and below an RI-implanted sample, the up/down ratio~$R_{\textrm{on}}$ of the $\beta$-ray yields is expressed as
\begin{eqnarray}
R_{\textrm{on}}=a \frac{(1+A_{\beta}P)}{(1-A_{\beta}P)} \simeq a(1+2A_{\beta}P),
\label{Eq:bNMR_R}
\end{eqnarray}
where~$a$ is a factor that takes into account the differences of the solid angles and efficiencies of the $\beta$-ray detectors. Note that an approximation $P^2$ $\simeq 0$ is used in this case.  When the polarization,~$P$, is altered due to the resonant spin reversal (see below for details of the spin reversal), a change appears in the $R_{\textrm{on}}$ ratio.  Thus, the resonance frequency is derived from the observed peak or dip in the $R_{\textrm{on}}$ spectrum measured as a function of the frequency of an oscillating magnetic field $B_1$. The $\beta$-NMR spectrum can also be expressed as a double ratio~$R_{\textrm{on}}/R_\textrm{off}$, in which the $R_{\textrm{on}}$ ratio measured by searching a certain frequency window is divided by the $R_\textrm{off}$ ratio measured without applying the $B_1$~field. Using Eq.~(\ref{Eq:bNMR_R}), the $R_{\textrm{on}}/R_\textrm{off}$ ratio is given as
\begin{eqnarray}
  R_\textrm{on}/R_\textrm{off} = \left.
  \frac{(1-A_{\beta}P)}{(1+A_{\beta}P)}
  \middle/
  \frac{(1+A_{\beta}P)}{(1-A_{\beta}P)} \right.
  \simeq 1-4A_{\beta}P.
  \label{eq:bNMR_on_off}
\end{eqnarray}
The value deviates significantly from unity when the resonance frequency is swept because the $R_\textrm{on}/R_\textrm{off}$ ratio offsets systematic errors related to detector efficiency. The statistical significance of this deviation is given by
\begin{eqnarray}
  \lvert 1-R_\textrm{on}/R_\textrm{off} \rvert= x \sigma .
  \label{Eq:bNMR_dRR}
\end{eqnarray}
The value of $\sigma$ is the standard deviation of $R_\textrm{on}/R_\textrm{off}$, and $x$ denotes the number of its repetitions. By using the error propagation rule, Eq.~(\ref{Eq:bNMR_dRR}) can be approximately transformed to
\begin{eqnarray}
  t_m \approx
  \frac{x^2}{{(A_{\beta}P)^2}I_{\beta}},
  \label{Eq:bNMR_FOM-T}
\end{eqnarray}
where $I_{\beta}$ is the average counting rate and $t_m$ is its measurement time.  Eq.~(\ref{Eq:bNMR_FOM-T}) reveals that the measurement time necessary to obtain a statistical accuracy of $x\sigma$ is inversely proportional to ${P^2}I_{\beta}$.  Hence, the figure of merit (FOM) of the $\beta$-NMR measurement is given by ${P^2}I_{\beta}$. All these considerations are valid under the assumption that nuclear spins are completely inverted while maintaining the magnitude of the spin polarization, {\it i.e}., the magnetization.

\paragraph{The adiabatic fast passage method}\label{Sec:AFP}
In conventional NMR, the distribution among magnetic substates is saturated by applying a $B_1$ field at the Larmor frequency, resulting in an unpolarized state. If this method is applied to $\beta$-NMR, the $R_\textrm{on}/R_\textrm{off}$ ratio is given by
\begin{eqnarray}
  R_\textrm{on}/R_\textrm{off}= \left.  1 \middle/
           \frac{(1+A_{\beta}P)}{(1-A_{\beta}P)} \right.
           \simeq 1-2A_{\beta}P .
           \label{Eq:bNMR_Rdepol}
\end{eqnarray}
A comparison of Eqs.~(\ref{Eq:bNMR_R}) and (\ref{Eq:bNMR_Rdepol}) shows that for the conventional technique, the FOM, ${P^2}I_{\beta}$, is reduced by a factor of four compared to the case discussed above, assuming a spin reversal. For this reason, in $\beta$-NMR experiments, the adiabatic fast passage~(AFP) technique~\cite{abra61} is used to achieve spin reversal. In this technique, instead of a single-frequency RF field as in a conventional NMR measurement, an RF field with a continuously swept frequency through the resonance region is applied.

The classical motion of the magnetization $\bm{M}$ under a static magnetic field $\bm{B}$ is described using the Bloch equation as~\cite{abra61}
\begin{eqnarray}
\frac{d\bm{M}}{dt} = \gamma_\textrm{R} \bm{M} \times \bm{B},
\label{Eq:bNMR_Bloch0}
\end{eqnarray}
where $\gamma_\textrm{R}$ is the gyromagnetic ratio.  The definition of $\gamma_\textrm{R}$
is  $g{\mu}_N/\hbar$, where $g$ is the $g$~factor of a relevant nuclear state, $\mu_N$ is the nuclear magneton, and $\hbar$ is the reduced Planck's constant.  The relaxation term is not considered.  The magnetization $\bm{M}$ corresponds to the sum of the nuclear magnetic dipole moments per unit volume, that is, $\bm{M}$ $=$ $\bm{\mu}{n}{P}$, where $\bm{\mu}$ is the nuclear magnetic dipole moment, $n$ is the number of nuclei per unit volume, and $P$ is the spin polarization.  Considering a system rotating around a static field $\bm{B}$ with angular velocity $\bm{\omega}$, the equation of the $\bm{M}$ motion  in this rotating frame of reference can be written as
\begin{eqnarray}
  \frac{d\bm{M}}{dt} =
    {\gamma{_R}}{\bm{M}}{\times}
    \left( {\bm{B}}+\frac{\bm{\omega}}{\gamma_\textrm{R}} \right).
    \label{Eq:bNMR_Bloch-omg}
\end{eqnarray}
This transformation corresponds to the replacement of the static field $\bm{B}$ with ${\bm{B}^{\prime}} = \bm{B}+\bm{\omega}/{\gamma_\textrm{R}}$ in Eq.~(\ref{Eq:bNMR_Bloch0}). The coordinate transformation to a rotating frame under the condition that $\bm{\omega}_\textrm{L}$ $=$ ${-}{\gamma_\textrm{R}}{\bm{B}}$ leads to ${{B}^{\prime}}$ $=$ $0$, where the magnetization is stationary. In the laboratory frame, the magnetization $\bm{M}$ is rotating around the static field with an angular velocity $\omega_\textrm{L}$, {\it i.e.}, Larmor precession.

\begin{figure}[htbp]
  \begin{center}
    \includegraphics[width=0.95\linewidth]{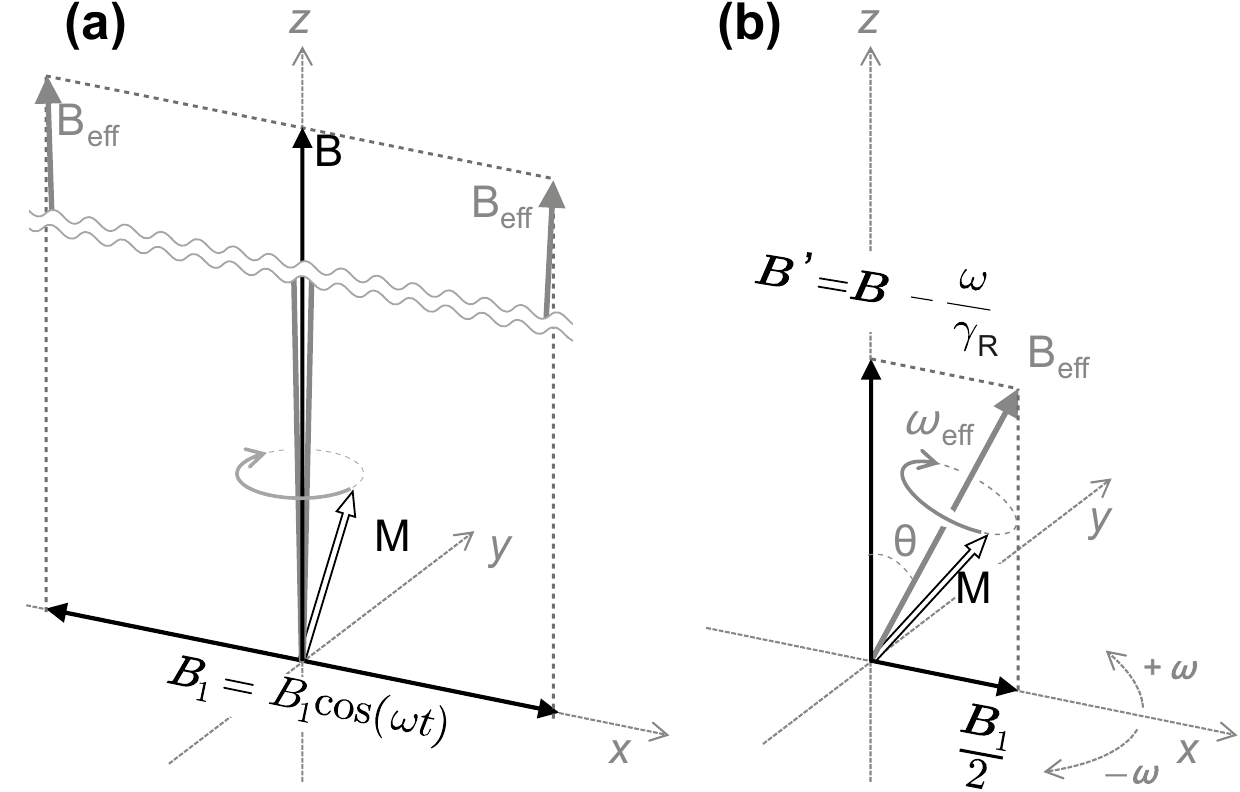}
  \end{center}
\caption{Relationship between the magnetic fields $\bm{B}$ and $\bm{B}_1$ and magnetization  $\bm{M}$ in (a) the laboratory frame and (b) the rotating frame with an angular velocity $-\omega$ after the transformation described in the text.  The direction of the Larmor precession is depicted considering positive magnetization. The effective magnetic field  $\bm{B}_\textrm{eff}$ is also displayed.
}
\label{FIG:bNMR_B0B1}
\end{figure}
Next, we consider the case when the RF field, $\bm{B}_1$, is applied perpendicular to the static field, $\bm{B}$. The relationship between the two magnetic fields in the laboratory frame is shown in Fig.~\ref{FIG:bNMR_B0B1}(a).  The $\bm{B}_1$ field has an angular velocity $\omega$  $=$ $2\pi\nu$, where $\nu$ is its oscillation frequency. It can be represented by a superposition of magnetic fields rotating in opposite directions to each other with angular velocity ${\pm}{\omega}$ as
\begin{eqnarray}
  \bm{B}_1 =
  {B_1}\cos({\omega}{t}) =
  \textrm{Re}\left(
  \frac{1}{2}{B_1} e^{i{\omega}t}
  + \frac{1}{2}{B_1} e^{-i{\omega}t}
\right).
\label{Eq:bNMR_B1-RotField}
\end{eqnarray}
Considering the positive values of $\gamma_R$ and $\omega$, in a system rotating at an angular velocity $-{\omega}$, the magnetization $\bm{M}$ precesses around an effective magnetic field
\begin{eqnarray}
  \bm{B}_\textrm{eff} =
    \left(
      {\bm{B}}-\frac{\bm{\omega}}{\gamma_\textrm{R}}
    \right)
    + \frac{\bm{B}_1}{2}
\label{Eq:bNMR_Beff}
\end{eqnarray}
at an effective angular velocity related to the Larmor frequency
\begin{eqnarray}
  {\omega}_\textrm{eff} =
  \sqrt{\left( {\omega_\textrm{L}}-{\omega} \right) ^2
    + \left( \frac{\omega_1}{2}\right) ^2},
\label{Eq:bNMR_omgeff}
\end{eqnarray}
where ${\omega_1}$ $=$ ${\gamma_\textrm{R}}{B_1}$.  The relationship between the magnetic fields and magnetization after transformation to the rotating coordinate system is shown in Fig.~\ref{FIG:bNMR_B0B1}(b).

In the NMR method based on the fast AFP technique (AFP-NMR method) described here, a reversal of the magnetic moment is introduced by effectively changing the right-hand term in Eq.~(\ref{Eq:bNMR_Beff}).  In practice, $\omega$, or more properly the oscillation frequency $\nu$, is increasingly swept to reach the resonance frequency.  Note that the effect caused by the other term in Eq.~(\ref{Eq:bNMR_B1-RotField}), {\it i.e}., the counter-rotating magnetic field not involved in resonance, causes a shift, called the Bloch--Siegert shift, which is due to a high-frequency magnetic field with an angular frequency twice the Larmor frequency~\cite{bloc40,abra61}. This shifts the resonance frequency by approximately $(B_1/2B)^2$, which is negligibly small in the case which is considered here.

When the speed at which $\bm{B}_\textrm{eff}$ tilts due to a frequency sweep in $\omega$, $d\theta/dt$, is sufficiently slow compared to the angular velocity $\omega_\textrm{eff}$ of the Larmor frequency under an effective magnetic field.  That is, when the condition
\begin{eqnarray}
  \frac{d\theta}{dt} \ll \omega_\textrm{eff}
    \label{Eq:bNMR_AFP0}
\end{eqnarray}
is met, the system changes adiabatically without changing the spin polarization, where $\theta$ is the angle between $\bm{B}$ and $\bm{B}_\textrm{eff}$ (see Fig.~\ref{FIG:bNMR_B0B1}(b)).  Given a positive frequency sweep $\dot{\omega} = d\omega/dt$, ${d\theta}/{dt}$ is obtained from Eqs.~(\ref{Eq:bNMR_Beff}) and (\ref{Eq:bNMR_omgeff}) as
\begin{eqnarray}
  \tan{\theta} =
  \frac{{B_1}/2}{B-\omega/\gamma_\textrm{R}} =
  \frac{{\omega_1}/2}{\omega_\textrm{L}-\omega} \nonumber \\
  \implies \frac{d\theta}{dt}
  = \frac{\omega_1}{2\,{\omega_\textrm{eff}}^2}\,\dot{\omega}.
\label{Eq:bNMR_dThdT}
\end{eqnarray}
From Eqs.(\ref{Eq:bNMR_AFP0})~and~(\ref{Eq:bNMR_dThdT}), we obtain
\begin{eqnarray}
  \dot{\omega} \ll
  \frac{2\,{\omega_\textrm{eff}}^3}{\omega_1},
\label{Eq:bNMR_AFPomg}
\end{eqnarray}
for the adiabatic condition. As shown in Eq.~(\ref{Eq:bNMR_omgeff}), $\omega_\textrm{eff}$ has the minimum value at the resonance point with $\omega_\textrm{eff}$ $=$ ${\omega_1}/2$, giving the most severe adiabatic condition.  Under this condition, Eq.~(\ref{Eq:bNMR_AFPomg}) becomes
\begin{eqnarray}
  \sqrt{\dot{\omega}} &\ll& \frac{\omega_1}{2}\nonumber \\
  \implies \frac{2\sqrt{\dot{\omega}}}{\gamma_R} &\ll& B_1.
\label{Eq:bNMR_AdiabaticCond}
\end{eqnarray}
From Eq.~(\ref{Eq:bNMR_AdiabaticCond}) it follows that the adiabatic condition is more easily met by applying a stronger RF field, $\bm{B}_1$, with a slower frequency sweep. Computer simulations of the spin-motion time evolution based on the Bloch equation, Eq.~(\ref{Eq:bNMR_Bloch0}), demonstrate that $B_1$ $\sim5.4~(3.4)\times\sqrt{\dot{\omega}}/\gamma_R$ is required to reverse more than 99\%~(90\%) of the spins.

In reality, the magnetization tends to return to a thermal equilibrium state. Therefore, a relaxation term must be considered.  Then, Eq.~(\ref{Eq:bNMR_Bloch0}) takes a more general form
\begin{eqnarray}
  \frac{d\bm{M}}{dt} = \gamma_\textrm{R} \bm{M} \times \bm{B}
  - \frac{{M_\textrm{x}}{\bm{e}_\textrm{x}}+
    {M_\textrm{y}}{\bm{e}_\textrm{y}}}{t_\textrm{ss}}
  - \frac{({M_\textrm{z}}-{M_0})\bm{e}_\textrm{z}}{t_\textrm{sl}},
\label{Eq:bNMR_Bloch-T1T2}
\end{eqnarray}
where $M_\textrm{i}$ and $\bm{e}_\textrm{i}$ (i~$=$~x,\,y,\,z) are the projected components of magnetization and unit vectors in the laboratory frame, respectively, and $M_0$ is the thermal equilibrium magnetization.  The values of $t_\textrm{sl}$ and $t_\textrm{ss}$ are the longitudinal (spin-lattice) and vertical (spin-spin) relaxation times, respectively.  Because the condition $t_\textrm{ss}$ $\ll$ $t_\textrm{sl}$ usually holds in solid samples, it is the $t_\textrm{ss}$ relaxation time that is relevant for constraining the AFP technique.  This means that if the spin reversal time is slower than $t_\textrm{ss}$, the spin polarization will be disturbed because the internal equilibrium among the spins will be established first.  To prevent this, the spin reversal must be performed quickly to satisfy the ``fast passage'' condition given by
\begin{eqnarray}
  \tau_\textrm{p} \simeq \frac{{B_1}/2}{\dot{B}^{\prime}} \ll t_\textrm{ss}
  \nonumber \\
  \implies B_1 \ll \frac{2\dot{\omega}t_\textrm{ss}}{\gamma_\textrm{R}},
\label{Eq:bNMR_FastCond}
\end{eqnarray}
where $\dot{B}^{\prime} = d{B}^{\prime}/dt$ and $\tau_\textrm{p}$ is the time needed that $\bm{M}$ passes through the resonance
point.  Finally, in the AFP technique, the RF field, $\bm{B}_1$,
should be applied in such a way as to meet the condition
\begin{eqnarray}
  \frac{2\sqrt{\dot\omega}}{\gamma_\textrm{R}} \ll
   B_1 \ll \frac{2\dot{\omega}t_\textrm{ss}}{\gamma_\textrm{R}},
\label{Eq:bNMR_AFPCond}
\end{eqnarray}
combining the ``adiabatic''  and ``fast passage'' conditions obtained from Eqs.(\ref{Eq:bNMR_AdiabaticCond})~and~(\ref{Eq:bNMR_FastCond}), respectively.

Another constraint on $\beta$-NMR measurements arises from the longitudinal spin-lattice relaxation time, $t_\textrm{sl}$. Since the $\beta$-ray up/down asymmetry is significantly reduced, if the spin polarization is not preserved during the $\beta$-ray measurement, the host crystal into which the guest RI is implanted should be chosen to provide a long~$t_\textrm{sl}$ over the $\beta$-decay lifetime of the guest RI.

\subsubsection{ {The \texorpdfstring{$\beta-$}{}NQR technique}}
\label{subsub:bNQR}
 {
The essential difference between the $\beta$-NMR and the $\beta$-NQR techniques is that the nuclear states of interest are subjected not to a magnetic field but to an electric field gradient (EFG), which is obtained through their implantation in a material with a non-cubic lattice structure. The interaction of the oriented nuclear states with the EFG leads to the observation of the quadrupole frequency and, eventually, extracting the quadrupole moment Q of the state of interest. Due to the specificity of the quadrupole interaction, Q can be determined only for states with spin $I \ge 1$. }

 {The lifetime range of the application of the $\beta$-NQR technique is the same as for the $\beta$-NMR, though the precision usually obtained for the quadrupole moments is rarely better than a few percent. This uncertainty is largely defined by the accuracy with which the EFG can be determined (see Sect. \ref{EFG_calibration}). }

\paragraph{Interaction between a quadrupole moment and an electric field gradient in the presence of an external magnetic field}

As discussed in Sect.~\ref{sec:EFG}, a nucleus placed in a crystal lattice interacts with the crystalline electric field gradient (EFG), $V_{ij}$, whose largest component is defined as $V_{\textrm{zz}}=eq$. The $\beta$-NQR technique uses both an external magnetic field and EFG. The interaction Hamiltonian of the nuclear electric spectroscopic quadrupole moment $Q_s$ under the combined Zeeman and quadrupole interactions is given by
\begin{eqnarray}
\hat{\mathcal{H}}_{\mu Q} =
   -\bm{\mu}\cdot\bm{B} +
    \frac{e^2qQ_s}{4I(2I-1)}
    \left(
       3(\hat{I}_\textrm{z}^2 - \hat{I}^2 )
       + \frac{1}{2}{\eta}({\hat{I}_+}^2 + {\hat{I}_-}^2)
    \right),
\label{Eq:bNQR_Hml}
\end{eqnarray}
where $I$ is the nuclear spin, $\eta$ is the asymmetry parameter of the EFG tensor in the principal axis system fixed on the nucleus, $V_\textrm{zz} = eq$ is as defined above, $\hat{I}_\textrm{z}$ is the projection of the spin operator $\hat{I}$, and $\hat{I}_{+}$ and $\hat {I}_{-}$ are the spin ladder operators, cf. Eq.~(\ref{eq:Q_tensor}).  The axes $x$, $y$, and $z$ are the principal axes of the EFG tensor.  In the first-order perturbation theory, the energy of the magnetic sub-state $m$ is given by
\begin{eqnarray}
E_m = -\mu{B}\frac{m}{I}
  +e^2qQ_s\cdot
  \frac{
    3{\cos{\theta}^2}-1+{\eta}\textrm{sin}^{2}{\theta}\cos{2\phi}
  }{
      8I(2I-1)
  }\nonumber \\
    \times \left( {3m^2}-I(I+1) \right),
\label{Eq:bNQR_Em}
\end{eqnarray}
where $\theta$ and $\phi$ represent the polar and azimuthal angles of the static magnetic field, $\bm{B}$.  As shown in Eq.~(\ref{Eq:bNQR_Em}), due to the quadrupole interaction, the magnetic substates are not equally spaced. The energy splitting of the levels is shown on the right-hand side of Fig.~\ref{fig:zeeman_q_b}.  The resonance frequency $\nu_{m,{\,}m+1}$ between magnetic substates $m$ and $m+1$ is given by
\begin{eqnarray}
\nu_{m,{\,}m+1} = \nu_{\rm L} - \nu_\textrm{Q} \cdot
  3(3\textrm{cos}^{2}{\theta}-1)
  \frac{2m+1}{8I(2I-1)},
\label{Eq:bNQR_NuQ}
\end{eqnarray}
where $\nu_{\rm L}$ $=$ ${{\mu}B}/{hI}$ denotes the Larmor frequency and $\nu_\textrm{Q}$ $=$ $e^2qQ_s/h$ the nuclear quadrupole coupling constant ($h$ is Planck's constant).  Hereafter, we consider an axially symmetric EFG tensor with $\eta=0$ for simplicity.  Provided the $eq$ value is known, $\nu_\textrm{Q}$, i.e., the $Q_s$ moment can be determined using $\beta$-NMR spectroscopy through a measurement of $\nu_{m,{\,}m+1}$.  In cases where the $\beta$-NMR FOM is sufficiently favorable, it is possible to determine $\nu_{\rm Q}$ from the resonance frequencies of satellite transitions as in conventional NMR.  However, in actual RI-beam experiments, especially for RIs far from stability, such favorable conditions are rarely available.  Therefore, as described below, several experimental techniques have been developed to obtain an optimum FOM by using the AFP method described in Section~\ref{Sec:AFP}.

NQR generally refers to a method for observing only the quadrupole resonance without applying a magnetic field.  Therefore, resonance frequency observations based on Eq.~(\ref{Eq:bNQR_NuQ}) should be mentioned as a variant of NMR spectroscopy.  However, this method is sometimes conventionally called $\beta$-NQR, a notation accepted within this paper.

\paragraph{Implementation of the adiabatic fast passage technique}

\begin{figure}[htbp]
  \begin{center}
    \includegraphics[width=0.95\linewidth]{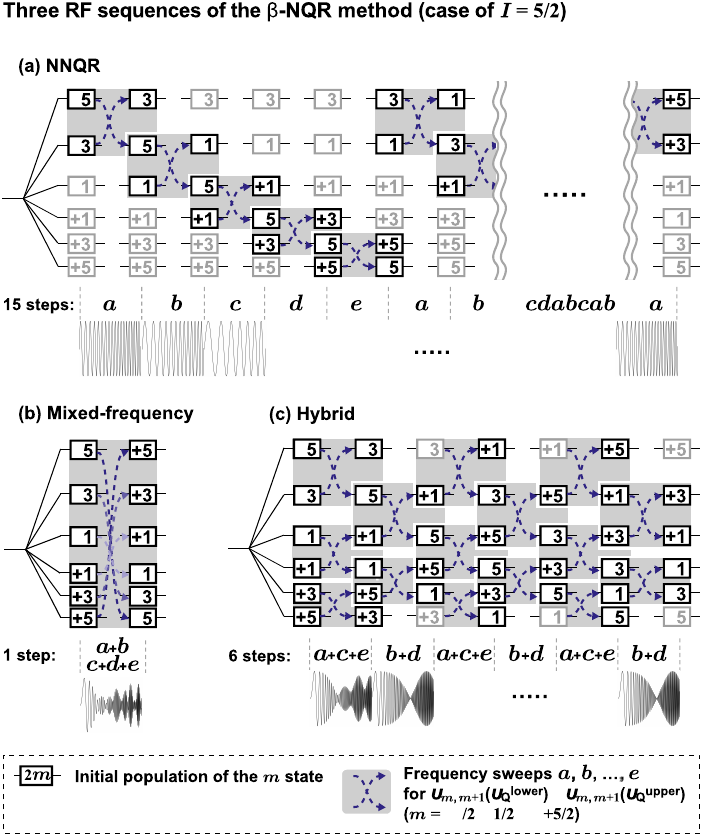}  \end{center}
\caption{Comparison of spin reversal using (a)~the NNQR, (b)~mixed-frequency, and (c)~hybrid methods in the case of nuclei with $I=5/2$. Solid squares labeled with $2m$ represent the population of the initial $m$~states, and gray squares represent the population exchange between $m$~levels by the AFP technique. The number of steps in the $B_1$-field application sequence required for the spin reversal is also shown.  See text for details on the frequency sweep settings of the $B_1$ field indicated by  \textit{a}, \textit{b}, $\cdots$, \textit{e}.
}
\label{FIG:bNMR_MixRF}
\end{figure}

Similarly to the AFP-NMR method, it is possible to exchange the populations between substates $m$ and $m+1$ by sweeping the RF field, $B_1$, across a resonant frequency $\nu_{m,{\,}m+1}({\nu_\textrm{Q}})$ while meeting the adiabatic condition given by Eq.~(\ref{Eq:bNMR_AFPCond}).  Furthermore, it is possible to induce spin reversal in the $\beta$-NQR method, enabling efficient measurements with a four-fold larger FOM compared with methods leading to saturation.  The newly developed nuclear quadrupole resonance (NNQR)~\cite{mina92} induces a full spin reversal by applying an RF field based on a sequence of stepwise reversals between two contiguous substates.  This requires a sequence comprising $I(2I+1)$ steps. The frequency sweep varies stepwise from $\nu_{m,{\,}m+1}({\nu_\textrm{Q}}^{\rm lower})$ to $\nu_{m,{\,}m+1}({\nu_\textrm{Q}}^{\rm upper})$, with ${\nu_\textrm{Q}}^{\rm lower}$ and ${\nu_\textrm{Q}}^{\rm upper}$ denoting the lower and upper bounds of the covered quadrupole frequency 
window. In Fig.~\ref{FIG:bNMR_MixRF}(a), an example for $I=5/2$ nuclei is provided. The $B_1$ field was applied in $I$($2I+1$) $=15$ steps in a sequence {\it abcdeabcdabcaba}, where \textit{a}, \textit{b}, $\cdots$, \textit{e} are the set of the $\nu_{m,{\,}m+1}$ frequencies for $m=-5/2$, $-3/2$, $\cdots$, $+5/2$, respectively.

This technique is highly effective and has led to many successful quadrupole-moment measurements of RIs.  However, for RIs with significantly short lifetimes or a large spin, the increase in the total RF-field application time $t_\textrm{RF}$, which is due to the multi-step sequence, makes it difficult to meet the adiabatic conditions. A typical example is $^{17}$B having a short halflife of $T_{1/2}$ $=5.08(3)$~ms. The ground~state magnetic moment of $^{17}$B was measured using the $\beta$-NMR method with a $B_1$ field of $t_\textrm{RF}$ $=$ 1~ms~\cite{ueno96}.  Because the spin of $^{17}$B is $I=3/2$, the NNQR method requires a total application time of $t_\textrm{RF} = 6$~ms under the same experimental conditions, which exceeds the half-life.

For this reason, another AFP-NQR method, the mixed-frequency technique, has been developed~\cite{ogaw03}. In this case, the frequency sweeps $\nu_{m,{\,}m+1}({\nu_\textrm{Q}}^{\rm lower})$ $\rightarrow$ $\nu_{m,{\,}m+1}({\nu_\textrm{Q}}^{\rm upper})$ for all $m$~substates are performed simultaneously by applying a frequency-mixed RF field, as shown in Fig.~\ref{FIG:bNMR_MixRF}(b).  In fact, for the quadrupole-moment measurement of $^{17}$B, the spin reversal was achieved with the total RF-field application time of $t_\textrm{RF}$~$=$~2~ms~\cite{ogaw03}.  Note that when using such a composite RF field, the $B_1$ strength that can be delivered to one transition is reduced by a factor of $1/(2I)$.  In the $I=5/2$ example shown in Fig.~\ref{FIG:bNMR_MixRF}, the duration of the $B_1$ field in the mixed-frequency method was $1/15$ shorter than that in the NNQR, but the $B_1$ field strength distributed to each transition was attenuated by $1/5$, which may significantly narrow the
quadrupole frequency window, thus failing to meet the adiabatic condition.  To overcome this limitation, another technique, the hybrid method, which limits the number of $\nu_{m,{\,}m+1}(\nu_\textrm{Q})$ components to be combined at a time, as shown in Fig.~\ref{FIG:bNMR_MixRF}(c), can be applied.  In the sequence of the hybrid method, the number of steps is $2I+1$, which is $1/I$ of the steps required for the NNQR method.  Notably, in the hybrid method, a simultaneous sweep of two adjacent transitions leads to an AFP-NMR between three levels, with no population change for the middle $m$ sub-state.

\paragraph{Spin-polarization measurements with the adiabatic field rotation technique}

In the $\beta$-NMR/NQR method, only after observing the resonance, we can know that an NMR resonance frequency was scanned and that spin-polarized RIs were produced.  Conversely, when the resonance is not observed in a measurement, it is not possible to separately investigate whether there is a problem with the frequency scan or with the production of spin polarization.  This is especially problematic in experiments using fragmentation-induced spin polarization, where the production of spin-polarized RI beams involves numerous parameters, such as target dependence, fragment momentum, and scattering angle, among others.  For this reason, the adiabatic field rotation (AFR) method was developed to confirm the spin polarization of RIs before scanning for resonance using $\beta$-NMR measurements~\cite{ogaw99,ishi13}.  In this method, the holding magnetic field is slowly half-rotated while maintaining the $\beta$-ray detector in place, causing a spin reversal relative to the detector system.  Thus, by measuring the change in the $\beta$-ray up/down asymmetry before and after the spin reversal, $A_{\beta}P$ can be determined from Eq.~(\ref{Eq:bNMR_R}) based on the same principle as the $\beta$-NMR method.

\begin{figure}[htbp]
  \begin{center}
    \includegraphics[width=0.95\linewidth]{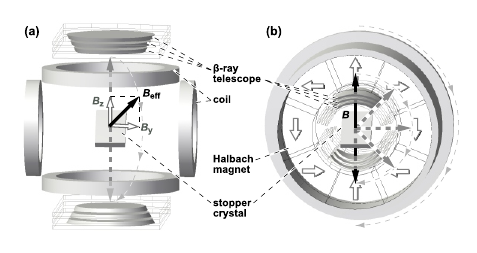}  \end{center}
\caption{Two types of $\beta$-AFR apparatus. The half rotation of the $B$ field is performed by (a)~controlling the composite field with two air-core coils and (b)~physically rotating a Halbach-type permanent magnet. The white arrows in (b) indicate the direction of the magnetic field in each magnetic domain.
}
\label{FIG:bNMR_AFR}
\end{figure}

The method was first implemented in experiments with $^{18}$N~\cite{ogaw99}. Since a strong static magnetic field was not required in this experiment to preserve the spin polarization, a holding field $B$~$=$~39~mT was applied. Two air-core coils were used to avoid hysteresis problems in magnetic field rotation.  The composite magnetic field, which is generated by the two coils as a function of time,
\begin{eqnarray}
  \bm{B}_\textrm{eff}(t)=
\bm{B}_\textrm{z}\cos(\frac{\pi t}{t_\textrm{hr}})+
\bm{B}_\textrm{y}\sin(\frac{\pi t}{t_\textrm{hr}}),
\label{eq:bNQR_comp}
\end{eqnarray}
is shown in
Fig.~\ref{FIG:bNMR_AFR}(a). It was half-rotated with a duration of ${t_\textrm{hr}}$~$=$~20~ms. Here, $\bm{B}_\textrm{z}$ and $\bm{B}_\textrm{y}$ are the components of the generated holding field, $z$ is the direction of the initial holding field (vertical), and $y$ is the horizontal direction perpendicular to the beam axis.  The rotational angular velocity of the $\bm{B}_\textrm{eff}(t)$ field must be sufficiently slower than the precession of the spins and should fulfill the adiabatic condition
\begin{eqnarray}
  \frac{\left( B_\textrm{y} \right) ^{2}}{B_\textrm{z}}
  \gg \frac{\pi}{t_\textrm{hr}\gamma_\textrm{R}}.
  \label{Eq:bNMR_AFRcond1}
\end{eqnarray}
It is also necessary that ${t_\textrm{hr}}$ is sufficiently shorter than
the $\beta$-decay lifetime.

Further experiments were carried out in the light nuclei covering the mass region from the $p$-shell to the $sd$-shell. These require a stronger holding field, with strengths in the order of several hundred~mT.  The air-core coil described above is advantageous for rotating an effective magnetic field in a short time because no hysteresis remains. However, it is difficult to generate such a strong magnetic field because this system needs to be incorporated into an NMR apparatus.  For this purpose, an AFR system using Nd permanent magnets has been developed ~\cite{ishi13}.  A circular magnet was assembled to form a Halbach array~\cite{halb80}, as shown in Fig.~\ref{FIG:bNMR_AFR}(b). It is designed to generate a strong longitudinal magnetic field of $B$ $\gtrsim$ 300~mT but with a small moment of inertia.  The system was efficiently used for studies of $sd$-shell nuclei.  However, it requires $t_\textrm{hr}$~$\sim$~150~ms for half rotation and can only be applied to RIs with longer $\beta$-decay lifetimes. The adiabatic condition in this case is
\begin{eqnarray}
  {B} \gg \frac{\pi}{t_\textrm{hr}\gamma_\textrm{R}}.
  \label{Eq:bNMR_AFRcond2}
\end{eqnarray}

\subsubsection{Experimental studies with the
  \texorpdfstring{$\beta-$}{}NMR/NQR method}\label{sec:NMRNQR_exp}

As described above, the $\beta$-NMR/NQR method was developed as an ultra-sensitive NMR specialized for RIs, and a large number of ground-state nuclear electromagnetic moment measurements have been performed for RIs in the low-mass region.

Since the late 1960s, nuclear-moment measurements have been performed utilizing spin polarization produced by low-energy direct reactions, such as the transfer reaction and polarized proton/neutron-capture reactions, using Van~de~Graaff accelerators.  In addition to the studies that were conducted actively at Osaka University, measurements were also performed at Princeton, Argonne, Stanford, Brookhaven National Laboratory (BNL), Rochester, TU M\"{u}nchen, and other accelerator facilities.  The measured nuclear electromagnetic moments are listed in Table~\ref{tab::moments_capture}. Note that the  $g$ factor of $^{15}$O was first measured using the atomic beam resonance method~\cite{comm63}, and the $\beta$-NMR experiment was subsequently carried out to obtain a more precise value~\cite{tani93}. A $\beta$-NMR/NQR apparatus was also assembled at the China Institute of Atomic Energy, and a few measurements were carried out~\cite{zhou03,zhou09}.

\begin{table}[]
\caption{Moment measurements in low-energy direct reactions. The references for the measured $g$ factors and quadrupole moments, $Q$, are provided.}
\begin{tabular}{|c |l c |l c|}
\hline
 $^ZA$ & $g$ & ref.  & $Q \; [b]$ & ref. \\
\hline
 $^8$Li & 0.826676(9) & \cite{winn78} & 0.0317(4) & \cite{dubb77} \\
 $^9$Li & 2.2927(4) & \cite{corr83} & 0.032(7) & \cite{corr83} \\
 $^8$B & 0.51775(15) & \cite{mina73} & 0.068(2); 0.0645(14) & \cite{mina92,sumi06} \\
 $^{12}$B & 1.003(1) & \cite{sugi68,mina73} & 0.0134(14); 0.01321(26) & \cite{mina78,ohts93} \\
 $^{12}$B & 1.0002(28) & \cite{zhou03} & &  \\
 $^{13}$B & 2.11808(34) & \cite{will71} & 0.0365(8) & \cite{naga04} \\
 $^{12}$N & 0.4571(5) &\cite{sugi68} & 0.0098(9) & \cite{mina98} \\
 $^{16}$N & 0.9930(5) & \cite{mats01} & 0.0179(17) &\cite{mats01} \\
 $^{15}$O & 1.43902(24) & \cite{tani93} & &\\
 $^{19}$O & 0.6128(3) & \cite{mina99} & 0.0037(4) & \cite{mina99} \\
 $^{17}$F & 1.8890(5); 1.8885(1)& \cite{sugi66,mina93} & 0.10(2) & \cite{mina74} \\
 $^{20}$F & 0.5235(5) & \cite{tsan63} &  & \\
 $^{19}$Na & -3.77084(16) & \cite{maca82} & & \\
 $^{24\textrm{m}}$Na & -1.928(3) & \cite{murn79} & &  \\
 $^{23}$Mg & 1.0728(6) & \cite{fuku93} & 0.114(3) & \cite{mats99a} \\
 $^{25}$Al &  1.4582(5) & \cite{mina76} & & \\
 $^{28}$Al & 1.0807(17) &\cite{mina81} & 0.172(12) & \cite{stoc78} \\
 $^{27}$Si & (-) 0.34216(16) & \cite{hugg84} & & \\
 $^{31}$S & 0.97586(16) & \cite{mina76b} & & \\
 $^{33}$Cl & 0.502(1);0.5033(2) & ~\cite{roge86,mats04} & & \\
 $^{29}$P & 2.4692(6) & \cite{zhou09} & & \\
 $^{39}$Ca & 0.6811(1)& \cite{mina76c} & 0.036(7) & \cite{mats99a} \\
 $^{41}$Sc & 1.5516(5) & \cite{mina90} & 0.120(6) & \cite{mina90a} \\
\hline
\end{tabular}
\label{tab::moments_capture}
\end{table}

With the construction of RI beam facilities worldwide in the 1980s, measurements have been extended to RIs of neutron-rich nuclei far from the $\beta$-decay stability line. At the isotope mass separator on-line facility (ISOLDE) at CERN, collinear fast beam laser spectroscopy (CFBLS) has been widely used for spectroscopic studies of alkaline (-earth) elements, taking advantage of high-quality low-energy beams, {\it i.e.}, low-emittance and narrow momentum-spread beams, which can be achieved at an ISOL facility. There, $\beta$-NMR/NQR measurements were also performed in combination with highly spin-polarized RI beams using the laser optical pumping method.  The RIs for which nuclear moments were determined using the $\beta$-NMR/NQR technique include:
$^{8}$Li ($Q$~\cite{borr05}),
$^{9}$Li ($g$~\cite{borr05}),
$^{11}$Li ($g$~\cite{arno87,neug08}, $Q$~\cite{arno92,neug08}),
$^{11}$Be ($g$~\cite{geit99}),
$^{17}$Ne ($g$~\cite{baby04}),
$^{20}$Na ($g$~\cite{neye05,kowa08}),
$^{21}$Mg ($g$~\cite{kram09}),
$^{29}$Mg ($g$~\cite{kowa08}),
$^{31}$Mg ($g$~\cite{neye05,kowa08}), and
$^{33}$Mg ($g$~\cite{yord07}).
The nuclear-moment measurements of $^{11}$Li~\cite{arno87,arno92,neug08} and $^{11}$Be~\cite{geit99} attracted much attention since they were related to the understanding of neutron halos in atomic nuclei. The nuclear-moment measurements of the Mg isotopes, in particular of $^{31}$Mg~\cite{neye05,kowa08} and $^{33}$Mg~\cite{yord07}, are noteworthy achievements.  These experiments were important for the investigation of the $N=20$ ``island of inversion'', see Sect.~\ref{sec:N20}.  The quadrupole-moment measurements of the $^{26\mbox{\scriptsize --}29}$Na were first performed using the CFBLS method and then with the $\beta$-NMR/NQR method~\cite{keim00}. At TRIUMF several quadrupole-moment measurements were carried out, {\it e.g.}, for $^{9}$Li ($Q$~\cite{voss11}), $^{20}$Na($Q$~\cite{mina04,mina09}), and $^{21}$Na($Q$~\cite{keim00,mina09}) using the $\beta$-NQR technique. In the case of $^{21}$Na, the quadrupole moment was first measured using the laser optical spectroscopy method~\cite{touc82}.

The milestone in these studies was the discovery of the spin-polarization phenomenon in the projectile-fragmentation reactions~\cite{okun94}, which paved the way for $\beta$-NMR/NQR measurements at RI-beam facilities based on the fast-fragmentation scheme.  This opened the avenue for nuclear moment measurements with RIs other than alkaline (-earth) elements.  Since the 1990s, nuclear-moment measurements have been actively conducted at RIKEN by the Tokyo Tech and Osaka University research groups.  The RIs for which nuclear moments were measured are listed in Table~\ref{tab::moments_fast_beams}. The results obtained from these measurements include the observation of the isospin dependence of effective charges based on the quadrupole moment measurements of the boron isotopes, the determination of anomalous ground-state spin/parity assignments in the carbon isotopes, and the abnormally large isoscalar spin expectation value of $^{9}$C in the  $^{9}$C--$^{9}$Li mirror pair, see Sect.~\ref{sec:9Li9C}.  This anomaly was of great interest, and a subsequent measurement of the $g$ factor of $^{9}$C at the National Superconducting Cyclotron Laboratory~(NSCL), Michigan State University~(MSU)~\cite{huht98} provided the same result, thus reconfirming the anomaly. Mirror moments are also discussed in Sect.~\ref{Sec:MEC}.  In this period, at GSI, the Helmholtzzentrum f\"ur Schwerionenforschung, a measurement of the $g$ factor of the $^{35}$K ground state was carried out~\cite{scha98}.  This experiment was performed at higher energies, $E/A\sim500$~MeV, than those used at RIKEN and Grand Acc\'el\'erateur National d'Ions Lourds (GANIL), typically $E/A\sim70$~MeV.  All these studies have deepened the understanding of fragmentation-induced spin polarization.

\begin{table}[ht!]
\caption{Nuclear moment measurements of exotic nuclei produced in projectile-fragmentation reactions. The references for the measured $g$ factors and quadrupole moments, $Q$, are provided, together with the laboratory where the experiments were done and the method used.}
    \centering
    \begin{tabular}{|c|c|c|c|c|}
    \hline
      $^AZ$ & $g$ & $Q$ & Lab & method \\
      \hline
       $^{13}$B  & 2.1185(3) \cite{naga04} & 0.0366(8) \cite{naga04} & QST, HIMAC & $\beta$-NMR/NQR \\
       $^{14}$B  & 0.5925(25) \cite{okun95} & 0.02984(75) \cite{izum96} & RIKEN & $\beta$-NMR/NQR \\
       $^{15}$B  & 1.77(1) \cite{okun95} & 0.038(1) \cite{izum96} & RIKEN & $\beta$-NMR/NQR \\
       $^{17}$B  & 1.697(13) \cite{ueno96} & 0.0386(15) \cite{ogaw03} & RIKEN & $\beta$-NMR/NQR \\
       $^{9}$C  & 0.9276(3) \cite{mats95} &  & RIKEN & $\beta$-NMR \\
                & 0.9307(20)\cite{huht98} &  & NSCL, MSU &  $\beta$-NMR \\
       $^{15}$C  & 3.44(2) \cite{asah02} &  & RIKEN & $\beta$-NMR \\
       $^{17}$C  & 0.5054(25) \cite{ogaw02} &  & RIKEN & $\beta$-NMR \\
       $^{17}$N  & 0.704(4) \cite{ueno96} &  & RIKEN & $\beta$-NMR \\
                 & 0.7102(8)\cite{rydt09} &  & GANIL & $\beta$-NMR \\
       $^{18}$N  & 0.3279(13) \cite{ogaw99} & 0.0123(12) \cite{ogaw99} & RIKEN & $\beta$-NMR/NQR \\
                 & 0.3273(4) \cite{rydt09} &  & GANIL & $\beta$-NMR \\
       $^{19}$N  & 0.61(3) \cite{kame04} &  & RIKEN & $\beta$-NMR \\
       $^{13}$O  & 0.926(2) \cite{mats96} & 0.0110(13) \cite{mats99} & RIKEN & $\beta$-NMR/NQR \\
       $^{21}$O  & (-)0.6036(14) \cite{ishi23} & \cite{ishi23} & RIKEN & $\beta$-NMR/NQR \\
       $^{21}$F  & 1.57(2) \cite{okun93} &  & RIKEN & $\beta$-NMR \\
       & 1.5678(5) \cite{mats99a} & 0.110(22) \cite{mats99a} & QST, HIMAC & $\beta$-NMR/NQR \\
       $^{22}$F  & 0.6736(1) \cite{miha10} & 0.003(2) \cite{miha10} & QST, HIMAC & $\beta$-NMR/NQR \\
       $^{27}Na$ & 1.558(2) \cite{borr02a} & & GANIL & $\beta$-NMR \\
       $^{23}$Al  & 1.557(88) \cite{ozaw06} & 0.163(47) \cite{naga10} & RIKEN & $\beta$-NMR/NQR \\
       $^{24\textrm{m}}$Al  & 2.99(9) \cite{nish07} &  & RIKEN & $\beta$-NMR \\
       $^{25}$Al  &  & 0.24(2) \cite{mats07} & QST, HIMAC & $\beta$-NQR \\
       $^{31}$Al  & 1.517(20) \cite{borr02} & 0.1340(16) \cite{rydt09a} & GANIL & $\beta$-NMR/NQR \\
       &   & 0.1365(23) \cite{heyl16} & GANIL & $\beta$-NMR/NQR \\
                  & 1.532(2)  \cite{himp06} & 0.112(32) \cite{naga09} & GANIL & $\beta$-NMR/NQR \\
       $^{32}$Al  & 1.959(9) \cite{ueno05} & 0.024(2) \cite{kame07} & GANIL & $\beta$-NMR/NQR \\
                  & 1.9516(22) \cite{himp06} & 0.0255(3) \cite{xu19} & GANIL & $\beta$-NMR/NQR \\
       $^{33}$Al  & 1.635(2) \cite{himp06} & $\sim$ 0.130; 0.132(16) \cite{naga09a,shim12} & GANIL & $\beta$-NMR/NQR \\
                  &  & 0.141(3) \cite{heyl16} & GANIL & $\beta$-NMR/NQR \\
       $^{34}$Al  & 0.539(4) \cite{himp08} &  & GANIL & $\beta$-NMR \\
       $^{34\textrm{m}}$Al  & 1.757(17) \cite{xu19} & 0.038(5) \cite{xu19} & GANIL & $\beta$-NMR/NQR \\
       $^{27}$Si  &  0.3461(1) \cite{mats99a} & 0.060(13) \cite{mats99a} & QST, HIMAC & $\beta$-NQR \\
       $^{35}$Si  & (-)0.468(1) \cite{neye07} &  & GANIL & $\beta$-NMR \\
       $^{28}$P  & 0.103(2) \cite{zhou07} &  & QST & $\beta$-NMR \\
         & 0.103(3) \cite{zhen09} &  &  HIMAC & $\beta$-NMR \\
        & 0.1036(11) \cite{mats10} & $\sim 0.127$ \cite{mats10} & QST, HIMAC & $\beta$-NMR/NQR \\
       $^{32}$Cl  & 1.114(6) \cite{roge00} &  & NSCL, MSU & $\beta$-NMR \\
       $^{44}$Cl & (-) 0.1375(1)\cite{rydt10} &  & GANIL & $\beta$-NMR \\
       $^{35}$Ar & 0.4215(1) \cite{mats02} & & QST, HIMAC & $\beta$-NMR \\
       $^{35}$K  & 0.24(2) \cite{scha98} &  & GSI & $\beta$-NMR \\
            & 0.261(47) \cite{mert06} &  & NSCL, MSU & $\beta$-NMR \\
       $^{37}$K &  & 0.0106(4) \cite{mina08} & NSCL, MSU & $\beta$-NQR \\
       $^{55}$Ni  & 0.65(2) \cite{berr09} &  & NSCL, MSU & $\beta$-NMR \\
       $^{57}$Cu  & 1.33 (3) \cite{mina06} &  & NSCL, MSU & $\beta$-NMR \\
    \hline
    \end{tabular}
    \label{tab::moments_fast_beams}
\end{table}

Since the 2000s, at GANIL and RIKEN, measurements on RIs in the $sd$~shell have been performed, focusing primarily on Al isotopes.  The RIs for which measurements were conducted are listed in Table~\ref{tab::moments_fast_beams}. The central theme here was the study of nuclear structure at the $N=20$ ``island of inversion'' through nuclear electromagnetic moment measurements, a topic which is discussed in Sect.~\ref{sub::islands_of_inversion}.  Within this series of experiments, $^{34}$Al was produced from a primary beam of $^{36}$S utilizing a neutron pickup reaction.  The production of spin polarization in intermediate-energy nucleon-pickup reactions was reported in previous studies~\cite{groh03,turz06}. Measurements were also carried out at QST-HIMAC and MSU-NSCL. For example, a measurement of the ground state $g$ factor of $^{35}$K was carried out at NSCL, MSU~\cite{mert06}, confirming the GSI result~\cite{scha98}. Note that the ground-state $Q$ moments in the $sd$~shell, including $\beta$-NMR/NQR measurements, are summarized and evaluated in a study by De~Rydt~et~al.~\cite{rydt13}.

\paragraph{Applications of the \texorpdfstring{$\beta-$}{}NMR technique}
The $\beta$-NMR method finds application in other research fields. The first example is the search for the $G$-parity violation of weak nucleon current in nuclear $\beta$ decay~\cite{sumi08,mina98a,mina01,mina11}. This study aimed to measure the spin alignment correlation term in the $\beta$-decay angular distribution from purely spin-aligned mirror-pair nuclei and determine the $G$-parity violating induced tensor term in the weak nucleon current.  The sophisticated method of repeated population inversion between the magnetic substates embedded in the $\beta$-NQR method enables measurements that eliminate systematic errors.  So far, measurements have been reported for the mass $A$ $=8$~\cite{sumi08}, 12~\cite{mina98a,mina01,zhen10}, and 20~\cite{mina11} systems.

The second example is related to $\beta$-delayed neutron and $\gamma$-ray spectroscopy from spin-polarized RIs~\cite{miya03,hira05,ueno13,hira15,nish17,nish19,nish20}. The large $Q_{\beta}$~windows of neutron-rich nuclei render $\beta$-delayed spectroscopy a useful tool.  Provided that the parent nucleus is spin-polarized, the asymmetry parameter $A_{\beta}$ of Eq.~(\ref{Eq:bNMR_Wtheta}) can be determined for each Gamow--Teller transition.  Being more specific, $A_{\beta}$ approximately selects discrete values $1$, $1/(I+1)$, $-I/(I+1)$ for the spin change ${\Delta}{I}$ $=$ $-1$, $0$, and $+1$ from the initial spin $I$, respectively, such that the spin parity of the connected state in the daughter nuclei can be determined through the determination of $A_{\beta}$ based on the $\beta$-NMR technique.  The proof-of-principle experiment was a study of $^{15}$C in the $\beta$~decay of spin-polarized $^{15}$B~\cite{miya03}. After that, the method was applied for the study of
$^{17}$C~\cite{ueno13} at RIKEN and
$^{9}$Be~\cite{hira15},
$^{11}$Be~\cite{hira05},
$^{30}$Mg~\cite{nish20}, and
$^{31}$Mg~\cite{nish17,nish19} at TRIUMF.

The $\beta$-NMR method was applied as well to interdisciplinary studies. At TRIUMF, material science research is carried out utilizing highly spin-polarized, high-quality, low-energy beams. In the experiments, $^{8}$Li nuclear probes are implanted at controlled depth on surfaces or interfaces and measured with a $\beta$-NMR system~\cite{koum15}. At ISOLDE, a high-accuracy liquid-sample $\beta$-NMR system, which aims at measurements related to chemistry and biology problems, was developed~\cite{croe21}. A proof-of-principle experiment for the measurement of the $g$~factor of $^{26}$Na with part-per-million accuracy was carried out at ISOLDE~\cite{hard20}. The ions of interest were implanted in room-temperature ionic liquids. In a conventional NMR experiment, a reference measurement of the magnetic moment of the stable isotope $^{23}$Na was measured in a NaCl water solution. For the improvement of the accuracy of the reference magnetic moment, {\it ab initio} calculations of NMR shielding constants were performed, thus removing a large systematic error. As a result, the accuracy of the measured ground state magnetic moment of $^{26}$Na was improved by two orders of magnitude, {\it i.e.}, $\mu = 2.849390(20)~\mu_{\mathrm{N}}$, compared to the old value $\mu = 2.851(2)~\mu_{\mathrm{N}}$.

These experiments demonstrate the potential to transform $\beta$-NMR spectroscopy into a more widely applicable technique, based on a multitude of ultrasensitive $\beta$-NMR probes with accurate magnetic moments. Problems that range from the neutron distribution in exotic nuclei to interactions of metal ions with biomolecules will become possible in the future.


\subsection{Nuclear moment studies of microsecond isomeric states using the TDPAD technique}
\label{sec:isomers}
\begin{figure}[ht]
    \centering
    \includegraphics[width=0.9\linewidth]{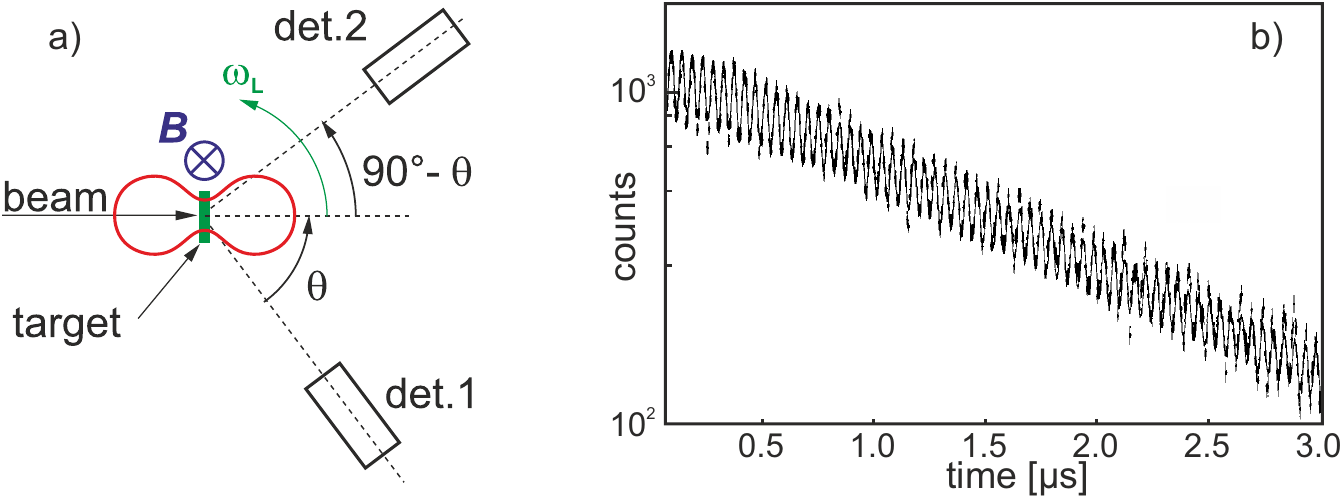}
    \caption{a) Principle scheme of a TDPAD setup.  {b) Typical decay curve of a single $\gamma$-ray detector (for a multipolarity 3 transition), including the modulation of the $\gamma$-ray intensity, due to the rotation of the spin ensemble. The figure is taken with modification from Ref. \cite{stuc87a}.}}
    \label{fig:TDPAD_scheme}
\end{figure}

For isomeric states with lifetimes in the range from about one nanosecond to a few microseconds, the Time Dependent Perturbed Angular Distribution (TDPAD) method appears to be the most suitable approach for nuclear moment studies. In such experiments, a full rotation of the spin ensemble within the nuclear lifetime may be observed.  {In certain cases, many full rotations can be detected, as depicted in Fig. \ref{fig:TDPAD_scheme} b). The typical precision of a TDPAD measurement is of the order of a few percent. }

The TDPAD technique is based on the rotation of the nuclear spin ensemble immersed in a magnetic field, usually perpendicular to the spin-alignment axis.  {
The angular distribution of the $\gamma$-rays is observed as a function of time, using detectors positioned in a plane perpendicular to an external magnetic field (see Fig. \ref{fig:TDPAD_scheme} a)). Using detectors positioned at $90^{\circ}$  with respect to each other allows the construction of an $R(t)$ function (see Eq.~(\ref{eq:Rt_90})) from which the $g$ factor of the isomeric state can then be extracted, provided the applied magnetic field is known.
}

The short-lifetime limit of applicability of the TDPAD technique is determined both by the production mechanism of the state of interest and by the possibility of obtaining a sufficiently strong magnetic field. For example, an isomeric state with a \textit{g} factor, $g = 1$, needs to be exposed to a magnetic field of about 130 Tesla to induce a full precession of its spin within one nanosecond. It is worth mentioning that the observed oscillation frequency in the TDPAD technique is twice that of the Larmor precession, {\it i.e.}, the period is $T = \pi/\omega_L$.

The long-lifetime range of applicability of the TDPAD technique is determined mostly by the relaxation effects or by the correlation time between the creation of the isomeric state, {\it e.g}., by a nuclear reaction or its implantation in the host and its decay. The relaxation times depend strongly on the probe nucleus and on the implantation host. They can vary between a few nanoseconds, {\it e.g.}, for rare-earth nuclei, up to a few (tens of) microseconds, when implanting in a liquid host or at temperatures very close to the melting point of the host. In the very long-lifetime regime (many microseconds), the Stroboscopic Observation Perturbed Angular Distribution (SOPAD) technique \cite{chri70} may be a better approach.

For many years, this method has been applied to in-beam experiments using fusion-evaporation or transfer reactions. This implied a beam pulsing with a pulse width of typically 1 -- 2 ns and a period that is considerably longer than the lifetime of the isomer, $T_{1/2}$, ($T \geq 3\:T_{1/2}$). In all those experiments the spin-alignment axis is parallel to the beam direction, {\it i.e.}, the spins are in the plane perpendicular to the beam, the applied magnetic field $\bm{B}$ is perpendicular to the beam axis and the $\gamma$-ray detectors are in a plane perpendicular to the magnetic field vector, see Fig. \ref{fig:TDPAD_scheme}. The $\gamma$-ray detectors are usually positioned at $90^{\circ}$ with respect to each other and, in the general case, at an angle $\theta$ with respect to the beam axis. Often $\theta = 45^{\circ}$ or 135$^{\circ}$. The $t=0$ reference is given by the beam pulse, and the $\gamma$-ray intensities are monitored as a function of time between the beam bursts:
\begin{equation}\label{eq:int_theta_t}
  I_\gamma(\theta,t,B) = I_0 e^{-t/\tau}W(\theta,t,B)
\end{equation}
\begin{equation}\label{eq:Rt_90}
  R(\theta,t,B) = \frac{I_\gamma(\theta,t,B) - I_\gamma(\theta+\pi/2,t,B)}{I_\gamma(\theta,t,B) + I_\gamma(\theta+\pi/2,t,B)},
\end{equation}
or, an alternative ratio function can be formed using the same detector at two different directions of the magnetic field:
\begin{equation}\label{eq:Rt_B}
  R(\theta,t,\pm B) = \frac{I_\gamma(\theta,t,+B) - I_\gamma(\theta,t,-B)}{I_\gamma(\theta,t,+B) + I_\gamma(\theta,t,-B)},
\end{equation}
where $+B$ and $-B$ indicate the two opposite directions of the magnetic field.

The Lamor frequency can be extracted from the $R(t)$ function by substituting the angular distribution function from Eq.~(\ref{eq:w_distribution_axial}) into Eq.~(\ref{eq:Rt_90}) giving
\begin{equation}\label{eq:Rt_gen}
  R(\theta,t,B) = \frac{[(3/2)A_2B_2+(5/8)A_4B_4] \cos[2(\theta-\omega_Lt)]}{2+(1/2)A_2B_2+(9/32)A_4B_4+(35/32)A_4B_4 \cos[4(\theta-\omega_Lt)]}.
\end{equation}
 Terms up to $k_{\rm max} = 4$ are considered, which is sufficient for multipolarities up to quadrupole transitions. In most of the cases, $A_4B_4\ll A_2B_2$. Thus Eq.~(\ref{eq:Rt_gen}) can be reduced to
\begin{equation}\label{eq:Rt_2}
  R(\theta,t,B) = \frac{3A_2B_2 \cos[2(\theta-\omega_Lt)]}{4+A_2B_2}.
\end{equation}
When the $\gamma$-ray detectors have considerably different characteristics, it might be advantageous that the $R(t)$ function is constructed using the same detector. For this purpose, the magnetic field direction could be flipped as mentioned above. In that case the $R(t)$ function, considering only $k = 2$ terms, becomes:
\begin{equation}\label{eq:Rt_2pm}
  R(\theta,t,\pm B) = \frac{3A_2B_2 \sin(2\theta)\sin(2\omega_Lt)}{4+A_2B_2+3A_2B_2 \cos(2\theta) \cos(2\omega_Lt)}
\end{equation}
The symmetry properties of the above equation apply conditions on the positioning of the $\gamma$-ray detectors. For example, for in-beam experiments, when the spin-alignment axis is parallel to the beam direction, the best position of the detector used in a magnetic-field flip measurement is at $\pm$ $45^{\circ}$. In that case, the last term in the denominator becomes zero, and one obtains the maximum amplitude of the $R(t)$ function $(3A_2B_2/(4+A_2B_2))$. On the contrary, for detectors at $0^{\circ}$ and at $90^{\circ}$, the $R(t)$ is strictly zero.

The fusion-evaporation reactions have the advantage of producing a sizable amount of spin alignment. They can readily be used for populating isomeric states on the proton-rich side of the nuclear chart. However, in order to study the properties of neutron-rich nuclei, one needs to apply other production mechanisms, such as {\it e.g.} projectile-fragmentation reactions. As discussed earlier (see Sect.~\ref{sec:orientation}), nuclear spin orientation has been observed in projectile fragmentation, and it has been utilized in a number of cases. The main difference between in-beam (fusion-evaporation or transfer reactions) experiments and those using projectile fragmentation is the decoupling between the position at which the states of interest are populated and where they are observed. This certainly brings the advantage of cleaner experimental conditions (no prompt or short-lived radiation produced by the nuclear reactions) but limits the investigations to lifetimes longer than the flight time through the spectrometer (usually longer than 100 ns).

In addition to the lifetime limitations for TDPAD studies in projectile-fragmentation, there are two more issues to be taken into account:
 \begin{description}
   \item[(i)] the orientation of the nuclear spin ensemble has to be preserved during the flight of the ions through the spectrometer;
   \item[(ii)] the orientation axis of the spin ensemble is not necessarily parallel to the beam axis at the implantation point.
 \end{description}

One of the main reasons for spin deorientation for ions flying in vacuum is the coupling of the nuclear and the electron spins. The straightforward way to cope with this is by selecting only fully stripped ions to be transported through the spectrometer. At the fragmentation energies (a few hundreds of MeV/u), this is usually easy to obtain for the lower to medium $Z$ nuclei, up to $Z \sim$ 50. However, for higher-$Z$ nuclei, towards the Pb region, the fraction of fully-stripped ions, even above 500 MeV/u, is quite small, and one is forced to look for other possibilities, {\it e.g.}, selecting well-defined charge states with zero electron spins. More detailed discussion on the interaction of the nuclear and atomic spins of ions moving in vacuum can be found in Sect. \ref{sect:RIV-method} and those that follow on the Recoil in Vacuum (RIV) method.

\begin{figure}
  \centering
  \includegraphics[width=0.8\linewidth]{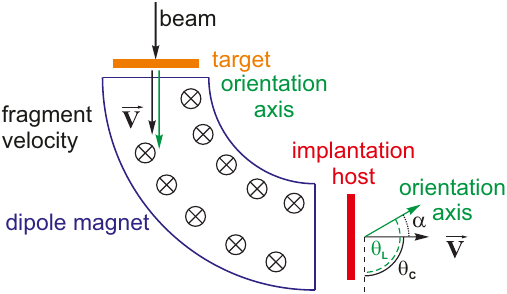}
  \caption{Deviation between the beam axis and the spin-orientation axis after passing of the ions through a magnetic spectrometer, see explanations in the text. }\label{fig:phase_alpha}
\end{figure}

\paragraph{ {Sign determination of \texorpdfstring{$g$}{} and \texorpdfstring{$Q$}{} in TDPAD/TDPAC measurements}}

Here, we assume that only fully stripped ions are selected, and we consider the interaction of the nuclear spins with the electromagnetic fields of the spectrometer. During the passage of the ions through a dipole magnetic field, their spins are rotated at an angle $\theta_L$ with the Larmor frequency $\omega_L = - g\mu_NB/\hbar$ while their trajectories are deviated with the cyclotron frequency $\omega_\mathrm{c} = qB/A$, where $g$ is the $g$ factor of the nuclear state, $A$ is the mass of the ion, and $q$ is its charge, which is equal to $Z$ for fully stripped ions. Therefore, at the exit of a magnetic spectrometer, the ions will be deviated at an angle $\theta_\mathrm{c}$, and the angle between the beam direction and the axis of the nuclear spin ensemble will be (see Fig.~\ref{fig:phase_alpha})
\begin{equation}\label{eq:dev_angle}
  \alpha = \theta_L - \theta_\mathrm{c} = - \theta_\mathrm{c} \left( 1 - \frac{gA}{2q}\right).
\end{equation}

 This results in a modification of Eq.~(\ref{eq:Rt_2}):
\begin{equation}\label{eq:Rt_2_frag}
    R(\theta,t,B) = \frac{3A_2B_2 \cos[2(\theta-\alpha-\omega_Lt)]}{4+A_2B_2}.
\end{equation}
It is worth mentioning that the ratio function depends on the $g$ factor of the isomeric state both through the observed frequency and the initial phase of the oscillations. In the particular case when the sum of the deviation angle of the spectrometer and the position of the first detector ($\theta_\mathrm{c} + \theta$) is a multiple of $\pi/2$, ($\theta_\mathrm{c} + \theta) = n\pi/2$, Eq.~(\ref{eq:Rt_2_frag}) can be rewritten in the form
\begin{equation}\label{eq:Rt_2_frag_pi}
    R(n\frac{\pi}{2} - \theta_\mathrm{c},t,B) = \pm \frac{3A_2B_2}{4+A_2B_2} \cos\left\{ 2g\left(\frac{\theta_\mathrm{c}A}{2q} - \frac{\mu_NBt}{\hbar}\right)\right\},
\end{equation}
where the positive sign corresponds to even $n$ and the negative one to odd $n$ values. The cosine is an even function and, therefore, the sign of the $g$ factor cannot be determined in this specific configuration.

While discussing the possibility, or not, to determine the sign of the $g$ factor in a TDPAD/TDPAC measurement using a magnetic interaction, it might be worth mentioning a peculiarity of the quadrupole interaction (QI) in spin-precession techniques. In the general case, a spin-oriented ensemble is obtained either through $\gamma\gamma$ correlations (TDPAC) or in the reaction mechanism itself (TDPAD). In a parity-conserving process, such as a $\gamma$-ray emission, this orientation is revealed in the form of a spin alignment if the circular polarization of the photon is not detected. Thus, the spin ensemble can be considered as having $p(m) = p(-m)$, while $p(m_1) \ne p(m_2)$ for $m_1 \ne m_2$. If the spin ensemble is subjected to a QI with a direction of the EFG parallel to the spin-orientation axis, the entire system is stationary, and no modification of the angular anisotropy is observed. However, if the two symmetry axes of the system, namely of the QI interaction (the EFG) and the spin orientation, do not coincide, this will cause a time-dependent modulation of the population of the different $m$ substates, which depends on the quadrupole coupling frequency $\omega_Q$, see Eq.~(\ref{eq:quadrupole_frequency}). This is observed as a coherent superposition of different frequencies (quantum beats) in the $\gamma$-ray angular anisotropy. The modulation can either $i)$ cause an oscillation of the spin alignment, while keeping the interaction only between even-tensor terms (alignment part); or $ii)$ convert the alignment to polarization by coupling odd and even tensor terms (polarization part). This second effect is unique for the quadrupole interaction and is not observed in a magnetic interaction, which does not modify the spin alignment. It comes from the fact that the QI makes the $+m$ and the $-m$ substates precess in opposite directions.

This behavior can be expressed mathematically by looking at, {\it e.g.}, the $\beta - \gamma$ angular correlations:
\begin{equation}
    \label{eq:QI_pol}
    W(t) = \sum_{k_1, k_2,\newline N_1,N_2} \frac{(-1)^{k_1 + k_2}A_{k_1}(\beta)A_{k_2}(\gamma)}{\sqrt{(2k_1+1)(2k_2+1)}} \\
     \times G_{k_1  k_2}^{N_1 N_2}(t)Y_{k_1}^{N_1*}(\theta_1, \phi_1)Y_{k_2}^{N_2}(\theta_2, \phi_2),
\end{equation}
where $A_k$ are the angular correlation coefficients (see Eq.~(\ref{eq:angular_distribution_coeff})); $Y_k^q$ are the spherical harmonics; $\theta$ and $\phi$ are the emission angles of the detected particles. The indices  $k_1$, $k_2$, $N_1$ and $N_2$ are all integers with $-k \le N \le +k$. Since the circular polarization of the $\gamma$-rays is not measured, $k_2$ is restricted to even integers (0, 2, 4, etc.) while $k_1$, due to the parity non conservation of the $\beta$ decay, can take all values (0, 1, 2, etc.). The properties of the angular correlation coefficients $A_k$ lead to a maximum value of $k_{2 {\rm max}} \le {\rm min}(2I,2L)$, where $I$ is the spin of the state and $L$ is the multipolarity of the transition. The limit on $k_1$ depends on the $\beta$ decay. The $G_{k_1  k_2}^{N_1 N_2}(t)$ are the  perturbation factors. For an axially symmetric EFG (for example, in a hexagonal crystal), the QI is diagonal, resulting in $N_1 = N_2=N$ and
\begin{equation}
    \label{eq:G_k_correlations}
    \begin{aligned}
    G_{k_1 k_2}^{NN} &= \sum_{n \ge 0}S_{nN}^{k_1k_2}\:\cos(n\omega_0 t) \;\;& \mathrm{for } \; k_1 + k_2 = \mathrm{even} \\
    G_{k_1 k_2}^{NN} &= (-i) \sum_{n \gt 0}S_{nN}^{k_1k_2}\:\sin(n\omega_0 t)  \;\; &\mathrm{for } \; k_1 + k_2 = \mathrm{odd.}
    \end{aligned}
\end{equation}
Here $\omega_0$ is related to the quadrupole frequency $\omega_Q = e^2qQ/4I(2I-1)\hbar$, Eq.~(\ref{eq:quadrupole constatnt}), as
\begin{equation}\label{eq:omega0}
    \begin{aligned}
    \omega_0 &= 3\omega_Q \quad \mathrm{for\;\; integer } \;I \\
    \omega_0 &= 6\omega_Q \quad \mathrm{for \;\; half-integer } \;I.
    \end{aligned}
\end{equation}
More information on the $S_{nN}^{k_1k_2}$ factors can be found, e.g., in Ref. \cite{ragh75}, and they are tabulated in Ref. \cite{wata91}. In the case of a TDPAD measurement, the angular correlation coefficients for $\beta$-decay $A_{k_1}(\beta)$ need to be replaced with the orientation parameters $B_k(I)$ (Eq.~(\ref{eq:B_orientation})).

The key feature of the quadrupole interaction becomes evident by the second term of Eq.~(\ref{eq:G_k_correlations}). If odd $k_1$ values are allowed, either by the nature of the first radiation, {\it e.g.}, $\beta$ decay, or by the reaction mechanism populating the spin ensemble (through $\rho ^k_q$), the anisotropy of the angular distribution becomes sensitive to the sign of the quadrupole moment. Otherwise, only its absolute value can be determined.

Further details on this subject can be found in Ref. \cite{ragh75}. In addition to calculations and specific examples for angular anisotropies in $\beta - \gamma$ correlations, Raghavan et al. discuss more options on how the sign of the nuclear quadrupole moment can be derived, {\it e.g.}, by starting from an initially spin-aligned ensemble and having a detection sensitive to the spin polarization, e.g., $\beta$ asymmetry or $\gamma$-ray circular polarization. Here we will limit ourselves only by giving two examples, namely of a $\beta-\gamma$ measurement using a combined magnetic and quadrupole interaction of Rots et al. \cite{rots75} and of a sign of the quadrupole moment study of the $10^+$ isomeric state in $^{54}\mathrm{Fe}$, using a spin polarized ensemble by means of the tilted-foils technique (TFT) by Hass et al.~\cite{hass84}.

The special interest in the work of Rots et al.~\cite{rots75} comes from the assumption that no magnetic moment information can be obtained in $\beta - \gamma$ correlations. This is indeed correct if the spin ensemble is subjected only to a magnetic interaction. However, Rots et al. have used a combined magnetic plus quadrupole interaction, with the second one much weaker, thus treated as a perturbation. This has allowed them to observe a clear beating pattern, dominated by the Larmor frequency of $\mathrm{In}\underline{\mathrm{Ni}}$. This measurement opens the possibilities for magnetic moments studies in $\beta - \gamma$ correlations in the special cases when the isomeric state of interest is directly populated by $\beta$ decay but no $\gamma - \gamma$ path is observed. Another interesting point to mention about the work of Rots et al. \cite{rots75} is that a quadrupole interaction, although much weaker than the magnetic one, is observed following the implantation of Te into Ni.

In the work of Raghavan et al. \cite{ragh75}, there are two methods considered for obtaining a spin-polarized ensemble through $\beta$ decay or after a Coulomb excitation, which limits the states that can be studied. The approach of Hass et al. \cite{hass84} relies on polarizing the states of interest after their population in, {\it e.g.}, a fusion-evaporation reaction, in a fast and chemically non-selective manner. The TFT transfers the atomic polarization, obtained by the interaction of ions passing through a stack of carbon foils at oblique angles, to the nuclear spin ensemble. Studies of the obtained nuclear spin polarization as a function of the nuclear spin and the number of the foils are presented in Ref. \cite{hass84}. For example, for the high-spin ($10^+$) isomer, the nuclear polarization has been measured at 8\% (for a stack of 13 foils) and up to 18\% for 17 foils. A similar level of spin orientation should be suitable for nuclear moments studies. Further specificities of the TFT can be found in the review of Berry et al. \cite{berr82}.

To summarize, the signs of the magnetic dipole and the electric quadrupole moments can be obtained in the spin-rotation technique, although this might require some specific arrangements. The fine control of the nuclear spin orientation with the transfer from alignment to polarization and {\it vice versa} is a subject that can be further explored and provide new opportunities.

The greater part of the TDPAD and TDPAC results, published in the period between 2000 and 2025, will be covered in the discussion of the scientific results in Sect. \ref{sec:results}. However, some of the physics cases go beyond the scope of the present review. Among those one can mention the TDPAD measurements in the mass $A \sim 90$ region using transfers reactions populating, {\it e.g.}, isomeric states in $^{86}\mathrm{Y}$ \cite{ione00,rusu10}, $^{84}\mathrm{Y}$ \cite{ione05}, and in  $^{93}\mathrm{Sr}$ \cite{sasa04}. In the latter case, the online TDPAC technique was applied. Both time-integral and time-dependent perturbed angular correlation methods have been applied. For example, in the mass $A\sim140$ region, these include results on the very short-lived isomeric states in $^{132}\mathrm{I}$ ($T_{1/2} = 1.120(15)$~ns) \cite{tani09} and in $^{135}\mathrm{I}$ (calculated $T_{1/2} \sim 2$~ns) \cite{good08}, as well as the $15/2^-$ state in $^{137}\mathrm{Xe}$ \cite{liu10} and the $4^+$ isomer in $^{140}\mathrm{Ce}$ \cite{ohku13}. Time integral $\gamma$-$\gamma$ angular correlations, using a static external field of up to 5 Tesla, have been applied in the mass $A \sim 170$ mass region. These allowed determining the $g$ factors of the $2^+_1$ states in $^{164}\mathrm{Yb}$ \cite{bera04}, $^{160}\mathrm{Er}$ \cite{wolf05}, $^{170}\mathrm{Hf}$ \cite{wolf07}, $^{172}\mathrm{Hf}$ \cite{bera09} and $^{168}\mathrm{Hf}$ \cite{wolf12}, having half lives between 0.8~ns and 1.3~ns with a precision of about 20\%. In the mass $A \sim 190$ region IPAD measurements were reported in $^{190 - 192}\mathrm{Pt}$~\cite{kovg01} and in $^{190 - 194}\mathrm{Pt}$ and $^{196-198}\mathrm{Hg}$~\cite{levo06}.

\subsection{Short-lived excited states: An overview}
\label{sect:short-lived-meth}

Measurements of the magnetic moments of excited states with lifetimes below about one nanosecond require the application of a magnetic field to the nucleus with a magnitude of at least a few hundred tesla. Such large magnetic fields can be produced by hyperfine interactions between the nucleus and its local electronic environment.

 {
The experimental precision achieved for short-lived states depends on the size of the experimental effect being measured and statistics. However, precision of the order of a few percent is typically reported for the transient-field method (see Sect. \ref{sect:TF-method}) and the recoil-in-vacuum technique (Sect. \ref{sect:RIV-method}). For these methods a dominant contribution to this uncertainty arises from the limited knowledge and calibration of the hyperfine fields employed.
}

Three types of hyperfine magnetic field are commonly used to measure the $g$~factors of short-lived states. These are: (i) the static hyperfine magnetic field that acts on the nuclei of impurity atoms at rest within a ferromagnetic host (usually on a substitutional lattice site), (ii) the transient magnetic field that acts on the nuclei of ions swiftly traversing a polarized ferromagnetic medium, and (iii) the hyperfine interactions of free ions in vacuum, often referred to as `Recoil in Vacuum' or RIV.

The static hyperfine field was discussed in sect.~\ref{sect:SF} and will not be considered here in any detail. However, before turning to focus on the transient field and RIV, it is worth noting that the static field is useful not only for short-lived states, but also for longer-lived states for which the TDPAD method may be used (see sect.~\ref{sec:isomers}). In contrast, the transient field and the free-ion fields can be applied to the nucleus only fleetingly.

The static and transient hyperfine fields act like externally applied classical magnetic fields, which means, in practical terms, that the nuclear precession is independent of the nuclear angular momentum and that the effect of the field is manifested as a rotation of the decay radiation pattern around the magnetic field direction.

The hyperfine fields of free ions are more complex in that the perturbation of the nuclear angular momentum depends on its value, as well as on the details of the electronic configuration of the ion. Ions are often highly charged, and there can be a distribution of charge states, each potentially with a distribution of excited atomic states, which may also change with time.

As discussed in Sect. \ref{sec:isomers}, the RIV interaction can be a cause of the loss of spin alignment in TDPAD measurements with fast radioactive beams, where longer-lived (microsecond) states are transported between the production target and the point of observation, and the nucleus of interest carries one or more electrons during the transportation. However, for short-lived states with lifetimes on the order of picoseconds, the RIV interaction is useful to measure the magnetic moment and has some advantages over the static- and transient-field methods for application to radioactive beams.

The static hyperfine field strength is typically on the order of 10 to 100 times weaker than the transient field, whereas free-ion hyperfine fields can be much stronger than both.
 {It is worth noting that the RIV technique is insensitive to the sign of the $g$ factor, whereas the transient-field and static-field methods allow its determination.
}

The following discussion of $g$-factor measurements on short-lived states will focus first on transient-field measurements and then review RIV measurements.

\subsection{The transient-field method}
\label{sect:TF-method}

 {
\subsubsection{Introduction to the transient-field method}
}
\label{sect:TF-intro}
As noted above, the \textit{transient field} is an intense hyperfine magnetic field that acts on the nuclei of ions while they are in motion through a polarized ferromagnetic medium; for transient-field $g$-factor measurements, typical ion velocities are of the order of a few percent of the velocity of light. The transient field was discovered in the late 1960s \cite{borc68}. Also important was the discovery by Eberhardt et al. \cite{eber75a} in 1975 that the transient field increases in strength with ion velocity, not inversely with ion velocity as was believed at the time.

Having a typical strength of several kilotesla, and acting like a classical macroscopic magnetic field in the direction of the magnetization of the host, the transient field has proven to be very useful for measuring the magnetic moments of excited nuclear states with lifetimes typically on the order of picoseconds. In fact, the shortest-lived state for which transient-field precessions have been measured \cite{kumb79,spei81,spei87a} is the first excited state of $^{12}$C with a mean lifetime of $\tau=65$~fs. In this case, the measurements were performed to study the characteristics of the transient field rather than to measure the $g$~factor (which is presumed to have a value very near $g=0.5$). The precession angles are less than one milliradian (on the order of 0.05$^{\circ}$).

Measurements of $g$~factors have been performed on a number of excited states with lifetimes of about 100 fs; examples include excited states in light nuclei such as $^{20}$Ne \cite{lesk03} and $^{24}$Mg~\cite{spei83,spei84}, the 2$^+_1$ and 2$^+_2$ states in $^{90}$Zr \cite{jako00} and $^{92}$Zr \cite{wern08}, and the first-excited 2$^+$ states in the $N=82$ isotones $^{140}$Ce, $^{142}$Nd and $^{144}$Sm \cite{bazz91}.

 {
\subsubsection{Reviews of the transient-field method}
}
\label{sect:TF-prev-reviews}

There have been several reviews concerning transient-field $g$-factor measurements. Benczer-Koller, Hass and Sak \cite{benc80} covered the discovery of the transient field, its characterization and atomic origin, and discussed moment measurements until 1980.
Rud and Dybal \cite{rud86} in 1986 collated data on the transient-field strength for light ions and correlated them with measured K-vacancy fractions to describe the atomic origin of the transient field in terms of the polarization of electrons bound to the moving ion. Overall, in the period since, more attention has been given to using the transient field as an empirical tool for magnetic-moment measurements than to understanding the ion-solid interactions that give rise to it.

In 2002, Speidel, Kenn, and Nowacki \cite{spei02} reviewed advances in transient-field measurements, mainly focusing on the use of inverse kinematics (projectile excitation) to study nuclei near shell closures and compare with large basis shell model calculations.
Five years later, in 2007, Benczer-Koller and Kumbartzki \cite{benc07} reviewed technical advances and showed selected highlights of measurements in the previous decade, including examples of applications of the transient-field technique to moment measurements on excited radioactive beams.

Here, an overview of the technique is given with a review primarily of research published since the year 2000. The prospects and challenges for future research are noted.

 {
\subsubsection{The experimental method}
}
\label{sect:TF-experimental-method}

\begin{figure}[ht]
  \centering
  \includegraphics[width=0.8\linewidth]{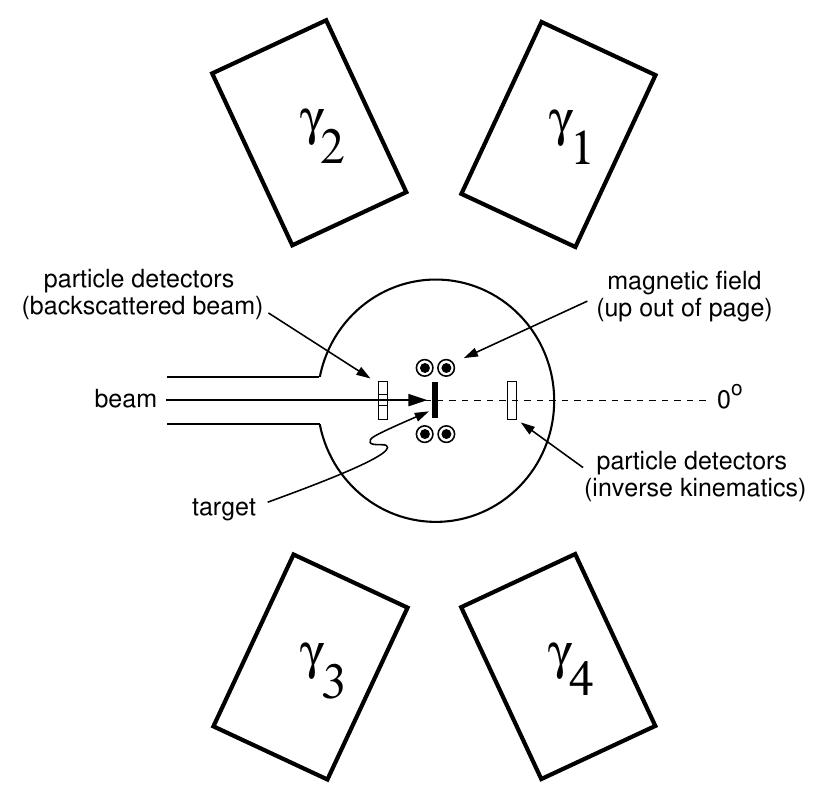}
  \caption{Plan-view sketch of the apparatus used for transient-field $g$-factor measurements. The $\gamma$-ray detectors are in the horizontal plane through the beam axis, and the magnetic field applied to the target is in the vertical plane. The figure is taken from Ref.~\cite{stuc20}.}
\label{fig:TF_det_schematic}
\end{figure}

To describe the experimental method, it is convenient to consider the case of Coulomb excitation, which has been used to excite the states of interest in the majority of the transient-field $g$-factor measurements to date. A sketch of the apparatus, including the particle and $\gamma$-ray detectors and showing the magnetic field applied to the target, is shown in Fig.~\ref{fig:TF_det_schematic}.

The `classic'  { transient-field} measurements, performed since the 1960s, used `conventional kinematics', i.e., a lower-mass beam was employed to Coulomb excite a heavier target nucleus. De-excitation $\gamma$ rays were detected in coincidence with backscattered beam ions detected in an annular counter placed around the beam axis \cite{benc80}. More recently, there has been a shift to `inverse' kinematics in which a heavier beam ion is Coulomb-excited on a lower-mass target with the knock-on target ions detected at forward angles \cite{benc07}. These `conventional' and `inverse' kinematics experiments both correspond to near head-on collisions in the center-of-mass frame. The advantages of inverse kinematics over conventional kinematics include (i) the improved sensitivity achieved by virtue of the forward-focusing of the reaction products in the laboratory frame (the particle detector therefore can cover a larger solid angle in the center-of-mass frame), and (ii) the applicability of inverse kinematics to experiments on nuclei produced as radioactive beams, which open up new regimes for nuclear structure studies \cite{kumb04,benc08,kumb12a,illa14}. A third alternative for the reaction is to excite the beam ions in glancing collisions on heavier target nuclei \cite{spei91,cub92,cub93,stuc05,davi06,stuc06,fior12}. This case has been used to measure the properties of the transient field for ions with $K$-shell electrons \cite{spei91,cub92,cub93,stuc05}, and for $g$-factor measurements on radioactive beams produced by projectile fragmentation \cite{davi06,stuc06,fior12}.

\begin{figure}[ht]
  \centering
  \includegraphics[width=0.8\linewidth]{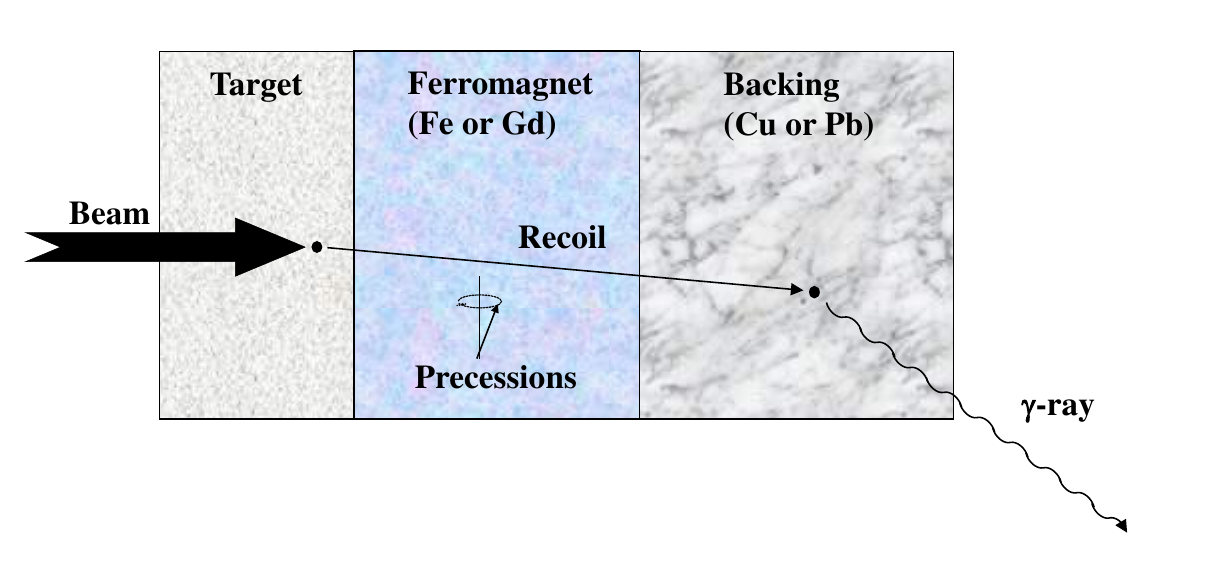}
  \caption{Representation of the triple-layered target used in conventional ‘thin-foil’ transient-field $g$-factor measurements. The target nuclei of interest are Coulomb-excited by the beam and recoil through the ferromagnetic layer, where they experience the transient field, before coming to rest in a non-magnetic backing layer (often copper or lead). The figure is taken from Ref.~\cite{stuc20}.}
\label{fig:TF_target}
\end{figure}

Essentially all transient-field measurements use a triple-layer target like that represented in Fig.~\ref{fig:TF_target}. The nuclei of interest are excited by a chosen reaction in the first layer of the target and then recoil into a polarized ferromagnetic host (usually the second layer of the target). As they move through this ferromagnetic layer, the transient field acts on the nuclei of the ions, causing the nuclear spin to precess about the direction of the magnetic field applied to the target. The ions slow and then stop in the third, nonmagnetic, layer of the target, where the nuclear decay (predominantly) takes place. The precession of the nuclear spin is observed via the perturbed particle-$\gamma$ angular correlation of the de-excitation $\gamma$ rays.

The effect of the transient field is simply to rotate the angular correlation pattern $W(\theta)$ through an angle $\Delta \theta$ about the applied field direction. For a {\em thin-foil} transient-field measurement, using a target as represented in Fig.~\ref{fig:TF_target}, the precession angle is
\begin{equation}\label{eq:dtheta}
\Delta \theta = - g \frac{\mu_{\rm N}}{\hbar} \int^{t_e}_{t_i}
B_{\rm TF}(t) e^{-t/\tau} dt,
\end{equation}
where $g$ is the nuclear $g$~factor, $\mu_{\rm N}$ is the nuclear magneton, $\hbar$ is the reduced Planck constant, and $B_{\rm TF}$ is the transient-field strength, which depends on the time $t$ through its dependence on the velocity of the ion as it slows in the ferromagnetic layer of the target. The time at which the ion enters the ferromagnetic layer is $t_i$ and the time at which it exits is $t_e$. The factor $e^{-t/\tau}$ accounts for nuclear decays that occur during transit through the ferromagnetic foil, but typically remains near unity because the excited nucleus usually traverses a thin ferromagnetic foil in a time ($ < 1$~ps) that is short compared to the mean life of the excited state of interest. (The cases noted above with $\tau \approx 100$~fs are an exception to this usual case.)

For the magnetic field in the `up' (`down') direction, the perturbed angular correlation is
\begin{equation}
\label{eq:W_up_down}
W^{\uparrow(\downarrow)}(\theta_{\gamma}) = W(\theta_{\gamma} \mp \Delta \theta) \simeq
W(\theta_{\gamma}) \mp \Delta \theta \frac{dW}{d \theta},
\end{equation}
where the negative sign applies for field up ($\uparrow$) and the positive sign applies for field down ($\downarrow$). The meaning of `up' and `down' is defined in Fig.~\ref{fig:TF_det_schematic}. The precession angle is determined experimentally by placing a pair of $\gamma$-ray detectors at angles $\pm \theta_\gamma$ with respect to the beam axis, and forming a double ratio
\begin{equation} \label{eq:rho}
\rho_\textrm{TF} =
\sqrt{\frac{N_\gamma^{\uparrow}(+\theta_\gamma)}{N_\gamma^{\downarrow}(+\theta_\gamma)}
\frac{N_\gamma^{\downarrow}(-\theta_\gamma)}{N_\gamma^{\uparrow}(-\theta_\gamma)}},
\end{equation}
where, for example, $N_\gamma^{\uparrow}(+\theta_\gamma)$ is the number of counts recorded in the detector at $+\theta_\gamma$ for field up. Two such pairs of detectors are shown in Fig.~\ref{fig:TF_det_schematic}: $\gamma_1$ and $\gamma_4$ are one pair; $\gamma_2$ and $\gamma_3$ are a second pair.

Defining
\begin{equation} \label{eq:epsilon}
\epsilon_\textrm{TF} = \frac{1 - \rho_\textrm{TF}}{1 + \rho_\textrm{TF}},
\end{equation}
the experimental precession angle is given by
\begin{equation}\label{eq:delta-theta}
\Delta \theta = \epsilon_\textrm{TF}/S,
\end{equation}
where
\begin{equation}\label{eq:S}
S = S(\theta_{\gamma}) = \left . \frac{1}{W}\frac{dW}{d\theta} \right \vert _{\theta_\gamma}.
\end{equation}
In other words, $S$ is the logarithmic derivative of the angular correlation evaluated at the $\gamma$-ray detection angle $+\theta_{\gamma}$. It is sometimes referred to as the `slope'. The `effect' $\epsilon_\textrm{TF}$ is formally equivalent to
\begin{eqnarray}\label{eq:epsilon1}
\epsilon_\textrm{TF} &=&  \frac{W^{\downarrow}(+\theta_\gamma)-W^{\uparrow}(+\theta_\gamma)}{W^{\downarrow}(+\theta_\gamma)+W^{\uparrow}(+\theta_\gamma)},
\end{eqnarray}
which is similar in form to the ratio defined in the TDPAC/D method; see Eq.~(\ref{eq:Rt_B}).

The particle-$\gamma$ angular correlation in general takes the form (see Refs.~\cite{alde75,stuc02a,stuc03}
and references therein)
\begin{equation}
W(\theta_p, \theta_\gamma, \Delta \phi) = \sum_{k q} B_{k
}^{q}(\theta_p) Q_k A_k D^{k *}_{q 0}(\Delta \phi, \theta_\gamma,
0), \label{eq:pacTF}
\end{equation}
where $(\theta_p, \phi_p)$ and $(\theta_\gamma, \phi_\gamma)$ are the spherical polar angles corresponding to particle and $\gamma$-ray detection, respectively, with the $z$-axis along the beam direction, and $\Delta \phi = \phi_\gamma - \phi_p$. The statistical tensor $B_{k}^q(\theta_p)$ defines the spin alignment of the initial state. $A_k$ represents the usual combination of $F$-coefficients for the $\gamma$-ray transition \cite{yama67}; it is defined in Eq.~(\ref{eq:angular_distribution_coeff}) and depends on the initial and final level spins and the (possibly mixed) multipolarity of the transition. $Q_k$ is the attenuation factor for the finite size of the $\gamma$-ray detector \cite{rose53} and $D^{k *}_{q 0}(\Delta \phi, \theta_\gamma, 0)$ is the Wigner rotation matrix. The majority of cases concern either $E2$ transitions or mixed $M1+E2$ transitions for which $k =0,2,4$.

In the case of azimuthal symmetry (i.e., cylindrical symmetry around the beam axis), the angular correlation reduces to the form
\begin{equation}
W(\theta_\gamma) = 1 + \sum_{k =2,4} a_k P_k(\cos \theta_\gamma), \label{eq:ac-sym}
\end{equation}
where $a_k = B_{k} Q_k A_k$, and $P_k$ is a Legendre polynomial (see Eq.~(\ref{eq:w_distribution_axial})).

A typical transient-field $g$-factor measurement requires a relatively long run with a pair (or pairs) of $\gamma$-ray detectors at angles $\pm \theta_\gamma$ (and $180^\circ \pm \theta_\gamma$) to determine the `effect' $\epsilon_\textrm{TF}$, along with a somewhat shorter run to determine the `slope' $S(\theta_{\gamma})$. For `safe' Coulomb excitation measurements \cite{CLINE1986}, and with the excited nuclei implanted into a medium that does not perturb the spin alignment (typically a cubic metal), the angular correlation, and hence $S(\theta_{\gamma})$, can be calculated precisely; for examples, see Refs.
\cite{bolo83,stuc85,stuc88,lamp89,stuc91,stuc91a,stuc92,stuc94,lamp94,ande95a,stuc98,robi99,stuc00,beza00,mant01,east09,cham09,cham11}. One caveat on this statement is that $S(\theta_{\gamma})$ is a very sensitive function of $\theta_{\gamma}$ when the detectors are placed near the maximum slope of the angular correlation, as is done for this type of precession measurement. In such cases, if there is an offset in the $\gamma$-ray detector angle from its nominal value, the calculated slope will be inaccurate. Such inaccuracies largely factor out if the experiment is designed to measure the relative precessions of two or more states simultaneously under the same experimental conditions; however, the detector angle must be known accurately to calculate $S(\theta_{\gamma})$ reliably in cases where the absolute magnitude of the precession angle is required. It can be checked via angular correlation measurements, as was done in Refs.
\cite{bolo83,stuc85,stuc88,lamp89,stuc91,stuc91a,stuc92,stuc94,lamp94,ande95a,stuc98,robi99,stuc00,beza00,mant01,east09,cham09,cham11}. Another exception is for longer-lived states of paramagnetic ions, often rare earth ions, for which hyperfine interactions can cause a reduction in the anisotropy relative to the calculated angular correlation. An example is the $2_1^+ \rightarrow 0_1^+$ transition in $^{152}$Sm reported by Byrne et al. \cite{byrn87}.

Some  {authors} have preferred not to measure the full angular correlation, but rather to deduce the required slope from a more limited number of measurements. The $\gamma$-ray detector angles are shifted, and the slope is determined from the change in $\gamma$-ray intensity, without measuring the whole angular correlation. The shift in detector angle can be small to determine $\frac{dW}{d \theta} \approx \frac{\Delta W}{\Delta \theta}$. Alternatively, Ernst et al. \cite{erns00} describe a more elaborate procedure to determine the slope from measured double ratios similar to Eq.~(\ref{eq:rho}), but with field-up and field-down data added, and the detectors in the forward quadrant first at $+50^{\circ}$ and $-80^{\circ}$ and then reversed to $+80^{\circ}$ and $-50^{\circ}$. This analysis yields a single parameter that accounts for the finite size of the particle detector such that the full angular correlation function in the form of Eq.~(\ref{eq:ac-sym}) can be evaluated. Numerical calculations, like those in Refs.\cite{bolo83,stuc85,stuc88,lamp89,stuc91,stuc91a,stuc92,stuc94,lamp94,ande95a,stuc98,robi99,stuc00,beza00,mant01,east09,cham09,cham11}, verify this method.

In many measurements, several excited states of the nucleus may be populated, and the transient-field precessions that are observed for the lower states are affected by the feeding from higher states that feed into them. The formulation to correct for these feeding effects was first discussed by H\"ausser et al. \cite{haus80}. As procedures for feeding corrections have been developed and reported in many subsequent works (e.g., \cite{stuc85,byrn87,robi99,stuc07}), no detail will be given here.

Once the experimental precession angle $\Delta \theta$ has been obtained from Eq.~(\ref{eq:delta-theta}), the experimental $g$~factor can be obtained from Eq.~(\ref{eq:dtheta}), provided that the transient-field strength is known under the conditions of the experiment. Unfortunately, the physical processes that give rise to the transient field are too complex and too poorly understood to allow the calculation of its strength from first principles; the transient field has to be calibrated based on independently known $g$~factors. Herein lies a problem that is not entirely solved: Generally, for nuclei with $ 8 \leq Z \leq 12$ and $42 \lesssim Z \lesssim 82$, there are reliable independent calibration $g$~factors to provide a reference for the transient-field strength. However, for the intermediate region $12 \lesssim Z \lesssim 42$, independently determined, precise, and accurate calibration $g$~factors are not available.

\begin{figure}[ht]
  \centering
  \includegraphics[width=0.85\linewidth]{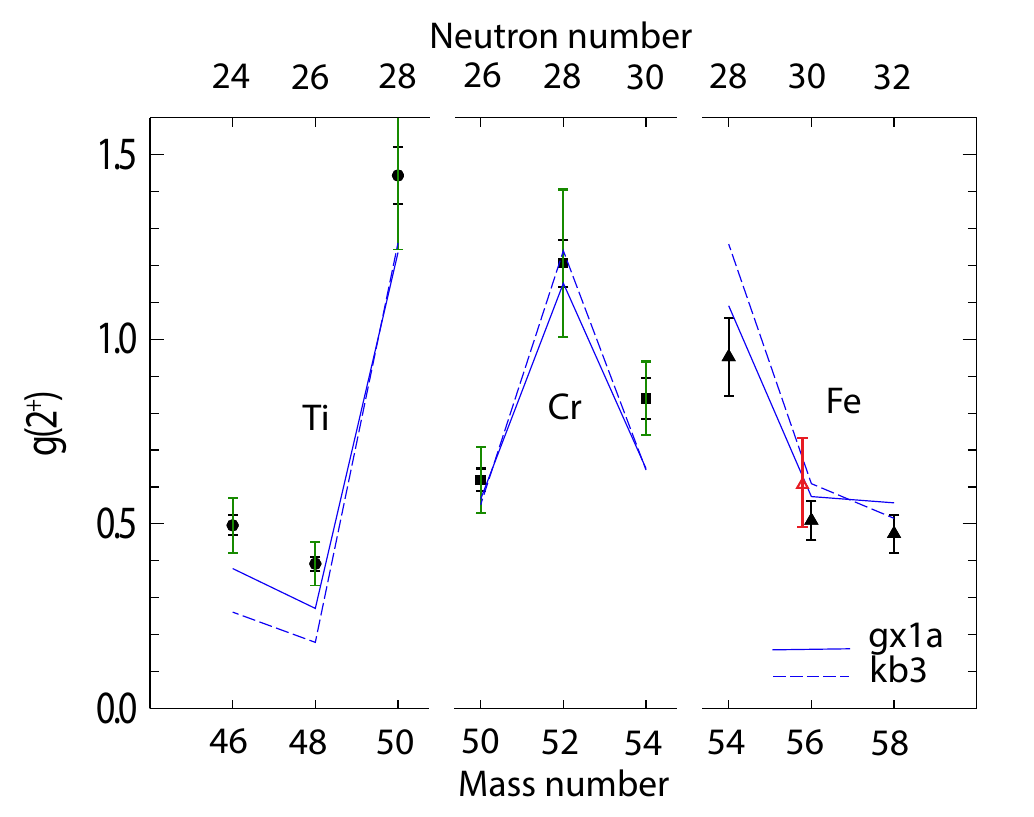}
  \caption{Experimental $g$~factors of the 2$^+_1$ states in the stable isotopes of Ti, Cr, and Fe \cite{erns00,erns00a,spei00,wagn01,east09,east09a,ston20} compared with shell model calculations in the $fp$ shell. Calculations for two interactions are shown: \texttt{gx1a} \cite{honm04,honm05} and \texttt{kb3} \cite{pove01}. Note that the full name for \texttt{gx1a} is \texttt{GXPF1A}. The smaller black error bars are those originally published, whereas the larger green uncertainties from the table of recommended values \cite{ston20} are intended to reflect uncertainties in the strength of the transient field. The $g$~factors of the Fe isotopes \cite{east09,east09a} were measured relative to the independently known $g$~factor of the 136-keV 5/2$^-$ state in $^{57}$Fe \cite{fahl79}. Also shown by the open red triangle is the $g$~factor of $^{56}$Fe that was used in the global fits aiming to parametrize the transient-field strength \cite{eber75,eber77,eber79,shu80}.
  }\label{fig:fp-gfactors}
\end{figure}

The calibration problem is illustrated in Fig.~\ref{fig:fp-gfactors} wherein the experimental $g$~factors of the 2$^+_1$ states in the stables isotopes of Ti, Cr and Fe
\cite{erns00,erns00a,spei00,wagn01,east09,east09a,ston20}
are compared with large-basis shell model calculations. For the Ti and Cr isotopes, two sets of error estimates are shown. The smaller uncertainties (black) indicate the $g$~factors as reported, whereas the larger (green) uncertainties include the additional uncertainty due to uncertainties in the transient-field calibration that is added in the most recent compilation of recommended values by Stone \cite{ston20}. For the Fe isotopes, the transient-field data (black filled triangles) are normalized to the $g(2^+_1)$ value of $^{56}$Fe obtained by East et al. \cite{east09} relative to the independently determined and precisely known \cite{fahl79} $g$~factor of the 5/2$^-$, 136-keV state in $^{57}$Fe. Unfortunately, the precision of the measurement of East et al. was hampered by the long lifetime of the 136~keV state in $^{57}$Fe ($\tau = 12.6$~ns), which underwent a significant degree of precession in the externally applied polarizing field relative to the transient-field contribution. It proved impossible to isolate the transient-field precession precisely. Nevertheless, an improved value for the $g(2^+_1)$ value in $^{56}$Fe was obtained. Also shown in Fig.~\ref{fig:fp-gfactors} (red open triangle) is the historic value for $^{56}$Fe, originally used to calibrate the transient field in this region \cite{eber75,eber77,shu80}.

\subsubsection{Parametrization of the transient field }
\label{sect:TF-param}

The evidence accumulated over the past few decades suggests that there is no simple universal empirical parametrization of the transient-field strength. Nevertheless, from the beginning, there have been efforts to find a global parametrization in terms of the ion velocity and atomic number.

 {
\paragraph{Linear-velocity parametrization}
}

\begin{figure}[ht]
  \centering
  \includegraphics[width=0.95\linewidth]{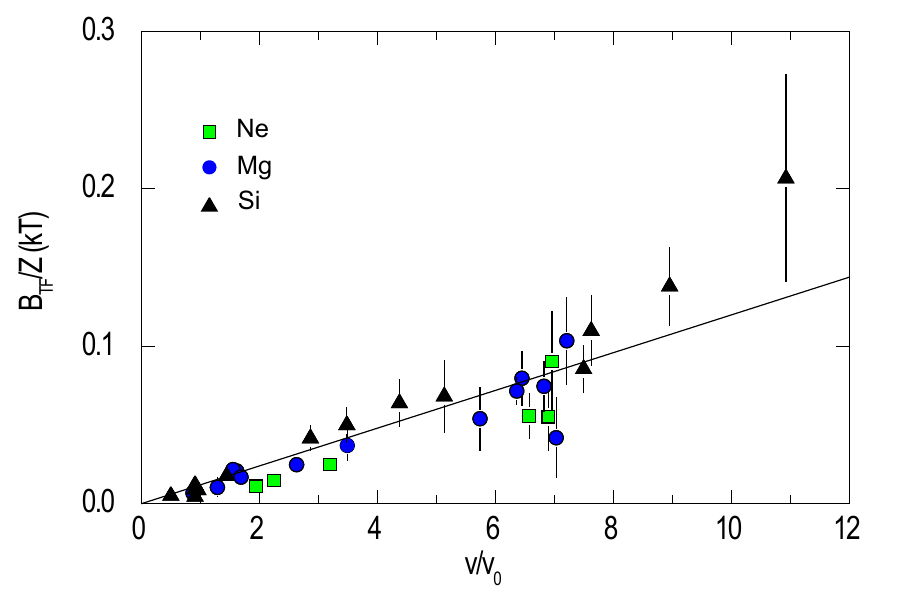}
  \caption{Experimental transient-field strengths for Ne, Mg, and Si ions in iron compared with the linear velocity-dependent parametrization of Eq.~(\ref{eq:lin-param}). The data are from \cite{eber74,eber75a,eber77,zalm78,dera79,spei80,spei81,spei83,bazz86,spei89,spei94}. The adopted $g(2^+_1)$ factors for $^{20}$Ne and $^{24}$Mg are 0.54(4) and 0.538(13), respectively. For $^{28}$Si, no experimental value is available, so the USDB \cite{brow06} shell model value, $g=0.541$, is used. To show the velocity dependence of the transient field more clearly, the data with a thick iron foil in which the recoiling ions stop have been used twice; the observed values are plotted, and also differences that reveal the field strengths at higher velocity. Specifically, the precession measured with a lower initial velocity is subtracted from that measured with a higher initial velocity to reveal the transient-field strength corresponding to the velocity range between the two initial velocities. Note that no fitting of data has been performed in the preparation of this figure: it aims to give a visual comparison of the data with the proposed parametrization, taking $a=12$ tesla.
\label{fig:Btf-low-Z}}
\end{figure}

Along with their discovery that the transient field increases with ion velocity,
Eberhardt et al. \cite{eber77,eber79} introduced the parametrization
\begin{equation} \label{eq:lin-param}
B_{\rm TF}(v,Z)=a Z R (v/v_0),
\end{equation}
where $a$ was an adjustable parameter and $R=1+(Z/84)^{2.5}$ is a relativistic factor, which is near unity for light nuclei. Subsequent work usually omitted it. The velocity is given relative to the Bohr velocity, $v_0=c/137$, where $c$ is the speed of light. This parametrization gives a good description of transient-field strengths for Ne, Mg, and Si ions traversing iron hosts with velocities up to the $K$-shell electron velocity ($Zv_0$), as illustrated in Fig.~\ref{fig:Btf-low-Z}.

The fit performed by Eberhardt et al. \cite{eber77,eber79} included cases with $Z$ from across the nuclear chart, but in most cases the ion velocity was relatively low, $v \lesssim 2v_0$ for the higher-$Z$ ions, and, in addition,  frequently the probe ion stopped in the ferromagnetic host, requiring the subtraction of a static-field contribution, a procedure that may not be reliable \cite{ande95,stuc99}. Moreover, the case of Pt in iron was included, which subsequently proved to be a special case \cite{stuc81,stuc83,stuc94}.

Due to a lack of data, it is not clear how high in $Z$ and ion velocity this parametrization, in which $B_{TF} \propto v$ up to $v = Z v_0$ applies; $Z < 20$ seems to be a reasonable limit. Certainly, departures from the parametrization of Eq.~(\ref{eq:lin-param}) are found for higher-$Z$ ions and higher velocities (well below $Zv_0$). The measurements showing this departure employed the thin-foil technique (i.e., three-layer targets as in Fig.~\ref{fig:TF_target}) so that possible ambiguities associated with a contribution from the static hyperfine field are avoided. The behavior of the transient field at $v=Zv_0$ as a function of the atomic number $Z$ is discussed below in Sect. \ref{sect:TF-physics}.

The Bonn group has ``maintained" the linear-velocity parametrization, most often applying it for nuclei with $Z \lesssim 30$, and recommending $a \approx 12$~T for iron hosts and $a \approx 17$~T for gadolinium, with $R=1$ \cite{spei91a}.

In cases where a heavy beam impinges on the target, the Bonn group introduces a multiplicative correction factor, $G_{\rm beam}$, which is unity for the lightest beams ($Z \lesssim 16$) but decreases with increasing mass. It is apparently related to the energy loss ($dE/dx$) of the beam ion in the ferromagnetic host, and was attributed to a dynamic demagnetization of the host induced by the beam.  An empirical formula for $G_{\rm beam}$, for $Z_{\rm beam} \gtrsim 16$, has been proposed  \cite{spei89}:
\begin{equation}\label{eq:Gbeam}
G_{\rm beam} = 1 - \alpha \frac{1 - \exp(-\lambda_r t_{\rm eff})}{\lambda_r t_{\rm eff}},
\end{equation}
where  $t_{\rm eff}$ is the effective interaction time of the beam in the ferromagnetic host (determined by the $dE/dx$, reaction kinematics and target properties). The parameters $\alpha$ and $\lambda_r$ have the values $\alpha=1.0(1)$, $\lambda_r^{-1} =0.4(1)$~ps for iron, and $\alpha=0.5(1)$, $\lambda_r^{-1} \gtrsim 1.0$~ps for gadolinium. It is not very clear from the papers of the Bonn group reporting $g$-factor measurements exactly how they determine $G_{\rm beam}$ in each case, but the adopted value is always given with an assigned uncertainty, so the transient-field parametrization and its uncertainty are made clear. Whether or not one agrees with the Bonn group's interpretation of the physical mechanism that determines $G_{\rm beam}$, their procedure to calibrate the transient-field strength for their $g$-factor measurements is reasonable. The weakness, which applies to all empirical parametrizations, is the quality and consistency of the calibration data.\\

To sum up: a current assessment of the proposed linear velocity dependence of the transient field is that it is a good approximation for low-$Z$ ions ($10\lesssim~Z~\lesssim20$) and for ion velocities up to about the $K$-shell electron velocity ($Zv_0$), as illustrated in Fig.~\ref{fig:Btf-low-Z}.\\

 {
\paragraph{Rutgers parametrization}
}
The Rutgers group generalized the form of Eq.~(\ref{eq:lin-param}) and included a specific dependence on the host magnetization \cite{shu80}:
\begin{equation}\label{eq:RUparam}
B_{\rm TF}(v,Z) = a (v/v_0)^{p_v} Z^{p_Z} \mu_B N_p,
\end{equation}
where $a$, $p_v$, and $p_Z$ are dimensionless parameters, and $\mu_B N_p$ is the host magnetization in tesla (expressed as the product of the Bohr magneton and the volume density of polarized electrons in the ferromagnetic host; there is an implicit scaling by the permeability of vacuum, $\mu_0$, so that the magnetization is expressed in tesla).

A fit to data for ions ranging from O ($Z=8$) to Sm ($Z=62$), mainly in iron hosts, but including some limited data (for $^{82}$Se; $Z=34$) on gadolinium hosts,  yielded $a=96.7 \pm 1.6$, $p_v = 0.45\pm 0.18$, and $p_Z = 1.1 \pm 0.2$. The uncertainties given on these parameters appear to be correlated, so a quantitative estimate of the uncertainty in the transient-field strength in any particular case is not easily determined; this difficulty may explain why measured $g$~factors evaluated with the Rutgers parametrization are often given without an assigned uncertainty in the transient-field strength.

The main difference between this parametrization and that of Eq.~(\ref{eq:lin-param}) is that the velocity dependence is now nearer to the square root than linear. With the benefit of hindsight, this difference comes about because the Rutgers parametrization includes fits to additional transient-field data on heavier nuclei and traversing the ferromagnetic hosts with higher velocities. Also, data on Pt ($Z=78$), which showed a clear linear velocity dependence \cite{haus79a,kali80,stuc81}, were excluded from the Rutgers fit. The reason for this different behavior of Pt to somewhat lighter ions was identified later \cite{stuc83}.

It can be noted that for nuclei up to the $A \sim 80$ region and ion velocities up to about 5$v_0$, the Rutgers parametrization and the linear-velocity parametrization with the attenuation factor $G_{\rm beam}$ included agree quite well.

\begin{figure}[ht]
  \centering
  \includegraphics[width=0.8\linewidth]{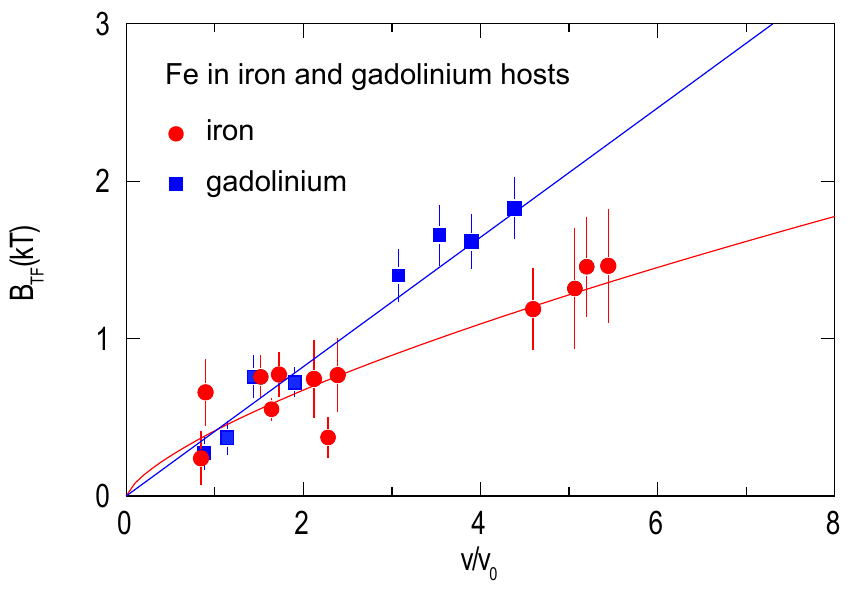}
  \caption{Comparison of transient-field strengths for $^{56}$Fe and $^{54}$Fe in iron and gadolinium hosts. The experimental data are compiled (see text) from \cite{bren77,hass76,hubl74,shu80,stuc87} for iron and \cite{fahl77,fahl78,fahl78a,erns00,east09a} for gadolinium. In this plot, the relative transient-field strengths are independent of the absolute magnitudes of the $g$ factors.
  }
\label{fig:Btf_Fe-Gd}
\end{figure}

The Rutgers parametrization assumes that the transient fields for iron and gadolinium hosts are the same, apart from the difference in the host magnetization. This assumption seems to be valid in some cases (see e.g. \cite{beza00}), but it cannot be relied upon. For example, Chamoli et al. \cite{cham11} observed that the transient-field strength for $^{108}$Pd in gadolinium is a factor of 1.4 times stronger than predicted by the Rutgers parametrization. As another example, Fig.~\ref{fig:Btf_Fe-Gd} compares the transient-field strengths for $^{54}$Fe and $^{56}$Fe in iron and gadolinium hosts. The experimental data are compiled from \cite{bren77,hass76,hubl74,shu80,stuc87} for iron and \cite{fahl77,fahl78,fahl78a,erns00,east09a} for gadolinium. Much of the data displayed for iron hosts, particularly at low ion velocities, is from thick-foil measurements in which the ions stop in the ferromagnetic host and experience the static field as well as the transient field. As described in relation to Fig.~\ref{fig:Btf-low-Z}, these data for thick iron hosts have been subtracted in several combinations to yield equivalent thin-foil data, thus removing the static-field contribution. The transient-field strength for Fe in gadolinium is clearly greater than that for Fe in iron for ion velocities around 4$v_0$. The situation around 2$v_0$ is less clear because there are discrepant low-velocity data for iron hosts. Moreover, caution is required concerning the lower-velocity points for gadolinium hosts (those below 2$v_0$), which are taken from the thick-foil measurements of Fahlander et al. \cite{fahl77,fahl78,fahl78a} assuming that the static field is $B_{\rm SF}=0\pm 1.6$~T. Thus, the comparison at the lower velocities may have some bias. Nevertheless, the data strongly suggest that the transient fields for Fe ions in iron and gadolinium hosts differ by more than the 5\% to 10\% that would be the case if they scaled only with the bulk magnetization of the ferromagnetic host.

 {
\paragraph{Chalk River parametrization}
}

The Chalk River group proposed parametrizations of the form
\begin{equation}\label{eq:CRparam}
B_{\rm TF}(v,Z)= a Z (v/v_0) \exp(-\beta (v/v_0)),
\end{equation}
where $a$ and $\beta$ are parameters. These parameters were allowed to vary with the ferromagnetic host and with the atomic-number regime of the moving ion. The interest was particularly in high-spin spectroscopy of rare-earth and heavier nuclei, so fits were performed for iron \cite{andr82} and gadolinium  \cite{haus83} hosts in the rare-earth region, and for gadolinium hosts at $Z=82$ \cite{haus84}.

The parameters obtained were $a=19.0 \pm 0.5$~T for $\beta = 0.12$ (iron hosts, rare earth ions, mainly Tm ($Z=69$)); $a=29.0 \pm 1.8$~T for $\beta = 0.135$ (gadolinium hosts assuming a magnetization of 6.2 $\mu_B$ per atom, rare earth ions, mainly Tm ($Z=69$)); and $a=28.0 \pm 2.6$~T for $\beta = 0.135$ for $^{207}$Pb ($Z=82$) in gadolinium. Thus, within uncertainties,  the parametrization for Pb ($Z=82$) is similar to that obtained for Tm ($Z=69$) in gadolinium hosts. Note that $a$ and $\beta$ are highly correlated in fits to the data. The procedure adopted was to set $\beta$ to a fixed ``best'' value and then determine $a$ and its uncertainty. This approach provides a reasonable and simple means to give an estimate of the uncertainty on the transient-field strength without performing a detailed analysis of the uncertainties on the correlated parameters.

The Chalk River parametrization was determined for the rare earth region and heavier nuclei, but it has sometimes been used well away from its region of calibration, such as in the $A \sim 80$ region \cite{zhus00}. While the use of a parametrization outside the regime in which it was derived is not recommended, the Chalk River and Rutgers parametrizations happen to give similar field strengths in the $A \sim 80$ region.

The above parametrizations were based on limited data and have not been updated since the 1980s, which constitutes a potential weakness of transient-field measurements that rely on one of these empirical parametrizations to set the absolute magnitudes of the $g$~factors.

\begin{figure}[ht]
  \centering
  \includegraphics[width=0.55\linewidth]{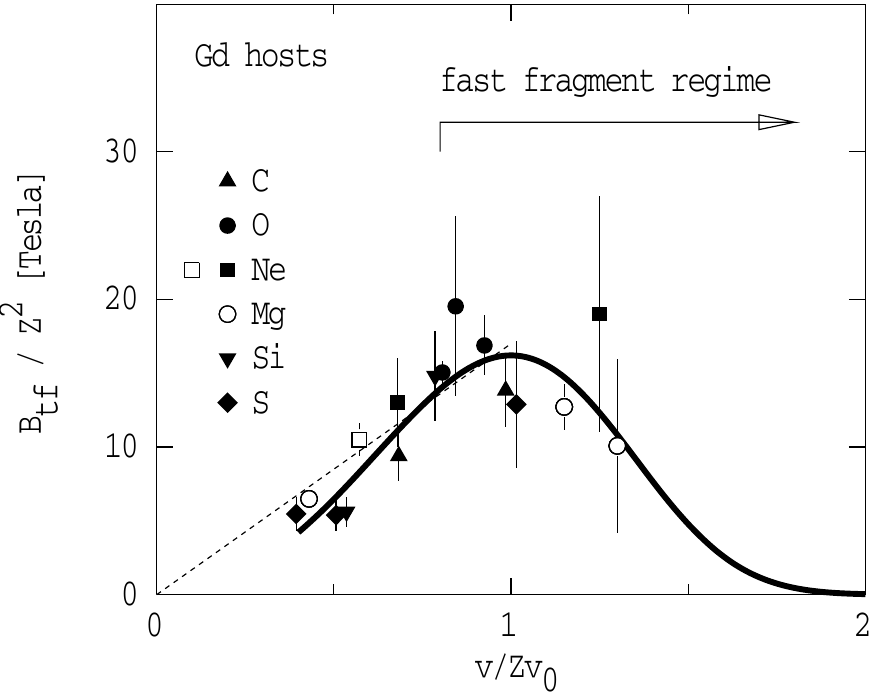}
  \caption{Transient-field strengths for high-velocity light ions, $6 \leq Z \leq 16$, in gadolinium hosts. The bold line represents a parametrization of the field strength for $v \gtrsim Zv_0/2$:  $B_{\rm TF}(v,Z,{\rm Gd} ) = 26.7 Z^2 (v/Zv_0)^2 {\rm e}^{- \frac{1}{2}(v/Zv_0)^4}$ T. The dotted line shows the linear parametrization applicable for $v \lesssim Zv_0$: $B_{\rm TF}(v,Z,{\rm Gd} )  = 16  Z (v/v_0)$ T. The figure is taken from Ref.~\cite{stuc05a}.}
\label{fig:HVTF-param}
\end{figure}

 {
\paragraph{High-velocity parametrization}
}
More recently, to enable transient-field $g$-factor measurements on radioactive beams produced by projectile fragmentation \cite{davi06,stuc06}, a model-based transient-field parametrization was proposed for light ions at high velocity \cite{stuc04}:
\begin{equation}
B_{\rm TF}(v,Z) = A Z^P (v/Zv_0)^2 {\rm e}^{-
\frac{1}{2}(v/Zv_0)^4}, \protect \label{eq:aes-param}
\end{equation}
where fitting the available data
\cite{spei81,brig84,spei86,bazz86,trol86,spei87,simo87,reut89,spei89,spei91,cub93},
for gadolinium hosts gave $A=26.7 \pm 1.1$~T for $P=2$, and for iron hosts \cite{zalm79,dybd79,beck81,spei81,trol86,spei86,spei83,bazz86,spei87,simo87,spei89,spei94}
 $A=1.82 \pm 0.05$ T for $P=3$. The parametrization for gadolinium hosts is compared with data in Fig.~\ref{fig:HVTF-param}.

 {
\paragraph{Summary}
}

 {
In summary, this section has provided a historical survey and critical appraisal of attempts to parametrize the transient-field strength in terms of the ion velocity and atomic number. The evidence is that there is no simple universal empirical parametrization.
}

 {
A current assessment of the originally proposed \cite{eber77,eber79} linear velocity dependence of the transient field is that it is a good approximation for low-$Z$ ions ($10\lesssim~Z~\lesssim20$) and for ion velocities up to about the $K$-shell electron velocity ($Zv_0$), as illustrated in Fig.~\ref{fig:Btf-low-Z}.
}

 {
The Rutgers parametrization \cite{shu80} was based on a fit to data for ions ranging from O ($Z = 8$) to Sm ($Z = 62$), mainly in iron hosts. Despite the inclusion of data on $Z=8$, it should not be used for low-$Z$ ions, say $Z \lesssim 20$. It has been used to calibrate many data on ions with $Z \geq 20$ since the year 2000. See further discussion in section~\ref{sect:TF-gfactors-Z.le.50}.
}

 {
The Chalk River group were focused on rare earth and heavier nuclei \cite{andr82,haus83,haus84}.  For such nuclei, their parametrizations for iron and gadolinium hosts are recommended; however, in such cases, the Chalk River and Rutgers parametrizations often agree.
}

 {
As noted, the above parametrizations were based on limited data and have not been updated since the 1980s. The region where the calibration of the transient field is most uncertain is between $Z=12$ and $Z=46$, where suitable independent measurements of $g$ factors are not available. By happenstance, the linear (with beam induced attenuation factors \cite{spei89}), Rutgers, and Chalk River parametrizations tend to agree for nuclei near $A=80$. Of course, this accidental agreement does not mean that the field calibration is correct. Current effort is directed towards obtaining an independent $g$-factor measurement in this region to solve the transient-field calibration problem. Specifically, Time Dependent Recoil in Vacuum on Na-like $^{76}$Ge ions is being pursued \cite{mcco21}. See also the discussion in sections \ref{sect:TF_prospects}, \ref{sect:RIV-TDRIV}, and \ref{sect:TF-gfactors-Z.le.50}. \\
}

\subsubsection{Physical origins of the transient field }
\label{sect:TF-physics}

The transient field arises from polarized electrons bound to the moving ion. Somehow, the polarization of the ferromagnetic host is transferred to unpaired electrons bound to the moving ion. As the largest hyperfine fields produced at the nucleus by atomic electrons are those in $s$ orbitals, the transient field may be expressed in terms of the contact fields present at the nucleus from unpaired electrons in $s$ orbitals. The model proposed by Eberhardt et al. \cite{eber77} writes the transient-field strength as:
\begin{equation}\label{eq:BTF_model}
B_{\rm TF} = \sum_{n} \xi_{ns}(v,Z)F^1_{ns}(v,Z)B_{ns}(Z),
\end{equation}
where $B_{ns}$ is the contact field from a single electron in the $n^{th}$ $s$ orbit, $F^1_{ns}$ is the fraction of ions carrying a single electron in the $ns$ orbital, and $\xi_{ns}(v,Z)$ is the degree of polarization of these vacancies. Note that $n$ is the principal quantum number ($n=1$ for the $K$~shell, $n=2$ for the $L$~shell, etc.).

The most detailed and assumption-free modeling of the transient-field strength is that reviewed by Rud and Dybdal \cite{rud86}, wherein the transient-field for light ions ($ 6 \leq Z \leq 16 $) is modeled based on measured $K$-vacancy fractions, i.e., measured $F^1_{1s}$ values as a function of ion velocity \cite{dybd87}. For $K$-shell (1$s$) electrons, the contact fields can be evaluated quite accurately by assuming the ions are hydrogen-like, in which case
\begin{equation}\label{eq:B_ns}
B_{ns}=16.7 (Z/n)^3 R(Z),
\end{equation}
in tesla units, where $R(Z)$ is the relativistic correction factor as in Eq.~(\ref{eq:lin-param}). Rud and Dybdal found that observed transient fields for these light ions in iron hosts could be explained by assuming a constant value of  $\xi_{ns} = 2.2/16 \simeq 0.14$, where the fraction represents the average number of polarized electrons in the outer-shell $3d$ and $4s$ orbits of iron.
The model of Eq.~(\ref{eq:BTF_model}) has also been used semi-quantitatively for heavier ions \cite{haus84,stuc94}.

By definition, the single vacancy fraction for an $s$ orbit has a maximum value of $F^1_{ns}=0.5$. It is generally accepted that this maximum value of $F^1_{ns}$ occurs when the ion velocity $v$ matches the $ns$ electron velocity, which for $K$-shell electrons is $v=Zv_0$.
This property was invoked to determine the high-velocity parametrization of the transient field given in Eq.~(\ref{eq:aes-param}) and Fig.~\ref{fig:HVTF-param}, along with the assumption that the velocity dependence of $F^1_{ns}$ follows the shape of the hydrogen-like charge distribution. The polarization degree $\xi(v,Z)$ was parametrized and determined by a fit to data.

By measuring the transient-field strength at the $K$-shell electron velocity for several ions, the Bonn group has determined the polarization degree of $K$-shell vacancies at the $K$-shell electron velocity for ions up to $Z =24$, mainly in gadolinium hosts
\cite{spei81,spei86,spei87,simo87,simo88,spei91,cub93,grab97,grab97a}.

\begin{figure}[ht]
  \centering
  \includegraphics[width=0.6\linewidth]{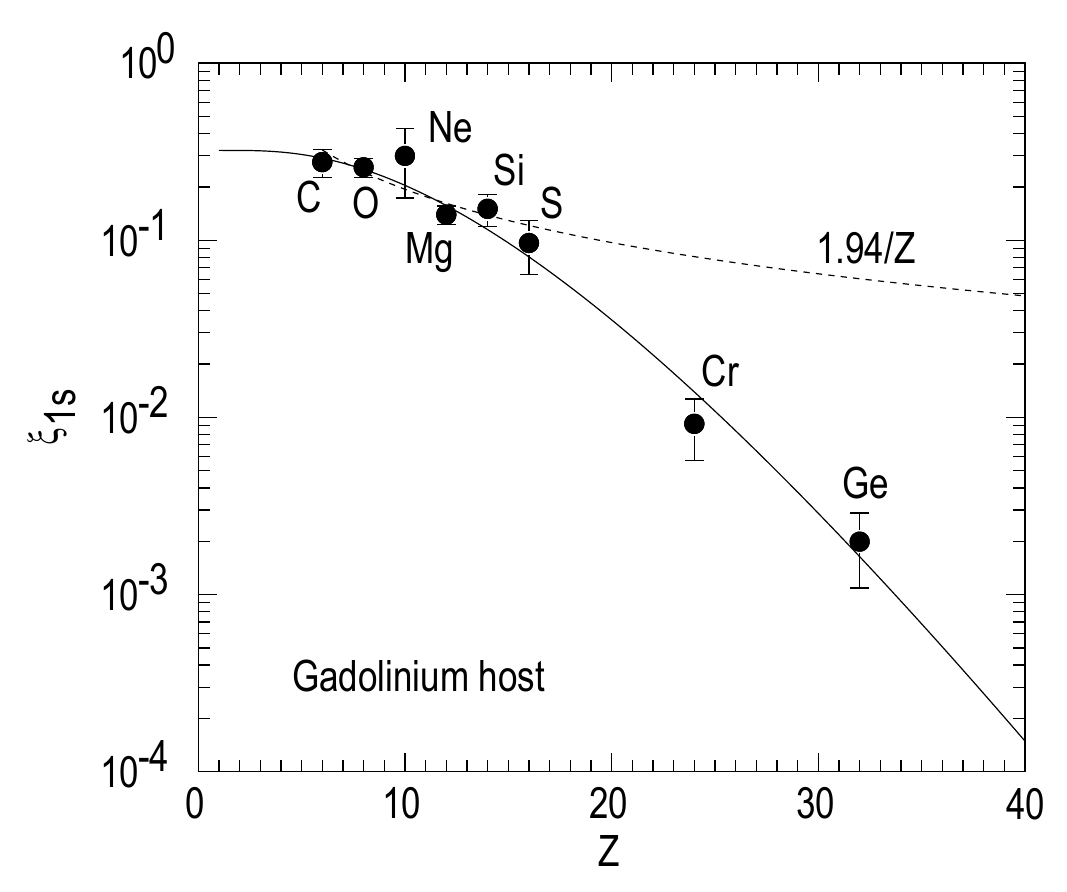}
  \caption{ $K$-shell polarization in transient-field measurements on ions traversing magnetized gadolinium hosts with velocities near the $K$-shell electron velocity, $Zv_0$. The dashed line shows the $1/Z$ dependence, which was used to describe the low-$Z$ data in Fig.~\ref{fig:HVTF-param}. The solid curve is the fit to Eq.~(\ref{eq:p1s-param}). The figure is taken from Ref.~\cite{fior12}.}
\label{fig:TF-p1s}
\end{figure}

An additional data point for Ge ($Z=32$) was obtained by Fiori et al. \cite{fior12}. The combined data are plotted in Fig.~\ref{fig:TF-p1s}. There is a clear decrease in the polarization degree at $Zv_0$ as $Z$ increases.

The solid curve in Fig.~\ref{fig:TF-p1s} is an empirical fit to the function
\begin{equation} \label{eq:p1s-param}
\xi_{1s} = a \exp[-(\ln Z)^6/b],
\end{equation}
where the best fit parameters are $a=0.32(3)$ and $b=328(25)$. At face value, this procedure assumes that the average value of $\xi_{1s}$ near $Zv_0$ does not depend strongly on velocity, but varies with $Z$. However, because the cases considered are constrained by the requirement that $v \sim Zv_0$, the changes in $\xi_{1s}$ required to fit the data can equivalently be considered to represent a dependence of $\xi_{1s}$ on ion velocity. Whichever interpretation is adopted, these data indicate that high-velocity $v \sim Zv_0$ transient-field measurements are not possible for heavy ions, $Z \gtrsim 30$, which limits the application of the transient-field technique at facilities that use projectile fragmentation to produce radioactive beams. This point is discussed further in section \ref{sect:TF-gfactors-Z.le.50}, in relation to measurements on radioactive beams of $^{72}$Zn.

 {
\subsubsection{Doppler broadened line shape lifetimes from transient-field \textit{g}-factor measurements}
\label{sect:TF-DBLS}
}

 {
Due to the small precession angles, the transient-field method requires measurements with high statistics, and often the $\gamma$-ray spectra show clear Doppler broadened line shapes that may be analyzed to determine the lifetimes of excited nuclear states. Given that the time taken by the ions under study to stop in the target material is of the order of a picosecond, this  Doppler Broadened Line Shape (DBLS) approach is relevant for lifetimes up to a few picoseconds \cite{schw68}. The extracted lifetime depends on the stopping powers adopted to evaluate the slowing of the ion in the multi-layer target. The Melbourne-ANU collaboration developed a computer code to analyze these data during the late 1980s \cite{dora88} and reported lifetime measurements in $^{155}$Gd \cite{stuc98} and $^{197}$Au \cite{stuc88}. The standard procedure in these cases was to check and adjust the stopping powers, if necessary, to reproduce a known lifetime included in the measurement.
}

 {
Starting around the year 2000, there have been a few measurements of lifetimes based on the application of the DBLS method to data taken during a transient-field $g$-factor measurement \cite{kenn01,jung11,benc16}. Unfortunately, some of those have subsequently been shown to be incorrect by precision Coulomb excitation studies \cite{allm14,allm15,gray22}. In the case of the Ni isotopes, apart from $^{62}$Ni, the results of Kenn et al. \cite{kenn01} do not agree with either the trend or magnitude of other measurements (see \cite{allm14} and references therein). In the case of $^{106}$Cd \cite{benc16}, the lifetime of the 2$^+_1$ state from the transient-field measurement is about 30\% shorter than other measurements, while the simultaneously measured lifetime of the 4$^+_1$ state is almost twice as long as expected.
}

 {
The reason for the discrepancies is not clear, but the following observations must be considered: (i) The targets employed for transient-field measurements consist of multiple layers; are the layer thicknesses well known? (ii) How well are the stopping powers known for each of the layers in the targets?  (iii) If target layers separate, would this be clear in the line shape, and could it lead to a false lifetime analysis? (iv) The discrepant measurements all used a modified version of the LINESHAPE code of Johnson and Wells \cite{well91}, which was originally written for the analysis of high-spin states populated by heavy ion fusion-evaporation reactions and did not account for angular variation in the $\gamma$-ray intensity. Is the reaction kinematics for Coulomb excitation included appropriately? How do angular correlation effects change the line shape?
}

 {
The $\gamma$-ray intensity can change significantly across the face of the detector in transient-field precession measurements because the detectors are placed near the maximum slope of the angular correlation. The typical angular correlation for a $2^+_1 \rightarrow 0^+_1$ transition following Coulomb excitation increases the portion of the line shape associated with higher $v/c$, which would tend to reduce the extracted lifetime or increase the implied $B(E2)$. This effect might contribute to the reduced lifetime obtained for the 2$^+_1$ state in $^{106}$Cd, but the majority of the discrepant DBLS data correspond to lifetimes that are too long.
}

 {
There has been much discussion \cite{allm15,spie18,kund19,kund21,beus24} of the lifetime measurements of Jungclaus et al. \cite{jung11} on the $^{112-124}$Sn isotopes, based on data taken during transient-field $g$-factor measurements at GSI \cite{walk11} and ANU \cite{east08}. In that case, the $B(E2)$ values are on average $84\pm2$\% of those obtained in the precise Coulomb excitation study of Allmond et al. \cite{allm15}. The {\em trend} in $B(E2)$ values agrees well with the Coulomb excitation measurements, despite the combination of data sets from two laboratories, one using inverse kinematics ($^{112,114,116,122}$Sn beams on a gadolinium-backed carbon target) and the other using conventional kinematics ($^{58}$Ni beams on an iron-backed natural Sn target). Although the reason for the 84\% factor cannot be identified with certainty, an increase in the stopping powers by 15-20\% would account for the discrepancy.
}

 {
To summarize, level lifetimes on the order of picoseconds can be deduced from the Doppler-broadened line shapes observed in transient-field $g$-factor measurements. However, care must be taken to assess all possible sources of systematic error. In particular, stopping powers can depend on the method of target preparation and should be calibrated either by direct measurement (cf. \cite{allm14a,stuc96a}) or by reference to an independently known lifetime present in the data set.
}\\

\subsubsection{Prospects and challenges}
\label{sect:TF_prospects}

A most pressing challenge for transient-field $g$-factor measurements is to obtain a robust calibration of the transient-field strength for probe ions in the range between $Z \sim 12$ and $Z \sim 42$. This part of the nuclear landscape has been the focus for much of the recent work and is reviewed in section \ref{sect:TF-gfactors-Z.le.50} below. It also covers much of the range of neutron-rich nuclei that are accessible for study with newer radioactive beam facilities such as FRIB. As pointed out above in section \ref{sect:TF-method} and in the discussion of the Ge isotopes in section \ref{sect:TF-gfactors-Z.le.50},  the problem is that there are no suitable independently known $g$~factors among this group of nuclei that can serve as a precision calibration point. It is not sufficient that a state have a precisely known $g$~factor; the state must also be suitable for a precise transient-field measurement. As shown by East et al. \cite{east09}, the 136-keV state in $^{57}$Fe with a $g$~factor measured precisely by the TDPAD technique \cite{fahl79}, cannot be used for a high-precision transient-field measurement because it is too long-lived; it undergoes a precession in the applied magnetic field of sufficient magnitude to obscure the transient-field precession.

As an alternative to seeking an already known independently measured $g$~factor (of which no suitable candidates have been identified), there is a possibility to measure the $g$~factors of picosecond states directly with precision by the time-dependent recoil in vacuum (TDRIV) technique based on free ion hyperfine fields as described below in section \ref{sect:RIV-TDRIV}. Selected measurements on say $^{46}$Ti, $^{56}$Fe and $^{76}$Se would solve the transient-field calibration problem. This direction is currently being explored \cite{mcco20,mcco21,mcco26}.

 {Nuclear structure insights from selected transient-field $g$-factor measurements are reviewed in chapter \ref{sect:short-lived}.}
The discussion here now turns to the possibilities for future transient-field $g$-factor measurements.

The foregoing review of the transient-field method,  {along with the examples reviewed in chapter \ref{sect:short-lived}} demonstrate that it is best suited to measuring first-excited-state $g$~factors following Coulomb excitation of even-even stable nuclei. These measurements have been performed for the majority of such nuclei. Exceptions are extremely low abundance isotopes like $^{162}$Er (0.14\%), $^{168}$Yb (0.13\%) and $^{184}$Os (0.029\%). (The abundances are given in parentheses.) Other exceptions are some of the rare earth nuclei where M\"osbauer measurements already exist, and so transient-field measurements would not yield new information.

There are a few cases of stable even-even nuclei where low-precision data from the early 1980s may be worth revisiting to improve the measurement of $g(2^+_1)$ but the general conclusion is that  $g(2^+_1)$ systematics for the stable nuclides are essentially complete.
In section \ref{sect:TF-gfactors-Z.le.50} below, this review emphasizes recent measurements on higher-excited states, $2^+_2$, $4^+_1$, and $3^-$. The overview presented indicates that these measurements are most instructive in nuclei near closed shells, where the $g$ values can vary considerably with the single-particle structure and relatively low-precision measurements can lead to decisive nuclear structure insights.

For stable nuclei, future applications of the transient-field technique must seek high-precision measurements of excited states above the 2$^+_1$ state. This will require considerable dedicated beam time, so study cases must be chosen carefully. One direction for further investigation concerns the nature of notional vibrational states, which is being questioned. The Cd isotopes have long been cited as perhaps the best examples of nearly spherical vibrational nuclei, however, recent work has proposed an alternative explanation in terms of multiple shape coexistence and rotational degrees of freedom \cite{garr19,garr20,sici21}. Despite the fact noted above that $g$~factors might not be very sensitive to shape coexistence, there is an intriguing result for $^{106}$Cd, namely $g(4^+_1)/g(2^+_1) \sim 0.58$, which cannot be explained by collective models. An alternative interpretation is that single-particle structure is in play, and $^{106}$Cd is accessible to shell model calculations. However, the shell model calculations in Ref.~\cite{benc16} cannot explain this $g$-factor ratio. Further experimentation is called for.

The discussion about the nature of notionally vibrational states in the near-spherical Cd isotopes may lead to questioning of the nature of $\gamma$-vibrational states in deformed nuclei. For example, these states might have an origin in triaxial shapes rather than surface vibrations. There are data on the ratio of the $\gamma$-band $g$~factor to the ground-state-band $g$~factors in $^{160,162,164}$Dy and $^{164,166,168}$Er which compellingly show $g(2_{\gamma})/g({\rm gsb}) \sim 1.2$ in all of these nuclei \cite{bran96,bran99}. This ratio implies a difference between the $g_K$ value for the $\gamma$~band and the rotational $g$~factor of the ground-state band, $g_R$, which agrees with an analysis of the mixing ratios inferred from $\gamma$-ray branching ratios; see Eq.~(\ref{eq:gnils}) through Eq.~(\ref{eq:gKgRQ0}) above. However, the reason for the difference has not been explored.

Several successful $g$-factor measurements have been performed on radioactive beams. These measurements are very challenging. In heavier nuclei near $^{132}$Sn (see section \ref{subsubsec:Te_RIV} below), where both transient-field and RIV measurements have been published, the RIV technique has yielded higher precision for the magnitude of the $g$~factor. The transient field will be necessary, however, in cases where a measurement of the sign of the $g$~factor is critical. Moving forward, it will be useful to explore with stable beams the feasibility of performing combined transient-field and RIV measurements: the beam ion would be excited on a polarized ferromagnetic target that also can induce a transient-field precession that determines the sign of the $g$~factor, and the ion would then recoil into vacuum, where the magnitude of the moment could be measured with higher precision. Whether or not a transient-field precession is to be measured, future RIV measurements on radioactive beams will likely require empirical calibrations based on stable isotopes with known $g$~factors.  In many cases, those stable-beam $g$~factors will have been determined by the transient-field technique.

Reaching up to even heavier nuclei around $^{208}$Pb, there are a number of nuclides with long half-lives (days to years) that could be produced as radioactive beams and studied by the transient-field technique, almost in the same experimental conditions as stable-beam measurements. With inverse kinematics and gadolinium as the ferromagnetic host, precession angles of more than 100 mrad can be anticipated. Such measurements appear promising in terms of both the feasibility and the possibility to expose features of low-excitation nuclear structure near $^{208}$Pb.

As a final example of possible future applications of the transient-field technique, there are opportunities to study nuclear physics at high spin. Although this area is out of fashion at present, contemporary $\gamma$-ray detector arrays together with techniques developed at the turn of the century \cite{stuc02a,robi02a,robi01,robi02} suggest new opportunities to probe nuclear structure at high excitation energy and high spin. Depending on the feeding time to the discrete levels after heavy-ion-induced reactions, the average $g$~factor can be determined in the quasicontinuum above the discrete states (long feeding time) or of discrete states (short feeding time). The former tends to apply in the rare earth region and heavier nuclei (examples: \cite{mahn84,haus84a,tara85,hass91a,weis96,weis98,maye98,weis99a,robi02}) whereas the latter applies near $A=80-100$ and in lighter nuclei (examples: \cite{ward81,barc86,kuch89,moun92,pako92,bill93,pako93,pako94,moun95,zhu01}).

\subsection{The Recoil In Vacuum method }
\label{sect:RIV-method}

\begin{figure}[ht]
\begin{center}
\includegraphics[width= 0.65\columnwidth]{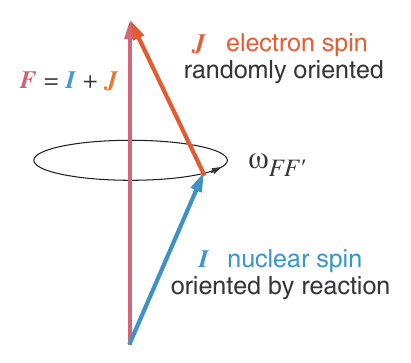}
\caption{\label{fig:free-ion} Vector model of the
free-ion hyperfine interaction on which the Recoil in Vacuum
technique is based. The figure is taken from Ref.~\cite{stuc13a}.}
\end{center}
\end{figure}

When a free ion moves through vacuum, the magnetic hyperfine interaction couples the atomic spin $\bm J$ to the nuclear spin $\bm I$, and together they precess about the total spin ${\bm{ F = I + J}}$, as illustrated in Fig.~\ref{fig:free-ion}. The precession frequency $\omega_{F,F^\prime}$ is proportional to the nuclear $g$~factor and the magnitude of the hyperfine magnetic field at the nucleus. To measure the $g$~factor, the nuclear state of interest is excited by a suitable reaction and then allowed to recoil into vacuum. The effect of the hyperfine interaction is observed via the perturbation of the angular correlation or angular distribution of the $\gamma$-rays de-exciting the state.

 {
In practice, experiments may measure how the shape of the radiation pattern varies with time, or may simply measure the difference between the perturbed and unperturbed radiation patterns. In either case, assuming axial symmetry for simplicity, the angular correlation (or distribution) has the form given by Eq.~(\ref{eq:W_theta_t}),
\begin{equation}
    \label{eq:W_theta-Gk}
    W(\theta) = \sum_{k} A_{k}B_{k}G_k P_{k}(\cos \theta),
\end{equation}
where $A_k$ and $B_k$ are as defined in section~\ref{sec:orientation}, and $G_k$ is the vacuum deorientation coefficient, which is unity if there is no hyperfine field present. In the presence of the free-ion hyperfine fields, $G_k$ departs from unity and depends on the magnitude of the nuclear $g$~factor. Thus, the strategy is to measure the $G_k$ coefficients and from them determine the nuclear magnetic moment.
}

Recoil-in-vacuum (RIV) can refer to two quite distinct experimental techniques, depending on whether the ion has a very simple, few-electron configuration, or whether it has a complex many-electron configuration \cite{gold82}.
 {
More specifically, on one hand, we consider ions with a dominant `simple' H-like, Li-like or Na-like configuration. In this case, the hyperfine interaction is well controlled and time-dependent RIV measurements are performed to identify the nuclear precession frequency. On the other hand, complex ions, with a range of many-electron configurations and no dominant precession frequency, require a more empirical approach and generally, the time-integral attenuation effect is measured.
}

In the following subsection, the foundational concepts and formalism of RIV, which apply in all cases, are reviewed first. Recent developments in relation to simple, generally low-$Z$ ions are then discussed before turning to review the RIV method as applied to slower-moving many-electron heavier ions, particularly radioactive ions produced by the ISOL method. This chapter is concerned primarily with the experimental methods. The outcomes for the structure of nuclei near $^{132}$Sn are discussed in section~\ref{subsubsec:Te_RIV}.

\subsubsection{Formalism and foundational concepts }
\label{sect:RIV-concepts}

This section reviews the formalism and develops the foundational concepts needed to evaluate the free-ion hyperfine interactions associated with the RIV method. Additional details may be found in the review by Goldring \cite{gold82}, and more recently in Stuchbery and Stone \cite{stuc07a} and Stuchbery \cite{stuc13a}.

As indicated by the vector model in Fig.~\ref{fig:free-ion}, the free-ion hyperfine interaction is determined by angular momentum coupling. The starting point for calculating the $\gamma$-ray angular correlation is the statistical tensor of the nuclear state, which specifies the alignment of the nuclear angular momentum. To evaluate the effect of the hyperfine interaction, the general case begins with the statistical tensor of the coupled (nuclear + electronic) system, which evolves in time as
\begin{equation}\label{eq:rhoFFt}
\rho_{q}^{k}(F F^\prime;t) = \rho_{q}^{k}(F F^\prime;0) {\rm e}^{- i \omega_{F F^\prime} t} ,
\end{equation}
where $\omega_{F F^\prime} = (E_F - E_{F^\prime})/\hbar$. Since we are concerned with magnetic dipole interactions
\begin{equation}\label{eq:omegaFF}
\omega_{F F^\prime} = \frac{A_J}{2 \hbar} [ F(F+1) -
F^\prime(F^\prime + 1)],
\end{equation}
where the hyperfine interaction constant $A_J$ is
\begin{equation}\label{eq:AJ}
A_J=\frac{ g \mu_{\rm N}  B_\textrm{H}}{J},
\end{equation}
and $g$ is the nuclear $g$~factor, $\mu_{\rm N}$ is the nuclear magneton, $B_\textrm{H}$  is the hyperfine magnetic field at the nucleus and $J$ is the angular momentum of the electronic configuration. The hyperfine interaction constant $A_J$ is usually given as a frequency in MHz (i.e., as $A_J / h$). The hyperfine field in tesla is related to $A_J$ in MHz by
\begin{equation}\label{eq:BH}
B_\textrm{H} = 0.1312 \frac{J}{g} A_J.
\end{equation}

The initial statistical tensor of the nucleus is determined by the nuclear excitation mechanism, and in the case of Coulomb excitation, it can be calculated accurately. If the atomic system is initially randomly oriented, as is usually the case, it follows that the time dependence of the nuclear tensor is given by
\begin{equation}\label{eq:rhot}
\rho_{q}^{k}(I;t) = \rho_{q}^{k}(I;0) G_k(t) ,
\end{equation}
where the time-differential attenuation coefficient is
\begin{equation}\label{eq:Gk}
G_k(t) = \sum_{F,F^\prime}
\frac{(2F+1)(2F^\prime + 1)}{2J+1} \left
\{ \begin{array}{rrr} F & F^\prime & k \\
I & I & J \\ \end{array} \right \}^2
\cos( \omega_{F F^\prime} t).
\end{equation}
Further details of the relation between Eq.~(\ref{eq:rhoFFt}) and Eq.~(\ref{eq:rhot}) can be found in Ref.~\cite{stuc13a}.

Some experiments observe the average deorientation effect averaged over the decay of the nuclear level. In this case, the time-integral attenuation factors are
\begin{equation}\label{eq:Gkinf}
G_k^\infty(\tau) = \int_0^\infty G_k(t) {\rm e}^{-t/\tau} {\rm d}t/\tau,
\end{equation}
where $\tau$ is the mean life of the nuclear state.

\begin{figure}[t]
\includegraphics[width=\columnwidth]{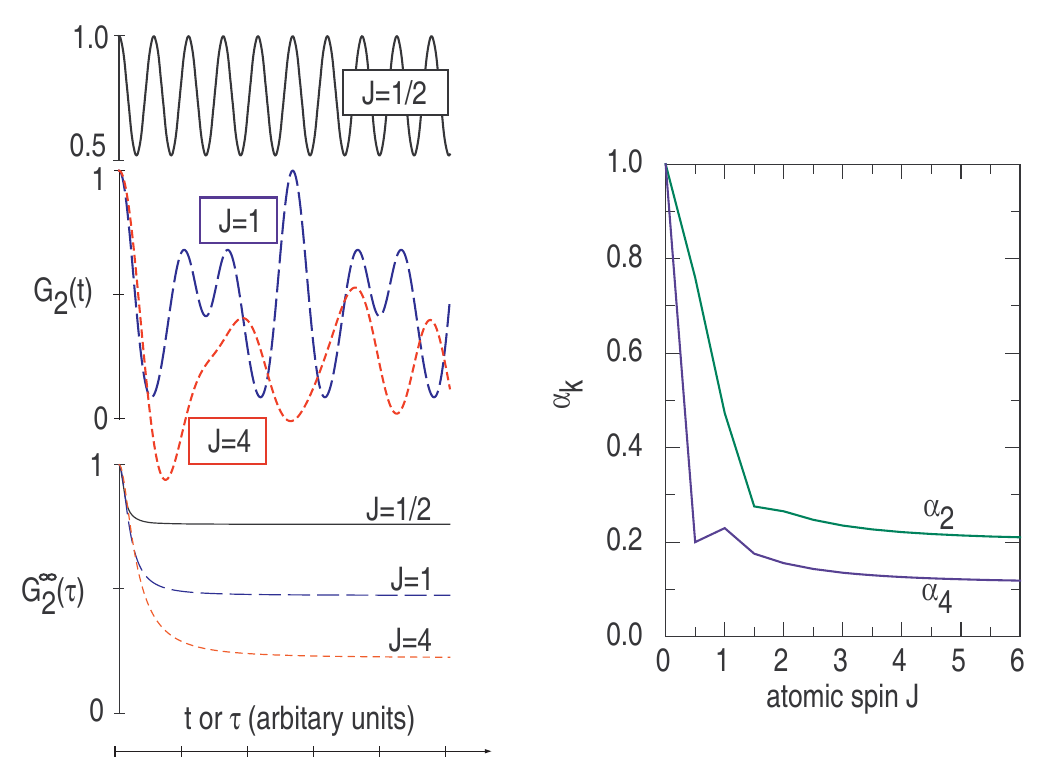}
\caption{Left: Examples of time-differential and time-integral attenuation coefficients for $I=2$. Right: $J$ dependence of the hard-core vacuum-deorientation coefficients for $I=2$. The figure is taken from Ref.~\cite{stuc13a}.}
\label{fig:Gkexamples}
\end{figure}

Examples of $G_k(t)$ and $G_k^\infty(\tau)$ for $I=2$ and the same hyperfine field strength are shown in Fig.~\ref{fig:Gkexamples}. A single frequency is present for $J=1/2$. As $J$ increases, the number of different frequencies present increases and the $G_k(t)$ function becomes more complex.

For a given hyperfine field strength $B_\textrm{H}$, the time-integral coefficients for the different $J$ values all decrease from unity at a similar rate, but they approach different `hard core' values. The hard core, often denoted $\alpha_k$, is given when $F = F^\prime$, or $\omega_{F,F^\prime} = 0$, in Eq.~(\ref{eq:Gk}). Specifically,
\begin{equation}\label{eq:alphak}
\alpha_k = \sum_{F}
\frac{(2F+1)^2}{2J+1} \left
\{ \begin{array}{rrr} F & F & k \\
I & I & J \\ \end{array} \right \}^2.
\end{equation}

The right side of Fig.~\ref{fig:Gkexamples} shows how the hard-core parameters depend on the atomic angular momentum $J$ (for $I=2$). The magnitude of the hard core is a very strong function of $J$ for $J < 2$, but it becomes almost independent of the atomic angular momentum for $J \geq 2$. This behavior has implications for the application of RIV to nuclear moment measurements to be discussed in the next two subsections.

\subsubsection{Simple ions: Time dependent recoil in vacuum }
\label{sect:RIV-TDRIV}

The Time-Dependent Recoil-In-Vacuum (TDRIV), or `plunger', technique was employed for measuring the $g$~factors of excited states in light nuclei with simple electronic configurations (ideally H-like) mainly in the 1970s and early 1980s. The $g$~factors of $\sim 20$ excited states in nuclei ranging from $^{13}$C to $^{24}$Mg were measured. The states studied had spin-parity $I^\pi = 1^-$, $2^\pm$, $5/2^+$, and $3^-$. Their meanlives ranged from $\sim 1$~ps to $\sim 3$~ns \cite{ston14}; with this method, it is possible to observe the time-dependence of the nuclear spin rotation over one or more periods and hence perform precise and accurate $g$-factor measurements on states with lifetimes of a few picoseconds \cite{gold82}.

The method has been revived with modifications for applications to radioactive beams \cite{stuc05}. In this `radioactive beam geometry' it has yielded a new measurement of the $g$~factor of the first-excited state of $^{24}$Mg \cite{kuso15,kuso15a}, for the first time achieving the accuracy and precision needed to test state-of-the-art shell model calculations, which predict departures from $g=0.5$ at the level of about 10\% for $N=Z$ nuclei in the $sd$ shell \cite{rich08}.

\begin{figure}[t]
\includegraphics[width=\columnwidth]{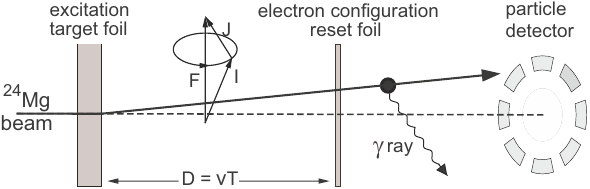}
\caption{The modified TDRIV technique, designed to measure excited-state $g$~factors of fast radioactive beams, has the `stopper' of the traditional plunger technique replaced by a thin foil that resets the electron configuration of H-like ions. As illustrated here, the method was tested via a high-precision measurement on stable $^{24}$Mg~\cite{kuso15,kuso15a}. The particle detector, with segmentation around the beam axis, is located downstream of the $\gamma$-ray detectors. The figure is taken from Ref.~\cite{kuso15}.}
\label{fig:Mg24fig1}
\end{figure}

The experimental method is based on the observation of the precession of the nuclear moment as hydrogen-like ions fly through vacuum. As illustrated in Fig.~\ref{fig:Mg24fig1} for the  $^{24}$Mg experiment, excited nuclei emerge from a target foil as ions carrying one electron.

As noted above, the hyperfine interaction couples the atomic spin to the nuclear spin, and together they precess about the total spin with a frequency proportional to the nuclear $g$~factor. For the ensemble of H-like ions, the nuclear spins are rotated in random directions but at the same frequency. After one period, the nuclear spins all return to their initial orientation. Thus, the orientation of the nuclear spin is periodically reduced and restored during the flight through vacuum. As a consequence, the angular intensity pattern of the $\gamma$ rays emitted by the nuclei varies periodically, in step with the orientation of the nuclear spin. In the traditional recoil-in-vacuum `plunger' technique \cite{gold82}, the ions travel a set distance through vacuum before being stopped in a thick `stopper' foil, which immediately quenches the hyperfine interaction and freezes the orientation of the nuclear spin. The nuclear precession frequency is determined by observing changes in the radiation pattern as the flight time is varied by changing the distance between the target and `stopper' foils.

An important advantage of this method is that the hyperfine fields are well defined. Specifically, the hyperfine field at the nucleus due to a $1s$ electron is
\begin{equation}\label{eq:B1s}
B_{1s} = 16.7 Z^3 R(Z) \;\; {\rm T},
\end{equation}
where $R(Z) \simeq [1+(Z/84)^{2.5}]$ is a relativistic correction factor (also introduced in relation to the transient field, Eq.~(\ref{eq:B_ns})). More precise evaluations of the relativistic correction may be found in ref.~\cite{pyyk71} or by evaluation with the General Relativistic Atomic Structure Package \cite{froe19}. Relativistic effects for $Z=12$ are of order 1\%; in general, the uncertainty in $B_{1s}(0)$, which stems from the accuracy of the relativistic correction, is negligible in comparison to other experimental uncertainties.

Ignoring the relativistic correction, the resulting nuclear precession frequency for an H-like ion in its $^{2}S_{1/2}$ ground state is
\begin{equation}\label{eq:omegas}
\omega_{\rm s} = g (2I+1) \frac{\mu_{\rm N}}{\hbar} B_{1s} = g
(2I+1) 800 Z^3 \;\; {\rm MHz},
\end{equation}
where $g$ is the nuclear $g$~factor and $I$ is the nuclear spin.

In the presence of vacuum deorientation, the time-dependent particle-$\gamma$ angular correlation takes the form (see e.g.~Ref.~\cite{stuc03} and references therein, and cf. Eq.~(\ref{eq:pacTF}))
\begin{equation}
W(\theta_p, \theta_\gamma, \Delta \phi,t) = \sum_{k q} a_{k}^q(\theta_p) G_k(t) D^{k *}_{q 0}(\Delta \phi, \theta_\gamma,0), \label{eq:pacRIV}
\end{equation}
where $\theta_p$ and $\theta_\gamma$ are the polar detection angles for particles and $\gamma$ rays, respectively. $\Delta \phi = \phi_\gamma - \phi_p$ is the difference between the corresponding azimuthal detection angles,  $ a_{k}^q(\theta_p) = B_{k}^q(\theta_p) Q_k F_k $, where $B_{k}^q(\theta_p)$ is the statistical tensor, which defines the spin orientation of the initial state. $F_k$ represents the ordinary $F$-coefficient (Eq.~(\ref{eq:f_ordinary})) for the $\gamma$-ray transition, and $Q_k$ is the attenuation factor for the finite size of the $\gamma$-ray detector. $D^{k *}_{q 0}(\Delta \phi, \theta_\gamma, 0)$ is the Wigner-$D$ matrix. For $E2$ excitation, $k= 0,2,4$ and $-k \leq q \leq k$. The attenuation coefficients, $G_k(t)$, specify the time-dependent vacuum deorientation effect.

For H-like $J=1/2$ configurations the $G_k(t)$ are cosine functions,
\begin{equation} \protect{\label{eq:GK}}
G_k(t) = 1 - b_k(1 - \cos \omega_{\rm s} t),
\end{equation}
where $b_k = k(k+1)/(2I+1)^2$ ; see Fig.~\ref{fig:Gkexamples}, top left. In the traditional technique, the hyperfine interaction is quenched, thus freezing the orientation of the nuclear spin, by {\em stopping} the excited ions in a thick metallic foil after a flight time $T_\textrm{f}=D/v$, where $D$ is the flight distance (or plunger separation). The perturbation factor for the stopped ions is therefore \cite{gold82}
\begin{equation} \protect{\label{eq:stopped}}
G_k^{\rm stopped}(T_\textrm{f}) = G_k(T_\textrm{f}),
\end{equation}
while the ions that decay in flight have an average deorientation coefficient:
\begin{eqnarray} \protect{\label{eq:flight}}
 G_k^{\rm flight} (T_\textrm{f}) &=& \int_0^{T_\textrm{f}} G_k(t) e^{-\lambda t} \lambda
{\rm d}t /\int_0^{T_\textrm{f}} e^{-\lambda t}
\lambda {\rm d}t\\
&=& 1 -b_k(1-F(T_\textrm{f}))
\end{eqnarray}
where
\begin{equation}\label{eq:FT}
F(T_\textrm{f}) = \frac{1 - e^{-\lambda T_\textrm{f}} \left ( \cos \omega_{\rm s} T_\textrm{f} -
\omega_{\rm s} \tau \sin \omega_{\rm s} T_\textrm{f} \right ) } {\left ( 1 +
\omega_{\rm s} ^2 \tau ^ 2 ) (1 - e^{-\lambda T_\textrm{f}} \right ) }.
\end{equation}
(We have corrected a typographical error in Ref.~\cite{stuc05}.)
In the limit that $T_\textrm{f} \rightarrow \infty$, the integral
perturbation factors are obtained,
\begin{equation}\protect{\label{eq:Ginfty}}
G_k(\infty) \equiv G_k^{\rm flight} (\infty) = 1 - b_k \left (
\frac{\omega_{\rm s}^2 \tau^2}{1 + \omega_{\rm s}^2 \tau^2} \right
).
\end{equation}\label{eq:Gk_infinity}
If $\omega_{\rm s} \tau >> 1$, the integral attenuation coefficients approach their hard core values, $G_k({\rm h.c.}) = 1 - b_k = \alpha_k$. For $I=2$ and H-like ions with $J=1/2$, $G_2({\rm h.c.}) = 0.76$, so the nuclear spin can retain a significant level of alignment despite the hyperfine-induced deorientation.

The novel feature for applications to radioactive beams is to replace the thick `stopper' foil by a thinner foil that simply {\em resets} the electron configuration. This change allows projectile-excitation experiments in which the radioactive beam ion is detected at forward angles out of the view of the $\gamma$-ray detectors. In the following, ions that decay between the target and the reset foil are designated as {\em fast}, whereas those that decay after the reset foil are called {\em slow}.

With the thick stopper foil replaced by a much thinner foil that serves to reset the electronic state, the perturbation factor for the nuclei that decay before reaching the `reset' foil, i.e. the {\em fast} component, is unchanged compared with the traditional technique, $ G_k^{\rm fast} (T_\textrm{f}) =  G_k^{\rm flight} (T_\textrm{f})$. The average perturbation factor for the {\em slow} decays that occur beyond the reset foil is just the product of $G_k(T_\textrm{f})$ and the appropriate $G_k(\infty)$ value. In other words, the physical process is identical to the traditional technique up until the ion strikes the reset foil. At that point, the electronic configuration is reset randomly, and the nucleus experiences further perturbations identical in effect to those of the flight peak for an infinite flight path. Assuming, for example, that the ion which enters the reset foil is H-like and emerges from it again as an H-like ion (after several electron exchanges within the foil) the perturbation factor is
\begin{equation}\label{eq:Gkslow}
G_k^{\rm slow}(T_\textrm{f}) = G_k(T_\textrm{f}) \cdot G_k(\infty).
\end{equation}
Should the ion become fully stripped after the reset foil, then $G_k^{\rm slow}(T_\textrm{f}) = G_k(T_\textrm{f})$. Thus for $I=2$ and $J=1/2$, $G_2^{\rm slow}(T_\textrm{f}) = 0.76 G_2(T_\textrm{f})$ in the worst case.

The TDRIV method does not require that the $\gamma$-rays emitted from the {\em fast} and {\em slow} ions be separated in the observed energy spectrum. In such cases, the net angular correlation shows damped oscillations, with the rate of damping determined by the nuclear lifetime \cite{gold82,stuc05}.

\begin{figure}
\begin{center}
\includegraphics[width=0.8\columnwidth]{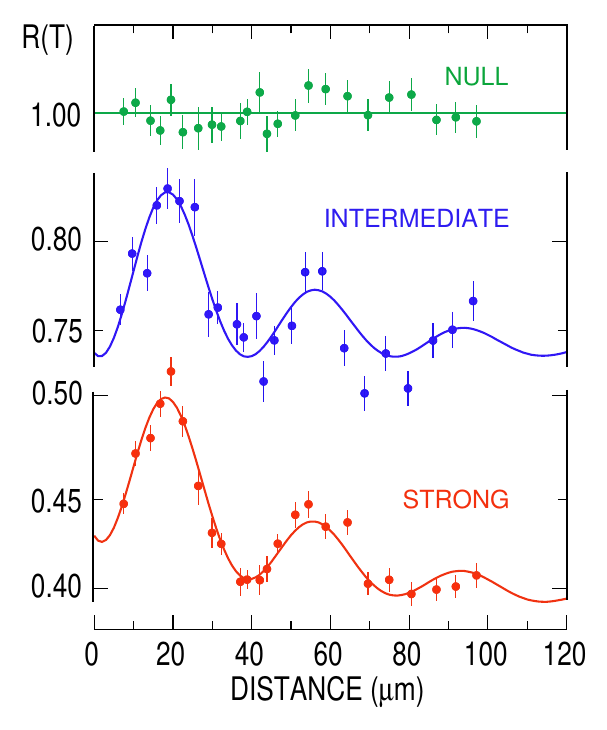}
\caption{Ratios of angular correlations, $R(T_\textrm{f})$, versus flight distance for $^{24}$Mg \cite{kuso15}. The distance is the separation of target and reset foils ($22.4 \mu {\rm m} = 1$~ps flight time). The frequency of the oscillation determines the $g$~factor. The figure is taken from Ref.~\cite{kuso15}.}
\label{fig:Mg24RT}
\end{center}
\end{figure}

Figure~\ref{fig:Mg24RT} shows data from the $^{24}$Mg measurement, which used the Orsay gamma (ORGAM) array of Compton-suppressed germanium $\gamma$-ray detectors and an eight-fold segmented scintillator particle detector \cite{kuso15,kuso15a}. The data were analyzed in groups according to the expected amplitude of the oscillations, $R(T_\textrm{f})$, in ratios of angular correlations observed at different particle and $\gamma$-ray angles as a function of flight time, $T_\textrm{f}$. Due to the symmetry of the particle- and $\gamma$-detector arrays, certain particle-$\gamma$ detector combinations should show the same angular correlation at all times. Ratios of such combinations should show a null effect and serve as a check on the data and analysis procedures. An example is shown in Fig.~\ref{fig:Mg24RT}, labeled `null', along with $R(T_\textrm{f})$ data having strong and intermediate amplitude oscillations.

In this case, the {\em statistical} uncertainty in the $g$~factor from the strong $R(T)$ data is about $\pm 2.4\%$.
Sources of {\em systematic} error arise from:
\begin{enumerate}
\item [(i)] uncertainty in the ion velocity, $v/c$, which affects the flight time;
\item  [(ii)] uncertainty in the distance from the target to particle-detector face, which affects the path length and hence the flight time;
\item  [(iii)] uncertainty in the nuclear lifetime, which affects the damping of the oscillations in the $R(T_\textrm{f})$ function; and
\item [(iv)] uncertainties in the distribution of hyperfine fields, which affect the relationship between the observed frequency and the $g$~factor.
\end{enumerate}

The systematic uncertainties were of similar magnitude to the statistical uncertainty in the case of the $^{24}$Mg measurement \cite{kuso15}: $g= 0.538 \pm 0.011$ {\rm (statistical)} $\pm 0.007$ {\rm (systematic)}.

In some historical TDRIV data, the H-like charge state does not dominate. An example is the measurement on $^{24}$Mg by Horstman et al. \cite{hors75}, where the H-like fraction is only 10\%. Zero-field contributions (such as the He-like ground-state) contributed 36\%. Such contributions reduce the amplitude of the observed oscillations, which reduces the precision of the measurement, but have little impact on systematic uncertainties. Of greater concern are non-zero contributions such as those from Li-like ions, of the order of 38\% in the historical $^{24}$Mg case, which can introduce additional frequency components into the observed $R(T_\textrm{f})$ data and lead to systematic uncertainties in the extracted $g$~factor. The presence of additional non-zero fields can limit the accuracy of TDRIV measurements and must be evaluated at least as a contribution to the systematic uncertainties.

Of at least equal concern for the accuracy and precision of TDRIV measurements is the determination of the zero of the plunger separation and whether the $R(T_\textrm{f})$ data clearly show at least one oscillation period. If a full oscillation period is not observed with some precision, the uncertainty in the zero-distance position of the plunger can limit the accuracy of the $g$-factor measurement. An example is again provided by Horstman et al. \cite{hors75}, in the case of their $^{20}$Ne measurement. In that case, the $R(T_\textrm{f})$ data rise from $R \approx 0.2$ at $T_\textrm{f}=0.5$~ps to a maximum of $R \approx 0.3$ beyond $T_\textrm{f}=1.5$~ps, with little evidence of a second minimum. The measurement of the precession period then relies on an accurate knowledge of both the `zero distance' and the slope of the $R(T_\textrm{f})$ data. In the $^{20}$Ne case, the uncertainty from the zero distance was comparable to the statistical uncertainty (see Table 3 of \cite{hors75}). These remarks are also relevant to the first TDRIV measurement on a radioactive beam of $^{28}$Mg performed at ISOLDE, to be discussed below.

Before coming to the measurement on $^{28}$Mg, it is useful to review the three TDRIV measurements of the magnitude of the $g$~factor of the 351-keV 5/2$^+$ state in $^{21}$Ne published in 1977-78. The results were: $\vert g\vert=0.196(14)$ \cite{rowe78}, $\vert g\vert=0.28(3)$ \cite{beck77}, and $\vert g\vert=0.35(8)$ \cite{angh78}. Of these three measurements, only that of Rowe et al. \cite{rowe78} compellingly observes oscillations (three peaks, two troughs), and yields a $g$-factor magnitude close to the {\tt USDB} shell model \cite{brow06,rich08} value of $g = -0.220$. These examples illustrate the quality of the $R(T_\textrm{f})$ data required to obtain a reliable TDRIV $g$-factor measurement.

The first TDRIV $g$-factor measurement on a radioactive beam was performed successfully on $^{28}$Mg at ISOLDE in November 2017. However, it has not yet been published due to uncertainties in the plunger zero distance. To determine the zero distance, a TDRIV measurement was performed on $^{22}$Ne, a standard beam used to set up Miniball, and having precise values for $g(2^+_1)$ in the literature, namely $\vert g\vert =0.326(12)$ \cite{hors77} and $\vert g\vert =0.36(3)$ \cite{bohm76}. However, the new TDRIV data in the `radioactive beam geometry', and at higher recoil velocity, suggest a higher value for this $g$~factor, closer to the {\tt USDB} shell model value of $g=0.379$ than the smaller literature value. However, the data were not taken with sufficient precision to determine precisely both the $^{22}$Ne $g$~factor and the zero distance. Thus, the final result from the $^{28}$Mg measurement awaits the analysis of a new precise measurement of  $g(2^+_1)$ in $^{22}$Ne, which has recently been completed at GANIL.

Other insights for future TDRIV measurements on radioactive beams follow from the fact that the $^{28}$Mg measurement was rate-limited by the decay of $^{28}$Mg. The rate limit can be reduced by opening out the reaction geometry, so that the beam does not stop within view of the $\gamma$-ray detectors, e.g., in the collimator before the plunger and by ensuring that the particle detector is located far enough downstream. Also, if the beam intensity to be delivered is known accurately enough in advance of the experiment, the thickness of the target and reset foils can be optimized to maximize yield while keeping multiple scattering to an acceptable level.

A new, precise measurement of the $^{22}$Ne $g$~factor has recently been performed at GANIL, using a 121 MeV $^{22}$Ne beam, the plunger device OUPS (Orsay Universal Plunger System \cite{ljun12}), the EXOGAM $\gamma$-ray detector array \cite{azai99} and the OPSA (Orsay Particle Scintillator Array) particle detector. Online analysis shows 3 clear oscillation periods, and detailed data analysis is underway. This case is an example where precision data on stable beams is essential for radioactive beam studies. In this case, the connection is technical, but the same principle holds concerning the need for precise systematic data on stable beams to interpret the structure of nuclei away from the valley of stability.

The experience with OPSA for the $^{22}$Ne measurement in comparison with the DSSD (double-sided silicon strip detector) in the $^{28}$Mg run suggests that a scintillator array is better suited to these measurements as it can take a higher rate; in these measurements there is always a high Rutherford scattering rate at the forward angles in competition with the sought Coulomb excitation of the beam.

A limitation of the TDRIV method with H-like ions is that the H-like field increases with $Z^3$ and so the precession frequency, $\omega_L$, becomes too short to measure as $Z$ increases; for $g=0.5$, the period of the oscillations is 3.1~ps at $Z=10$ and 0.38~ps at $Z=20$. Thus, the method with H-like fields becomes impractical beyond $Z \sim 20$, depending somewhat on the magnitude of the $g$~factor.

To apply the TDRIV method to higher-$Z$ nuclei would require the use of weaker fields of shielded electrons, such as Li-like ions (2s electron) or Na-like ions (3s electron). Applications of the method with Li-like and Na-like ions are under investigation. There are a number of indications from measurements of charge-state distributions and integral recoil in vacuum measurements on relatively low-$Z$ ions that atomic ground states and low-excitation atomic states dominate the free-ion interactions at least after a few picoseconds \cite{stuc13a,stuc07a}. However, a Na-like ion, for example, has many more excited states than an H-like ion, with the potential to wash out the unique frequency of the atomic ground-state configuration. Moreover, for ions with $Z \sim 26$, it is not possible to get more than about 28\% of the ions in the Na-like charge state. Even so, Sprouse et al. \cite{spro76,spro76a} and Lin et al. \cite{Lin1978}  have reported evidence of population of Na-like $^{41}$Ca ions via the observation of a spin precession at the expected frequency in transverse decoupling experiments on the 3.830-MeV, $I=15/2^+$, $g=+0.29 \pm 0.02 $, $\tau = 4.7 \pm 0.2$ ns, level.

\begin{figure}
 \includegraphics[width=0.85\columnwidth]{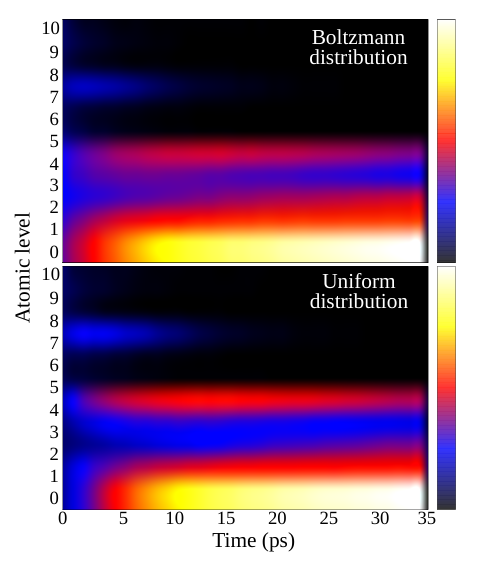}
\caption{Monte Carlo simulation of Na-like atomic-state distributions
for Fe ions, showing an initial Boltzmann ($T = 100$ eV)
vs uniform distribution. The $y$ axis shows the energy-ordered
level index. Level 0 is the $3s_{1/2}$ ground state. The temperature scale of the `heat map' is set relative to the most intense
population on the plot. Higher-excited states, which feed
the lower states, are included in these calculations but not shown.
Due to binning and fast transitions, the time $t = 0$ distribution
does not show the true initial distribution shape. The figure is taken from Ref.~\cite{mcco20}.}
\label{fig:NaLikeFeMC}
\end{figure}

These conclusions are supported by detailed Monte Carlo calculations based on level energies and lifetimes from the general relativistic atomic structure package (GRASP) \cite{froe19}. An example is shown in Fig.~\ref{fig:NaLikeFeMC}.

Further experimentation on Na-like ions of $fp$ shell nuclei is in progress. A conventional TDRIV measurement was performed on the $^{56}$Fe $2^+_1$~state at the Orsay Tandem accelerator during the Miniball campaign in 2014. A 130~MeV $^{56}$Fe beam impinged upon a 0.30~mg/cm$^2$ carbon target backed by 0.67~mg/cm$^2$ of nickel. After excitation and recoil into vacuum, the $^{56}$Fe ions travelled a known distance before stopping in a second 5.8~mg/cm$^{2}$ nickel foil.

The beam energy was chosen based on charge-state-distribution measurements performed at the Australian National University Heavy Ion Accelerator Facility (HIAF), so that the charge-state distribution of $^{56}$Fe ions exiting the target was centered on the Na-like charge state. The Orsay Universal Plunger System (OUPS)~\cite{ljun12} was used to vary the distance between the target and secondary (stopping) foil, thus measuring the time-dependence of the hyperfine interaction.

Oscillations were observed that can be associated with the Na-like $3s_{1/2}$. A contribution from the F-like $2p_{1/2}$, which gives rise to a similar precession frequency, was also in evidence. Despite the complexity of the hyperfine-field distribution, the value of the $g$~factor extracted was robust, varying by at most a few percent under alternative fitting assumptions.

This measurement can remove the uncertainties assigned by Stone \cite{ston20} to the recommended transient-field $g$~factor measurements in the $fp$ shell. It is therefore important that it be confirmed, not necessarily by repetition of the $^{56}$Fe case, but by measurements on other suitable neighboring nuclei. One lesson from the $^{56}$Fe study is that Mg-like and Al-like ions are preferable to F-like ions in the charge-state distribution as they contribute only a smooth variation to the $R(T)$ function, whereas F-like ions can introduce a frequency near that of the sought Na-like ground state.

Another approach to obtaining accurate $g$~factors of short-lived states in the $fp$ shell is being pursued through TDRIV experiments on Li-like Ti ions at Jyv\"askyl\"a (experiment J27). As the number of species in the charge-state distribution is smaller than for Na-like ions, the number of hyperfine fields acting on the nucleus is reduced, which may simplify the interpretation of the data and reduce the scope for systematic uncertainties stemming from the hyperfine interactions.

\subsubsection{RIV with complex ions }
\label{sect:RIV-complex-ions}
 {
\paragraph{Introduction}
}
The RIV technique has proved a powerful method to measure the $g$~factors of excited states of neutron-rich nuclei produced as radioactive beams, particularly in the tin and tellurium isotopes near the neutron-rich doubly magic nuclide $^{132}$Sn \cite{ston05,stuc07a,stuc13,allm13,allm15,allm17,stuc17}.

One of the method's advantages is that the $g$~factor of the 2$^+_1$ state can be measured simultaneously with $B(E2; 0^+ \rightarrow 2^+)$ and $Q(2^+)$
\cite{stuc13,allm11,allm13,allm15,allm17,stuc17}.

Although the RIV method gives only the magnitude of the $g$~factor, it has proven to give it more precisely \cite{ston05,stuc07a,allm13} than the transient-field method \cite{benc08,kumb12a} for radioactive beam measurements. The advantage of the transient-field method, however, is that it can determine the sign.

The primary examples discussed here concern the application of the RIV technique to many-electron radioactive ions as produced by the Isotope Separation On Line (ISOL) type facilities like the Holifield Radioactive Ion Beam Facility (HRIBF) at Oak Ridge National Laboratory, USA. In these $g$-factor measurements, the experimental procedures are identical to those in a measurement of the reduced transition probability, or $B(E2)$, by Coulomb excitation. The focus of the analysis, however, is on the angular correlation pattern of the $\gamma$ radiation de-exciting the state of interest rather than its total intensity.

\begin{figure}
\includegraphics[width=\columnwidth]{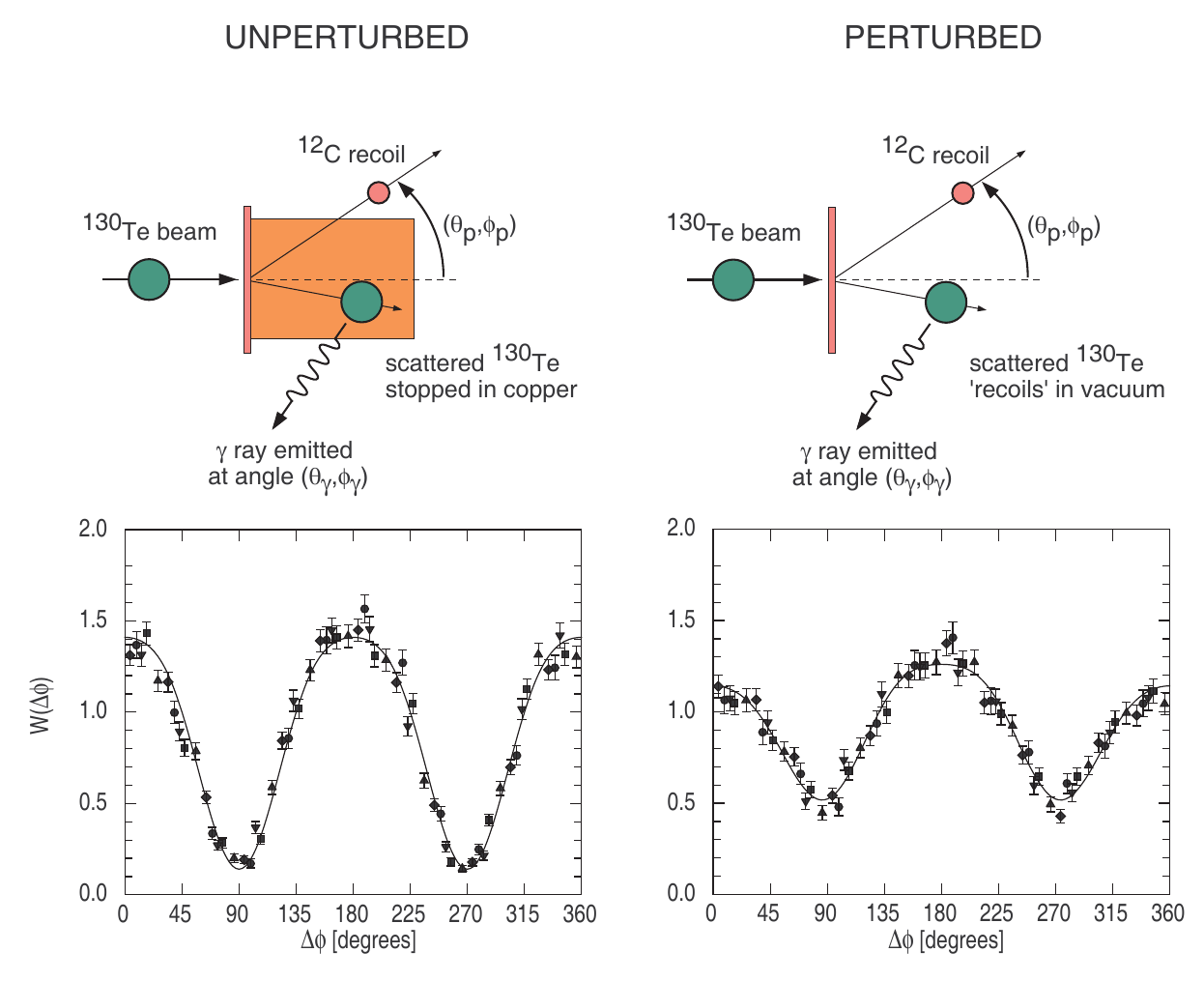}
\caption{Unperturbed and perturbed angular correlations for the $2^+_1 \rightarrow 0^+_1$ transition in $^{130}$Te \cite{ston05}. Left: Unperturbed angular correlations following implantation into copper. Right: Perturbed angular correlations with reduced anisotropy resulting from vacuum deorientation of the nuclear spin. The figure is taken from Ref.~\cite{stuc13a}.}
\label{fig:riv-concept}
\end{figure}

To measure the $g$~factor, the nuclear state of interest is excited by a suitable reaction and then allowed to recoil into vacuum. The effect of the hyperfine interaction is observed via the reduced anisotropy of the angular correlation of the $\gamma$-rays de-exciting the state. The concept is illustrated in Fig.~\ref{fig:riv-concept} based on data from HRIBF \cite{ston05}. The left side of Fig.~\ref{fig:riv-concept} shows the measurement of the unperturbed angular correlation by implantation of (stable) $^{130}$Te into a copper host. The right side shows the equivalent attenuated angular correlation observed when the $^{130}$Te ions recoil into vacuum. The $g$~factor is determined from the difference between these angular correlations.

The angular correlation formalism following Coulomb excitation was given in Eq.~(\ref{eq:pacRIV}), but now the vacuum attenuation coefficients are for a complex electronic configuration and are integrated over the nuclear lifetime; $G_k(t)$ in Eq.~(\ref{eq:pacRIV}) is replaced by $G_k^\infty(\tau)$. The first step to determine the $g$~factor is to determine the vacuum deorientation coefficients, $G_k=G_k^\infty(\tau)$, from the measured angular correlations. Once these have been determined, the $G_k$ values must be related to the $g$~factor and the level mean life $\tau$. There are three possible ways to do this: (i) make an essentially empirical parametrization of $G_k$ versus $\vert g \vert \tau$ or $g^2 \tau$ \cite{stuc07a,allm17,stuc17}, (ii) seek a model-based parametrization of $G_k$ versus $\vert g \vert \tau$ \cite{ston05,stuc07a,stuc13,allm13,allm15}, or (iii) perform a first-principles calculation of the $G_k$ coefficients based on atomic physics calculations \cite{chen13,ston10,ston15}. As discussed by Stuchbery et al. \cite{stuc17}, despite considerable progress in calculations from first principles, an empirical foundation is still needed to relate the $G_k$ coefficients to the $g$~factor.

\begin{figure}
\includegraphics[width=0.95\linewidth]{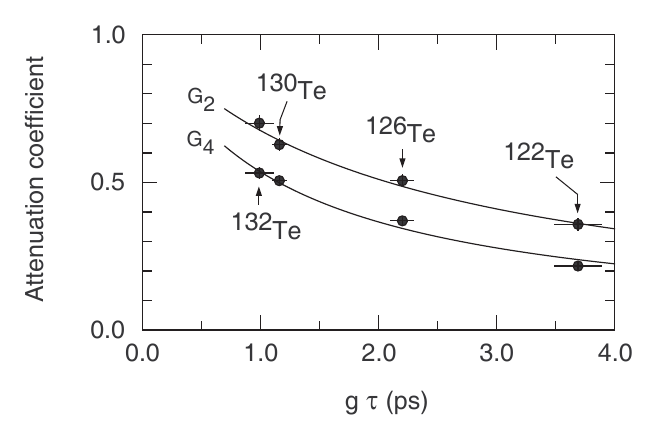}
\caption{Time-integral attenuation coefficients for the Te isotopes. The solid line is the best fit to Eq.~(\protect \ref{eq:Gkemp}) from which $g(2^+)$ in $^{132}$Te was determined \cite{stuc07a}. The figure is taken from Ref.~\cite{stuc07a}.}
\label{fig:Gkplot-emp}
\end{figure}

 {
\paragraph{Determining $g$~factors from attenuation coefficients}
}

Fig.~\ref{fig:Gkplot-emp} illustrates the empirical procedure used to determine the $g(2^+_1)$ value in neutron-rich $^{132}$Te based on a calibration using stable $^{122}$Te, $^{126}$Te and $^{130}$Te, as described in Ref. \cite{stuc07a}. The measured attenuation factors are plotted versus the product $\vert g \vert \tau$. The $g$~factors for the stable isotopes were measured precisely at the Australian National University \cite{stuc07}. The time-dependent attenuation coefficients were assumed to decay exponentially with time to a hard core value $\alpha_k$ at the limit of long times
\begin{equation}\label{eq:lorentz-Gkt}
G_k(t) = \alpha_k + (1-\alpha_k) e^{-\Gamma_k t},
\end{equation}
where $\Gamma_k$ is the relaxation constant, related to the $g$ factor as discussed below. A time-independent (static) distribution of hyperfine fields with a Lorentzian frequency distribution gives rise to such an exponential time-dependence of the attenuation coefficients.

The integral attenuation coefficients corresponding to Eq.~(\ref{eq:lorentz-Gkt}) have the form
\begin{equation}
\label{eq:lorentz-Gktau}
G_k^\infty(\tau) = \alpha_k + (1- \alpha_k)\frac{1}{1+ \lvert \Gamma_k \rvert \tau}.
\end{equation}
The dependence on the $g$~factor is given by
\begin{equation}\label{eq:Ck-defn}
\Gamma_k = \lvert g^n \rvert /C_k ,
\end{equation}
where $C_k$ is the parameter that determines the strength of the interaction, and hence deorientation. The static limit (no atomic transitions) corresponds to $n=1$ \cite{stuc07a} whereas the fluctuating limit (continuous atomic transitions) corresponds to $n=2$  \cite{abra53}.

Fits to the $G_k$ values for the stable isotopes $^{122,126,130}$Te in Fig.~\ref{fig:Gkplot-emp} favored $\alpha_2 = \alpha_4 = 0$. These `hard core' parameters were therefore set to zero for subsequent fits. Acceptable fits could be obtained with $n=1$ or $n=2$. However the case of $n=1$, corresponding to the static limit, was favored \cite{gold78a,andr78,gold85,bill86a}. Thus the fit in Fig.~\ref{fig:Gkplot-emp} assumes that
\begin{equation} \label{eq:Gkemp}
G_k^\infty(\tau) = C_k/(C_k +  \lvert g \rvert \tau).
\end{equation}
In this purely empirical approach, the parameters for $G_2$ and $G_4$ were determined independently of each other.

When dealing with low-statistics data such as those typically obtained with radioactive beams, it is important to minimize the number of fit parameters and strive to determine the $g$~factor from fits to the angular correlation data with $ \lvert g \rvert $ as the only free parameter. Strategies to achieve this goal will now be discussed, first in terms of the relation between theoretical attenuation coefficients and the $g$~factor, and then in relation to eliminating normalization factors from the experimental angular-correlation data.

Eq.~(\ref{eq:Gk}) indicates that $G_2$ and $G_4$ are intimately related to each other; for a single electronic configuration, the only difference is the value of $k$ in the six-$j$ symbol. However, in the case of a complex superposition of static and time-varying hyperfine interactions, the relationship between the resultant $G_2$ and $G_4$ values is not readily calculated from first principles.

\begin{figure}[hb]
\includegraphics[width=\columnwidth]{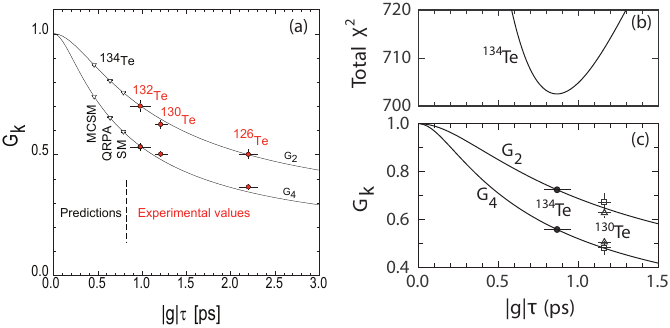}
\caption{\label{fig:Gkplot} (a) Attenuation factors versus $\vert g \vert\tau$ for several isotopes of Te. Data are from \cite{ston05}. For $^{134}$Te, the open points indicate the attenuation factors corresponding to predicted $g$~factors: MCSM \cite{shim04}, QRPA \cite{tera02}, SM \cite{brow05}. (b) Total $\chi^2$ versus $ \vert g \vert \tau$ for $^{134}$Te from \cite{stuc13}. (c)  $G_k$ versus $ \vert g \vert \tau$ calibration curves from \cite{stuc13}. The best fit $\vert g \vert \tau$ value for $^{134}$Te, and its uncertainty, is projected onto the curves (filled circles). Also shown are $^{130}$Te data (squares) from \cite{stuc13} and \cite{ston05} (triangles) which, together with additional measurements \cite{allm13}, define the calibration curves \cite{stuc07a}. Part (a) of the figure is from Ref.~\cite{stuc13a}; parts (b) and (c) are taken from Ref.~\cite{stuc13}.
 }
\end{figure}

For the $Z=50$ region, a pure static-model approach has been used to relate $G_2$ and $G_4$ and successfully describes a large fraction of the data. As an example, Fig.~\ref{fig:Gkplot} again plots $G_2$ and $G_4$ versus $\vert g \vert \tau$ for the $2^+_1$ states in Te isotopes (cf. Fig.~\ref{fig:Gkplot-emp}), but now the curves represent a static-model-based fit to the calibration data \cite{stuc07a,allm13,stuc13a,allm15}.
Specifically, in the static model of the RIV interaction, as described by Stuchbery and Stone \cite{stuc07a} and applied to excited 2$^+$ states near $^{132}$Sn with lifetimes up to a few picoseconds, the observed $G_k$ value is modeled as a superposition of the attenuation coefficients for a distribution of many individual electronic configurations with spin $J_i$ and hyperfine field $B_j$, denoted $G_k(J_i,B_j,  \lvert g \rvert \tau)$. The distribution of such atomic states is parametrized by normalized Gaussian distributions, $w_J$ and $w_B$, which represent the distribution of electronic spins and hyperfine fields, respectively. Thus
\begin{equation}\label{eq:staticM}
G_k = \sum_{i,j} w_J(J_i) w_B(B_j) G_k(J_i,B_j, |g|\tau).
\end{equation}
For nuclear spin $I=2$, the $G_k$ become insensitive to the precise value of $\bar{J}$  once $\bar{J} > 2$ \cite{stuc12,stuc13a}. As a consequence, for nuclei in the vicinity of $Z=50$, the average atomic spin can be set to $\bar{J}=4.5$ \cite{stuc07a}. In general, there is little sensitivity to the width of the spin distribution, making it possible to set $\sigma_J = 1$. Moreover, the data are well described by setting the width of the field distribution $\sigma_B$ equal to the mean field, $\bar{B}$; see Refs.~\cite{stuc07a,stuc12,stuc13a}. It has thus proved possible to describe much of the RIV data near $Z=50$ in terms of the single parameter $\bar{B}$. In other words, setting the average hyperfine field strength can describe the $g \tau$ dependence of both $G_2$ and $G_4$ in many cases.

Fig.~\ref{fig:Gkplot}(a) shows the locations of points on the $G_k$ versus $g\tau$ curve as predicted by various nuclear models for $g(2^+_1)$ in $^{134}$Te \cite{shim04,tera02,brow05}. The results of the subsequent measurement are shown to the right in Fig.~\ref{fig:Gkplot}(b). It is evident that the range of predicted $G_k$ values for semimagic $^{134}$Te is near the maximum slope of the $G_k$ vs $g\tau$ curve, and hence the RIV measurement \cite{stuc13} had the sensitivity required to distinguish between the theoretical models.

The above discussion has assumed that the RIV interaction is primarily due to a superposition of static electronic configurations throughout the lifetime of the nuclear state. However, this assumption is not generally applicable. In the subsequent analysis of the RIV interaction to determine $g(2^+_1)$ in $^{136}$Te, which has a lifetime of $\tau = 27.5(23)$~ps, i.e., an order of magnitude longer than the cases considered previously, it was necessary to include calibration data on suitably long-lived states in $^{125}$Te. From these data, it was found that the $G_k(\tau)$ data were correlated with $g^2\tau$ rather than $\lvert g \rvert \tau$. Thus, atomic transitions must be playing a role, and the static model-based calibration used for the shorter-lived states in the isotopes $^{132,134}$Te \cite{ston05,stuc07a} and the sequence of Sn isotopes \cite{allm13,allm15} could not be used. The analysis therefore reverted to an empirical approach based on Eqs.(\ref{eq:lorentz-Gktau}) and (\ref{eq:Ck-defn}). In addition, to reduce the number of parameters in the fit to the low-statistics radioactive beam data, the calibration data were used to constrain the relation between $G_2$ and $G_4$ \cite{stuc17}. The fluctuating-field model of Abragam and Pound \cite{abra53} for pure magnetic interactions specifies a simple relationship between $G_2$ and $G_4$, namely,
\begin{equation} \label{eq:G2vG4-AP}
G_4 = \frac{0.3 G_2}{1-0.7G_2}.
\end{equation}
To describe the RIV data, this form was generalized as
\begin{equation} \label{eq:G2vG4}
G_4 = \frac{a G_2^p}{1-(1-a)G_2^p},
\end{equation}
where $a$ and $p$ are parameters. This expression has the correct limits that $G_4 \rightarrow 0$ as $G_2 \rightarrow 0$ and $G_4 \rightarrow 1$ as $G_2 \rightarrow 1$.

The hyperfine interaction clearly depends on the charge states of the ions as they recoil into vacuum, generally becoming stronger as the charge state increases. Reaction kinematics determine the recoil velocity. Hence, the strength of the hyperfine fields varies with beam energy, target mass, and scattering angle. For beams near $^{132}$Sn, the change in velocity with scattering angle is small for C targets, but significant for Ti targets for which a separate $G_k$ vs $\vert g \vert \tau$ (or $g^2 \tau$) curve must be used for each beam, target, and particle detection angle. Likewise, the relation between $G_2$ and $G_4$ varies with ion velocity.

\begin{figure}[hb]
\centering
\includegraphics[width=0.95\columnwidth]{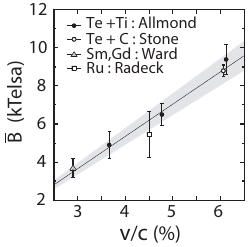}
\caption{\label{fig:BbarvsV}
Dependence of the average hyperfine field on velocity extracted from RIV data for Te ions recoiling into vacuum from Refs.~\cite{ston05,stuc07a} (Stone) and Ref.~\cite{allm13} (Allmond). Fields for Ru, Sm, and Gd ions were obtained from a re-analysis \cite{allm13} of the results of Radeck~et~al.~\cite{rade12} and Ward~et~al.~\cite{ward74}, respectively. The shaded region represents an uncertainty of $\pm 8\%$ assigned to $\bar{B}$ in Ref.~\cite{allm13}. The figure is taken and adapted from Ref.~\cite{allm13}.}
\end{figure}

For those cases where the static model described above is applicable, the strength of the hyperfine interaction as a function of ion velocity can be parametrized in terms of the effective average field at the nucleus, $\bar{B}$. As shown in Fig.~\ref{fig:BbarvsV}, for nuclei with $42 \leq Z \leq 64$, $\bar{B} \propto v/c$, where $3 \lesssim  v/c \lesssim 6$, and $v/c$ is the ion velocity relative to the speed of light.

The $Z$-dependence of the RIV fields in Fig.~\ref{fig:BbarvsV} is weak, as the data for $_{44}$Ru, $_{62}$Sm, and $_{64}$Gd all overlap the calibration curve for $Z=52$. This weak dependence on $Z$ comes from a trade-off: the increase in hyperfine field with increasing $Z$ for a given electronic configuration is offset by a change in charge state - the higher-$Z$ ion carries more electrons (for given $v/c$), which moderates the increase in hyperfine field. It is justified, therefore, to consider the average hyperfine fields for $_{50}$Sn and $_{52}$Te equal, well within the precision of the measurements.

For those cases where the static model is not applicable, the parameters in Eq.~(\ref{eq:lorentz-Gktau}) must be determined as a function of the ion velocity; see \cite{stuc17}. But, again, a weak dependence on the atomic number of the ion is expected.

 {
\paragraph{Instrumentation and angular correlation analysis}
}
The discussion now turns to the analysis of the angular correlation data, specifically the strategy adopted to eliminate normalization and efficiency factors in the data taken at the Holifield Radioactive Ion Beam Facility (HRIBF) at Oak Ridge National Laboratory with the BareBall (particle) \cite{gali10} and CLARION ($\gamma$-ray) \cite{gros00} detector arrays.
 {
This combination of particle and $\gamma$-ray detector arrays served as the platform for developing the RIV method for applications to radioactive beams. Moreover, they have proven near optimal for such measurements. For this reason, the following discussion is based on the specifics of these arrays.
}

The key feature is the use of {\em azimuthal} angular correlations around the beam axis, in terms of the difference in the particle and $\gamma$-ray detection angles, $\Delta \phi = \phi_{\gamma} - \phi_{p_i}$. This choice differs from the more familiar  $\gamma$-ray angular correlations following Coulomb excitation, which are usually measured with respect to the beam axis, i.e., with respect to $\theta_{\gamma}$.  Examples of the azimuthal angular particle-$\gamma$ correlations for $^{130}$Te are shown in Fig.~\ref{fig:riv-concept} above.

\begin{figure}[htb]
\centering
\includegraphics[width=0.95\columnwidth]{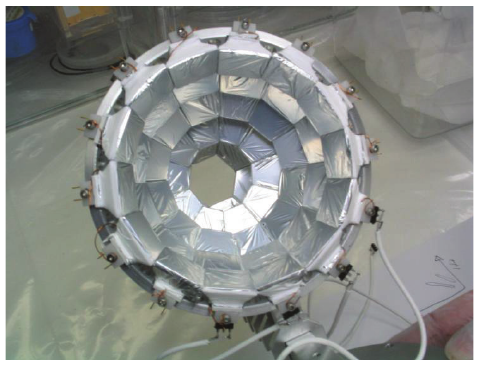}
\caption{\label{fig:BareBall} A view of BareBall, the particle detector array with 55 CsI(Tl) photodiode detectors arranged in rings around the beam axis. The figure is taken from Ref.~\cite{gali10}.}
\end{figure}

The BareBall detector is shown in Fig.~\ref{fig:BareBall}. Typical combined RIV and $B(E2)$ measurements have used three rings of BareBall (ring~2: $\theta_p = 14^\circ-28^\circ$, ring~$3=28^\circ-44^\circ$, and ring~$4=44^\circ-60^\circ$). In most cases, three rings of the CLARION array (five Compton-suppressed `Clover' detectors at $\theta_{\gamma} = 90^\circ$, four at $132^\circ$, and two at $154^\circ$) were used, thereby constructing nine  particle-$\gamma$ angular correlations in $\Delta\phi$. (Detection angles for each ring are given in the laboratory frame with respect to the beam axis.)

For a given BareBall and CLARION ring, the coincidence count for a particular BareBall segment and $\gamma$-ray detector was first normalized to the Rutherford scattering yield to account for any variations in efficiency of the BareBall elements. The efficiency of the $\gamma$-ray detector was factored out by normalizing the coincidence counts in a chosen pair of particle and $\gamma$-ray detectors, $N(\theta_p, \theta_\gamma,\phi_\gamma-\phi_{p_i})$, to the sum over all particle detector segments in the particular BareBall ring:
\begin{equation} \label{eq:wphinorm}
\frac{N_H N(\theta_p, \theta_\gamma,\phi_\gamma-\phi_{p_i})}{\sum_i
N(\theta_p, \theta_\gamma,\phi_\gamma-\phi_{p_i})}=
\frac{W(\theta_p, \theta_\gamma,\phi_\gamma-\phi_{p_i})}{
W(\theta_p, \theta_\gamma)}.
\end{equation}
$N_H$ is the number of detectors in the BareBall ring and $W(\theta_p, \theta_\gamma)$ is given by Eq.~(\ref{eq:pacRIV}) with $q\equiv 0$. By this procedure, the only free parameters required to fit the perturbed angular correlations are the vacuum attenuation factors, $G_2$ and $G_4$, which can be related to the value of $\lvert g \rvert$ as described above (assuming that $\tau$ is known).

The identity in Eq.~(\ref{eq:wphinorm}) is exact, which can be shown by the following: Beginning with the statistical tensor for the particle detector (or detector segment) at $\phi_p=0$ and denoting it $B_{k q}$, then the theoretical normalization factor for each $\gamma$-ray detector requires the evaluation of a sum of statistical tensors:
\begin{equation}
B_{k}^{q \rm(sum)} = B_{k}^{q} \sum_{j=1}^{N_{H}} {\rm e}^{i q (j-1) 2 \pi /N_{H}},
\end{equation}
where the sum is over the $N_H$ detectors in the particle-detector ring. If $q=0$, the sum is clearly just $N_{H}$. For $q \neq 0$ the sum is zero. The mathematical reason for it being identically zero stems from the fact that the terms in the sum are related to the $n$th complex roots of unity, which sum to zero. Thus we can write in general for each BareBall ring, that
\begin{equation}
B_{k}^{q{\rm(sum)}} = B_{k}^q {N_{H}} \delta_{q 0}.
\end{equation}
This result proves Eq.~(\ref{eq:wphinorm}). The interesting point is that it does not assume cylindrical symmetry for the individual detectors in the Bareball ring. Moreover, the statistical tensors to be used are just the $B_{k q}$ values evaluated for the single (finite) particle detector element at $\phi_p = 0$. This procedure is applicable to any particle detector array that has rings of identical equally-spaced segments (or elements) around the beam axis.

\subsubsection{Status of ab initio RIV calculations for complex ions}
\label{sect:RIV-abinitio}

In principle, the RIV hyperfine interactions can be calculated based on atomic structure calculations. The status of this approach as appraised by Stuchbery~et~al. in 2017 \cite{stuc17} largely remains current and will be reviewed here.

With the computer power available today, along with comprehensive atomic structure
codes such as the multiconfiguration Hartree-Fock (MCHF) Atomic Structure Package \cite{froe97,froe00} and the General Relativistic Atomic Structure Package (GRASP) \cite{froe19}, it has become reasonable to attempt ab initio calculations of the free-ion hyperfine interactions of relevance to RIV and magnetic moment measurements.

The simplest microscopic approach to model the RIV attenuation for many electron ions is to superimpose the deorientation coefficients for the calculated hyperfine interactions up to a cut-off in excitation energy, assuming a weighting factor of $(2J + 1)$ for each atomic state. Stone et al. \cite{ston10} have reported such calculations for Mo, Ru, and Pd ions recoiling into vacuum with velocity $v/c \sim 0.05$, and more recently for $^{54,56}$Fe ions at $v/c \sim 0.08$ \cite{ston15}. This static model can then be improved by including the effect of atomic transitions, based on the calculated atomic level lifetimes, which Stone et al. also explored \cite{ston10,ston15}. Chen et al. \cite{chen13} have taken the calculations further by implementing a Monte Carlo method to evaluate the effect of atomic transitions and applying it to the tellurium isotopes. In their work, rather than an energy cut-off, the maximum number of electrons excited from the ground-state configuration is treated as a free parameter. The physical justification for this choice is that it determines an excitation energy regime for the ions with a physical basis. However, whether this strategy is appropriate in other cases is not known.

An important outcome of the work of Chen et al. is the recognition, based on atomic calculations, that many electronic states may have lifetimes comparable to or shorter than the nuclear lifetime. Consequently, atomic transitions may contribute to the observed average hyperfine interaction. Indeed, the empirical modeling of the hyperfine interactions for Te ions recoiling into vacuum with $v/c \sim 0.06$ reported in Ref.~\cite{stuc07a} indicated that there is evidence for several atomic transitions taking place on the time-scale of $\sim 10$~ps (see Fig.~7 of \cite{stuc07a}).

\begin{figure}[htb]
\includegraphics[width=\columnwidth]{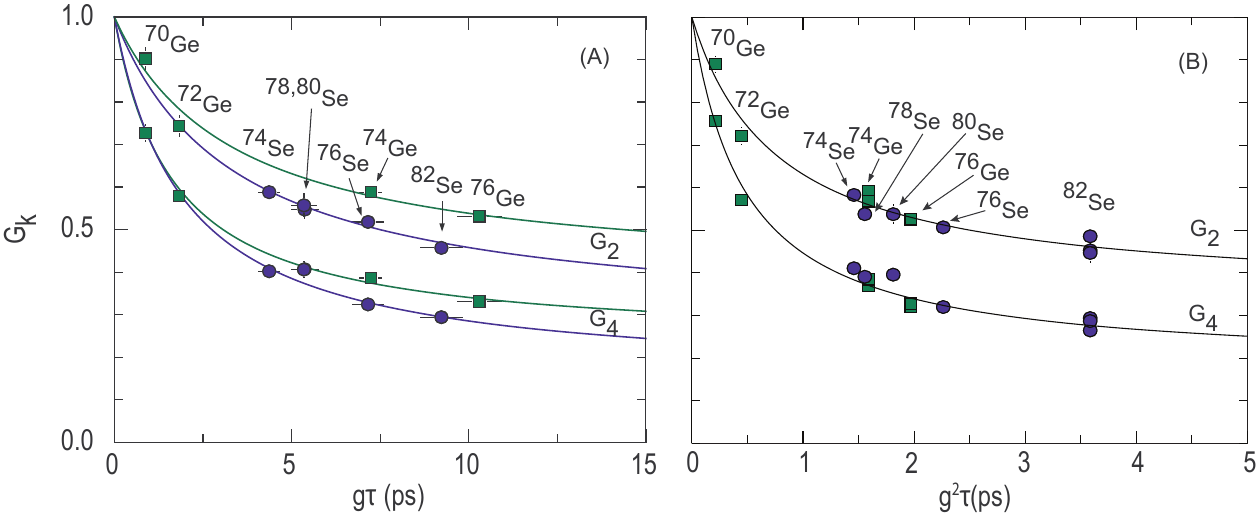}
\caption{\label{fig:GeSeGk}
Vacuum deorientation coefficients, $G_k(\infty)$, for Ge and Se ions with $v/c \sim 0.052$ recoiling in vacuum.  The solid lines are fits based on Eq.~(\ref{eq:lorentz-Gktau}). (A) $G_k(\infty)$ vs $g \tau$ as reported in 2012 \cite{stuc13a}. (B) $G_k(\infty)$ vs $g^2 \tau$ with updated $g$~factors from McCormick et al. \cite{mcco19}. The difference in behavior is due to plotting versus $g^2 \tau$ rather than $g \tau$, not to changes in the $g$~factors. The left panel (A) is taken and adapted from Ref.~\cite{stuc13a}.
}
\end{figure}

The main limitation that makes first-principles calculations impractical at present is that the initial population of atomic states, when the ion enters vacuum, is not well known. For example, the calculations described above do not account for the atomic structure effects that are seen in the free-ion hyperfine interactions for Ge and Se ions carrying $\sim 12 - 15$ electrons \cite{chen13a}, as reported in Ref.~\cite{stuc13a} and reproduced on the left side of Fig.~\ref{fig:GeSeGk}. The difference in the $G_k$ versus $g \tau$ dependence is prominent in the hard-core region, and the view in 2012 \cite{stuc13a} was that it apparently stems from a difference in the average atomic angular momentum in the range between $\bar{J} = 1$ and $\bar{J} = 2$ for Ge versus Se ions within about 10 ~ps of the ions entering vacuum. As evident in the right panel of Fig.~\ref{fig:Gkexamples}, the hard-core attenuation coefficients vary somewhat with atomic spin for $J < 2$ but become insensitive to $J$ for $J \geq 2$. Ref.~\cite{stuc13a} assumed that the interaction is primarily due to a static superposition of hyperfine fields and that there is a strong preference to populate lower-excited atomic states soon after the ions enter vacuum, which explained why, for these ions, that $\bar{J}$ is rather sensitive to the charge state of the ion. Further investigation and scrutiny was called for.

In 2018, McCormick et al. \cite{mcco19} reported new measurements of the $g$~factors of the 2$^+_1$ states in the stable Ge and Se isotopes by the transient-field technique. A primary motivation was to ensure that the differences observed between Ge and Se ions in Fig.~~\ref{fig:GeSeGk}a were not due to the adopted $g$~factors and any systematic difference between the previous measurements for Ge and Se ions. A simultaneous measurement on a cocktail beam of $^{74}$Ge and $^{74}$Se was performed to eliminate possible sources of systematic error.

McCormick et al. \cite{mcco19} confirmed the relative $g$~factors reported previously \cite{pako84,lamp87,gurd13,spei98} but proposed absolute values smaller than those used in Ref.~\cite{stuc13a} and hence in Fig.~\ref{fig:GeSeGk}A. The new values would contract the horizontal $g \tau$ scale somewhat, but the difference between the Ge and Se behavior would persist. In contrast, if the $G_k$ data are plotted versus $g^2 \tau$ rather than $g \tau$, as shown in Fig.~\ref{fig:GeSeGk}b, the difference between the behavior of the Ge and Se ions is no longer evident.

This dichotomy between a $g \tau$ versus a  $g^2 \tau$ dependence of the attenuation coefficients requires further investigation. The persistence of a hard-core in the $G_k$ values in Fig.~\ref{fig:GeSeGk} (evident for cases with longer nuclear lifetimes) requires a significant static component of the hyperfine interaction (unchanging electronic configurations), whereas a dependence on $g^2 \tau$ requires changes in the atomic configurations, i.e., atomic transitions. It is hardly surprising that the actual free-ion hyperfine interaction is neither static nor continually changing; excited ions leaving the target will decay, and transitions must cease when the ions reach the atomic ground state.

As noted above, precise modeling of the hyperfine interactions of free ions in vacuum is hampered by a lack of information on the excitation energies of the atomic configurations as the ion emerges from the target foil. A related complication is that charge-state distributions, which have been used as input for the calculations of the hyperfine interactions to date, are measured many nanoseconds, if not microseconds, after the ions emerge into vacuum. It is known from the strength of the transient field that ions in transit through a solid medium such as a target foil have deep inner-electron vacancies, and these must persist at least fleetingly after the ion emerges into vacuum, before the ion relaxes through a sequence of X-ray and Auger emissions. The measured charge state therefore corresponds to a time long after any charge-changing Auger emission cascade is complete, which could impact the effective hyperfine interaction by causing (i) a period of time during which atomic transitions must be considered, and/or (ii) a shift in the average charge state applicable in the period that the RIV hyperfine interaction takes place.

Thus, although there has been considerable progress, and some clear directions for improvement are identified, the point has not yet been reached whereby ab initio atomic calculations can reliably calculate the RIV hyperfine interaction for complex many-electron ions. Experimental data on similar atomic systems for isotopes with known $g$~factors are still needed to define the hyperfine interaction, and, in essence, the calibration procedures will remain largely empirical until ab initio calculations are demonstrated to account quantitatively for a more extensive set of data than has been examined to date. Fortunately, as demonstrated by the work to date, the RIV method based on empirical calibration procedures can yield measured $g$~factors of excited radioactive beams with sufficient precision to test nuclear models.

\subsubsection{Time-dependent RIV for complex ions}
\label{sect:TDRIV-complex}

Historically, measurements of time-dependent vacuum deorientation based on plunger measurements played an important role in determining the nature of the RIV interaction for complex many-electron ions  \cite{gold82,ward74,andr78}. So far, there have been few such measurements in the new millennium, however, the measurements of Radeck et al. \cite{rade12} on $^{98}$Ru and Naqvi et al. \cite{naqv14} on $^{138}$Ce and $^{142}$Ce demonstrate the potential of such measurements. Additional lifetime and RIV $g$~factor measurements using the plunger method have been performed on $^{120}$Te \cite{terr09}, $^{92,94}$Zr \cite{wern12} and $^{104,106,108}$Pd \cite{wern12a,wern12b}. These cases represent Coulomb excitation of stable nuclei and were directed towards the study of nuclear structure. For applications to $g$-factor measurements on radioactive beams, further such measurements on stable beams would be of considerable value in order to characterize the RIV interaction.

As will be noted in the next section \ref{sect:RIV-future}, the use of time-dependent RIV on radioactive beams carrying many electrons would be necessary in cases where the action of the hyperfine field on the nuclear state must be restricted so that the attenuation coefficient $G_k$ remains in the region sensitive to the nuclear $g$~factor, i.e., well away from the `hard-core' region. However, as also noted below, such measurements may be limited by available beam intensities.

As with the case of H-like ions discussed above in section \ref{sect:RIV-TDRIV},  the radioactive beam ions cannot be stopped in view of the $\gamma$-ray detectors. The use of an electron-configuration reset foil (or velocity degrader foil) is required. Experiments then become more challenging because the deorientation after the reset/degrader foil can be significant; it is expected to be very much reduced from the H-like case described above. The decays of the `slow' ions after the degrader are expected to be dominated by the hard-core value of $G_k$ and thus insensitive to the $g$~factor.

The only viable option may be to design the measurement so that the decays of the `fast' ions before the degrader can be separated (via the Doppler shift) from the decays of the `slow' ions after the degrader. In this way, the attenuation of the angular correlation of the `fast' component can be restricted to the time-regime where the RIV effect is most sensitive to the $g$~factor. For longer-lived nuclear states, the `fast' component may be a small fraction of the total decay intensity. Thus $\gamma$-ray detection that cannot distinguish the `fast' and `slow' components must be avoided.
For any given radioactive beam measurement, careful modeling of feasibility must be made before it is attempted. Pilot studies on stable beams of nearby isotopes, under similar measurement conditions, would be strongly recommended.

\subsubsection{Future work on RIV for complex ions }
\label{sect:RIV-future}

As RIV $g$-factor measurements can be performed in parallel with $B(E2)$ measurements by Coulomb excitation on beams of rare isotopes, there is scope for wide application of the technique. As it happens, the particle and $\gamma$-ray detector arrays (BareBall and CLARION) at Oak Ridge National Laboratory were near ideal for RIV measurements because the measurement of azimuthal angular correlations and the measurement relative to the summed intensity as represented in Eq.~(\ref{eq:wphinorm}) are important to factor out detector efficiencies and other possible systematic effects. With a view to future radioactive-beam measurements, primarily at the nuCARIBU-ATLAS facility of Argonne National Laboratory \cite{mcla22}, and the ReA3 facility of FRIB-MSU \cite{glas24}, a new particle-$\gamma$ detector array called CLARION2-TRINITY has been commissioned \cite{gray22}. These detectors improve upon CLARION-BareBall, but retain the features required for efficient RIV measurements.

\begin{figure}[htb]
\centering
\includegraphics[width=0.95\columnwidth]{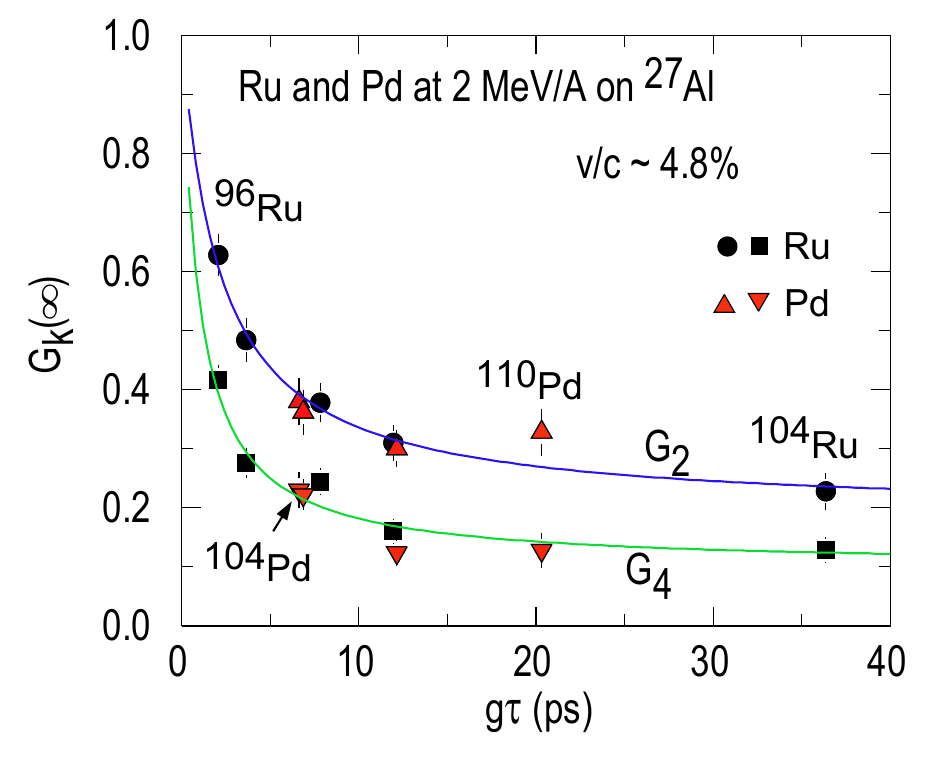}
\caption{\label{fig:RuPdRIV}  Vacuum deorientation coefficients ($G_k(\infty)$) for Ru and Pd ions with $v/c \sim 0.048$ recoiling in vacuum. To evaluate $g \tau$, the lifetimes were taken from Ref.~\cite{rama01} and the $g$~factors from Chamoli et al. \cite{cham11}. The solid lines are fits based on the empirical static model described in Sect. \ref{sect:RIV-complex-ions}, Eq.~(\ref{eq:staticM}).
The figure was prepared from data acquired for Ref.~\cite{lesl15}.
}
\end{figure}

To date, the RIV method on heavier radioactive beams has been applied primarily to nuclei near $^{132}$Sn. However, calibration data exist for $Z \sim 32$ (i.e., Ge and Se), as shown above in Fig.~\ref{fig:GeSeGk}, and $Z \sim 44$ (Ru and Pd), as shown in Fig.~\ref{fig:RuPdRIV}. Complementary time-dependent (plunger) data are also available for Ru and Pd ions \cite{rade12,wern12a,wern12b}.

\begin{figure}[htb]
\centering
\includegraphics[width=0.95\columnwidth]{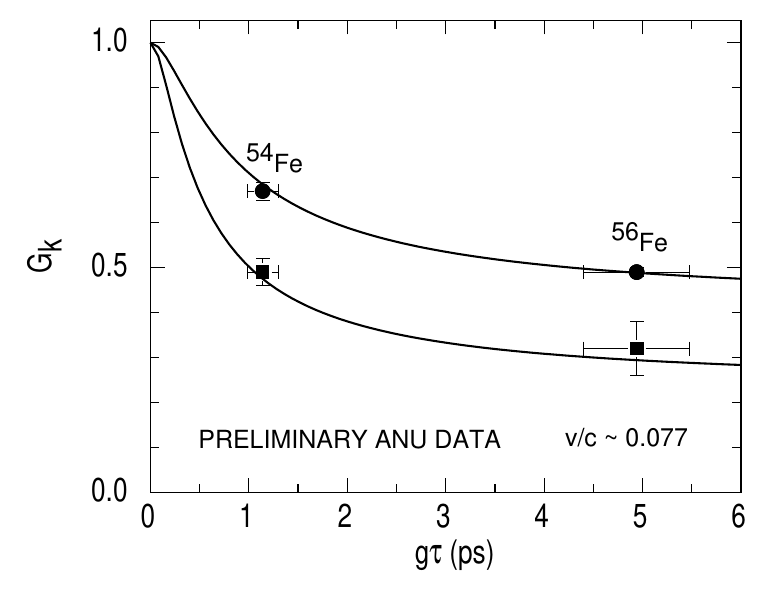}
\caption{\label{fig:GkFe} Vacuum deorientation coefficients, $G_k(\infty)$, for Fe ions with $v/c \sim 0.077$ recoiling in vacuum.
Lifetimes were taken from Ref.~\cite{rama01} and the $g$~factors from East et al. \cite{east09a}, giving $g \tau(^{54}{\rm Fe})=+1.14(16)$ and $g \tau(^{56}{\rm Fe})=+4.9(5)$. (The $g$~factors are known to be positive in this case.)
}
\end{figure}

To conclude this section, it is noted that all of the RIV $g$-factor measurements on radioactive beams to date have used reactions in inverse kinematics - a heavy beam is incident on a lower-mass target (often C) with the knocked-on target ions detected at forward angles. This reaction geometry corresponds to near head-on collisions in the center of mass frame. In many cases, it is more favorable to perform $B(E2)$ measurements by Coulomb excitation of the beam on a heavier target; the result is a glancing collision in both the center of mass and the lab frame. In anticipation of such measurements on radioactive beams near $Z=28$, the ANU group has performed some exploratory calibration measurements on beams of $^{54}$Fe and $^{56}$Fe at 200 MeV scattered on $^{\rm nat}$Ag. This beam energy corresponds to near-barrier (or near-grazing) collisions at the scattering angle. The Fe ions were scattered, leaving the target with $v/c \sim 0.077$, which corresponds to an average charge state of $\bar{q} \sim 19+$, or 7 electrons remaining on the ion. Results are shown in Fig.~\ref{fig:GkFe} along with an empirical static model fit as described above, see Eq.~(\ref{eq:staticM}). Stone et al. \cite{ston15} made schematic calculations for five-electron ions, which also reproduced the trend of the data. Under these conditions, the RIV measurement is sensitive to the $g$~factor if $\vert g \vert \tau \lesssim 2$. To make measurements for cases where $\vert g \vert \tau \gtrsim 2$, it will be necessary to reduce the recoil velocity (and hence the charge state) or otherwise restrict the time for which the hyperfine interaction acts on the nucleus. This might be possible, in principle, by use of a plunger device that slows but does not stop the ions. Count-rate considerations are likely to determine the feasibility of such measurements on rare isotope beams.

\section{ {Key results on nuclear structure of short-lived states} }
\label{sect:short-lived}

 {
In this section, key nuclear structure insights from $g$-factor measurements on short-lived states are reviewed. The focus is primarily on transient-field measurements on stable and near-stable nuclei, and at low spin.
Measurements on high-spin states populated in heavy ion reactions, which are often in the quasicontinuum above the near-yrast discrete states, are discussed in section~\ref{par::high-spin}. Superdeformed bands are discussed in section \ref{par::superdef}.
}

 {
Transient-field $g$-factor measurements following Coulomb excitation have been performed for stable nuclides across the nuclear chart. The data for even-even stable nuclei are almost complete (a few very low-abundance isotopes have not been measured). For stable odd-$A$ nuclei with $I=1/2$ ground states, measurements have also been completed; a few cases with $I=3/2$ have been measured, but to our knowledge, no cases with $I \geq 5/2$ have been published. The reason is that the particle-$\gamma$ angular correlation following Coulomb excitation becomes almost isotropic as the ground-state spin increases, as discussed at the end of section \ref{subsubsec:spin_orientations}.
}

 {
\subsection{An overview of transient-field \textit{g}-factor measurements in even-even nuclei with \texorpdfstring{$Z \leq 50$}{}}
\label{sect:TF-gfactors-Z.le.50}
}

 {
In this and the next subsections, the objective is to give an overview of transient-field $g$-factor measurements focused on those published from the year 2000 onward. Where relevant, earlier publications will be mentioned. The overview in this section includes measurements following Coulomb excitation of even-even nuclei up to $Z=50$ (i.e., Sn isotopes). Also included are the $\alpha$-transfer reaction cases, which use essentially the same technique but have the beam energy sufficiently close to the Coulomb barrier that $\alpha$ transfer occurs from a $^{12}$C target to the projectile nucleus.
Much of the experimental effort since the year 2000 is covered.
Measurements on higher-$Z$ nuclei, including the measurements on radioactive beams of $^{132}$Te \cite{benc08} and $^{126}$Sn \cite{kumb12a}, are discussed in section~\ref{sect:TF-gfactors-Z.gt.50} and in section~\ref{subsubsec:Te_RIV} .
}

 {
\paragraph{Transient-field \texorpdfstring{$g$}{}-factor measurements: \texorpdfstring{$10 \leq Z \lt 20$}{}}
The transient-field $g$-factor measurements performed on nuclei with $10 \leq Z \lt 20$ are displayed in Table~\ref{tab:TFZ=10-20}. This table includes results from before the year 2000 as well as more recent measurements. The aim is to show that a variety of procedures and parametrizations have been used to calibrate the transient-field strength. These procedures can be justified, but users of the data should be aware that relative $g$~factors may be more reliable than the absolute values. On the other hand, the precision of some of these measurements is limited, and concerns about the precise field calibration are secondary. Also, in cases like the 2$^+_1$ state of $^{40}$Ar, the experimental $g$~factor is consistent with zero, which challenges shell model calculations without the need for an accurate transient-field calibration.
}

\begin{table}[ht]
\caption{Transient-Field $g$-factor measurements $10 \leq Z < 20$.}
\label{tab:TFZ=10-20}
\begin{tabular}{|l|r|l|l|c|c|l|}
\hline
Nuclide & $E_x$  & $I^{\pi}$ & $g$ factor & ref. &  Host &TF calibration \footnotemark[1] \\
        & (keV) & & & & & \\
\hline

$^{20}$Ne  & 4247 & 4$^+$ &  $+0.38(8)$ & \cite{lesk03}  &  Fe & $g(2^+_1) = +0.54(4)$ \cite{hors75}\\
$^{22}$Ne  & 3357 & 4$^+$ &  $+0.55(14)$ & \cite{bazz84} &  Fe & $g(2^+_1) = +0.326(12)$ \cite{hors77} \\
$^{24}$Mg  & 4123 & 4$^+$ & $+0.4(3)$  & \cite{spei83} &  Fe & $g(2^+_1) = +0.51(2)$\cite{hors75}\\
$^{24}$Mg  & 4238 & 2$^+$ & $+0.6(2)$  & \cite{spei83} &  Fe & $g(2^+_1) = +0.51(2)$\cite{hors75}\\
$^{24}$Mg  & 6010 & 4$^+$ & $+0.5(4)$  & \cite{spei84} &  Fe & $g(2^+_1) = +0.51(2)$\cite{hors75}\\
$^{26}$Mg  & 1809 & 2$^+$ & $+0.86(10)$ & \cite{mcco18} &  Gd & $g(2^+_1;^{24}{\rm Mg}) = +0.538(13)$ \cite{kuso15} \\


$^{30}$Si  &  2235 & 2$^+$ & $+0.38(9)$ & \cite{zalm78} &  Fe & Local parametrization~\footnotemark[2]. \\

$^{31}$P  & 1266 & 3/2$^+$ & $+0.20(5)$  & \cite{holt82} &  Fe & $B_{\rm TF}=(238 \pm 10)v/v_0$~T~\cite{zalm78} \\
$^{31}$P  & 2234 & 5/2$^+$ & $+1.13(18)$ & \cite{holt82} &  Fe & $B_{\rm TF}=(238 \pm 10)v/v_0$~T~\cite{zalm78} \\

$^{32}$S  & 2230 & 2$^+$ & $+0.47(9)$  & \cite{zalm79} &  Fe & $B_{\rm TF}=(281 \pm 11)v/v_0$~T~\cite{zalm78} \\
$^{32}$S  & 2230 & 2$^+$ & $+0.44(10)$ & \cite{spei06} &  Gd & $B_{\rm TF}=(261 \pm 19)v/v_0$~T \footnotemark[3]\\

$^{32}$S  & 4459 & 4$^+$ & $+0.40(15)$ & \cite{simo88} &  Fe &  $g(2^+_1) = +0.47(9)$ \cite{zalm79} \\

$^{34}$S  & 2128 & 2$^+$ & $+0.50(8)$ & \cite{zalm79} &  Fe & $B_{\rm TF}=(281 \pm 11)v/v_0$~T~\cite{zalm78}  \\


$^{38}$S  & 1292 & 2$^+$ & $+0.13(5)$ & \cite{davi06,stuc06} &  Fe & HVTF \cite{stuc04} \\
$^{40}$S  & 904 & 2$^+$ & $-0.01(6)$ & \cite{davi06,stuc06} &  Fe &  HVTF \cite{stuc04} \\

$^{36}$Ar  & 1970 & 2$^+$ & $+0.52(18)$ & \cite{spei06} &  Gd &  $B_{\rm TF}=(294 \pm 21)v/v_0$~T \footnotemark[3] \\
$^{38}$Ar  & 2167 & 2$^+$ & $+0.24(12)$  & \cite{spei06} &  Gd & $B_{\rm TF}=(294 \pm 21)v/v_0$~T \footnotemark[3]\\
$^{38}$Ar  & 3937 & 2$^+$ & $+1.1(11) $  & \cite{spei06} &  Gd & $B_{\rm TF}=(294 \pm 21)v/v_0$~T \footnotemark[3]\\

$^{40}$Ar  & 1461 & 2$^+$ & $-0.1(1)$    & \cite{cub92} &  Gd &  $B_{\rm TF}=(4.9 \pm 1.5)$~kT\\
$^{40}$Ar  & 1461 & 2$^+$ & $-0.02(4)$ & \cite{stef05} &  Gd & RU \cite{shu80} \\
$^{40}$Ar  & 1461 & 2$^+$ & $-0.02(3)$    & \cite{spei08} &  Gd &  $B_{\rm TF}=(294 \pm 21)v/v_0$~T \footnotemark[3]\\

$^{39}$K   & 8030 & 19/2$^-$ & $+0.35(3)$ & \cite{pako92} & Gd & $g$(15/2$^+_1$; $^{41}$Ca) = $+0.296(17)$ \footnotemark[4] \\

 \hline
\end{tabular}

\footnotetext[1]{The reference $g$~factor is given in cases where the measurement was made relative to a known value in the same or a neighboring nucleus. For the linear parametrization and $10 \leq Z < 20$, see Eq.~(\ref{eq:lin-param}), there has been some variation in the adopted parameters, so the field calibration is indicated by $B_{\rm TF}=a(Z)v/v_0$, where the parameter $a(Z)$ depends on the atomic number. The high-velocity transient-field HVTF parametrization \cite{stuc04}, see Eq.~(\ref{eq:aes-param}), is designated HVTF.  The Rutgers parametrization \cite{shu80}, see Eq.~(\ref{eq:RUparam}), is designated by RU.
}

\footnotetext[2]{Empirical transient field (TF) calibration based on data for $^{20}$Ne, $^{24}$Mg, $^{28}$Si ions, assuming a theoretical value for $^{28}$Si: $g(2^+_1;^{28}{\rm Si}) = +0.53(2)$. Data from \cite{eber75} were re-evaluated. }
\footnotetext[3]{Evaluated from $B_{\rm TF}=a G_{\rm beam} Z (v/v_0)$ with $a=17(1)$~T and $G_{\rm beam}=0.96(4)$.}
\footnotetext[4]{ Average of $g$-factor values for the 3830-keV state in $^{41}$Ca from
\cite{youn75,uhrm75}.}

\end{table}

\begin{figure}[ht]
  \centering
  \includegraphics[width=0.65\linewidth]{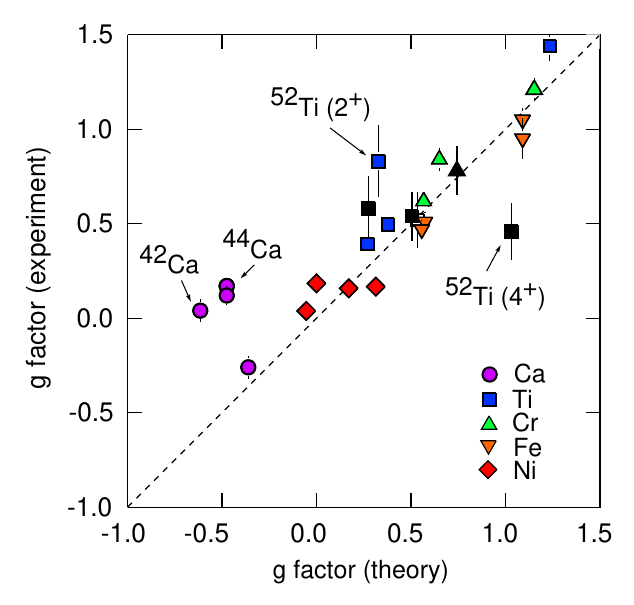}
  \caption{Transient-field $g$~factor measurements of 2$^+_1$ and $4^+_1$ states in the $fp$ shell compared with shell model calculations using the \texttt{GXPF1A} interaction \cite{honm04}. The $4^+$ $g$~factors are indicated by the black filled symbols. The data are from Table~\ref{tab:TFZ=20-28}. }
\label{fig:fp-gfac-th-vs-exp}
\end{figure}

 {
\paragraph{Transient-field \texorpdfstring{$g$}{}-factor measurements: \texorpdfstring{$20 \leq Z \leq 28$}{}}
Table \ref{tab:TFZ=20-28} lists the $g$~factors measured by the transient-field technique in even-even nuclei from $Z=20$ to $Z=28$. Whereas the valence protons in these nuclei span the $0f_{7/2}$ shell, the valence nucleons are within the full $fp$ shell: $0f_{7/2}$, $1p_{3/2}$, $0f_{5/2}$, and $1p_{1/2}$. With the exception of $^{52}$Ti, the $g(2^+_1)$ values of the Ti, Cr, and Fe isotopes have been plotted and compared with shell-model theory in Fig.~\ref{fig:fp-gfactors} above. In Fig.~\ref{fig:fp-gfac-th-vs-exp} the $g(2^+_1)$ and $g(4^+_1)$ values from Table~\ref{tab:TFZ=20-28} are compared with shell model calculations in the full $fp$ shell, using the \texttt{GXPF1A} interaction \cite{honm04,honm05}. The effective $g$~factors are $g_{\ell}(\pi) = +1.10$, $g_{s}(\pi) = +5.027$, $g_{\ell}(\nu) = -0.10$, and $g_{s}(\nu) = -3.443$. Note that the effective spin $g$~factors for both protons and neutrons are quenched to 0.9 times the free nucleon values. These effective $g$~factors were estimated by Honma et al. \cite{honm04} based on a least-squares fit that included precise measurements of ground-state moments. For the Ca isotopes, only the ground states of $^{47}$Ca and $^{49}$Ca were included.
}

 {
It is evident that the calculations generally give a reasonable description of the data, particularly for $^{44-50}$Ti, $^{50-54}$Cr, and $^{54-58}$Fe. The failure of the $fp$-shell model calculations for $^{42,44}$Ca is understood as due to mixing of the spherical shell model configuration with deformed states built on multiparticle-multihole excitations of the $^{40}$Ca core \cite{schi03a,tayl03,tayl05}. See also Ideguchi et al. \cite{ideg22} for additional discussion on shape co-existence in $^{40}$Ca. The review of Stuchbery and Wood \cite{stuc22} identifies the deformed bands in $^{42,44,46}$Ca and shows that in $^{42}$Ca the $g$~factor, $E2$ matrix elements, and one-neutron transfer cross sections all imply that the $2^+_1$ state has a near 50\% collective admixture.
}

 {
While the discrepancy between the shell model calculations in the $fp$ shell and the $g(2^+_1)$ values in the Ca isotopes is understood, the experimental $g(2_1^+)/g(4_1^+)$ ratio in $^{52}$Ti is contrary to theory, and this discrepancy is harder to understand. Speidel et al. \cite{spei06a} were able to account for it by raising the single particle energies of the neutron $1p_{3/2}$ and $1p_{1/2}$ levels in the \texttt{FPD6} interaction \cite{rich91}, however the implications of this change for other observables in $^{52}$Ti and neighboring nuclei were not fully explored. Further investigation is warranted.
}

\begin{table}[ht]
\caption{Transient-field $g$-factor measurements $20 \leq Z \leq 28$.}
\label{tab:TFZ=20-28}
\begin{tabular}{|l|r|l|l|c|c|l|}
\hline
Nuclide & $E_x$  & $I^{\pi}$ & $g$ factor \footnotemark[1]  & ref. &  Host &TF calibration \footnotemark[2] \\
        & (keV)  & & & & & \\
\hline

$^{42}$Ca  & 1525 & 2$^+$ &  $+0.04(6)$ & \cite{schi03a}  &  Gd & Bonn; $G_{\rm beam}$=0.90(5) \\

$^{44}$Ca  & 1157 & 2$^+$ &  $+0.17(3)$ & \cite{schi03a}  &  Gd & Bonn;  $G_{\rm beam}$=0.90(5) \\

           &      &       &  $+0.12(5)$ & \cite{tayl03}  &  Gd & RU \\

$^{46}$Ca  & 1346 & 2$^+$ &  $-0.26(6)$ & \cite{tayl05}  &  Gd & $g(2^+_1;^{46}\mathrm{Ti})=+0.496(27)$ \cite{erns00a} \\

           &      &       &  $-0.19(12)$ & \cite{spei03}  &  Gd & Bonn; $G_{\rm beam}$=0.90(5)  \\
  & & & & & &\\

 $^{44}$Ti  & 1083 & 2$^+$ &  $+0.52(15)$ & \cite{schi03}  &  Gd & Bonn; $G_{\rm beam}$=0.90(5)  \\
 $^{46}$Ti  & 889 & 2$^+$ &  $+0.496(27)$ & \cite{erns00a}  &  Gd & Bonn;  $G_{\rm beam}$=0.83(4)  \\
            & 2010 & 4$^+$ &  $+0.58(17)$ & \cite{erns00a}  &  Gd & Bonn; $G_{\rm beam}$=0.83(4)  \\
 $^{48}$Ti  & 984 & 2$^+$ &  $+0.392(19)$ & \cite{erns00a}  &  Gd & Bonn;  $G_{\rm beam}$=0.83(4)  \\
            & 2296 & 4$^+$ &  $+0.54(13)$ & \cite{erns00a}  &  Gd & Bonn; $G_{\rm beam}$=0.83(4)  \\
 $^{50}$Ti  & 1554 & 2$^+$ &  $+1.44(8) $ & \cite{spei00}  &  Gd & Bonn; $G_{\rm beam}$=0.83(4)  \\
 $^{52}$Ti  & 1050 & 2$^+$ &  $+0.83(19)$ & \cite{spei06a}  &  Gd & Bonn;  $G_{\rm beam}$=0.90(5)  \\
            & 2318 & 4$^+$ &  $+0.46(15)$ & \cite{spei06a}  &  Gd & Bonn;  $G_{\rm beam}$=0.90(5)  \\
  & & & & & &\\
$^{50}$Cr  & 783 & 2$^+$ &  $+0.619(31)$ & \cite{erns00a}  &  Gd & Bonn; $G_{\rm beam}$=0.83(4)  \\
           & 1881 & 4$^+$ &  $+0.78(13)$ & \cite{erns00a}  &  Gd & Bonn; $G_{\rm beam}$=0.83(4)  \\

$^{52}$Cr  & 1434 & 2$^+$ &  $+1.21(6)$ & \cite{erns00a}  &  Gd & Bonn;  $G_{\rm beam}$=0.83(4)  \\

$^{54}$Cr  & 835 & 2$^+$ &  $+0.84(6)$ & \cite{wagn01}  &  Gd & Bonn;  $G_{\rm beam}$=0.83(4)  \\
  & & & & & &\\
$^{54}$Fe  & 1408 & 2$^+$ &  $+1.05(6)$ & \cite{spei00}  &  Gd & Bonn; $G_{\rm beam}$=0.83(4)
\footnotemark[3]\\
           &      &       &  $+0.95(11)$ & \cite{east09a}  &  Gd &  $g(2^+_1;^{56}{\rm Fe})=+0.51(5)$ \cite{east09} \\

$^{56}$Fe  & 847 & 2$^+$ &  $+0.51(5)$ & \cite{east09}  &  Gd &  $g(5/2^-_1;^{57}{\rm Fe})=+0.374(4)$ \cite{fahl79} \\
$^{58}$Fe  & 811 & 2$^+$ &  $+0.47(5)$ & \cite{east09a}  &  Gd &  $g(2^+_1;^{56}{\rm Fe})=+0.51(5)$ \cite{east09} \\
  & & & & & &\\
$^{58}$Ni  & 1454 & 2$^+$ &  $+0.038(9)$ & \cite{kenn00,kenn01}  &  Gd &  Bonn; $G_{\rm beam}$=0.69(6) \\
$^{60}$Ni  & 1332 & 2$^+$ &  $+0.158(28)$ & \cite{kenn00,kenn01}  &  Gd &  Bonn; $G_{\rm beam}$=0.69(6) \\
$^{62}$Ni  & 1173 & 2$^+$ &  $+0.167(24)$ & \cite{kenn00,kenn01}  &  Gd &  Bonn;  $G_{\rm beam}$=0.69(6) \\
$^{64}$Ni  & 1346 & 2$^+$ &  $+0.184(31)$ & \cite{kenn00,kenn01}  &  Gd &  Bonn;  $G_{\rm beam}$=0.69(6) \\

 \hline
\end{tabular}

\footnotetext[1]{Apart from rounding, the values and uncertainties are quoted as in the original reference. }
\footnotetext[2]{The Rutgers parametrization \cite{shu80}, see Eq.~(\ref{eq:RUparam}), is designated by RU.  For the Bonn linear parametrization, $B_{\rm TF}=a G_{\rm beam} Z (v/v_0)$ with $a=17(1)$~T. The adopted $G_{\rm beam}$values are given. The reference $g$~factor is given in cases where the measurement was made relative to a known value.}
\footnotetext[3]{The transient-field strength calibration is stated to be the same as in Refs.\cite{erns00a,erns00}.}
\end{table}

 {
\paragraph{Preamble on Transient-field \texorpdfstring{$g$}{}-factor measurements for \texorpdfstring{$Z=30$ to $Z=48$}{}, Zn - Cd}
}

\begin{figure}
  \centering
  \includegraphics[width=0.7\linewidth]{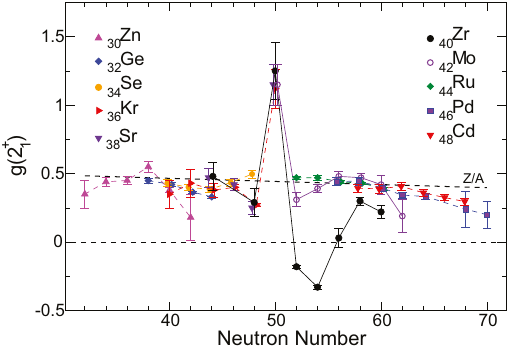}
  \caption{The $g(2^+_1)$ values for isotopic chains on both sides of the closed neutron shell $N = 50$. The exact $Z/A$ varies with $Z$; the dashed line shown represents an average value for the given range of nuclei. The figure is taken from \cite{kumb14}. A complete list of references for the data was not given there. Original references may be found in the present discussion on nuclei with $30 \leq Z \leq 48$.}
\label{fig:Zn-Cd-2plus}
\end{figure}

 {
The 2$^+_1$-state $g$~factors from transient-field measurements are displayed in Fig.~\ref{fig:Zn-Cd-2plus} for the even-even nuclides from $Z=32$ to $Z=48$. These data,  along with $g$-factor measurements on the higher-excited $2^+_2$, $4^+_1$ and 3$^-_1$ states where measured, will be discussed in the following paragraphs. A survey of the $g(2^+_1)$ systematics from Zn to Sr published by Mertzimekis et al. \cite{mert03} in 2003 found that simple collective models could describe the data for isotopes below $N=48$. For the $N=48$ and $N=50$ isotones, however, shell model calculations were required. This generalization remains valid, although, as will be discussed in the following paragraphs,  large basis shell model calculations are now feasible for most of the cases displayed in Fig.~\ref{fig:Zn-Cd-2plus}. The present survey also extends to collective nuclei above $N=50$.
}

 {
\paragraph{Transient-field \texorpdfstring{$g$}{}-factor measurements: \texorpdfstring{$Z=30$}{}, Zn}
}

 {
The even Zn isotopes have been studied extensively by the transient-field method:
$^{62}$Zn \cite{kenn02};
$^{64}$Zn \cite{kenn02,lesk05};
$^{66}$Zn \cite{kenn02,lesk06};
$^{68}$Zn \cite{kenn02,lesk05,lesk05b,bout07,mosc10};
$^{70}$Zn \cite{kenn02,much09};
$^{72}$Zn \cite{fior12,illa14}.
}

\begin{figure}[ht]
  \centering
  \includegraphics[width=0.5\linewidth]{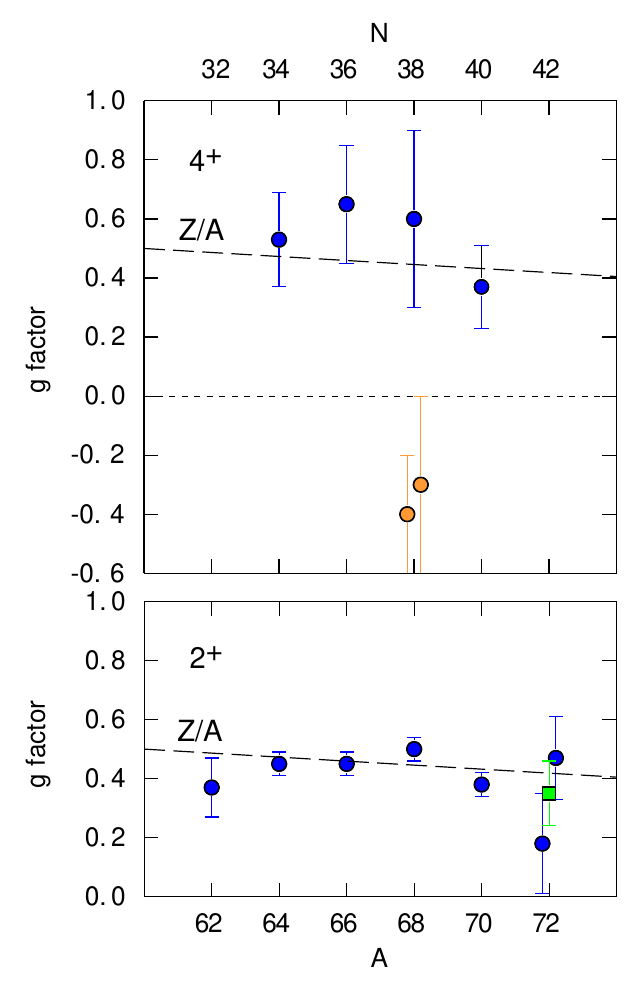}
  \caption{Transient-field $g$~factor measurements of the 2$^+_1$ and 4$^+_1$ states in Zn isotopes. The data are from \cite{kenn02,lesk05,lesk06,lesk05b,bout07,mosc10,much09,fior12,illa14}. Two measurements on $^{72}$Zn by the HVTF \cite{fior12} and LVTF \cite{illa14} methods are shown with their average (green square). The negative $g(4^+_1)$ values for $^{68}$Zn \cite{lesk05,lesk05b}, which do not fit the systematics and are difficult to explain theoretically, are discussed in the text.
}

\label{fig:ZnTFgactors}
\end{figure}

 {
A feature of these transient-field measurements on the Zn isotopes is the measurement of excited-state $g$~factors above the $2^+_1$ state. The $4^+_1$ level $g$~factors have been measured in all even-even stable isotopes (i.e., from $^{64}$Zn to $^{70}$Zn). The $g$~factors of the $2^+_1$ and $4^+_1$ states in the Zn isotopes are displayed in Fig.~\ref{fig:ZnTFgactors}. A surprise was the observation of near-constant $g$~factors as a function of mass number with values near the collective estimate $Z/A$. Specifically, there is no evidence of a neutron subshell closure at $N=40$, which would be expected to increase the $g(2^+_1)$ and $g(4^+_1)$ values in $^{70}$Zn. If anything, these $g$~factors at $N=40$ fall below the trend observed for the lighter isotopes.
}

 {
Shell model calculations have been performed in the $fp$-shell basis (i.e. a $^{40}$Ca core with $0f_{7/2}$, $1p_{3/2}$, $0f_{5/2}$, and $1p_{1/2}$ orbits for both protons and neutrons). These calculations are clearly inadequate for $N > 40$ and, as noted, also fail at $N=40$ for $^{70}$Zn \cite{kenn02}. Additional calculations with a $^{56}$Ni core and $1p_{3/2}$, $0f_{5/2}$, $1p_{1/2}$ and $0g_{9/2}$ orbits for both protons and neutrons were more successful \cite{kenn02}. It has been noted that calculations with a $^{48}$Ca core and $0f_{7/2}$, $1p_{3/2}$, $0f_{5/2}$ for protons, and $1p_{3/2}$, $0f_{5/2}$, $1p_{1/2}$ and $0g_{9/2}$ for neutrons might be more appropriate, but these were not pursued due to limits on computational power at the time  \cite{kenn02}.
}

 {
Two measurements \cite{lesk05,lesk05b} obtained a negative $g$~factor for the $4^+_1$ state in $^{68}$Zn, which was interpreted as evidence of a dominant neutron $\nu (0g_{9/2})^2$ contribution to its wavefunction \cite{lesk05b}. But this result was difficult to reconcile with shell model calculations and positive $g(4^+_1)$ values in the neighboring isotopes. A new measurement gave a positive $g$~factor $g(4^+_1) = +0.6(3)$ \cite{bout07}, and reanalysis of the earlier data gave $g(4^+_1) = +0.14(13)$ \cite{mosc10}. Transient-field $g$-factor measurements with sufficient precision give an unambiguous measurement of the sign. The problem here is likely due to statistical limits on the extraction of the so-called `effect', $\epsilon_{TF}$; see Eq.~(\ref{eq:epsilon}).
}

 {
The neutron-rich case of $^{72}$Zn was measured both by the high-velocity transient-field (HVTF) technique at GANIL \cite{fior12} and by the low-velocity transient-field (LVTF) technique at ISOLDE \cite{illa14}. As seen in Fig.~\ref{fig:ZnTFgactors}, the two measurements agree at the  $1\sigma$ limits, with an average value of $g(2^+_1)=+0.35(11)$. This average value is near that of $g(2^+_1)=+0.38(4)$ in $^{70}$Zn, which is consistent with the $g$-factor trend observed for these isotones in the Ge isotopes (see Fig.~\ref{fig:Ge-rel-g} below).
}

 {
An important outcome of the HVTF measurement \cite{fior12} on $^{72}$Zn was the clear evidence that the transient field does not increase as initially hoped for very high velocity heavy ions (due to reduced transfer of polarization from the ferromagnetic host to the rapidly moving ion, as discussed above in section \ref{sect:TF-physics} and as displayed in Fig.~\ref{fig:TF-p1s}). The subsequent LVTF measurement \cite{illa14} allows for a comparison of the two methods. In terms of the transient-field strength and consequent magnitude of the precession angle, HVTF and LVTF are roughly equivalent at around $Z=30$. The evidence from the comparison of these measurements is that LVTF is to be preferred for $Z \gtrsim 30$.
}

 {
\paragraph{Transient-field \texorpdfstring{$g$}{}-factor measurements: \texorpdfstring{$Z=32,34$}{}, Ge and Se}
}

 {
The $g$~factors of the Ge ($Z=32$) and Se ($Z=34$) isotopes most recently measured by the transient-field technique include the radioactive nuclide $^{68}$Ge \cite{lesk05a} populated by $\alpha$-transfer from a $^{12}$C target, and all of the stable even-even isotopes $^{70-76}$Ge \cite{lesk06a,bout07a,gurd13,mcco19}
and $^{74-82}$Se \cite{mcco19,spei98}.
All of the measurements on stable isotopes used inverse kinematics (Coulomb excitation of the beam) and, like for the Zn isotopes, there has been a focus on measurements of states above the 2$^+_1$ state \cite{gurd13,spei98}. The study of McCormick et al. \cite{mcco19} was not focused on higher-excited states, but rather on measuring the relative $g$~factors in the stable Ge and Se isotopes. To this end, it included a measurement using a cocktail beam of $^{74}$Ge and $^{74}$Se.
}

 {
McCormick et al. showed excellent agreement between their {\em relative} $g$~factors and the relative $g$~factors of all of the previous transient-field measurements on the Ge and Se isotope chains, as shown in Fig.~\ref{fig:Ge-rel-g} and Fig.~\ref{fig:Se-rel-g}.
However, there is a very significant difference between the absolute $g$~factors proposed by McCormick et al. \cite{mcco19} on one hand, and those of G\"urdal et al. \cite{gurd13} and Speidel et al. \cite{spei98} on the other. It may be relevant that McCormick et al. \cite{mcco19} used iron as the ferromagnetic host, whereas G\"urdal et al. \cite{gurd13} and Speidel et al. \cite{spei98} used gadolinium.
}

\begin{figure}
  \centering
  \includegraphics[width=0.65\linewidth]{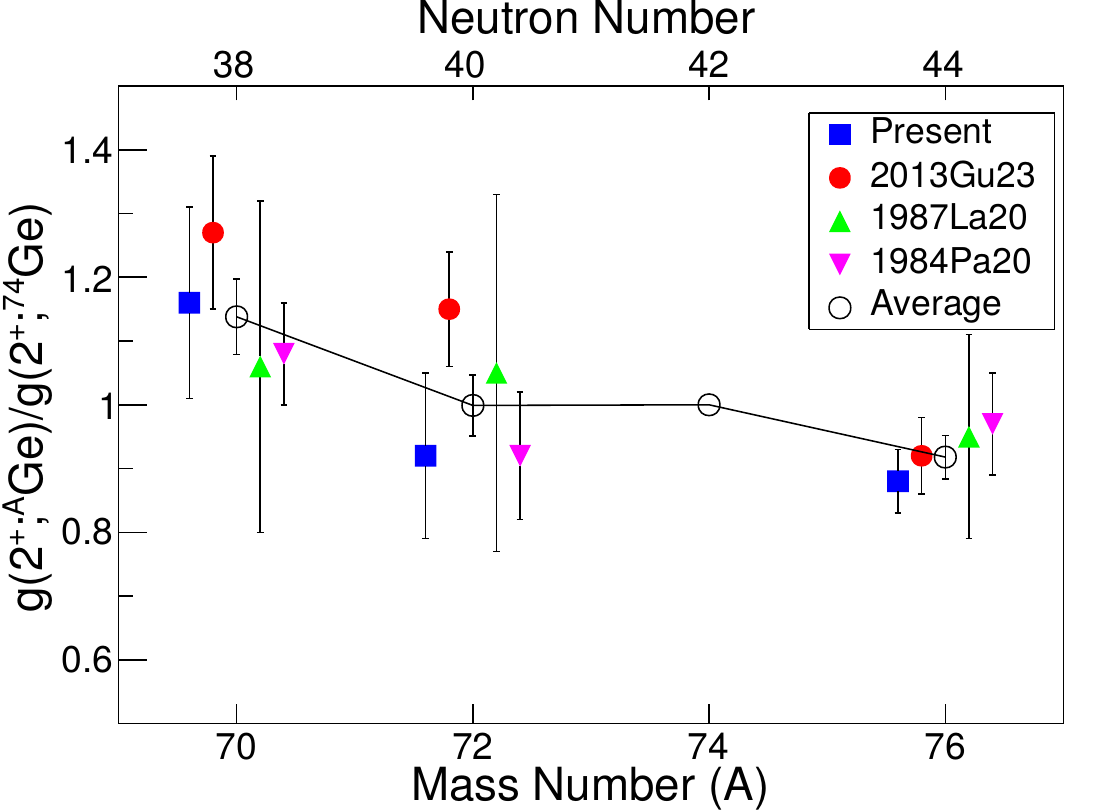}
  \caption{ Ratios of $g(2^+_1)$ in the Ge isotopes relative to $^{74}$Ge, from \cite{mcco19} (`Present'), 1984Pa20 \cite{pako84}, 1987La20 \cite{lamp87}, and
 2013Gu23 \cite{gurd13}.  The figure is taken from Ref.~\cite{mcco19}.}
\label{fig:Ge-rel-g}
\end{figure}

\begin{figure}
  \centering
  \includegraphics[width=0.65\linewidth]{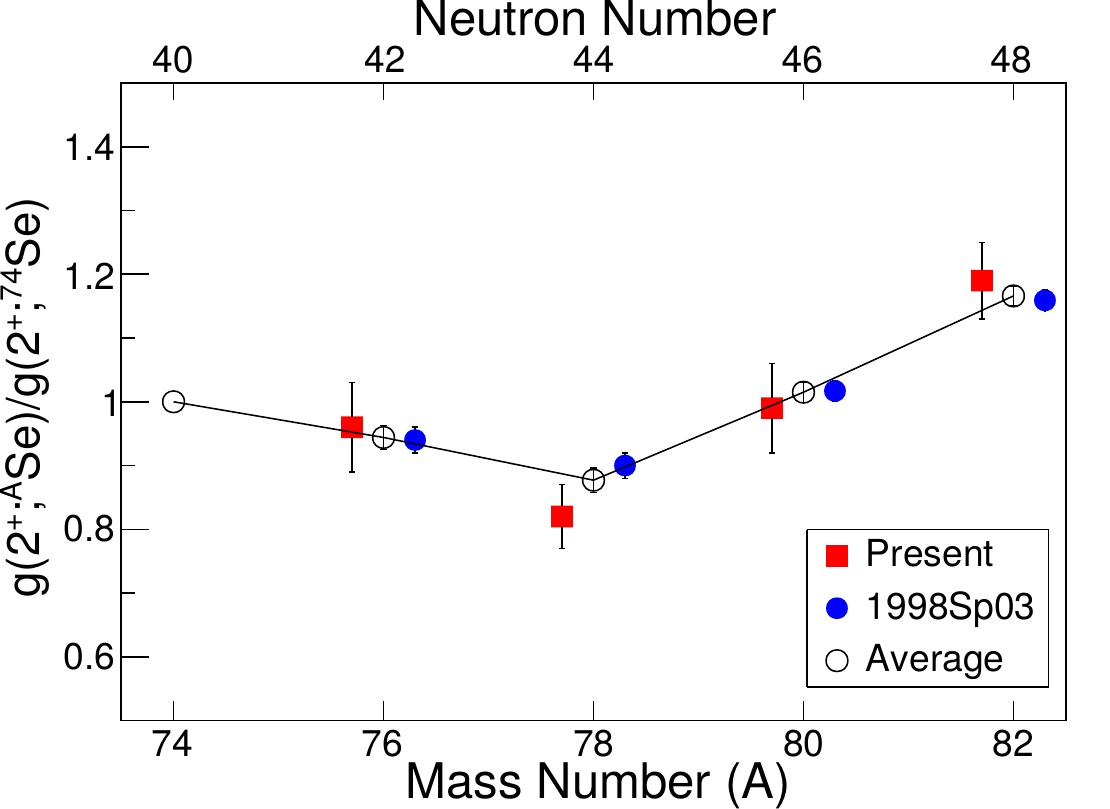}
  \caption{ Ratios of $g(2^+_1)$ in the Se isotopes relative to $^{74}$Se, from
 \cite{mcco19} (`Present')  and \cite{spei98} (1998Sp03).  The figure is taken from Ref.~\cite{mcco19}.}
\label{fig:Se-rel-g}
\end{figure}

 {
To set the absolute scale for both the Ge and Se isotopes, McCormick et al. performed a fit that included the Rutgers parametrization \cite{shu80} and the only independently determined $g$ factor among the cases under study, namely $g(2^+_1)$ in $^{76}$Se \cite{murr67}.  The $g$~factor from \cite{murr67} was adjusted for newer lifetime and hyperfine field values; see \cite{mcco19}. As it transpired, the resultant $g$ factor of $^{76}$Se agreed precisely with the magnitude of that from Ref.~\cite{murr67}.
}

 {
G\"urdal et al. \cite{gurd13} used the Rutgers parametrization to determine their absolute $g$~factors in the Ge isotopes. For the Se isotopes Speidel et al. \cite{spei98} normalized to $g(2^+_1;^{82}{\rm Se})=+0.496(29)$, citing Brennan et al. \cite{bren78} as the source of this calibration $g$~factor. However, in that 1978 publication the value is $g(2^+_1;^{82}{\rm Se})=+0.40(4)$, taken from an even older 1969 IMPAC measurement \cite{hees69}, where the value is given as $g(2^+_1;^{82}{\rm Se})=+0.42(12)$. It is not clear how the adopted value is derived from the original measurement.
}

 {
Uncertainties on the calibration of the transient field are most acute in this region of the nuclear chart. A precise independent measure of a suitable calibration $g$~factor in one of the Ge or Se isotopes is sorely needed.
As noted above in sections \ref{sect:TF-param}, \ref{sect:TF_prospects} and \ref{sect:RIV-TDRIV}, time-dependent recoil in vacuum measurements on Na-like Ge and Se ions appear to be the most promising path to address this problem. Preliminary data on $^{76}$Ge \cite{mcco21}, not specifically tuned to optimize Na-like ions, encourage this pursuit.
}

 {
Turning now to the nuclear structure of the Ge and Se isotopes, it can be noted that this region is challenging for the shell model because of its transitional nature and because shape-coexisting structures are present at low-excitation energies in some of the isotopes. Nevertheless, shell model calculations have been performed in the basis space of $1p_{3/2}$, $0f_{5/2}$, $1p_{1/2}$ and $0g_{9/2}$ for both protons and neutrons \cite{gurd13,mcco19}. Moreover, recent Coulomb excitation measurements \cite{ayan23} and electromagnetic properties deduced from $(n,n^{\prime})$ measurements \cite{mukh17} for excited states in $^{76}$Ge have shown good agreement with such shell model calculations. McCormick et al. found that the $g$~factors of the 2$^+_1$ states scale essentially with the total orbital angular momentum of the protons, a view that is consistent with collective models. Moreover, it was found that the minimum 2$^+_1$-state $g$~factor occurs at or near $N=44$, which is the midpoint of the combined $1p_{1/2}$-$0g_{9/2}$ shell ($38 \le N \le 50$) and agrees with experiment as seen most clearly for the Se isotopes (Fig.~\ref{fig:Se-rel-g}).
}

 {
The $g$~factors of the 4$^+_1$ and 2$^+_2$ states have been measured in all of the stable Ge and Se isotopes apart from $^{82}$Se, where the 2$^+_2$-state $g$~factor is yet to be measured. Unfortunately, the precision is limited; there is little evidence that the $g$~factors of the 2$^+_2$ and 4$^+_1$ states differ from those of the 2$^+_1$ states. However, the $g$~factors of these three low-excitation states must become essentially identical and near $Z/A$ as the excitations become collective. The data are therefore consistent with expectations.
}

 {
\paragraph{Transient-field \texorpdfstring{$g$}{}-factor measurements: \texorpdfstring{$Z=36-40$}{}, Sr, Kr and Zr}
}

\begin{figure}
  \centering
  \includegraphics[width=0.75\linewidth]{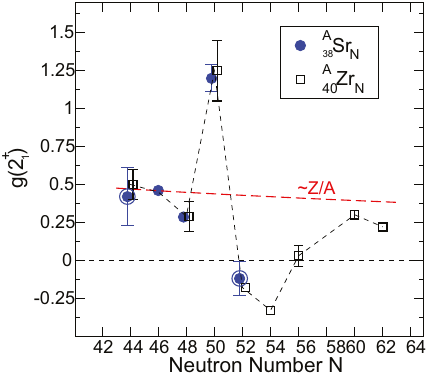}
  \caption{ Comparison of the 2$^+_1$-state $g$~factors in the Zr  and Sr isotopic chains. The figure is from \cite{kumb14} where new results for $^{82}$sr and $^{90}$Sr ($N=44,52$) were measured. These results are highlighted. The similarity of the $g$~factors in the Sr and Zr sequences is striking.}
\label{fig:ZrSrg}
\end{figure}

 {
The Kr, Sr, and Zr isotopes for which transient-field $g$-factor measurements have been performed span the $N=50$ shell closure, as shown in Fig.~\ref{fig:Zn-Cd-2plus}. These nuclei show less collectivity as $Z=40$ is approached, and the Zr and Sr isotopes show the most pronounced variations around $N=50$; see Fig.~\ref{fig:ZrSrg}.
}

 {
The following discussion in this section includes only measurements on low-excitation states in Kr, Sr, and Zr isotopes by Coulomb Excitation and $\alpha$-transfer reactions in inverse kinematics. There have been several transient-field measurements on high-spin states following heavy-ion-induced reactions in this region \cite{ward81,bill93,kuch89,yuan08a,fan09,fan10,yuan10,yuan05,moun92,zhus00,zhu01,moun95,weis95,yuan07,fan15,jung98,teic99}. These are discussed in section~\ref{par::high-spin}.
}

 {
The low-excitation $g$ factors of the stable even-even Kr isotopes $^{78}$Kr - $^{86}$Kr were measured by Mertzimekis et al. \cite{mert01}. As well as the 2$^+_1$ states, the $g$~factors of the 4$^+_1$ states in $^{78,80,82}$Kr and those of the 2$^+_2$ states in $^{78,80}$Kr were reported. The first radioactive beam transient-field $g$-factor measurement on $^{76}$Kr \cite{kumb04} included a more precise precession measurement on $^{78}$Kr. Additional $g$-factor measurements on $^{78}$Kr and $^{86}$Kr were reported as a by-product of a study on $^{82}$Sr and $^{90}$Sr using $\alpha$-transfer reactions \cite{kumb14}.
}

 {
Concerning the Sr isotopes, in addition to the study on $^{82}$Sr and $^{90}$Sr \cite{kumb14} just mentioned, Kumbartzki et al. \cite{kumb12} have measured $g$~factors of low-excitation states in $^{84,86,88}$Sr. The new measurements of $g(2^+_1)$ agree with the early measurements of Kucharka et al. \cite{kuch88} but have higher precision. As well as measurements for the 2$^+_1$-states, 4$^+_1$-state $g$~factors were measured in $^{82,86,90}$Sr, albeit with significant uncertainties.
}

 {
The transient-field method has provided measurements of excited state $g$~factors in $^{88}$Zr \cite{kumb12}, $^{90}$Zr \cite{jako00}, $^{92,94}$Zr \cite{jako99,wern08} and $^{96}$Zr \cite{kumb03}. In all cases, $g(2^+_1)$ was measured. In addition, $g(4^+_1)$ was measured in $^{88}$Zr, both $g(4^+_1)$ and $g(2^+_2)$ were measured in $^{92,94}$Zr, and $3^-_1$-state $g$~factors were measured in $^{90}$Zr and $^{96}$Zr \cite{jako99,kumb03}.
}

 {
As seen in Fig. \ref{fig:Zn-Cd-2plus} and Fig.~\ref{fig:ZrSrg} the 2$^+_1$-state $g$~factors of the $N=50$ isotones have values near $g = +1.25$, consistent with primarily $\pi (0g_{9/2})^2$ (i.e. seniority two) structure. At $N=52$ and $N=48$ the $g$~factor tends to drop well below the collective $Z/A $ value, going negative in $^{90}$Sr and $^{92}$Zr. Then, as successive pairs of neutrons are either added or removed, the $g$ factors trend toward $Z/A$. This behavior is most pronounced for the Zr isotopes, where in $^{92,94}$Zr $2^+_1$ and $4^+_1$ apparently have dominant neutron $\nu (1d_{5/2})^2$ character. The $g$~factors of the 2$^+_2$ states were measured to investigate their structure as possible mixed-symmetry states \cite{wern08}. It was striking to find that these states have a primarily proton character.
}

 {
This behavior of the lowest few states in the even-even nuclides adjacent to doubly magic (or approximately doubly magic) nuclei is strongly influenced by the relative excitation energies in the separate proton and neutron systems and the fact that the proton-neutron coupling is relatively weak in these nuclei.
}

 {
To illustrate the importance of the relative excitation energies of the parent proton and neutron excitations in the case of $^{92}$Zr consider it to have a $^{88}$Sr core as a first approximation and compare the proton excitation $E(2^+_1)=2186$ keV in $^{90}$Zr (= $^{88}{\rm Sr} + 2p$) with the neutron excitation $E(2^+_1)=832$ keV in $^{90}$Sr (= $^{88}{\rm Sr} + 2n$). If the interaction between the proton and neutron spaces is relatively weak, the $2^+_1$ state in $^{92}$Zr must be dominated by the neutron configuration, and the $2^+_2$ state must be dominated by the proton configuration. This mechanism as related to $g$ factors was noticed for the Mo \cite{mant01,stuc01} and Zr isotopes \cite{stuc01} prior to the measurements of Werner et al. \cite{wern08}. It is noteworthy that, despite being among the most challenging transient-field measurements yet attempted and having somewhat limited precision, the results of  Werner et al. show the power of the $g$ factors to reveal key elements of nuclear structure.
}

 {
The $g$~factors of the 3$^-_1$ states in $^{90}$Zr and $^{96}$Zr have values near unity, pointing to a proton-dominated structure. In fact the $g$-factor value suggests a $\pi  [1p_{3/2} 0g_{9/2}]_{3^-}$ configuration, in contrast to the $g(3^-_1)\sim Z/A \sim 0.4$ values observed in the Zn isotopes above. The 3$^-_1$-state $g$~factors in these Zr isotopes present a puzzle because they imply a single-particle configuration, whereas the experimental $E3$ transition rates imply collective structures that cannot be readily explained by shell model calculations. As a possible resolution of this problem, in Ref.~\cite{stuc04a} it was shown that a consistent description of the $g$~factors and $E3$ transition strengths in $^{90}$Zr and $^{96}$Zr can be obtained by coupling between the single-particle structure and a collective octupole vibration. The latest evaluated nuclear structure data \cite{basu20} reduce the $E3$ in $^{90}$Zr from that adopted in Ref.~\cite{stuc04a} to 8 W.u., which implies less collectivity and thus reduces the problem. However, the current data evaluation slightly {\em increases} the $E3$ strength in $^{96}$Zr to $B(E3;3^-\rightarrow 0^+)=57(4)$ W.u. \cite{abri08}, which is the largest $E3$ strength observed in all nuclei. To our knowledge, a microscopic resolution of this dichotomy between collective and single-particle contributions to the octupole states in the Zr isotopes is yet to be found. In any case, the experimental $g$~factors make it apparent that there are strong single-particle components in what otherwise would be interpreted as collective 3$^-$ states in $^{90}$Zr and $^{96}$Zr. The similarity of the $g$ factors in $^{90}$Zr and $^{96}$Zr suggests that further checking of the huge experimental $B(E3)$ value in $^{96}$Zr is required.
}

 {
\paragraph{Transient-field \texorpdfstring{$g$}{}-factor measurements: \texorpdfstring{$Z=42-48$}{}, Mo, Ru, Pd, Cd}
}

 {
As shown in Fig.~\ref{fig:Zn-Cd-2plus}, the 2$^+_1$-level $g$-factor measurements on Mo, Ru, Pd, and Cd isotopes span the region from $N=50$ toward the middle of the $50 \leq N \leq 82$ shell, thus reaching the region where collective structures emerge that have traditionally been classified as vibrational excitations. Recently, alternative interpretations of these nominally vibrational nuclei have been proposed in terms of multiple shape coexistence \cite{garr19,garr20,sici21}.
}

 {
A complete set of $g(2^+_1)$ values has been measured for the stable isotopes. Specifically, Mantica et al. \cite{mant01} studied the Mo isotopes, Taylor et al. \cite{tayl11} the Ru isotopes, and Chamoli et al. \cite{cham11} performed a comprehensive set of measurements on the Ru, Pd, and Cd isotopes. In addition, Torres et al. \cite{torr11} measured the $g$~factors of the 2$^+_1$ and 4$^+_1$ states in unstable $^{100}$Pd following $\alpha$ transfer to a $^{96}$Ru beam, which also yielded measurements of $g({2}_{1}^{+})$ and $g({4}_{1}^{+})$ in $^{96}$Ru \cite{torr12}. G\"urdal et al. \cite{gurd10} sought to test the vibrational model as applied to $^{106}$Pd by measuring the $g$~factors of the 2$^+_2$ and 4$^+_1$ states. The $g$~factor of the $4^+_1$ state in $^{106}$Cd has also been measured as a by-product of an $\alpha$-transfer reaction to measure $g$~factors in $^{110}$Sn \cite{kumb16,benc16}.
}

\begin{figure}
  \centering
  \includegraphics[width=0.95\linewidth]{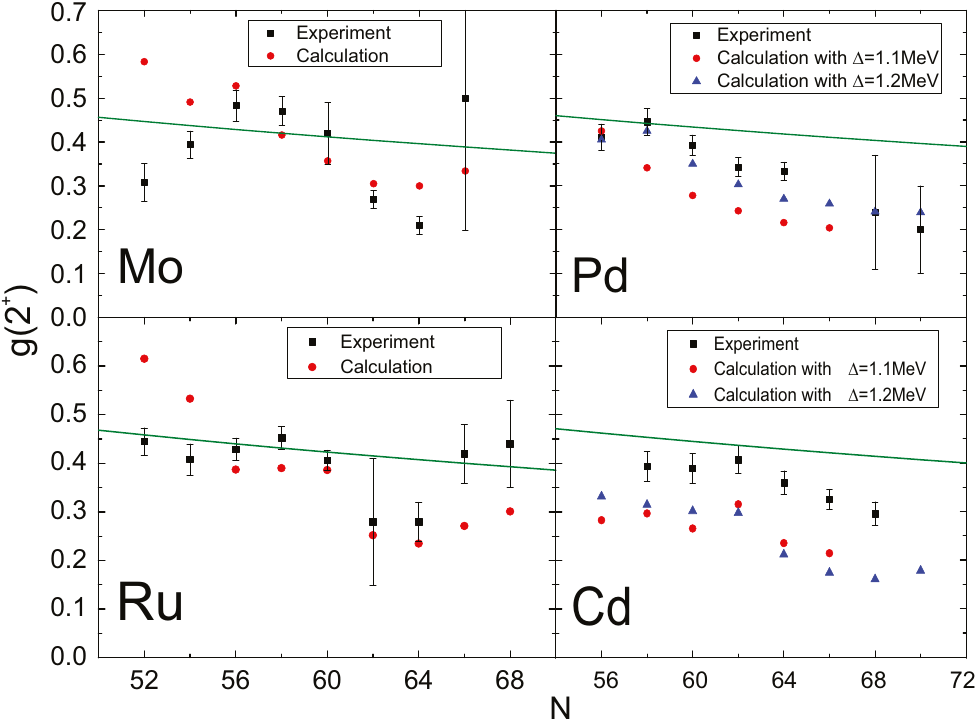}
  \caption{Tidal model calculations of $g(2^+_1)$ in Mo, Ru, Pd and Cd isotopes. The green line shows $Z/A$. The effect of the pairing gap parameter $\Delta$ is explored for Pd and Cd. Aside from the isotopes with $N=52$ and $N=54$, which may retain single-particle character, the tidal model describes the trends well and gives insights into the nucleonic structure causing the trends. In particular, the $\nu 0h_{11/2}$ orbit plays an increasing role as the Fermi surface moves beyond midshell at $N=66$. Data are from Refs.~\cite{cham11,smit04,smit05}. Figure adapted from Ref.~\cite{cham11}. }
\label{fig:RuPdCd}
\end{figure}

 {
Without going into details, it can be noted that the transient-field calibration in this group of nuclei is referenced either directly or indirectly to independently measured $g$~factors in the Pd isotopes, especially $^{106}$Pd. In addition to calibration data for the Pd isotopes \cite{stuc81a}, Mantica et al. \cite{mant01} also included some calibration data based on independently known $g$~factors in $^{103}$Rh \cite{lamp89}, but its inclusion or exclusion does not materially change their adopted $g$~factors. Having emphasized the concerns with transient-field calibration through the $20  \lesssim Z \lesssim 42 $ range, it is to be emphasized here that for $Z \gtrsim 42 $  the transient-field calibration can be soundly based on independent measurements.
}

 {
Along with their experimental results, Chamoli et al. \cite{cham11} presented an extensive set of calculations based on a version of the cranking model given the descriptive name of the `tidal wave' model, which views the nuclear excitations in weakly collective nuclei as tidal waves on the surface and allows for a microscopic description in terms of the nucleonic motion that can identify how the nucleus is carrying its angular momentum. A comparison of the tidal-wave theory and experiment is shown in Fig.~\ref{fig:RuPdCd}. For those cases with $N=52$ and $N=54$, which retain evidence of pre-collective single-particle behavior, the tidal wave model does not agree well with the data. However, once collectivity sets in, the mass dependence of the $g$~factors is well described as a trend even when it is not in excellent agreement with the magnitudes of the $g$~factors. Fig.~\ref{fig:RuPdCd} shows that the magnitudes of the $g$~factors are sensitive to the pairing strength.
}

 {
An examination of the angular momentum content of the 2$^+_1$ states showed that the tendency of all of the $g$ factors to decrease along an isotope chain as the neutron number increases is associated with increasing neutron $0h_{11/2}$ content of the wave function. These detailed calculations together with the detailed and precise data confirm the inference of Smith et al. \cite{smit04,smit05}, from static-field $g$-factor measurements on neutron-rich fission fragments, that $\nu 0h_{11/2}$ contributions cause departures from the trend suggested by the Interacting Boson Model, which predicts that the minimum $g$~factor along an isotopic sequence occurs at $N=66$, the midpoint of the major neutron shell.
}

 {
Thus, the $g$~factors in the even-even isotopes in this region help track the emergence of collectivity and develop microscopic descriptions of this emerging collectivity. However, they do not provide a very sensitive test for vibrational versus shape coexistence descriptions of the structure because the $g$~factor of a collective excitation in an even-even nucleus is near $Z/A$, largely independent of whether the state is vibrational or rotational, and largely independent of deformation. To make a critical assessment of the nuclear structure would require measurements of the 2$^+_2$ and 4$^+_1$ $g$~factors at a much higher level of precision than has been achieved to date \cite{gurd10}.
}

 {
Sensitivity to the vibrational versus rotational (deformed) character of an even-even nucleus can, however, be found in the $g$~factors of its odd-$A$ neighbors \cite{stuc16,coom19}.  In the cases of $^{111}$Cd and $^{113}$Cd, the $g$~factors of the ground and excited states differ significantly depending on whether the even-even core is vibrational or rotational. The reason is that the configuration mixing involving the odd neutron is very different for particle-vibration versus particle-rotation coupling to the core. This sensitivity, in turn, can be traced to the sensitivity of the $g_K$ of the particular Nilsson orbit to the deformation of the core.
}

 {
\paragraph{Transient-field \texorpdfstring{$g$}{}-factor measurements: \texorpdfstring{$Z=50$}{}, Sn}
}

\begin{figure}
  \centering
  \includegraphics[width=0.85\linewidth]{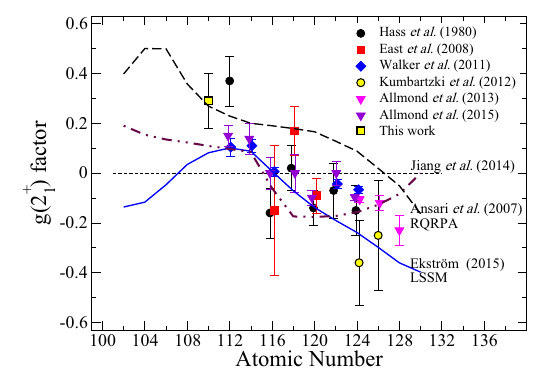}
  \caption{Compilation of experimental first-excited state $g$~factors for even-even Sn isotopes. The data references are: Hass et al. \cite{hass80}, East et al. \cite{east08}, Walker et al. \cite{walk11}, Kumbartzki et al. \cite{kumb12a}, Allmond et al. (2013) \cite{allm13}, Allmond et al. (2015) \cite{allm15}. ``This work" is \cite{kumb16}. Theory references are Jiang et al. \cite{jian14}, Ansari et al. \cite{ansa07}, and Ekstr\"om  who extended the calculations reported in  \cite{banu05,walk11}.
The figure is reproduced from \cite{kumb16}.}
\label{fig:Sngfactors}
\end{figure}

 {
The semimagic Sn isotopes have ongoing interest as the longest sequence of isotopes between two accessible double-magic shell closures at $^{100}$Sn and $^{132}$Sn. Tin also boasts the longest sequence of stable isotopes from $^{112}$Sn to $^{124}$Sn, for which several transient-field $g$-factor measurements have been performed \cite{hass80,east08,walk11}. In addition, measurements on radioactive $^{110}$Sn have been made after $\alpha$ transfer \cite{kumb16} to $^{106}$Cd, and a $g$-factor measurement on a radioactive beam of $^{126}$Sn has been performed \cite{kumb12a}. Complementing these transient-field studies are Recoil in Vacuum (RIV) measurements on the even-even isotopes between $^{112}$Sn and $^{128}$Sn \cite{allm13,allm15}.
}

 {
The $g(2^+_1)$ data and several theoretical calculations are displayed in Fig.~\ref{fig:Sngfactors}. The large-scale shell model calculation of Ekstr\"om, which is communicated as a private communication in \cite{kumb16} and extends the calculations reported in \cite{banu05,walk11}, tracks the trend in the data well. The trend can be understood in relation to the $g$~factors of the neutron orbits at the Fermi surface. Specifically, the orbits and their Schmidt $g$~factor estimates (in parentheses) in order from $N=50$ to $N=82$ are:
$1d_{5/2}$ $(-0.765)$;
$0g_{7/2}$ $(+0.425)$;
$2s_{1/2}$ $(-3.826)$;
$1d_{3/2}$ $(+0.765)$;
$0h_{11/2}$ $(-0.348)$:
$g(2^+_1$) is clearly positive when the Fermi surface is in the $0g_{7/2}$ orbit at $N=62,64$. Once the Fermi surface moves into the $2s_{1/2}$ and $1d_{3/2}$ orbits at $N=66,68$, the 2$^+_1$-state $g$~factors are essentially zero. Note that the $2s_{1/2}$ orbit cannot contribute to a 2$^+$ state on its own, but can couple to the $1d_{3/2}$ to produce the $[2s_{1/2} \otimes 1d_{3/2}]_{2^+}$ configuration with $g\simeq-0.38$. The $2s_{1/2}$ and $1d_{3/2}$ orbits are near degenerate, so it is reasonable to expect their contributions to $g(2^+_1)$ to cancel. The experimental $g$~factors reach negative values towards $^{132}$Sn once the Fermi surface moves into the $0h_{11/2}$ orbit.
}

 {
The experimental 2$^+_1$ $g$~factors are all small in magnitude compared with those of the single-particle orbits, indicative of configuration mixing, however, the trend tracks that expected by inspecting the $g$~factors of the single-particle orbits.
}

 {
In due course, it would be of considerable interest to extend the $g$~factor measurements below $^{110}$Sn. The experience to date indicates that the RIV method would give more precise magnitudes for the $g$~factor, however, the differences in signs for the alternative theoretical predictions in Fig.~\ref{fig:Sngfactors} indicate that a measurement of the sign is essential, and thus transient-field measurements would be required.
}

 {
\subsection{Transient-field \texorpdfstring{$g$}{}-factor measurements and emerging collectivity near \texorpdfstring{$^{132}$Sn}{} and \texorpdfstring{$^{146}$Gd}{}
}
\label{sect:TF-gfactors-Z.gt.50}
}

 {
Since the year 2000, transient-field measurements on heavy nuclei have focused on Te and Xe isotopes near doubly magic $^{132}$Sn and on the Nd isotopes near $^{146}$Gd, which, with $Z=64$ and $N=82$, can be considered approximately doubly magic. The measurements on Xe and Te isotopes near $^{132}$Sn are discussed below in section \ref{subsubsec:Te_RIV} together with the related measurements by the recoil in vacuum (RIV) method. As described there, the Te and Xe isotopes near $^{132}$Sn provide a fertile testing ground to map the path of emerging nuclear collectivity from seniority structures in the semimagic $N=82$  isotopes  $^{134}$Te and $^{136}$Xe towards the isotopes near the neutron mid-shell at $N=66$, which show vibrational level structures. For the Te and Xe isotopes, it is seniority-two proton ($0g_{7/2}$) structures that persist in the 4$^+_1$ and 6$^+_1$ states (see e.g. \cite{jako02,pete19,hick22,reec25}).
}

\begin{figure}
  \centering
  \includegraphics[width=0.5\linewidth]{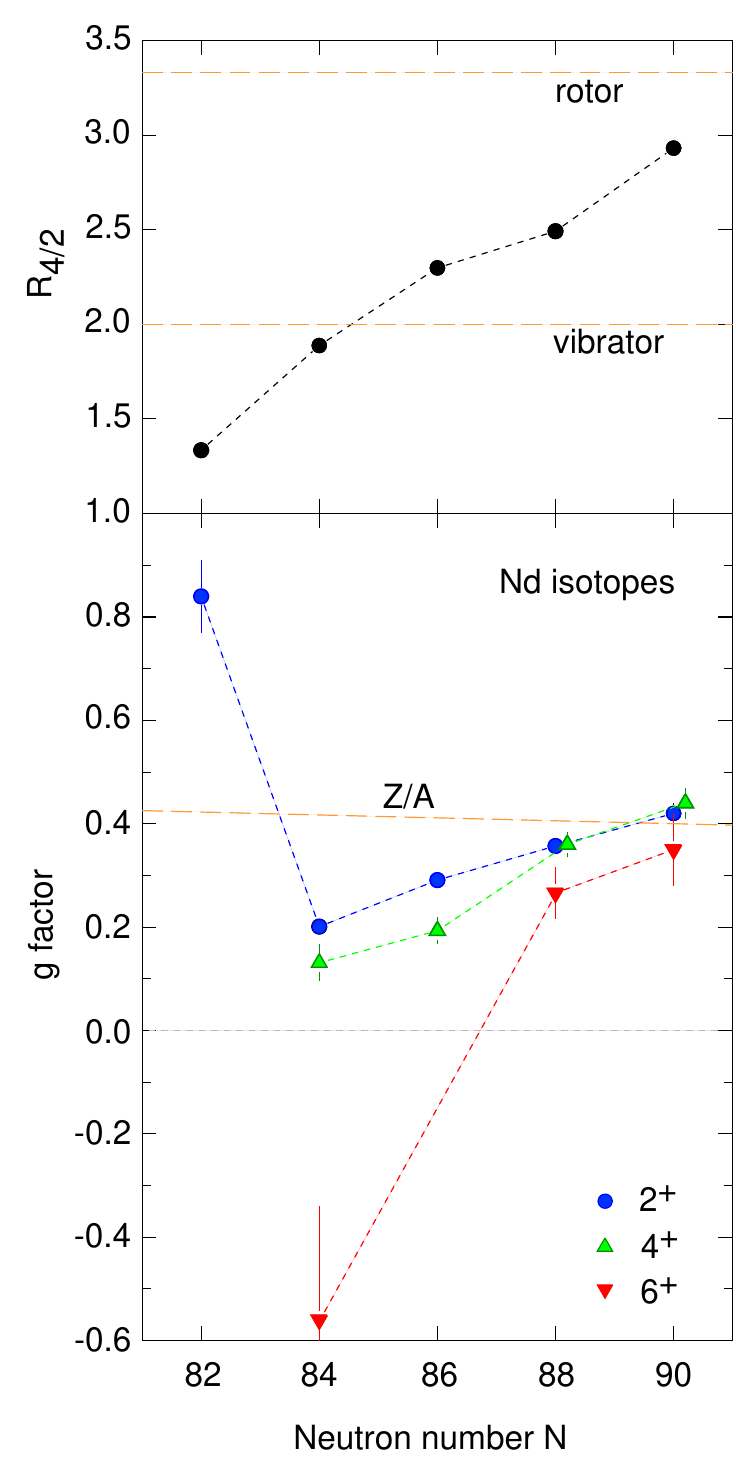}
  \caption{Mass and spin dependence of $g$~factors in the even Nd isotopes from semimagic $^{142}$Nd to $^{150}$Nd, which appears rotational. The $g(2^+_1) \sim + 0.8$ value in $^{142}$Nd is clearly associated with the lowest excitation in the open proton shell. Once a pair of neutrons is added, the $g$~factors fall well below $Z/A$. The $g(6^+_1) \sim -0.6$ in $^{144}$Nd is consistent with the Schmidt estimate for an $f_{7/2}$ neutron, the lowest shell model orbit above $N=82$. As additional pairs of neutrons are added, the $g$ factors of all of the low-excitation states converge towards the collective estimate, $g\sim Z/A$. Data are from \cite{bazz91,hold00,hold01}.}
\label{fig:Ndgfactors}
\end{figure}

 {
The $g$~factors of the stable neodymium isotopes, $^{142}$Nd ($Z=60$, $N=82$) to $^{150}$Nd, were measured by the transient-field technique \cite{bazz91,hold00,hold01} prior to the measurements on Xe \cite{jako00} and Te isotopes \cite{stuc07,coom20,coom22}. The measured Nd $g$~factors are displayed in Fig.~\ref{fig:Ndgfactors}. In this case, the nuclear structure transition is from a seniority structure in $^{142}$Nd ($N=82$) that can be associated with four proton holes in the $0g_{7/2}$ and $1d_{5/2}$ orbits below $Z=64$ to a rotational structure in $^{144}$Nd, which has an additional 8 neutrons. In contrast to the persistent proton $0g_{7/2}$ seniority structure in the Te and Xe isotopes,  in the Nd isotopes the $g$~factors point to persistence of neutron $1f_{7/2}$ contributions in the 4$^+_1$ states of $^{144,146}$Nd and in the 6$^+_1$ states of  $^{144,146,148}$Nd, and possibly also in $^{150}$Nd.
}

 {
The pattern shown by the Nd 2$^+_1$-state $g$~factors is similar to that found for the 2$^+_1$ states of the Mo isotopes \cite{mant01,stuc01}; see Fig.~\ref{fig:Zn-Cd-2plus}: At the neutron shell closure, $g(2^+_1) \sim 1$ originating from the open-shell proton excitation. However, with the addition of two neutrons to the closed shell, the $g$ factor falls to a value well below $Z/A$. This behavior points to a dominant neutron content of these $2^+_1$ states at $N=52$ and $N=84$ for the Mo and Nd isotopes, respectively.
}

\begin{figure}
  \centering
  \includegraphics[width=0.85\linewidth]{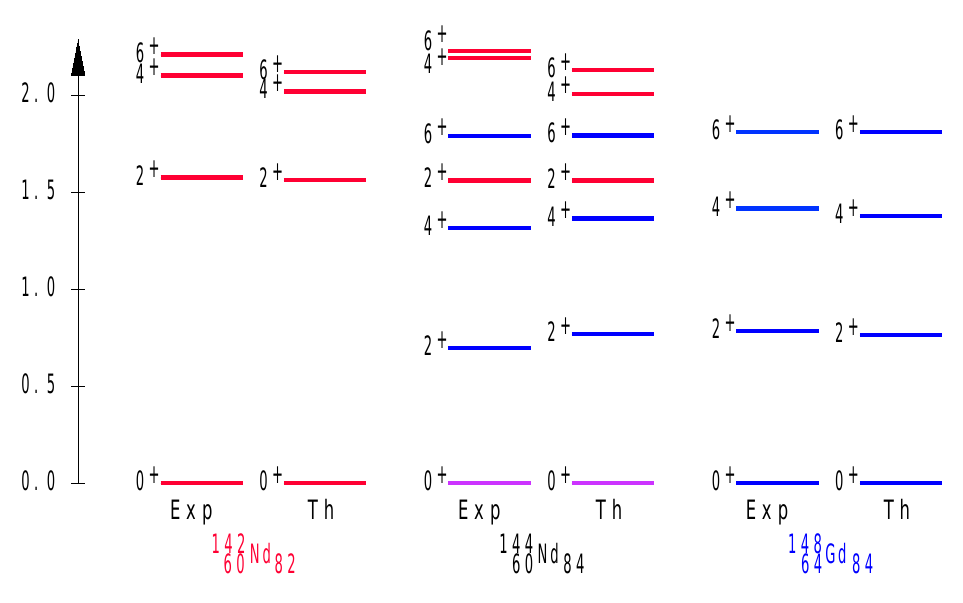}
  \caption{Level schemes illustrating the apparent weak-coupling of the proton and neutron excitations in $^{144}$Nd. The theoretical level schemes are from schematic shell model calculations based on surface-delta interactions and single-particle energies adjusted to describe nuclei near $^{146}$Gd. The figure is adapted from \cite{stuc01}.}
\label{fig:Nd144decomp}
\end{figure}

 {
A feature of the level spectra of both $^{94}$Mo and $^{144}$Nd is the apparent weak coupling of the proton and neutron valence spaces, as illustrated for $^{144}$Nd in Fig.~\ref{fig:Nd144decomp}. The level scheme of $^{144}_{\;60}$Nd$_{84}$ can be obtained approximately by superimposing the level schemes of $^{142}_{\;60}$Nd$_{82}$ and $^{148}_{\;64}$Gd$_{84}$, treating $^{146}_{\;64}$Gd$_{82}$ as a closed-shell nucleus.  Fig.~\ref{fig:Nd144decomp} suggests that neutron excitations must be prominent, if not dominant, in the low-excitation yrast states of $^{144}$Nd; however, it cannot be judged from the spectra alone whether the interaction between the proton and neutron spaces is actually weak or rather strong and state-independent. The $g$~factors are a sensitive measure of the coupling between the proton and neutron spaces, and hence play a key role in tracking the emergence of collectivity from the underlying single-nucleon structure. For example, the $g$~factor data in Fig.~\ref{fig:Ndgfactors} suggest that the coupling between protons and neutrons weakens with increasing spin up to 6$^+$. This feature is also found in the Te and Xe isotopes.
}

 {
As a final remark on the emerging collectivity in the Nd isotopes, it can be noted that the $g$~factor data show conclusively that the energy ratio $R_{4/2} = E_x(4^+_1)/E_x(2^+_1)$, which is shown in the upper panel of Fig.~\ref{fig:Ndgfactors}, cannot be used as a definitive signature of collective structure. For example,  $R_{4/2} \sim  2$ is often taken as evidence of anharmonic vibrational structure, but the $g$~factors show that this is not the case in $^{144,146,148}$Nd. This observation serves as a caution on over-interpreting limited information from excitation-energy patterns alone. \\
}

 {
\subsection{Odd-A nuclei}
\label{sec:oddA}
}

 {
The focus in this section is on magnetic moments of short-lived excited states of odd-$A$ nuclei. The measurements discussed primarily use the transient-field technique. In this class of transient-field measurement, which pertains to medium and heavy-mass nuclei, the transient-field strength can be calibrated accurately relative to independently known $g$~factors in the same nucleus or neighboring isotopes. While the transient-field measurements on the excited states are featured in the following discussion, it is important to note that these measurements are not interpreted in isolation: the nuclear structure interpretation includes studies of ground-state moments, quadrupole moments, and transition strengths in addition to the excited-state $g$~factors.
}

 {
\subsubsection{Rotational nuclei}
\label{sec:oddA-rotors}
}

 {
Expressions for the electromagnetic properties of single-quasiparticle rotational bands in odd-$A$ nuclei were summarized in section~\ref{subsub:nilsson}. That section concluded by noting that it is common practice to deduce $M1$ properties of these rotational bands from in-band $\gamma$-ray transition intensities assuming no Coriolis interactions and given reasonable estimates of the intrinsic quadrupole moment, $Q_0$, and rotational $g$~factor, $g_R$. The single-particle $g_K$ is deduced. The value of $g_K$ usually allows a unique assignment of a Nilsson configuration to the band.
}

\begin{figure}
  \centering
  \includegraphics[width=0.5\linewidth]{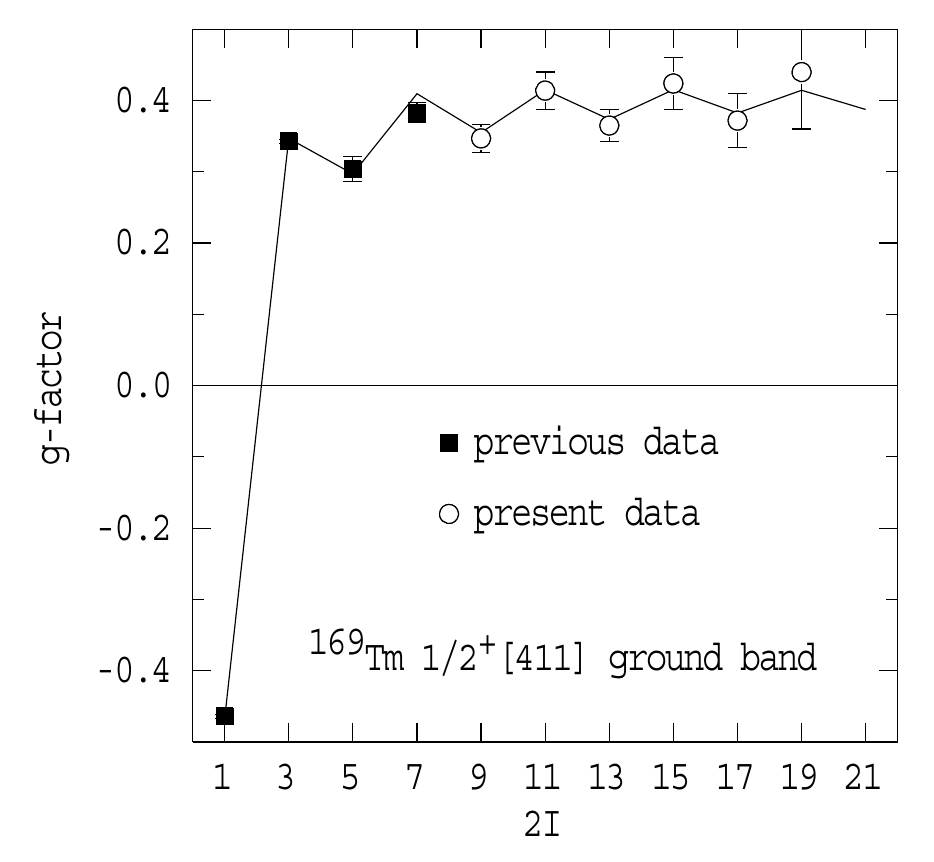}
  \caption{Experimental and calculated $g$~factors in the ground-state band of $^{169}$Tm, which has Nilsson assignment $1/2^+[411]$. The predicted $g$~factors are based on a fit to measured $M1$ observables performed by Taras et al. \cite{tara77}. From \cite{robi99}. }
\label{fig:Tm169-gfactors}
\end{figure}

\begin{figure}
  \centering
  \includegraphics[width=0.5\linewidth]{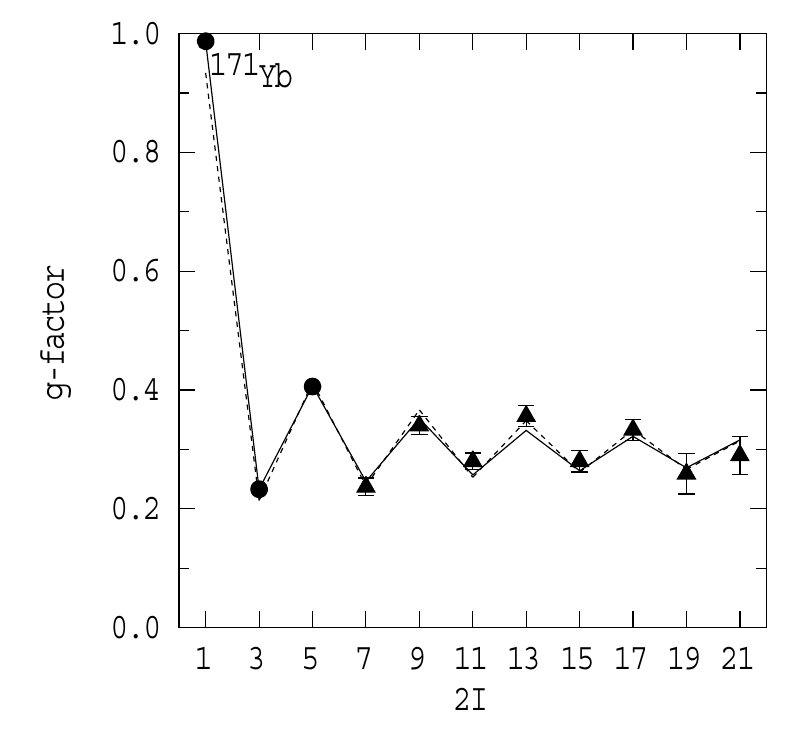}
  \caption{Experimental and calculated $g$~factors in the ground-state band of $^{171}$Yb, which has Nilsson assignment $1/2^-[521]$. The filled circles represent previous data, whereas the filled triangles are transient-field data from Ref.~\cite{stuc00}. The solid line results from a fit to the $g$~factors of the lowest three states assuming a pure $K=1/2$ band. The dotted line is a particle-rotor calculation that includes Coriolis interactions.
  From \cite{stuc00}. }
\label{fig:Yb171-gfactors}
\end{figure}

 {
Despite the wide use of this procedure, there have been few direct measurements to test the strong-coupling assumption, i.e. the presumption of a pure Nilsson band with no Coriolis interactions. There are four examples where detailed spectroscopy, along with $g$-factor measurements for several states in the rotational band, allow for a test of this analysis procedure based on the strong-coupling assumption. These nuclei are $^{155}$Gd \cite{stuc98}, $^{169}$Tm \cite{robi99}, $^{171}$Yb \cite{stuc00}, and $^{183}$W \cite{lamp92,lamp92a}. The $g$~factors of the ground-state band of $^{169}$Tm are shown in Fig.~\ref{fig:Tm169-gfactors}. A similar plot for $^{171}$Yb is shown in Fig.~\ref{fig:Yb171-gfactors}. These are $K=1/2$ bands, so there is a magnetic decoupling parameter, $b_0$, along with $g_K$ and $g_R$.
}

 {
In  Fig.~\ref{fig:Tm169-gfactors} the line is a prediction based on $g_K$, $g_R$ and $b_0$ values previously obtained by Taras et al. \cite{tara77}, which included the $g$~factors shown as filled squares and several $B(M1)$ transitions in the ground band of $^{169}$Tm.
For the case of $^{171}$Yb in Fig.~\ref{fig:Yb171-gfactors}, the solid line results from a fit to the $g$~factors of the lowest three states assuming a pure $K=1/2$ band. The dotted line is a particle-rotor calculation including Coriolis interactions.
}

 {
The conclusion from the detailed comparisons of theory and experiment in Refs.~\cite{stuc98,robi99,stuc00} is that the assumption of strong coupling when evaluating $g_K$ to assign a Nilsson orbit to a rotational band in an odd-$A$ nuclide appears to be justified, even when Coriolis mixing is present. The reason the simplified analysis works apparently stems from the fact that the first-order effect of the Coriolis interactions on the $g$~factors is to renormalize $g_R$. Thus, an appropriate choice of $g_R$ can give an accurate estimate of $g_K$, which may uniquely identify the Nilsson orbit on which the band is based.
}

 {
The above examples are becoming dated, however, recent advances in calculations of moments of odd-$A$ nuclei using Density Functional Theory (DFT) have brought them back into view \cite{doba26,doba26b}.
Traditionally, DFT (or Hartree-Fock) calculations evaluated only the ground-state moments. However, the methods described in Ref.~\cite{doba26} calculate also the moments of excited states in the rotational bands of odd-$A$ rotors. This is an important development, as having both experimental and theoretical magnetic moments for the excited states of a band can help unravel the relative contributions to the moment from the odd nucleon versus the collective core. Specifically, if there is a difference between theory and experiment, and only the ground-state moment is measured, there is no way to unravel whether the deficiency in the theory stems from the odd-nucleon part, the core part, or both. In contrast, if some of the excited-state moments are known, there is a chance to examine these contributions in theory versus experiment. Table~\ref{tab:Hf177gKgR} gives an example from the recent work of Dobaczewski et al. \cite{doba26,doba26a}, namely comparisons of measurement and calculation for three magnetic dipole moments in the 7/2$^-$[514] band of $^{177}$Hf.
}

\begin{table}[htbp]
\begin{center}
\caption{Comparison of theory and experiment for members of the 7/2$^-$[514] band in $^{177}$Hf. The last two rows give $g_K$ and $g_R$ derived from Eq.~(\ref{eq:gNils1}) as described in the text.}
\label{tab:Hf177gKgR}
\begin{tabular}{cllll}
\hline
\multicolumn{1}{c}{$I^{\pi}$} &
\multicolumn{4}{c}{$g$~factor} \\
\cline{2-5}
 & Exp. & DFT & Nilsson & PR\\
 \hline
 7/2$^-$  & $+0.2267(2)$ & $+0.288$ & $+0.276(7)$  & $+0.267(7)$ \\
 9/2$^-$  & $+0.229(7)$  & $+0.312$ & $+0.290(15)$ & $+0.279(15)$ \\
 11/2$^-$ & $+0.27(9)$   & $+0.326$ & $+0.298(20)$ & $+0.285(20)$ \\
\hline
$g_K$: & $+0.225$  & $+0.268$ & $+0.265$ & $+0.258$ \\
$g_R$: & $+0.233$  & $+0.355$ & $+0.315(30)$ \footnotemark[1] & $+0.299$ \\
\hline
\end{tabular}
\footnotetext[1]{This is the experimental value of $g(2^+_1)$ in $^{176}$Hf~\cite{alft96}.}
\end{center}
\end{table}

 {
The DFT $g$~factors in the 7/2$^-$[514] band all exceed the experimental values by about 30\%. It is instructive to analyze these data in terms of the particle-rotor model, beginning with the expression for a pure Nilsson band with $K \neq 1/2$ (see also Eq.~(\ref{eq:gnils})):
\begin{equation}\label{eq:gNils1}
    g(I)=g_R + (g_K-g_R)\frac{K^2}{I(I+1)},
\end{equation}
where the rotational $g$~factor, $g_R$, and the projection of the single-nucleon magnetic moment on the symmetry axis of the nucleus, $g_K$, are treated as parameters.
In Table~\ref{tab:Hf177gKgR}, the experimental $g$~factors are compared with the DFT calculation as well as the Nilsson model (i.e. particle-rotor model with no Coriolis mixing) and a full particle-rotor (PR) calculation including (unattenuated) Coriolis mixing. The effective values of $g_K$ and $g_R$ for each case are shown in the last two rows.
The Nilsson and PR calculations set $g_R=0.315(30)$, which is the experimental $g(2^+_1)$ value of $^{176}$Hf~\cite{alft96}. The theoretical $g$~factors from the Nilsson and PR models are given in Table~\ref{tab:Hf177gKgR} with an estimate of the uncertainty arising from the uncertainty in the adopted $g_R$ value, which scales as $1-K^2/I(I+1)$.
See Ref.~\cite{doba26} for additional details of these calculations.
}

 {
In principle, this analysis separates the contributions to the magnetic dipole moment from the core and the odd neutron. There is close agreement between the DFT calculation and the Nilsson model for $g_K$. Whereas it is usually claimed that $g_R$ in the odd-$N$ nucleus is smaller than $g(2^+_1)$ in the even-even neighbor (see~\cite{bohr75}, pp. 256 and 303), the DFT predicts the opposite trend here. The PR calculation suggests that Coriolis interactions effect a small reduction in $g_R$, but the observed effective $g_R$ is smaller than that predicted by any of the models. The empirical $g_K$ is also smaller than that given (or adopted) by the models. The difference between the DFT and empirical effective $g$~factors is about 20\% for $g_K$ and about 50\% for $g_R$.
}

 {
Although this analysis may suggest that the difference between the DFT and experiment in the 7/2$^-$[514] band of $^{177}$Hf is primarily due to the core contribution, it would be premature to draw a firm conclusion.
These observations nevertheless indicate the value of precise data on excited states in rotational bands to resolve the origins of discrepancies between theory and experiment. The need for further measurements is evident.
}

 {
\subsubsection{Near-spherical weakly collective nuclei}
\label{sec:oddA-spherical}
}

 {
Excited-state $g$~factors have been measured in a number of near-spherical nuclei with mass numbers $100 \lesssim A \lesssim 200$: $^{103}$Rh \cite{benc88,lamp89}, $^{107,109}$Ag \cite{wood84,bazz84a,ball86}, $^{111,113}$Cd \cite{benc88,stuc16,coom19}, $^{123,125}$Te \cite{benc88,cham09}, $^{191,193}$Ir \cite{beza00}, $^{195}$Pt \cite{lamp94}, $^{197}$Au \cite{stuc88}. These transient-field measurements are all on stable nuclei, and all have ground-state spins of $I=1/2$, apart from  $^{191,193}$Ir and $^{197}$Au, which have $I=3/2$.
}

 {
Generally, the $g$~factors of the lowest few collective states of these nuclei can be described by the weak-coupling core-excitation model. In this model, the $g$~factors are given by the  additivity relation, Eq.~(\ref{eq:additivity}), where $I_1$ and $g_1$ correspond to the odd nucleon, and $I_2$ and $g_2$ are identified with the first-excited state of the core. It can be noted that the particle-vibration model is a generalization of the weak-coupling model in that the weak-coupling model states readily form the basis for more sophisticated particle-vibration model calculations.
A more sophisticated Interacting Boson-Fermion Model analysis was performed in several cases, e.g. $^{103}$Rh, $^{107,109}$Ag, and $^{197}$Au \cite{lamp89,wood84,stuc88}, which generally improved the description of the energy spectrum and electromagnetic properties. The IBFM can be seen as a similar approach to the particle-vibration model, but with extra flexibility to specify the Hamiltonian of the core.
Additional discussion of the IBFM in relation to magnetic moments is given in the next subsection \ref{sec:IBFM}. }

 {
A critique of these models, particularly the core-excitation model, is that the $g$~factors of the odd nucleon and the core were often treated as parameters. For example, the $g$~factor of the odd nucleon was often identified with the measured $g$~factor of the ground state, without an attempt to understand its value.
}

\begin{figure}
  \centering
  \includegraphics[width=0.7\linewidth]{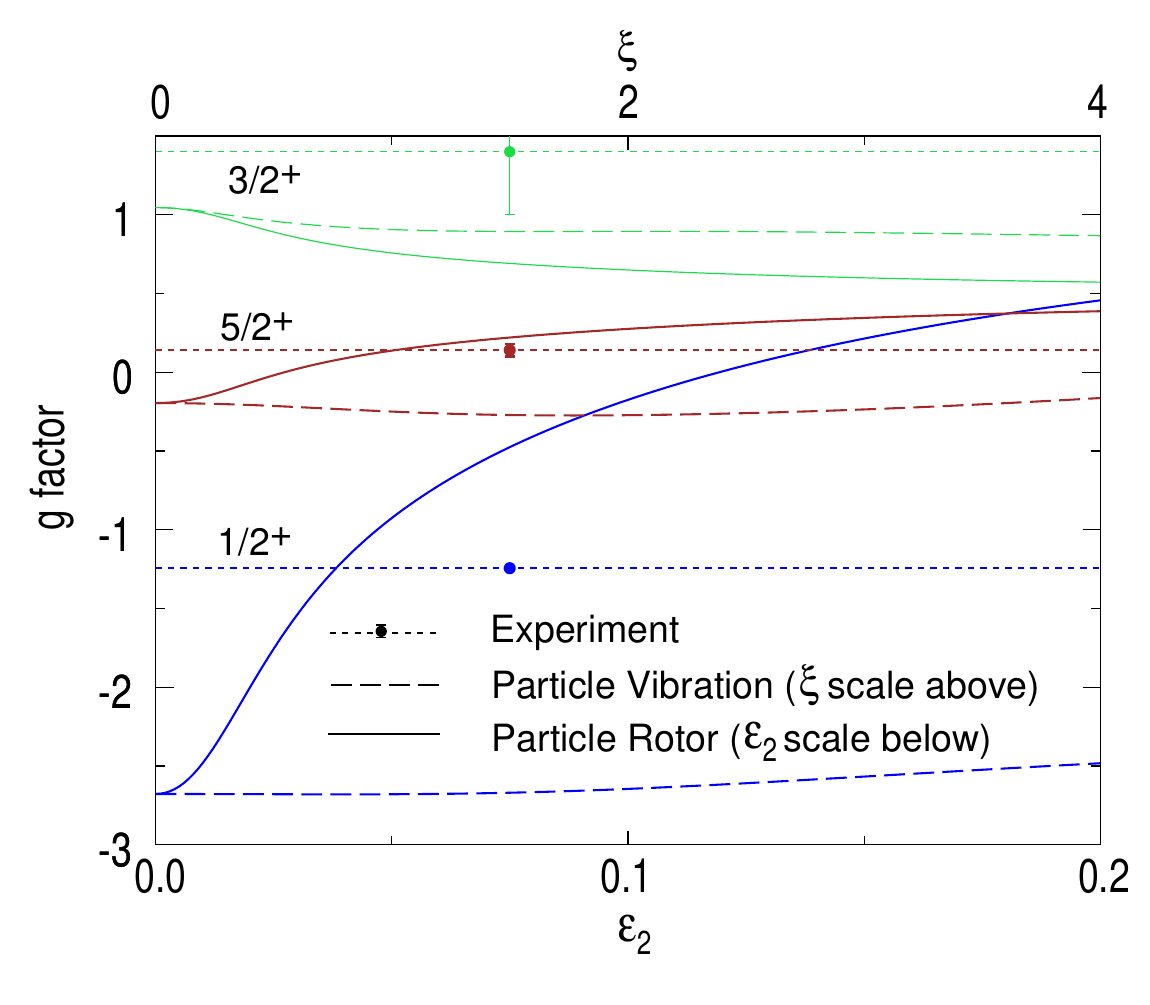}
  \caption{Comparison of g-factor variations in $^{113}$Cd with deformation or particle-vibration coupling strength. The parameters $\xi$ specified the particle-vibration coupling strength whereas $\epsilon_2$ is the Nilsson defomation parameter. The schematic particle-rotor calculations here ignore Coriolis interactions.
  From \cite{stuc16}. }
\label{fig:Cd-gfactors}
\end{figure}

 {
More recently, the cases of $^{111}$Cd and $^{113}$Cd have been re-examined \cite{stuc16,coom19} from the particle-vibration and particle-rotor perspectives.  As shown in Fig.~\ref{fig:Cd-gfactors}, the particle-vibration model, which assumes spherical core excitations, cannot explain the $g$ factors of the ground and low-excitation states in $^{113}$Cd, whereas a particle-rotor model with a small, nonzero core deformation does. Similar conclusions follow for $^{111}$Cd \cite{coom19}. The contrast of the two models is made stark by the fact that they begin from the same limiting g-factor values.
}

 {
It was shown in Ref.~\cite{stuc16} that when an odd nucleon occupies a spherical orbit with angular momentum $j = 1/2$, or a deformed orbit with j = 1/2 parentage, the particle-vibration model and the particle-rotor model both reduce to the same $g$-factor value in their respective limits of zero particle-vibration coupling or zero deformation. Thus it is demonstrated that the $g$~factors of low-excitation states in $^{111}$Cd and $^{113}$Cd are sensitive to the nature of the collectivity in these nuclei in ways that the electric quadrupole observables are not.
}

 {
The emergence of deformation and quadrupole collectivity away from double-magic shell closures remains a subject of on-going interest. Moreover, the true nature of weakly-collective nuclei remains open to debate. For example, the structure of the Cd isotopes near the middle of the neutron shell, which have long been interpreted as vibrators, has recently been interpreted in terms of multiple shape co-existence \cite{garr19,garr20,sici21}. Others, however, maintain the vibrational interpretation \cite{gavr23}. Moment measurements may help distinguish between, or otherwise challenge, these views.
}

 {
\subsubsection{Interacting boson-fermion models}
\label{sec:IBFM}
}
 {
The previous two sections described particle-core models broadly in terms of the coupling of a single nucleon to a core, with a focus mainly on cores described by the Bohr-Mottelson collective model. An analogous group of particle-core models can be obtained by coupling the odd fermion to a core represented by the Interacting Boson Model \cite{iach91}. The Hamiltonian for the Interacting Boson-Fermion Model (IBFM) has the standard particle-core coupling form
\begin{equation}\label{eq:HIBFM}
    \hat{\mathcal{H}}_{\mathrm{IBFM}}=\hat{\mathcal{H}}^{(B)} + \hat{\mathcal{H}}^{(F)}+\hat{V}^{(BF)},
\end{equation}
where $\hat{\mathcal{H}}^{(B)}$ is the Hamiltonian of boson core, $\hat{\mathcal{H}}^{(F)}$ is that of the added fermion, and $\hat{V}^{(BF)}$ is the boson-fermion interaction.
}

 {
A mention of the IBFM approach is included included in the previous section on near-spherical nuclei.
An extensive review of these models will not be discussed here. It is worth singling out, however, one novel outcome of the development of the interacting boson-fermion models, namely the idea that supersymmetry could be manifested in the low-excitation spectra of atomic nuclei \cite{iach80}. More specifically, the relationship between an odd-$A$ nuclide and its even-even neighbor can be set by supersymmetry requirements. In other words, the even-even nuclide and its odd-$A$ neighbor are described within the one supersymmetry theory. This approach largely avoids the criticism that the IBFM usually has a large number of free parameters. In fact, the wavefunctions are set by the supersymmetry requirement independent of Hamiltonian parameters.
}

 {
Initial supersymmetry schemes considered the coupling of a fermion in a specific single-particle orbit to a boson core represented by one of the symmetry limits of the Interacting Boson Model. The $U(6/4)$ supersymmetry scheme sought to describe odd-$Z$ nuclei in the Os-Pt region by coupling a fermion in the proton $d_{3/2}$ orbit \cite{bala81}.
}

 {
Initial tests of the $U(6/4)$ supersymmetry model for potential supersymmetry partners $^{190}$Os-$^{191}$Ir and $^{192}$Os-$^{193}$Ir compared excitation energies, binding energies, $E2$ transition rates and data from transfer reactions, but not $M1$ properties \cite{bala81}.
However, as emphasized by Kuyucak and Stuchbery \cite{kuyu93}, the $M1$ properties, including $g$~factors, in the odd-$A$ nuclide provide the most sensitive and parameter-free test of the supersymmetry model. Because the wavefunctions are set independent of Hamiltonian parameters, fits to energy levels provide a limited test of the model. As is the case in most particle-core coupling models, the $E2$ properties are determined primarily by the properties of the core; they therefore give limited insight into the particle-core properties prescribed by the supersymmetry. The utility of transfer reactions to test the supersymmetry models is limited by the fact that transfer reaction intensities require the introduction of additional parameters. In contrast, the $M1$ properties are sensitive to both the boson and fermion parts of the wavefunction and can be evaluated without the introduction of free parameters.
}

 {
Multi-$j$ supersymmetry models were developed subsequently, lifting the restriction on the single-$j$ assumption. For example, the $U(6/12)$ supersymmetry applicable to isotopes like $^{195}$Pt includes the $p_{1/2}$, $p_{3/2}$ and $f_{5/2}$ orbits, whereas $U(6/20)$ applicable to isotopes such as $^{191}$Ir and $^{193}$Ir includes $s_{1/2}$, $d_{3/2}$, $d_{5/2}$, and $g_{7/2}$ orbits.
}

 {
Transient-field $g$-factor measurements on $^{191}$Ir, $^{193}$Ir, $^{195}$Pt, and $^{197}$Au yielded data that were compared with these supersymmetry models \cite{beza00,kuyu93,lamp94,stuc88}.
}

 {
In the first published case of $^{197}$Au, the Spin(6) supersymmetry limit of the IBFM was evaluated along with a multi-shell IBFM. A fair description of the $g$~factors was obtained with the supersymmetry model, although it was concluded that inclusion of the $s_{1/2}$ orbit is important, and that the core nuclei in the Au region cannot be treated simply in a spherical limit of the IBM.
}

 {
By way of contrast, Kuyucak and Stuchbery \cite{kuyu93} concluded from parameter-free calculations of $g$~factors in the $U(6/12)$ supersymmetry that ``the level of agreement between theory and experiment is remarkable and, in general, supports the multi-$j$ supersymmetry model in $^{195}$Pt". It is worth noting that in the full account of the experimental work on $^{195}$Pt, which included more extensive comparisons with theory, a similar level of agreement was obtained by a particle-rotor calculation for states that could be assigned to the $K=1/2$ ground-state band, while the particle-rotor description of the $K=3/2$ band was relatively poor. The conclusion was that the multi-$j$ supersymmetry model provides a simple and successful description of the spectroscopy of the low-spin negative-parity states in $^{195}$Pt. The hope for future work was to explore whether this case is singular or whether supersymmetry models are a general feature of some classes of nuclei. To date, the case of $^{195}$Pt appears to be unique, notwithstanding extensive data and comparisons with $U(6/4)$ and $U(6/20)$ theory for $^{191}$Ir and $^{193}$Ir, which will now be discussed.
}

 {
Along with $U(6/4)$ and $U(6/20)$ supersymmetry theory, triaxial particle-rotor calculations were performed for $^{191}$Ir and $^{193}$Ir in a detailed experimental and theoretical study \cite{beza00}. Generally, a good description of electromagnetic observables was achieved by the triaxial particle-rotor calculations, whereas the $U(6/4)$ supersymmetry model was demonstrably inadequate. Also, the $U(6/20)$ model required modification to account for the data, the reason being that despite the inclusion of the $s_{1/2}$, $d_{3/2}$, $d_{5/2}$, and $g_{7/2}$ orbits, these are degenerate in the model. In Ref.~\cite{beza00}, the degeneracy was lifted by a BCS procedure and then included as a perturbation of the $U(6/20)$ Hamiltonian. Even so, it was concluded that this procedure could not mix the $s_{1/2}$ and $d_{3/2}$ fermion orbits with the precision required to explain the experimental data.
}

 {
While supersymmetry models have not been explored extensively in relation to nuclear moments into the 21st century, symmetry concepts and models based on them remain an important theme in nuclear structure research.
}

 {
\subsubsection{Effective field theories of odd-\textit{A} nuclei}
\label{sec:EFT-oddA}
}

 {
Coello P\'erez and Papenbrock have recently reviewed their effective field theory (EFT) approach to describing low-excitation collective modes in nuclei \cite{coel25}. An important feature of this approach is that it separates the energy scale of the low-lying collective states from a breakdown scale where non-collective excitations begin to appear. This separation allows for a quantification of theoretical uncertainties.}

 {
The model is akin to phenomenological particle-vibration and Interacting Boson-Fermion models. Although there are common concepts in the formulation of the model, it should not be confused with more microscopic models that determine an effective nucleon-nucleon interaction based on effective field theory. The use of effective field theory concepts means that the theory has an inbuilt uncertainty estimate determined by a ``cut-off'' in excitation energy. Of course, this theoretical uncertainty applies to the model itself - it is not related to the applicability of the model to the nucleus of interest, which may have structural features outside of the model.}

 {
Effective field theories of collective excitations have been used to compute spectra, transition rates, and other matrix elements of interest, including $g$~factors in odd-$A$ nuclei with $I=1/2$ ground states \cite{coel16} adjacent to notionally vibrational nuclei in the vicinity of $A=110$ \cite{coel15}. The agreement between theory and experiment for $^{103}$Rh, $^{109,111}$Ag, and $^{111,113}$Cd was considered satisfactory within experimental and theoretical uncertainties.
}

 {
It is relevant to note that in this EFT approach, the even-even core is strictly spherical. However, as shown by Stuchbery et al. \cite{stuc16}, the $g$~factors of low-excitation states in  $^{111}$Cd and $^{113}$Cd are strongly affected by nonzero core deformation. Thus, the limitations of the EFT in describing the $g$~factors may be due in part to the assumption of spherical cores, in addition to the quantified uncertainties in the EFT model itself.
}

 {
It would be of interest to extend the EFT approach to calculations of moments in deformed nuclei, which, to our knowledge, has not yet been done. \\
}

\section{Key results on nuclei at high angular momenta}
\label{sub::high-spins}
Studies of nuclear structure at high angular momentum were a central topic in nuclear-structure physics in the last decades of the $20^{th}$ century.  {In this period, a large number of excitations were discovered, which result from the evolution of nuclear structure with spin. These include the `backbending effect' in deformed nuclei, magnetic, antimagnetic and chiral rotational bands, superdeformation, {\it etc}. Isomers, built on different configurations, such as $K$ isomers or multi-quasiparticle high-spin isomers, were observed and their properties were investigated.} High-spin physics still attracts considerable interest, although the research focus of nuclear structure studies at the turn of the millennium has largely shifted to properties of neutron-rich nuclei.  {However, this chapter is far from being closed. Moreover, although nowadays nuclear structure research in `exotic' neutron-rich nuclei addresses the properties of low-lying states, with the advance of technologies, new experimental techniques are being developed for studies of the high-spin structure of nuclei lying far away from the valley of stability. Therefore, one should expect in the near future that new data on the high-spin structure of these nuclei will enlarge our understanding of excitations at high angular momenta.}

\begin{figure}[htbp]
	\centering
	\includegraphics[width=0.95\linewidth]{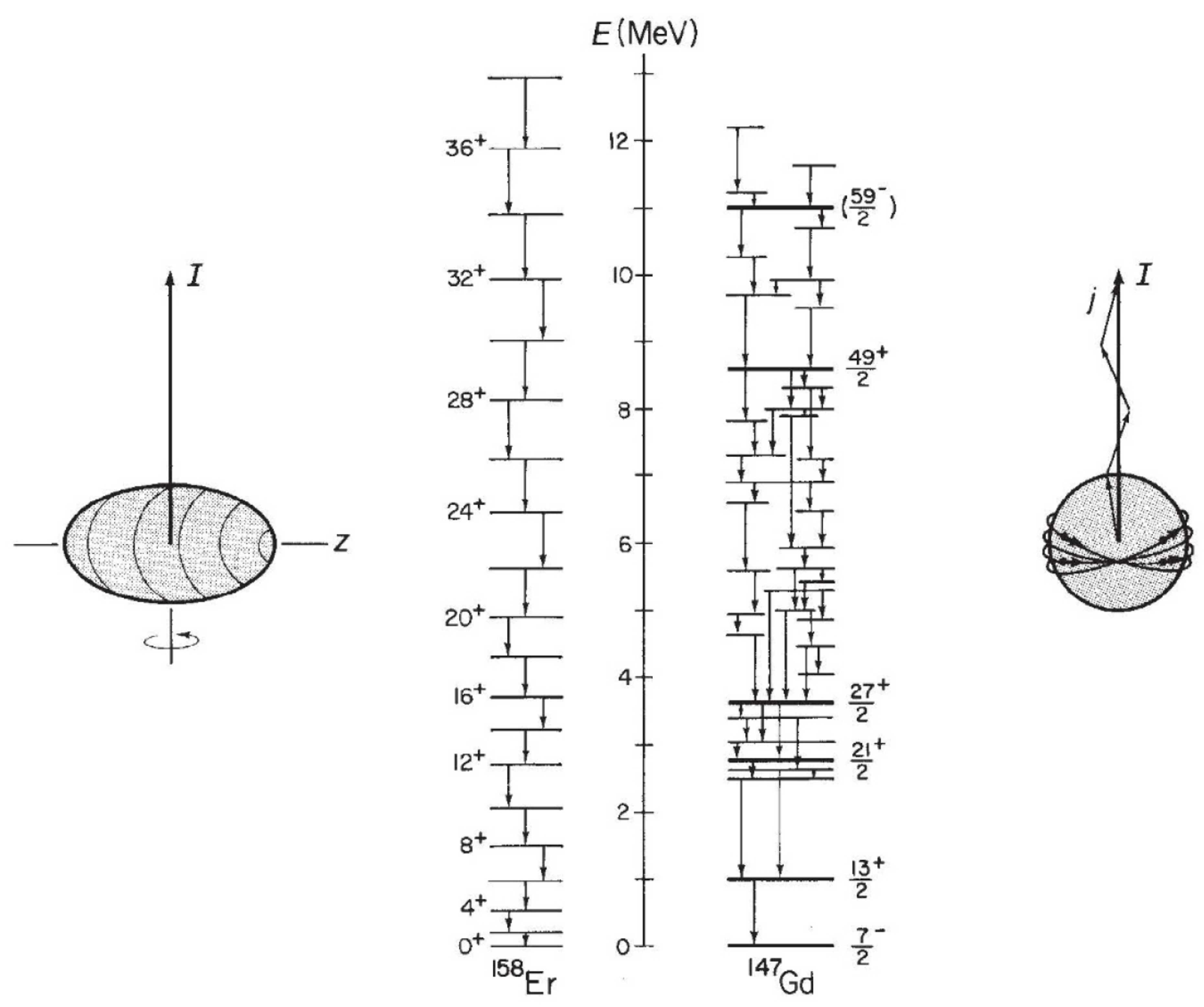}
	\caption{The lowest-energy high-spin states in $^{158}$Er and $^{147}$Gd. The sketches on both sides of the figure indicate the main mechanisms of accumulating nuclear angular momentum in each case, collective rotation (left) and single-particle alignment (right), respectively. The figure is taken from Ref.~\cite{diam84}.}
	\label{fig::31_spin_generation}
\end{figure}

There are two main mechanisms to generate high angular momentum in atomic nuclei, namely through collective rotation or by single-particle alignment, as illustrated in Fig.~\ref{fig::31_spin_generation}. The former mechanism, which is illustrated on the left-hand side of the figure for the ground-state rotational band in $^{158}$Er, appears in deformed nuclei, for which the spherical symmetry is broken. The single-particle alignment, schematically shown on the right-hand side of the figure for $^{147}$Gd, is a feature of nuclei that take spherical or close-to-spherical shapes.  In the figure, the isomeric states in $^{147}$Gd are indicated with thicker lines and with their spin-parity assignments shown.

The level structure of nuclei in the vicinity of the nuclear ground state is reasonably well understood at closed shells, where the collective motion is not very important, and the shell model accounts for the main features. In regions of well-deformed nuclei, the structure is also basically understood, as it is defined by the rotational motion of the system as a whole, superposed on a deformed shell-model level structure for the individual nucleons. Excited states in deformed nuclei are described by the rotational-energy formula

\begin{equation}\label{eq::K_rotational_energies}
E(I,K) = E_0 + \hbar^2\frac{ I(I+1) - K^2}{2\mathfrak{J}},
\end{equation}

\noindent where $E(I,K)$ is the energy of an excited state with spin $I$ and its projection on the nuclear symmetry axis $K$, $E_0$ is the band head energy, and $\mathfrak{J}$ is the nuclear moment of inertia. This formula reflects a basic concept of quantum mechanics, \textit{i.e.}, a perfect sphere cannot rotate because there is no point of reference against which its change in position can be detected. To observe the rotation of a quantal system, the spherical symmetry must be broken to allow an orientation in space to be defined. For ground-state bands in even-even nuclei, $K=0$, Eq.~(\ref{eq::K_rotational_energies}) takes the form
\begin{equation}\label{eq::rotational_energies}
E(I) = \frac{\hbar^2}{2\mathfrak{J}}I(I+1).
\end{equation}

Thus, at one limit, the nucleons act coherently, and collective bands develop that follow the $I(I+1)$ rotational pattern. In-band excited states are connected with stretched electric quadrupole, $E2$, transitions, such that the angular momentum of the nucleus increases with $\Delta I = 2$ at each step, as indicated on the left-hand side of Fig.~\ref{fig::31_spin_generation}. At the other limit, a few individual nucleons, which align their spins, generate all the angular momentum of a high-spin state. The nuclear moments of the long-lived states that are indicated on the right-hand side of Fig.~\ref{fig::31_spin_generation} have been measured~\cite{ston05a} and will be discussed in Sect.~\ref{subsub::spin_aligned}. Between these limits, more complex behavior is observed, as well as the coexistence of these limiting regimes. Studies of nuclei at high angular momenta address nuclear structure at these two limits and the interplay between them. Nuclear moment studies provide important information for the understanding of the structure in the high-spin regime. For such studies, it is necessary that the excited state of interest lives long enough to interact with an external magnetic dipole or electric quadrupole field. Thus, isomeric states are particularly amenable to both magnetic dipole and electric quadrupole moment measurements.

\begin{figure}[htbp]
	\centering
	\includegraphics[width=0.95\linewidth]{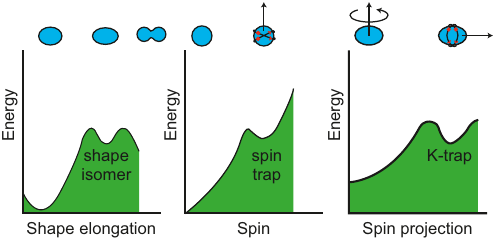}
	\caption{Excitation energy as a function of various nuclear variables. The secondary energy minima are responsible for the different kinds of isomers:
		(a) shape isomers; (b) spin isomers; (c) $K$ isomers. In each case, the relevant nuclear shapes are indicated in the upper part of the figure; where appropriate, angular momentum vectors are shown as arrows. For both the spin and the $K$ isomers, the angular momentum comes from a small number of orbiting nucleons, which are schematically illustrated in red in each case. The figure is taken from Ref.~\cite{walk99}. }
	\label{fig::31_isomer_generation}
\end{figure}

The decays of isomeric states are hindered because they require configuration changes of various types. Thus, nuclear metastable states are classified as shape, spin, seniority and $K$ isomers. The properties of the different types of isomers are systematized in Refs.~\cite{walk99,garg23}. The nuclear excitation energy {\it vs.} various nuclear variables that play a role in the appearance of isomers is presented in Fig.~\ref{fig::31_isomer_generation}. Recent reviews related to nuclear isomerism can be found in Refs.~\cite{drac16,walk20}.

 {In this chapter, we summarize the results in the field of high-spin nuclear physics obtained through studies of nuclear moments, which in many cases provided key information for the understanding of the underlying phenomena. In particular, we address nuclear moment studies of the structure of rotational bands in deformed and superdeformed nuclei, the structure of high-spin isomers and of magnetic and chiral bands.}

\subsection{Nuclear moments in axially deformed nuclei }\label{subsec:axially_def}
In deformed nuclei, measurements of nuclear magnetic moments often address the single-particle structure of the band-heads of the observed rotational bands,
while studies of quadrupole moments aim at the determination of the nuclear deformation. A compilation of the measurements of nuclear moments throughout the table of isotopes can be found in Refs.~\cite{ston05a,ston16}.

As noted and discussed in section~\ref{sec:oddA-rotors}, the usual approach in high-spin physics is to extract the ratio of the reduced transition probabilities $B(E2)/B(M1)$ and $g$ factors of the rotational bands from the energies and intensities of the in-band and intra-band transitions in odd-$A$ and doubly-odd nuclei. This is justified by the assumption that the nuclei of interest are well deformed and axially symmetric. For a given level of the rotational band with spin $I$, $B(E2)/B(M1)$ is given by Eq.~(\ref{eq::bm1be2}).

The mixing ratio can either be measured or can be determined from the branching ratio using Eq.~(\ref{eq::pureK-delta}). For strongly-coupled bands, the $B(M1)$ reduced transition probability is related to the $g$ factor through Eq.~(\ref{eq::bm1_strong_coupling}).

The $B(E2)$ reduced transition probability is related ~\cite{bohr75} to the intrinsic quadrupole moment, $Q_0$, by

\begin{equation}\label{eq::BE2_rotational}
B(E2; I \rightarrow I-2) = \frac{5}{16\pi} Q_0^2 \langle IK20 \vert I-2 K \rangle^2.
\end{equation}

\noindent Within the cranked shell model (CSM)~\cite{beng79a,beng79b}, which describes the nuclear excitations with respect to the rotational frame of reference, and making semiclassical approximations, the $B(M1)$ reduced transition probability can be defined as~\cite{dona83}

\begin{equation}\label{eq::BM1_CSM}
    \begin{aligned}
    B(M1; \Delta I = 1) &= \frac{3}{8\pi} K^2 (g_K – g_R)^2 \\
    &\times\bigg[\bigg(1 – \frac{K^2}{I^2}\bigg)^{1/2} – \frac{i}{I} + \frac{\Delta e^\prime}{\hbar \omega} – (g_{2qp} – g_R)\frac{i_{2qp}}{I}\bigg],
    \end{aligned}
\end{equation}

\noindent where the aligned angular momenta, $i$ and $i_{2qp}$, the signature splitting, $\Delta e^\prime$, and the rotational frequency, $\hbar\omega$, are defined with respect to the rotating frame of reference; $i$ denotes the aligned angular momentum of a single nucleon, and $i_{2qp}$ – the aligned angular momentum of an aligned pair of quasiparticles. The signature splitting is defined as the energy difference between two Routhians of different signature. The Routhian $e_{\omega}(Z,N,\beta)$ of a nucleus (Z, N) at frequency $\hbar\omega$ and deformation $\beta$ is the energy in a rotating frame of reference~\cite{beng79a,beng79b}.

In the following, we highlight experimental studies that were carried out to shed light on the structure of deformed nuclei. We discuss measurements related to the structure and deformations of the bandheads of rotational bands, properties of $K$ isomers, and studies related to superdeformed nuclear shapes.

\subsubsection{Band-heads of rotational bands }
\label{par::bandheads}
In even-even nuclei, the ground-state band is built on a $0^+$ state. The moments of two- or multi-quasiparticle states, which are the band-heads of excited rotational bands, are difficult to measure in even-even nuclei because they do not form isomeric states; high-$K$ isomers are an exception that will be discussed later in this section.

In odd-$A$ deformed nuclei, the band-heads of rotational bands are single-particle states, to which Nilsson-model configurations are assigned. As already mentioned, there is a wealth of information related to the $g$~factors and the quadrupole moments of such states~\cite{ston05a,ston16}. A special class of excitations are rotational bands built on intruder orbitals, {\it i.e.}, orbitals which originate from a higher shell and have unique parity, such as the $0g_{9/2}$ states in the $0f1p$ nuclei, the $0h_{11/2}$ states in the $1d2s$ nuclei, {\it etc}. Often, these are isomers, and in many cases their nuclear moments have been measured. Rotational bands, which are built on these configurations, benchmark the different regions of deformed nuclei, and their quadrupole and magnetic moments are used for comparison with the quantities extracted from the spectroscopic measurements of the in-band transitions and comparison with the theoretical models, see Eqs.~(\ref{eq::bm1be2},\ref{eq::pureK-delta},\ref{eq::bm1_strong_coupling},\ref{eq::BE2_rotational},\ref{eq::BM1_CSM}). However, a systematic survey of the existing information is still lacking.

Recently, the existing information about isomers was compiled and systematized~\cite{garg23,jain21}. For example, there are more than 70 $I^\pi = 9/2^+$ isomers, about 160 $I^\pi = 11/2^-$ isomers, and more than 90 $I^\pi = 13/2^+$ isomers, which can be associated with protons or neutrons occupying the $g_{9/2}$, $h_{11/2}$, and $i_{13/2}$ high-$j$ intruder unique-parity orbitals.

In doubly-odd nuclei, the rotational bands are built on two-quasiparticle configurations of the coupled proton and neutron. Similarly to odd-$A$ nuclei, the static moments of the band-heads provide crucial information for the characterization of these states.

The research efforts related to nuclear moments of low-lying states in exotic nuclei have employed a variety of techniques depending on their half-life. Ground state and isomeric moments were measured with the $\beta$-NMR technique (see Sect.~\ref{beta-NMR method}), by in-source and collinear laser spectroscopy, and using the NMR on-line (NMR-ON) method \cite{post86}. In the case of isomeric states, the TDPAD technique has also been applied, see Sect.~\ref{sec:isomers} for examples.  {A large amount of data related to band-head nuclear moments has been accumulated. The results are gathered in series of compilations~\cite{ragh89,ston05a,ston13a,ston14,ston15a,ston16,ston21,mert16a}. These data serve for comparison of the structure and deformations of these states. Although arguments based on systematics of nuclear moments are often used, a detailed study of the band-head properties throughout the Segr\'e diagram is still to be done.}

\subsubsection{High-spin states and backbending phenomena} \label{par::high-spin}

As discussed in Sect.~\ref{par::bandheads}, collectively rotating nuclei increase their angular momentum and excitation energy.  {At a certain rotational frequency, $\hbar\omega_c$, in deformed even-even nuclei, the ground-state band is crossed by a two-quasiparticle band, which becomes the yrast configuration. This phenomenon is called backbending~\cite{john71,john72} and is understood as due to the rotational alignment of a pair of quasiparticles~\cite{step72,step74}. The experimental observation is that a few rotational transitions from states with angular momenta of $I \gtrsim 10$ become lower in energy with increasing $I$, whereas transitions above and below this $I$-region have the normal rotational monotonic increase in energy as a function of angular momentum. In the adjacent odd-$A$ nucleus, the backbending results from the rotational alignment of the same pair of quasiparticles, which couple their spins to the angular momentum of the corresponding one-quasiparticle band. At the backbending frequency, $\hbar\omega_c$, it is more favorable to align the spin of two paired nucleons along the axis of rotation rather than increasing the rotational energy.} The backbending effect~\cite{john71,john72} was explained as the breaking of a specific pair of nucleons, moving in high-$j$ orbits, resulting in a jump of the angular momentum, when they aligned their internal motion to the external rotational field, {\it i.e.}, as due to rotational alignment~\cite{step72,step74}. In a more complete description, the ground-state band is crossed by a $2qp$ band, often called the s-band, which becomes yrast, and the backbending occurs in the band-crossing region. In general, in different parts of the Segr\'e diagram, rotational backbends are caused by the alignment of quasiparticles, either quasineutrons or quasiprotons, occupying intruder unique-parity orbitals.

Consequently, after the discovery of the backbending effect, rapid progress in the theoretical understanding of rapidly rotating nuclei took place, and by the end of the 1970s the Cranked Shell Model (CSM)~\cite{beng79a,beng79b,cwio80}  was developed, which created the foundation for understanding of the backbending phenomenon. Predictions were made of “higher order” backbends, rotational alignments, or band crossings at higher angular momentum values involving neutrons and protons, and situations where specific backbends might be “Pauli blocked.”

The $g$ factors of high-spin states beyond the backbending region depend mainly on the relative contributions of aligned protons and/or neutrons. The $g$ factor for states at and above the backbending, $g_{\rm S}$, with aligned angular momentum $i_S$, under the assumption that in even-even nuclei the $g$ factor of states in the ground-state band below the backbend is the rotational $g$ factor, $g_R$, is given as~\cite{frau81},
\begin{equation}\label{eq::g-average}
g_{\rm S} = g_R + (g_i - g_R) \frac{i_{\rm S}}{I_{\rm S}},
\end{equation}
\noindent where $g_i$ is the $g$ factor of the aligned nucleon, and $I_{\rm S}$ is the total angular momentum. The aligned spin $i_{\rm S}$ is determined from the separation of the two bands in the spin versus rotational frequency plot. From Eq.~(\ref{eq::g-average}) follows that the $g$ factors in the excited band increase when protons are aligned, while they decrease for neutron alignment.

\begin{figure}[htbp]
	\centering
	\includegraphics[width=0.7\columnwidth]{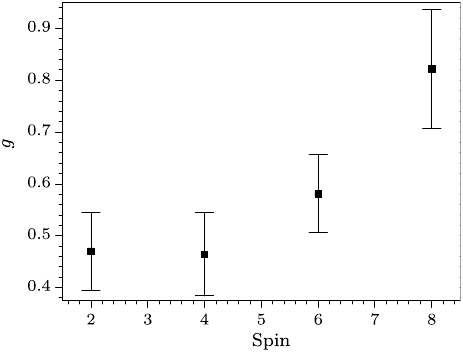}
	\caption{Measured $g$ factors of rotational states in $^{82}$Sr. The figure is taken from Ref.~\cite{yuan08a}.}
	\label{fig:g82Sr}
\end{figure}

In all deformed regions along the Segr\'e diagram, the band crossing is due to the rotational alignment of quasiparticles placed in intruder, {\it i.e.}, unique parity orbitals. Measurements of $g$ factors were carried out in several mass regions. In the mass $A = 80$ region, $g$ factors were measured in the yrast positive parity rotational band up to $I^\pi = 8^+$ in $^{82}$Sr using the transient-field (TF) technique~\cite{yuan08a,fan09}. The results are shown in Fig.~\ref{fig:g82Sr}. At low spins, the measured values $g \approx 0.46$ are consistent with the reported value for the $I^\pi = 2^+_1$ state in $^{84}$Sr, $g = 0.42(5)$~\cite{kuch88}. (See also Speidel et al. \cite{spei98} and \cite{mcco19} for more recent measurements of the 2$^+_1$-state $g$~factor in $^{84}$Sr.)
At $I^\pi = 6^+$, their values start to increase, indicating a $g_{9/2}$ proton alignment.

In the lanthanide nuclei, the backbending is understood as due to the rotational alignment of a $h_{11/2}$ quasineutron pair. The $g$ factors of discrete states in the ground-state band in $^{134}$Ce in the backbending region were carried out using both integral and time-dependent perturbed angular distribution measurements, with external fields and the static hyperfine field of Ce in iron and gadolinium hosts~\cite{gold80,zeme82}. In the experiment, the $g$ factor of the $10^+, \tau = 8(1)$~ps state was measured to be $g = -0.30(25)$, which is much lower compared to the states in the ground-state band, for which $<g> = 0.35$. This finding supports an $\nu h_{11/2}$ backbending.

In the rare-earth nuclei, the backbending is understood as due to the rotational alignment of an $i_{13/2}$ quasineutron pair. This suggestion was confirmed by a $g$-factor measurement of the  $I^\pi = 12^+_1$ state above the first backbend in $^{168}$W, $g(12^+_1) = -0.21(7)$. The negative sign and the magnitude of the $g$ factor are consistent with a stretched $i_{13/2}$ quasineutron configuration~\cite{bill86}. In the same experiment, the $g$ factors of the $2^+$ and $4^+$ states in the ground-state band were measured as $g(2^+) = +0.25(5)$ and $g(4^+) = +0.34(19)$, demonstrating a configuration change between the ground-state band and the s-band. This study used the internal (static) hyperfine field of W recoil-implanted into iron.

Multistep Coulomb excitation of $^{232}$Th and $^{238}$U coupled with the transient-field method provided the first direct evidence of $i_{13/2}$ {\em proton} alignment at high spin ($I=18$ - 24) in these two actinides, and evidence of neutron alignment below $I=16$ in $^{232}$Th~\cite{haus82a}.

An average high-spin $g$ factor for high-spin states in the quasi-continuum can also be measured by the transient-field technique following heavy-ion fusion–evaporation reactions in the rare-earth and higher-$Z$ regions. The measurements are usually performed with an array of Ge detectors. There have been relatively few such measurements, namely for rare earth nuclei: $^{152-156}\mathrm{Dy}$~\cite{haus84a,tara85,hass91a}, $^{162}\mathrm{Yb}$~\cite{mahn84}, $^{162–166}\mathrm{Hf}$~\cite{weis96,weis98}, and $^{168}\mathrm{W}$~\cite{bill86}; and in the mass $A \approx 180$ region for $^{180,182,184}\mathrm{Pt}$~\cite{robi02}. Similar measurements on Hg isotopes to measure $g$~factors in superdeformed bands are discussed below in Sect.~\ref{par::superdef}.


Robinson et al. measured the average quasi-continuum $g$ factors in $^{180,182,184}$Pt~\cite{robi02} by combining the transient-field technique with analysis of the perturbed $\gamma\gamma$ directional correlations observed in a multidetector array. The methodology of these experiments, which involves the interpretation of perturbed $\gamma \gamma$ angular correlations following the decays of reaction-aligned high-spin states, is discussed in detail in Refs.~\cite{stuc02a,robi02a}.

\begin{figure}[htbp]
	\centering
	\includegraphics[width=0.8\linewidth]{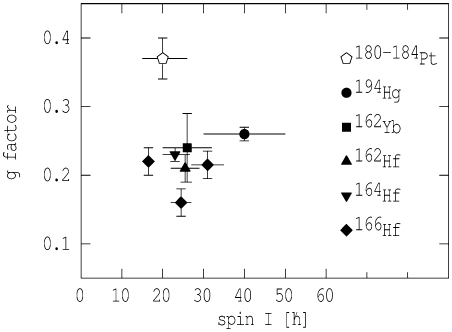}
	\caption{Comparison between the average quasicontinuum $g$ factors in even nuclei with $70 \leq Z \leq 80$. The horizontal lines associated with the data points indicate the approximate range of spins sampled. The figure is taken from Ref.~\cite{robi02}.}
	\label{fig::quasicontinuum-gfactors}
\end{figure}

In the Os-Pt nuclei with mass numbers $A \approx 180$, the competition between $\pi h_{9/2}$ and $\nu i_{13/2}$ alignment has attracted much attention~\cite{lara86,janz88,bala89,carp90,krei90,zhan90,kacz92,danc03}, a problem which is still not settled. For example, the first backbending in the $\pi h_{9/2}$ ground-state bands in $^{185,187}$Ir~\cite{andr77,bala89,danc03,moda10} is delayed compared to other rotational bands in the region. As demonstrated in Fig.~\ref{fig::quasicontinuum-gfactors}, the observed values of the average quasicontinuum $g$ factors in $^{180,182,184}$Pt~\cite{robi02} are higher than all measurements of average quasi-continuum $g$ factors in the rare-earth region. They are also larger than the measured $g(2^+_1)$ values in $^{180,182,184}$Pt~\cite{stuc96,bran98}. All these favor quasiproton alignment. However, a word of caution needs to be said, since there is only a subset of the states that experience the transient field, and hence the measured quasicontinuum $g$ factor cannot distinguish between details of the alignment scenarios. On the other hand, the aligned configurations observed in discrete spectroscopy must be present in the quasicontinuum, and, therefore, this measurement sheds additional light on understanding the problem.

 {In summary, in a series of nuclear moment measurements, the mechanism of the backbending effect was investigated in several mass regions of deformed nuclei within the Segr\'e diagram. The role of the different intruder orbitals, which are involved in the process, has been investigated. The results match the general understanding, which is based on the Nilsson model, except for the $A \approx 180$ region, where the data indicate that the first backbending is caused by the alignment of a $h_{9/2}$ quasiproton pair, while theoretical calculations favor the alignment of an $i_{13/2}$ quasineutron pair.}

\subsubsection{High-\textit{K} isomers }
\label{par::K_isomers}

 {The projection of the total angular momentum on the symmetry axis, $K$, is a good quantum number in axially deformed nuclei.} $K$ isomers arise from a sudden change in $K$ during the decay, see Fig.~\ref{fig::31_isomer_generation}. When the change of $K$, $\Delta K$, is larger than the multipolarity of the decaying transition, $\lambda$, the decay is hindered. The degree of hindrance of the corresponding $K$ isomer is $\nu = \Delta K - \lambda$. High-$K$ isomers are found in the mass $A \sim 180$ nuclei and in the superheavy nuclei. Details related to $K$ isomerism can be found in Refs.~\cite{drac05,kond15,drac16,walk24}. Here, we summarize the key features of such states. The conditions for their occurrence are axial symmetry, stable deformation, and the presence of high-$j$ orbitals near both proton and neutron Fermi surfaces, which can be coupled to form multi-quasiparticle states with high $K$ at relatively low energies and with limited decay paths. For example, in the hafnium nuclei, proton ($\pi^2$) and neutron ($\nu^2$) two-quasiparticle $2qp$ high-$j$ configurations are the building blocks of the observed $4qp$ high-$K$ isomers. The coupling of the $K = 6^+$, $\pi^2 5/2^+[402] \otimes 7/2^+[404]$ and $K = 8^-$, $\nu^2 9/2^+[624] \otimes 7/2^- [514]$ states, and of the $K = 8^{-}$, $\pi^2 9/2^+[514] \otimes 7/2^{-}[404]$ and $K = 8^-$, $\nu^2 9/2^+[624] \otimes 7/2^- [514]$ states results in the $K^\pi = 14^-$ and the $K^\pi = 16^+$ isomers in $^{178}$Hf. Here, the $2qp$ states are notated with their Nilsson-model quantum numbers.

According to the rotational model~\cite{bohr75}, the magnetic dipole moment of a state in a rotational band with spin $I$ built on a bandhead with a spin projection $K$ is
\begin{equation}\label{eq::mu_rotational_bandhead}
\mu = g_RI + (g_K - g_R)\frac{K^2}{I+1},
\end{equation}

\noindent which for the band head $K = I$ yields

\begin{equation}\label{eq::mu_rotational}
\mu = (g_KI + g_R)\frac{I}{I+1},
\end{equation}

\noindent where $g_K$ and $g_R$ are the single-particle and the rotational $g$ factors, respectively. There have been a number of experiments that aimed at measurements of magnetic dipole moments of high-$K$ isomers in the mass $A \sim 180$ nuclei. The magnetic moment measurements support the assigned multi-quasiparticle configurations, or help distinguish between different alternatives~\cite{hube75,walk80,aoki82,jain88,alde89,koni96,geor98,ione00a,drac00,thak06,muto14,rocc20}. Note that the deduced configurations are dependent on $g_K$ and $g_R$, which in the absence of a magnetic moment measurement can be obtained from the branching ratios of the associated rotational bands built on top of the isomers, provided the intrinsic quadrupole moment, $Q_0$ is known, or can be reliably estimated. Thus, the moment measurements place additional constraints on
$g_K$, $g_R$ and $Q_0$.

For estimation of the $g_K$ factors of the multi-quasiparticle configurations, the $g_K$ factors of the $1qp$ states are used, which are extracted from magnetic dipole moment measurements of ground state or isomeric states of the band heads of $1qp$ rotational bands having similar deformation. Thus, for a multi-quasiparticle state with $K = \sum K_i$, $g_K$ is defined as

\begin{equation}\label{eq::gK_additivity}
g_K = \frac{1}{K} \sum_i g_{K_i}K_i.
\end{equation}

\noindent The additivity of the $g_K$ values was demonstrated for a large number of high-$K$ isomers~\cite{ston13}. For example, for the $\pi^2\nu^2$ $4qp$, $K^\pi = 16^+$, 31~yr isomer in $^{178}$Hf, the estimated value based on the $g_K$ of the corresponding two-quasiproton and two-quasineutron configurations differs from the experimental value by $\sim 1$\%. It has been commented as well that the sum of the excitation energies for the two contributing $2qp$ states yields a value which differs from the excitation energy of the isomer by only 176~keV, a difference which is due to residual interactions~\cite{drac05,drac16}. All these observations demonstrate that the contributing quasiparticle excitations are approximately independent.

The spectroscopic quadrupole moment $Q$ is related to the intrinsic quadrupole moment $Q_0$ through Eq.~(\ref{eq:q_s-q_0}).
A few measurements of electric quadrupole moments of $2qp$, $3qp$, $4qp$, $5qp$ and $6qp$ $K$ isomers have been carried out~\cite{kain73,oert83,koni96,brou91,boos94,geor98,bala01,ione02,thak00,kuma07,biss07}. The results are summarized in Table~\ref{tab::Qmoments_Kisomers}. Different measurement techniques were used in the experiments, {\it e.g.}, the quadrupole moment of the $3qp$ $23/2^-$ isomer in $^{177}$Lu was measured using static nuclear orientation~\cite{oert83}, NMR on oriented nuclei~\cite{koni96}, and collinear laser spectroscopy~\cite{geor98,biss07}.

\begin{table}[h!]
\centering
\caption{Quadrupole moments of K isomers in the mass $A = 180$ region. The half-lives, $T_{1/2}$, spin-parity assignments, $K^\pi$, and configurations of the isomers are listed in the table.
}
\begin{tabular}{|c|c|c|c|r|}
\hline
   Nucleus  & $T_{1/2}$\tnote{a} & $K^\pi$ & Nilsson model configuration & $Q_s$~(eb) \\
\hline
  $^{177}$Lu & 160.4(1) d & $23/2^-$ & $\pi 7/2^+[404] \otimes$ & 5.71(5) \cite{geor98}\\
   & & & $\nu^2$($7/2^- [514]9/2^+[624])$ & +5.71(5) \cite{biss07} \\
  & & & & +5.2 (3) \cite{koni96} \\
  & & & & 4.23(67) \cite{oert83} \\
  $^{170}$Hf & 23(2) ns & $8^-$ & $\pi^2(9/2^-[514]7/2^+[404])$ & 4.91(17) \cite{kuma07} \\
  $^{171}$Hf & 18(2) ns & $23/2^-$ & $\pi^2(9/2^-[514]7/2^+[404])\otimes$ & 4.92(17) \cite{kuma07}\\
   & & &$\nu 7/2^+[633]$ &  \\
  $^{172}$Hf & 163(3) ns & $8^-$ & $\pi^2(9/2^-[514]7/2^+[404])$ & 5.40(19) \cite{kuma07}  \\
  $^{178}$Hf & 4.0(2) s & $8^-$ & $\nu^2$($9/2^-[514]7/2^{+}[404]$) & +4.99(4) \cite{biss07} \\
  & 31(1) y & $16^+$ & $\pi^2$($9/2^+[514]7/2^{-}[404]$)$\otimes$ & +6.00(7) \cite{boos94} \\
  & & & $\nu^2$($9/2^+[624]7/2^- [514]$) & \\
  $^{180}$Hf & 5.49(3) h & $8^-$ & $\pi^2$($9/2^+[514]7/2^{-}[404]$) & +4.6(3) \cite{kain73} \\
  $^{173}$Ta & 132(3) ns & $21/2^-$ & $\pi 9/2^-[514]\otimes$ & 6.23(18) \cite{thak00} \\
   & & & $\nu^2$($7/2^+[633]5/2^-[512])$  & \\
  $^{176}$W & 41(1) ns & $14^+$ & $\pi^2$($7/2^+[404]9/2^-[514]$)$\otimes$ & 5.99($^{+0.66}_{-0.82}$) \cite{ione02} \\
  & & & $\nu$($7/2^+[633]5/2^-[512]$) & \\
  $^{179}$W & 750(80) ns & $35/2^-$ & $\pi^2$($5/2^+[402]7/2^+[404]$)$\otimes$ & 4.00($^{+0.83}_{-1.06}$) \cite{bala01}\\
  & & & $\nu^3$($5/2^-[512]7/2^-[514]9/2^+[624]$) & \\
  $^{182}$Os & 150(10) ns & $25^+$ & $\pi^2$($9/2^-[514]11/2^-[505]$)$\otimes$  & 4.2(2) \cite{brou91}\\
  & & & $\nu^4$($9/2^+[624]7/2^+[633]7/2^-[503]7/2^-[514]$) & \\
\hline
\end{tabular}
\begin{tablenotes}
\item[($^a$)]~the values for the half-lives of the isomers are taken from Ref.~\cite{garg23}.
\end{tablenotes}
\label{tab::Qmoments_Kisomers}
\end{table}

The general understanding is that in well-deformed nuclei, the deformations of the ground state and the high-$K$ isomeric states are similar. The deformation parameter $\beta_2$ can be deduced from the measured spectroscopic quadrupole moment using Eqs.~(\ref{eq:q_s-q_0},~\ref{eq:q_moment_core_sharp}). Note that the nuclear radius is given as $R = r_0A^{1/3}$, and $r_0 = 1.2$~fm is related to the surface diffuseness. In most of the cases, the deduced deformation parameters for the ground states and isomers take close values, with three exceptions, $^{171}$Hf~\cite{kuma07}, $^{179}$W~\cite{bala01} and $^{182}$Os~\cite{brou91}, where the value of $\beta_2$ for the isomer is about 20\% lower than that of the ground state. The value of the quadrupole moment of the $35/2^-$ isomer in $^{179}$W deduced from the branching ratios of the cascade-to-crossover transition in the band built on top of the isomer~\cite{walk94} matches the direct measurement~\cite{bala01}. In the case of $^{182}$Os, the difference in deformation is understood to be caused by $\gamma$ softness~\cite{brou91}. On the other hand, the deformation of the $2qp$ $8^+$ isomer in $^{172}$Hf was found to be about 20\% higher compared to the ground-state deformation~\cite{kuma07}.

The excitation of multi-quasiparticle high-$K$ states provides a laboratory for studies of pairing in atomic nuclei. The idea is that as more orbitals lying close to the Fermi surface are occupied, the pairing will be reduced in a kind of controlled pattern, {\it i.e.}, the blocked orbitals are well characterized. Measurements of the $g$ factors of three-quasineutron $K^\pi = 21/2^-$ and the $5qp$ $K^\pi = 35/2^-$ isomers in $^{179}$W were discussed in terms of reduction of pairing, and it was concluded that static pairing is quenched with three quasineutron orbitals blocked~\cite{drac00}. The quadrupole moment of the $4qp$, $16^+$, $K$ isomer in $^{178}$Hf was measured using collinear laser spectroscopy (CLS)~\cite{boos94}. In the experiment, the mean-square charge radius, $\langle r^2\mathrm{(}^{178m}\mathrm{Hf)} \rangle $ was also measured.

\begin{figure}[htbp]
	\centering
	\includegraphics[width=0.8\linewidth]{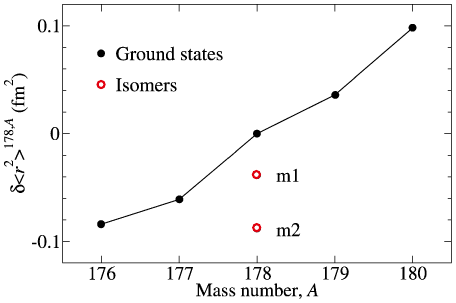}
	\caption{The isotope shifts of the mean-square charged radii in the Hf nuclei plotted with respect to the charge radius of the ground state in $^{178}$Hf. The isotope shifts of the $8^-$ and $16^+$ isomers in $^{178}$Hf are denoted as m1 and m2, respectively. The figure is taken from Ref.~\cite{biss07}.}
	\label{fig::31_Hf_shifts}
 \end{figure}

This isomer's mean-square charge radius appears to be smaller than that of the ground state, $\langle r^2\mathrm{(}^{178g}\mathrm{Hf)}\rangle $. The deduced difference of the isomeric shift $\delta \langle r^2 \rangle = -0.059(9)$~fm$^2$, is understood as quenching of pairing due to blocking of $4qp$, which results in a decrease of the surface diffuseness. In a follow-up CLS experiment~\cite{biss07}, the magnetic and quadrupole moments and the charge radius of the $8^-$ $2qp$ isomer in $^{178}$Hf were measured. The existing data for the isotope shifts in Hf nuclei~\cite{boos94,levi99,biss07} are presented in Fig.~\ref{fig::31_Hf_shifts}. Note that the value of the isomeric shift for the $^{178\mathbf{m}2}$Hf state (i.e., the 16$^+$ isomer) presented in the figure was recalibrated~\cite {biss07} in order to avoid the discrepancy between the isotope shifts, electronic factors, and charge radii, which is present in the data of Ref.~\cite{boos94}. An isomeric shift $\delta \langle r^2 \rangle = -0.035(6)$~fm$^2$ was observed also in $^{177}$Lu~\cite{geor98}, where $3qp$ excitations are blocked in the case of the $23/2^-$ isomer. This result was confirmed by a follow-up CLS experiment~\cite{biss07}.

\begin{figure}[htbp]
	\centering
	\includegraphics[width=0.8\linewidth]{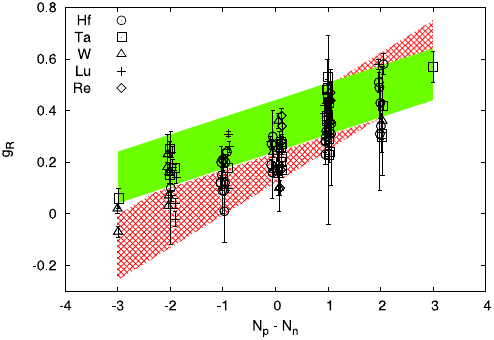}
	\caption{Plot of extracted $g_R$ values as a function of the difference between the number of proton and neutron quasiparticles, $N_p -N_n$. The colored and hatched bands represent alternative empirical model estimates which explore the degree to which the changes in $g_R$ in the multiparticle states can be treated as additive. The figure is taken from Ref.~\cite{ston13}.}
	\label{fig::31_gR_trend}
 \end{figure}

Two quantities are expected to be influenced by the reduction of the pairing, the moment of inertia, $\mathfrak{J}$, and the rotational $g$ factor, $g_R$. Changes in the nuclear moment of inertia can be deduced from the properties of the rotational bands, which are built on top of the high-$K$ states~\cite{drac00}. The systematic behavior of $g_R$ was studied in Ref.~\cite{ston13}. The demonstrated additivity of the $g_K$ values~\cite{ston13} provides an instrument to obtain values for $g_R$, based on the values of $\vert (g_K - g_R)/Q_0 \vert$ deduced from the branching ratios of the rotational bands. The results demonstrate a systematic behaviour of $g_R$ as a function of the difference between broken proton {\it vs} neutron pairs, $N_p - N_n$, taking lowest average values for the excess in numbers of broken neutron pairs, and highest values for the excess of broken proton pairs, as demonstrated in Fig~\ref{fig::31_gR_trend}.

This behavior is closely connected to the understanding of pairing and superfluidity of protons and neutrons in nuclei. Simple modeling connects the magnitude to $g_R$ to the partition of the collective motion between the protons and neutrons through their moments of inertia~\cite{ston13}, which is itself dependent on the magnitude of pairing. See also the discussion of deformation, pairing, moments of inertia, and the magnetic moments of rare earth nuclei in Ref.~\cite{stuc95}.

A recent measurement of the $g$ factor of the $K^\pi = 12^+$, $T_{1/2} = 124(8)$ isomer in $^{174}$W was reported to be $g = +0.304(11)$~\cite{rocc20}. By combining this value with the ratio $\vert (g_K - g_R)/Q_0 \vert = 0.04(1)$ deduced in Ref.~\cite{tand06}, an interesting interplay between values of the rotational $g$ factor and the intrinsic quadrupole moment, $Q_0$, was observed, namely if the deformation of the $K^\pi = 12^+$ isomer is similar to the one expected for the ground-state band in $^{174}$W, {\it i.e.}, $Q_0 \approx 7$~eb, $g_R$ is significantly smaller than that commonly assumed in the $A \approx 180$ mass region, {\it i.e.}, $g_R \approx 0.25$, and {\it vice versa}, if $g_R$ is close to this value, $Q_0$ would be smaller in the isomer than in the ground-state band. Clearly, more precise spectroscopic studies and direct measurements of the moments of $K$ isomers, as well as detailed theoretical studies, are needed to improve our knowledge of pairing and superfluidity in atomic nuclei.

High-$K$ configurations were proposed in the lanthanides as well. For example, $I^\pi = 8^-$ isomeric states have been systematically observed for the $N=74$ nuclei, {\it e.g.}, the properties of the band built on the $8^-$ isomer in $^{136}$Sm are consistent with the $7/2^+[404] \otimes 9/2^+[514]$ two quasi-neutron configuration~\cite{rega95}. The moments of the $I^\pi = 8^-$ isomer in $^{130}$Ba were measured using collinear laser spectroscopy~\cite{moor02}. The extracted values for the magnetic dipole and the electric quadrupole moment were found to be consistent with the proposed $2qp$  $7/2^+[404] \otimes 9/2^+[514]$ configuration. Also, the $g$ factor for the $8^-$ isomer in $^{128}$Xe was measured using the TDPAD method~\cite{lonn84}, $g= -0.036(9)$, which is consistent with the $\nu g_{7/2}h_{11/2}$ configuration.

A $3qp$ high-$K$ band was reported in $^{129}$Ba~\cite{byrn92}, which is built on the $23/3^+$, $T_{1/2}=27(2)$~ns isomer. The $g$ factor of this band head was measured using the TDPAD technique, $g= - 0.233(7)$~\cite{kaur13}, which demonstrates that it corresponds to the $5/2^+[402] \otimes 7/2^-[523] \otimes 11/2^-[505]$ configuration with an admixture of the $7/2^+[404] \otimes 7/2^-[523] \otimes 9/2^-[505]$ configuration.

 {In summary, $K$ isomers benchmark nuclear structure at high angular momenta in several mass regions of the Segr\'e diagram. Nuclear moment measurements have been carried out in the lanthanides and in the $A \approx 180$ rare-earth nuclei, which reveal their structure. The results are in agreement with the model expectations. A series of quadrupole moment measurements were performed aiming at testing the hypothesis that the ground-state and $K$ isomer deformations are similar. Few exceptions have been reported, which are attributed to different structural effects. However, the data are not sufficient for the creation of a consistent picture. $K$ isomers provide a playground for investigation of pairing in atomic nuclei. In general, the data seem to support the idea that pairing is reduced due to blocking of quasiparticle orbitals near the Fermi surface. It is worth noting that the topic still needs to be explored further. The existence of a multitude of $K$ isomers in some nuclei, {\it e.g.}, in $^{175,176}$Hf~\cite{kond15}, provides a possibility for systematic studies of pairing blocking effects through nuclear moment measurements.}

\subsubsection{Fission isomers and \textit{K} isomers in superheavy nuclei}
\label{par::superheavy}

Nuclear moment measurements for superheavy isotopes are scarce, see Ref.~\cite{ston14}. Moments in the heaviest isotopes have been measured for the ground states in $^{243,245,247}$Cm, $^{249}$Bk, and $^{253,254}$Es, as well as the $2^+$ isomer in $^{254}$Es. Most of these experiments were done in the 1970s. In the period covered by this review, Severijns et al.~\cite{seve09} measured the anisotropies of emitted $\gamma$-rays and $\alpha$ particles from oriented $^{250}$Bk, $^{253,254}$Es and $^{255}$Fm and determined the magnetic moment of the $(7^+)$ ground state in $^{253}$Es using the LTNO.

The second island of $K$ isomerism is found in the light superheavy nuclei in the mass $A \approx 250$ region, where $2qp$ and $4qp$ isomers were observed. Their properties are summarized in Ref.~\cite{kond15}. Their structure is still disputed, and more detailed spectroscopic studies, as well as nuclear moment measurements, are needed to shed light on the single-particle structure in this mass region. Of particular interest would be the $4qp$ states like the 2.928(3)~MeV $T_{1/2} = 184(2)$~$\mu$s isomer in $^{254}$No~\cite{clar10} and $T_{1/2} = 247(73)$~$\mu$s isomer in $^{254}$Rf~\cite{davi15}.

Fission isomers correspond to decays from states in the second well of the potential barrier; see Fig.~\ref{fig::31_isomer_generation}, left panel. These lie at the other extreme in terms of known properties, in that nothing is known of their spin, parity, or decay branches, except for the spontaneous fission half-lives. For reviews see, {\it e.g.}, Ref.~\cite{bjor80,meta80,kond15,drac16}.

Indications were reported of $2qp$ high-$K$ isomers in the second well of the even-even plutonium and curium isotopes~\cite{bjor80}. These states would take superdeformed shapes, as suggested by calculations~\cite{liu11} within the configuration-constrained potential energy surface model~\cite{xu98}. Detailed spectroscopy has not been done yet, and it is worth noting that such experiments would be very challenging. It is not known whether the isomers fission directly, or they decay to the corresponding second-well ground state, and then fission.

In a series of experiments in the 1970s the deformations of fission isomers in $^{236}$Pu~\cite{meta77}, $^{239}$Pu~\cite{habs77}, $^{238}$U~\cite{ulfe79}, $^{236}$U~\cite{meta80}, and $^{240}$Am~\cite{bemi79} were studied. The results are summarized in Ref.~\cite{meta80}. The obtained values for the quadrupole moments are, on average, about three times larger compared to ground-state measurements and correspond to a ratio of the major-to-minor axis of the nucleus $c/a \approx 2$, which confirms the suggestion that these states belong to the second well of the fission barrier and correspond to superdeformed shapes.

\subsubsection{Superdeformation }
\label{par::superdef}
Many nuclei take up extremely deformed shapes, corresponding to values of the deformation parameter $\beta_2 \approx 0.6$ at high angular momentum. Superdeformed nuclear shapes and the related rotational bands have been studied for many years, and a lot of their properties can be explained using the collective rotational model~\cite{bohr75}. They are understood to belong to the second well of the nuclear potential. Historically, such excitations were discovered in the actinide region near $A=240$, where fission (shape) isomers were identified and associated with the second potential well~\cite{poli62,poli73}. In the 1980s, with the development of the multi-detector $\gamma$-ray spectrometers, see, {\it e.g.}, Ref.~\cite{eber08} for a review of the field, discrete bands carrying large moments of inertia were discovered first in $^{152}$Dy~\cite{nyak84,twin86} and then in the neighbouring nuclei in the mass $A=150$ region. Nowadays, such excitations are established throughout the Segr\'e diagram. Their properties have been reviewed in a number of papers~\cite{nola88,twin90,jans91,bakt95}. These excitations are characterized by essentially constant energy spacing between transitions, {\it i.e.} picket-fence spectra,  strong population of superdeformed bands at high spins, and the lack of links between the superdeformed states and the yrast levels. A compilation of superdeformed bands and fission isomers can be found in Ref.~\cite{sing96,han96}.

\begin{figure}[htbp]
	\centering
	\includegraphics[width=0.95\linewidth]{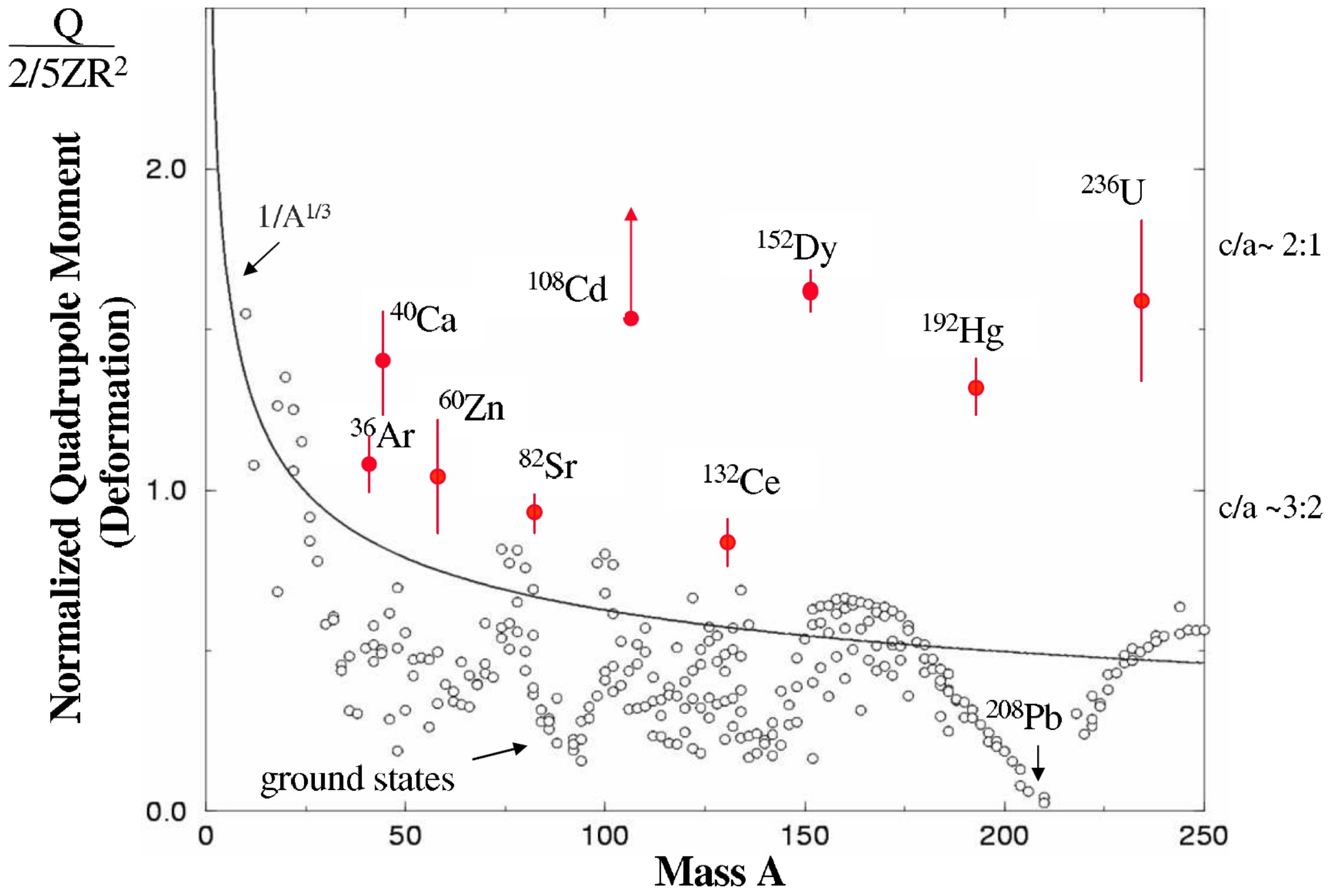}
	\caption{ Measured quadrupole moments scaled to remove the Z dependence for normal deformed (open circles) and superdeformed (filled circles) nuclei. The line gives the $A^{-1/3}$ dependence discussed in the text. Approximate prolate deformations given in terms of major-to-minor axis ratios are shown on the right. The figure is taken from Ref.~\cite{fall05}.}
	\label{fig::31_Q_superdef}
 \end{figure}

Transitional quadrupole moments of the superdeformed bands were extracted from lifetime measurements using the Doppler-shift attenuation method (DSAM)~\cite{nola79,dewa12}. In a heavy-ion fusion-evaporation reaction, the recoiling nuclei are usually produced with an initial recoil velocity of a few percent of the speed of light. They slow down in the target and backing materials and eventually come to rest. The DBLS measurement is based on the fact that the lifetimes of many nuclear states are of the same order of magnitude as this slowing-down time, {\it i.e.}, DSAM is used for measuring lifetimes in the range $10^{-13} > \tau > 10^{-15}$~s. Such measurements have provided the average transition quadrupole moments for superdeformed bands and verified that the nuclear deformation parameter, $\beta_2$, has a magnitude corresponding to superdeformation. The distinction between the quadrupole moments of normal and superdeformation excitations is illustrated in Fig.~\ref{fig::31_Q_superdef}.  The normalized quadrupole moment, $Q/(\frac{2}{5}ZR^2)$, for nuclei in the mass range $A \approx 10 - 200$ is plotted in the figure; data are from Ref.~\cite{rama87}. The deformation, expressed in terms of major-to-minor axis ratios of a prolate nucleus, is given on the right side. Bohr and Mottelson provide~\cite{bohr75} a simple estimate for the nuclear deformation in the absence of shell corrections, {\it i.e.}, the deformation $\beta_2$ is proportional to $A^{-1/3}$. The normalized quadrupole moments for selected superdeformed nuclei
are also shown in Fig.~\ref{fig::31_Q_superdef}. There are large deviations from the $A^{-1/3}$ trend
for nuclei with $A \sim 150, 190,$ and 240.

The average $g$ factors of high-spin, high-excitation energy, quasicontinuum structures were measured in $^{194}$Hg using the transient-field technique~\cite{maye98,weis99a}. By using a large Ge array and applying selected gating conditions to high-multiplicity events, it was possible to isolate specific decay paths in the $\gamma$ cascades and determine the average $g$ factors of the states leading to certain lower-energy discrete bands. The absolute values of the resulting $g$ factors depend on the parametrization of the transient field, but the relative $g$ factors of different ensembles of states are independent of the uncertainties in the transient hyperfine field. The experiment, which was analyzed in two independent ways, yields values for the average $g$ factor in the three superdeformed bands in $^{194}$Hg, $\langle g_{\rm {SD}} \rangle=+0.41(8)$ and $\langle g_{\rm {SD}} \rangle=+0.38(8)$, correspondingly. These values are significantly larger than the average $g$ factors of the normal-deformed states in $^{193,194}$Hg, which were measured with the same technique. Average quasicontinuum $g$ factors $\langle g\rangle = +0.19(1)$ and $\langle g \rangle = +0.26(1)$ were obtained for $^{193,194}$Hg, respectively. This result demonstrates that the nucleus in the superdeformed well behaves like a rigid rotor, {\it i.e.}, $g^{RR} = Z/A$. Note that the average $g$ factor reported for one of the superdeformed bands, SD$_3$, takes a larger value, $\langle g_{\rm {SD_3}} \rangle = +0.72(26)$, however, the authors do not consider it precise enough to draw firm conclusions.

 {Nuclear moment measurements of superdeformed states were used to provide estimates for the deformations of these excitations. The femtosecond half-lives of these states make the experiments difficult. Nevertheless, it was demonstrated that the extracted quadrupole moments correspond to large quadrupole deformations with a ratio of the long-to-short ellipsoid axes of $c/a \approx$~2:1 (see Fig.~\ref{fig::31_Q_superdef}). The measured average $g$ factors in superdeformed bands are in line with the idea that the nuclei are close to the rigid-rotor limit.}

\subsection{Spin-aligned isomeric states in nearly spherical nuclei}
\label{subsub::spin_aligned}
Spin-aligned states are observed in and near closed-shell nuclei. They take near-spherical shapes, and the alignment of the single-particle angular momenta often results in the appearance of isomeric states, called spin-traps, which are due to the fact that the decay of such states goes through transitions with high multipolarity, see Fig.~\ref{fig::31_isomer_generation}. Measurements of the moments of such states reveal their structure and deformations. Another class is seniority isomers, which are observed in semi-magic nuclei, having a closed proton or neutron shell, while the other type of particles are filling their shell. The seniority quantum number, $\nu$, refers to the number of particles that are not in pairs coupled to angular momentum $I = 0$. Seniority isomerism occurs because electric quadrupole $E2$ transitions between $\nu = 2$ states are small when the valence shell is close to half-filled. This result is a consequence of the fact that the matrix elements of the quadrupole operator between single-$j$ states with the same seniority vanish at the middle of the shell~\cite{talm93}. A compilation of spin traps and seniority isomers can be found in Refs.~\cite{jain21,garg23}. In this section, we discuss the structure of some multi-quasiparticle high-spin isomers, as well as the phenomena of magnetic and chiral rotation which occur in nearly-spherical or triaxial nuclei.

\subsubsection{Multi-quasiparticle high-spin isomers }
\label{par::multiparticle}
High-spin yrast trap isomers were first discovered in the Gd-Dy-Er nuclei with neutron number $N > 82$~\cite{pede77}. The mechanism of generating angular momentum is by aligning the angular momenta of individual particles, as illustrated in Fig.~\ref{fig::31_spin_generation}. Of special interest were the $N = 83$ isotopes $^{147}_{\;  64}$Gd, $^{149}_{\; 66}$Dy, and $^{151}_{\; 68}$Er, for which a series of yrast isomers were observed~\cite{garg23}. They decay via $E3$ transitions and are understood based on a particle-vibration coupling model. See Ref.~\cite{hama74} for a detailed discussion of the particle-vibration model applied to the Pb region. The interaction between the single-particle degrees of freedom and the octupole vibrational mode is the reason for the emergence of such isomers in nuclei not very far from the closed shell, which also explains the decay mode. For a recent review, see Ref.~\cite{drac16}.

Five isomers were observed in $^{147}$Gd, for which spin-parity was assigned as $13/2^+$, $21/2^+$, $27/2^-$, $49/2^+$, and $59/2^-$. The positive parity isomers decay via $E3$ transitions. Magnetic dipole~\cite{haus79,dafn87,hass89} and electric quadrupole moments~\cite{haus82,dafn85} were measured for all of them. The $g$-factor measurement for the $13/2^+$ isomer in $^{147}$Gd differs considerably from the value for the $\nu i_{13/2}$ orbital~\cite{rafa84}, which is explained by coupling with the $\nu f_{7/2} \otimes 3^-$ state. In fact, the dominant configuration must be $[\nu f_{7/2} \otimes 3^-]_{13/2^+}$. Oblate deformation was found for these states based on the quadrupole moment measurements~\cite{haus82,dafn85}. In one of the experiments, the tilted-foil TDPAD technique was used~\cite{dafn85}, which allowed a direct measurement of the signs of the deformation parameters. An increase of the oblate deformation with the increase of the angular momentum was observed, which is understood as due to a polarization effect of the aligned valence particles on the core~\cite{haus82}.

It is interesting to compare these findings with measurements in the other $N=83$ isotones, $^{149}$Dy, and $^{151}$Er. It is worth noting that the $g$-factor measurements in the rare-earth nuclei are rather difficult because of the large ambiguity of the effective magnetic field caused by the paramagnetic effect. So far, there is a single measurement in these nuclei, {\it i.e.}, of the $g$ factor of the $E_{\mathrm{ex}} = 8.522$~MeV, $I = (49/2)$, $T_{1/2} = 28$~ns isomer in $^{149}$Dy for which the observed value is $g=+0.41(6)$~\cite{wata03}. This value is similar to the $g$ factor of the $49/2^+$ isomer in $^{147}$Gd, $g=0.446(8)$, for which the $\pi(h^2_{11/2}d(^{-2}_{5/2})_0)) \otimes \nu(f_{7/2}h_{9/2}i_{13/2}(d^{-2}_{3/2})_0)$ configuration was suggested~\cite{haus79}. Exactly the same particles, {\it i.e.}, two protons and three neutrons in exactly the same orbitals, are suggested to form the configuration of the $(49/2)$ isomer in $^{149}$Dy, $\pi(h^2_{11/2}) \otimes \nu(f_{7/2}h_{9/2}i_{13/2}(d^{-2}_{3/2})_0)$. The only difference is that the configuration of $^{147}$Gd has two proton holes in the $d(^{-2}_{5/2})_0$, orbital which are coupled to zero, {\it i.e.}, the $^{147}$Gd isomer involves both the $Z = 64$ proton and $N = 82$ neutron core excitations, whereas only neutron particle–hole excitations across the $N = 82$ shell occur at the high-spin $^{149}$Dy isomer. The results are in good agreement with deformed independent particle model (DIPM)~\cite{doss81,neer81} calculations.

Next, the $g$ factor of the $I^\pi = 27^-$ $T_{1/2} = 1.6$~ns isomer in $^{152}$Dy was measured~\cite{fuji04} using the TIPAD technique, $g = +0.087(42)$. The known $g$ factor of the $I^\pi = 21^-$ isomer, which was populated simultaneously in the experiment, was used to deduce the effective magnetic field. While DIPM calculations indicate that the configuration of the isomer should be $\pi(h^2_{11/2}) \otimes \nu(f_{7/2}^2h_{9/2}i_{13/2})$, the value of the measured $g$ factor favors the $\pi(h_{11/2}d_{3/2}) \otimes \nu(f_{7/2}h_{9/2}i_{13/2}^2)$ configuration.

Another island of high-spin isomers was observed in the trans-lead region with neutron numbers $N \approx 126$ around $^{212}$Rn~\cite{garg23}. High-spin isomers in this mass region were first observed in the late 1970s~\cite{horn77,horn79}. Later, the At, Rn, and Fr nuclei were studied systematically in Canberra (for a review, see Ref.~\cite{drac16}). Their configurations correspond to spin-aligned high-$j$ valence protons and neutron-core excitations into the $g_{9/2}$, $i_{13/2}$, and $j_{15/2}$ orbitals. The neutron-core excitations are the building blocks in the configuration of high-spin isomers with $I$ above about $20 \hbar$. The decay of these yrast isomers is characterized by enhanced $E3$ transitions. These are understood as due to the coupling to the $3^-$  {octupole} vibration of the $^{208}$Pb core. Such correlations occur in the presence of pairs of single-particle orbitals with $\Delta \ell = 3$ and $\Delta j^{\pi} = 3^-$, such as the $\pi i_{13/2}$ -- $\pi f_{7/2}$ and $\nu j_{15/2}$ -- $\nu g_{9/2}$ pairs for the nuclei with $Z > 82$ and in the vicinity of $N = 126$. Thus, enhanced $E3$ transitions in the At-Rn-Fr nuclei are often related to $\pi (i_{13/2} \rightarrow f_{7/2})$ or $\nu (j_{15/2} \rightarrow g_{9/2})$ changes of configuration.

The $g$ factors of the high-spin isomers in the Rn region were measured for the $^{209-212}$At~\cite{haus76,sjor76,rahk78,sjor79,berg85}, $^{210-213}$Rn~\cite{horn77,horn79,maie81,pole86,stuc88a,stuc88b}, and $^{210-214}$Fr~\cite{horn79,byrn86,byrn94} nuclei. As noted in Ref.~\cite{drac16}, they are given by $g \approx (1.1 I_{\pi} - 0.1 I_{\nu})/I$, where $I$ is the spin of the isomer and $I_{\pi}$ and $I_{\nu}$ are the angular momenta of the valence protons and neutron-core excitations, respectively.
Often, the isomeric states have configurations that are mixed through coupling with the 3$^-$ core vibration of $^{208}$Pb. A multiparticle octupole coupling model, which explicitly accounts for the octupole coupling, is able to account for both $g$~factors and the observed $B(E3)$ transition strengths \cite{pole86,byrn86,stuc88a,stuc88b,stuc92a,stuc93,stuc98a}.

Another question that has been posed concerns the role of deformation in the formation of the isomers. The ground-state quadrupole moments in $^{211-213}$Fr nuclei~\cite{coc85} correspond to small oblate deformations. Quadrupole moment measurements of high-spin isomers were performed for the $^{209-211}$At~\cite{mahn83,mahn87,sche91}, $^{210-212}$Rn~\cite{dafn85a,berg85,berg86}, and $^{211-214}$Fr~\cite{byrn90,hard91a,neye95} nuclei. The data for the quadrupole moments in the $^{209-211}$At and $^{211-214}$Fr nuclei are listed in Table~\ref{tab::Qmoments_yrast}. Results for $^{211-212}$Fr, which were measured using the TDPAD and LEMS~\cite{hard91} techniques, were cross-compared in Refs.~\cite{byrn90,hard91a}. The calculated value, $Q_s^{\rm calc} = -0.70(7)$~barn, for the quadrupole moment of the $I^\pi = 29/2^+$ isomer in $^{213}$Fr was adopted for calibration. Similarly, for the LEMS measurements at the At nuclei, the measured quadrupole moment of the $29/2^+$ isomer in $^{211}$At was used as the calibration standard.

It is expected that alignment of the spins of the valence nucleons would result in concentration of valence nucleons in the equatorial plane, which in turn leads to a polarization of the core. The configurations of the lowest-lying $21/2^-$ and $29/2^+$ isomers in the odd-$A$ nuclei and the $11^+$ and $15^-$ isomers in the even-$A$ nuclei are described by the coupling of $h_{9/2}$ and $i_{13/2}$ protons to the corresponding Pb core. An increase of the quadrupole moments with the increasing removal of neutrons from the $\nu p_{1/2}$ orbital was observed in the $^{209-211}$At nuclei~\cite{mahn83} (see Table~\ref{tab::Qmoments_yrast}). The observed larger polarization was interpreted as an increased softness, {\it i.e.}, preference for deformation~\cite{mahn83,mahn87}. Follow-up measurements on higher-spin isomers addressed the role of neutron-core excitation from the $3p_{1/2}$ orbital to the $2g_{9/2}$ orbital, which softens the core and is related to an increase of the quadrupole moments of the $(19)^+$ and ($39/2^-$) isomers in $^{210,211}$At~\cite{sche91}. The extracted core deformation parameters, $\beta$, increase with the increase of the spin of the isomers, and are in agreement with calculations within the deformed independent-particle model (DIPM)~\cite{mats78}.

The quadrupole moments of the proton-aligned isomers in $^{208,210,211}$Rn were measured~\cite{berg86}. An increase is observed for the quadrupole moments of states with the same proton configuration when the number of neutron holes increases. The effect is more pronounced than the increase observed for At nuclei. Dafni {\it et al.} reported a low value of the quadrupole moment for the $I^\pi = (63/2^-)$ isomer in $^{211}$Rn~\cite{dafn85}, in contrast with the observations with the DIPM predictions and the observations in the neighbouring At nuclei, where the core-polarization effect holds up to the highest yrast isomers for which quadrupole moments were measured. The extracted core deformation $\beta = - 0.03(1)$ is similar to the ground-state deformation and is much lower than the DIPM prediction of $\beta = - 0.1$~\cite{mats78}.

The core-softness effect was also addressed for the Fr nuclei~\cite{hard91a}. Since there were no DIPM calculations for the Fr nuclei, it was assumed that the expected deformation for the ($27^-$) and ($65/2^-$) should be similar to that of the ($63/2^-$) isomer in $^{211}$Rn, a value that is much higher than the experimental values. On the other hand, the quadrupole moments measured in $^{212}$Fr were found to be in agreement with shell-model calculations which include the octupole vibrational coupling~\cite{byrn90}.

\begin{table}[h!]
\centering
\caption{Quadrupole moments of yrast isomers in the At and Fr nuclei. The spin-parity assignments, $I^\pi$ and the configurations are listed.}
\begin{tabular}{|c|c|c|c|l|}
\hline
   $N$  & $I^\pi$\tnote{a} & $Q_s$~(eb) $_{85}$At & $Q_s$~(eb) $_{87}$Fr & configuration \\
\hline
 124 & $21/2^-$ & 0.78(8)~\cite{mahn83} & & $\pi h^3_{9/2} \nu     (p^{-2}_{1/2})_{0^+}$ \\
      & & 0.78(8)~\cite{sche91} & & \\
     & $29/2^+$\tnote{b} & 1.50(15)~\cite{mahn83} & & $\pi h^2_{9/2}i_{13/2} \nu (p^{-2}_{1/2})_{0^+}$ \\
     & & 1.50(15)~\cite{sche91} & & \\
     & & & -- 1.07(18)~\cite{hard91a} & $\pi h^4_{9/2}i_{13/2} \nu (p^{-2}_{1/2})_{0^+}$ \\
     & ($45/2^-$) & & -- 1.98(56)~\cite{hard91a} & $\pi h^3_{9/2}i^2_{13/2} \nu (p^{-2}_{1/2})_{0^+}$ \\
\hline
 125 & $(11)^+$ & 0.65(8)~\cite{mahn83} & & $\pi h^3_{9/2} \nu p^{-1}_{1/2}$   \\
      & $15^-$\tnote{c} & & 1.22(12)~\cite{mahn83} & $\pi h^2_{9/2}i_{13/2} \nu p^{-1}_{1/2}$  \\
      & & & 0.84(13)~\cite{byrn90} & $\pi h^4_{9/2}i_{13/2} \nu p^{-1}_{1/2}$\\
      & & & -- 0.80(12)~\cite{hard91a} & \\
      & $(19)^+$ & 2.20(5)~\cite{sche91} & & $\pi h^2_{9/2}i_{13/2} \nu (p^{-2}_{1/2})_{0^+} g_{9/2}$  \\
      & ($27^-$) & & 1.74(33)~\cite{byrn90} & $\pi h^3_{9/2}i^2_{13/2} \nu (p^{-2}_{1/2})_{0^+}g_{9/2}$ \\
\hline
 126  & ($21/2^-$) & 0.53(5)~\cite{mahn83}\tnote{c} & & $\pi h^3_{9/2} \nu p^{-2}_{1/2}$  \\
      & ($29/2^+$) & 1.01(19)~\cite{mahn83}\tnote{d} & & $\pi h^2_{9/2}i_{13/2} \nu p^{-2}_{1/2}$  \\
      & ($39/2^-$) & 1.91(25)~\cite{sche91} & & $\pi h^2_{9/2}i_{13/2} \nu p^{-1}_{1/2}g_{9/2}$  \\
      & ($65/2^-$) & &  -- 2.19(53)~\cite{hard91a} & $\pi h^3_{9/2}i^2_{13/2} \nu (p^{-2}_{1/2})_{0^+} g_{9/2}i_{13/2}$ \\
\hline
 127 & ($11^+$) & &  0.82(22)~\cite{neye95} & $\pi
     h^4_{9/2}i_{13/2} \nu g_{9/2}$ \\
     & ($33^+$) & &  2.15(54)~\cite{neye95} & $\pi h^3_{9/2}i^2_{13/2} \nu p^{-1}_{1/2}g_{9/2}i_{13/2}$\\
\hline
\end{tabular}
\begin{tablenotes}
\item[($^a$)]~the $I^\pi$ values of the isomers are taken from ENSDF~\cite{tuli96}
\item[($^b$)]~tentative spin-parity assignment for the isomer in $^{211}$Fr
\item[($^c$)]~tentative spin assignment for the isomer in $^{210}$At
\item[($^d$)]~value derived from the $B(E2; 21/2^- \rightarrow 17/2^-)$ value
\item[($^e$)]~values used as a calibration standard for the LEMS measurements
\end{tablenotes}
\label{tab::Qmoments_yrast}
\end{table}

 {Nuclear moments of yrast trap isomers were studied for the $N > 85$ Gd-Dy-Er rare-earth nuclei and the $N \approx 126$ trans-lead nuclei. They are understood as due to the coupling of aligned particles and octupole vibrations. Oblate quadrupole deformations were measured for the yrast traps in $^{147}$Gd~\cite{haus82,dafn85}. The configurations of these isomers are understood as based on spin-aligned high-$j$ particles coupled to core excitations at higher angular momenta. The measured $g$-factors are described within the deformed independent particle model (DIPM) for the rate-earth isomers and the multiparticle octupole coupling model in the trans-lead nuclei. Electric quadrupole moments were measured to understand the effect of core polarization. Deformation parameters were extracted from the experimental data and compared to DIPM calculations. In general, a consistent description of the structure and decays of these states has been achieved~\cite{drac16}. }

\subsubsection{Magnetic and anti-magnetic rotation}
\label{par::magnetic_bands}
In the early 1990s, very regular $I(I+1)$ patterns of $\gamma$-rays were detected in the decays of nuclei that were known to be almost spherical~\cite{fant91,hube92,bald92,clar92}. The angular distribution and polarization of the $\gamma$-rays showed that they were not electric quadrupole $E2$ transitions, but magnetic dipole $M1$ transitions. In 1993, Stefan Frauendorf suggested that the patterns were due to a new excitation mode of the nucleus, \textit{i.e.}, that the angular momentum could be generated by just a few of the protons and neutrons, with the remaining nucleons being passive observers~\cite{frau93}. In this picture, the angular momentum of quasiprotons (quasineutrons) is along the deformation axis, and the angular momentum of quasineutrons (quasiprotons) is along the rotational axis, then the total angular momentum is generated by the gradual alignment of momentum vectors of the deformation-aligned and the rotation-aligned quasiparticles. This mechanism resembles the closing blades of a pair of shears, and thus, the mechanism has been referred to as the shears mechanism. Consequently, these bands were called magnetic or `shears' bands. The bands consist of strong $M1$, $\Delta I=1$ intraband transitions with only weak $E2$ crossovers, thus resulting in large ratios of the reduced transition probabilities $B(M1)/B(E2) \geq 20$~$ \mu_N^2/$(eb)$^2$. The angular momentum increases mainly due to the shears effect, which is a step-by-step alignment of high-$j$ proton and neutron orbitals into the direction of the total angular momentum. These excitations were reviewed in Ref.~\cite{hube05}. A compilation of the observed magnetic bands can be found in Refs.~\cite{amit00,kuma23,teng24}.

Another possibility of generating angular momentum in a near-spherical nucleus is by coupling of high-$j$ deformation-aligned proton holes (in time-reversed orbits) with rotation-aligned neutron particles. In this picture, two shears-like subsystems can be considered whose magnetic moments are antialigned. Hence, the net magnetic moment is zero, resulting in the absence of dipole transitions between the levels of the band. The rotational structure is characterized by weak $E2$ transitions with decreasing $B(E2)$ rates with increasing spin and by a large $\mathfrak{J}^{(2)}/B(E2) \sim 100~\mathrm{MeV}^{-1}(\mathrm{eb})^{-2}$ ratio, where $\mathfrak{J}^{(2)} \approx 4/\delta E_{\gamma}$ is the dynamical moment of inertia and $\delta E_{\gamma}$ is the difference between two successive $\gamma$-ray transition energies in the band. This phenomenon has been termed as antimagnetic rotation, and the bands are called antimagnetic bands.

At the time of writing, more than 200 magnetic rotational bands have been identified in 114 isotopes, and 37 anti-magnetic rotational bands in 27 nuclei~\cite{kuma23}. These are described in terms of the tilted-axis cranking (TAC) model~\cite{frau93,frau01,frau18}. Magnetic bands were reported in several mass regions in the vicinity of closed shells, {\it e.g.}, in the Pb region, where 88 bands have been observed in 33 nuclei, where they are most systematically studied. Anti-magnetic bands have been observed in the Pd-Cd-In nuclei.  {This is an active field of research nowadays, and large amounts of data are being generated by the INGA collaboration~\cite{pali25}, as well as by a few Chinese nuclear spectroscopy groups~\cite{teng24}.}

\begin{figure}[tp]
	\centering
	\includegraphics[width=0.95\linewidth]{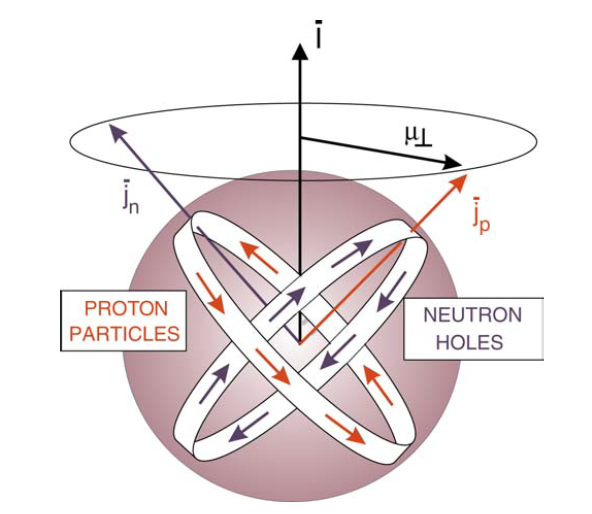}
	\caption{Schematic presentation of the spin-coupling scheme of a shears state. The perpendicular coupling of the proton-particle and neutron-hole orbitals, having angular momenta $\Vec{j_p}$ and $\Vec{j_n}$, respectively, results in a large transverse component of the magnetic moment vector, $\mu_{\perp}$, that rotates around the total angular momentum, $\Vec{I}$, and creates the enhanced $M1$ transitions between the shears states. The figure is taken from Ref.~\cite{hube05}.}
	\label{fig::31_magnetic_vectors}
 \end{figure}

The $\Delta I=1$ bands observed in the Pb isotopes have been assigned weakly oblate configurations consisting of the $K^\pi = 11^-$ excitation arising from two-proton particles in $\pi h_{9/2} \otimes i_{13/2}$ orbitals coupled to low-$\Omega$ one/three quasi-neutrons (in odd-$A$ nuclei) or two/four quasi-neutrons (in even-$A$ nuclei) dominated by $i_{13/2}$ neutron holes; see Table~\ref{tab::moments_blades}, where the configurations of the lowest-lying shears states in $^{191-197}$Pb are summarized. It should be noted that the role of high-$j$ orbitals is crucial in these structures as it is in many other rotational phenomena. The measured values of the $g$ factors and quadrupole moments for the blades of the shears are included in the table. The coupling of the proton and neutron angular momentum vectors at the bandhead of a shears band is shown schematically in Fig.~\ref{fig::31_magnetic_vectors}.

 A measurement of the quadrupole moment of the $K^\pi = 11^-$, $T_{1/2} = 72(4)$~ns isomer in $^{196}$Pb has been performed~\cite{vyve02}. This state, with configuration $\pi(2s_{1/2}^{-2}0h_{9/2}0i{13/2})_{11^-}$, is identified as the ``proton blade" of the shears band. The level-mixing spectroscopy (LEMS) technique~\cite{hard91} was used. Since the LEMS curve is sensitive to the ratio of the quadrupole interaction frequency, $\nu_Q = eQV_{zz}/h$, to the magnetic interaction frequency, $\nu_{\mu} = \mu B/h$, it is necessary to know the magnetic moment, $\mu$, the electric field gradient of the host lattice, $V_{zz}$, and the value of the applied magnetic field, $B$. The ``neutron blade" of the shears is the $I^\pi = 12^+$, $T_{1/2} = 270(4)$~ns isomer. The $11^-$ isomer decays to the $12^+$ isomer through a 497.8~keV $E1$ transition. The adopted values used to derive the nuclear moments for the blades of the shears in $^{194,196}$Pb are summarized in Ref.~\cite{vyve04a}.

The data were analyzed in two ways, initially by assuming that the Larmor frequency for the lower-lying $12^+$ isomer is not affected significantly by the higher-lying $11^-$ isomer~\cite{vyve02}, and then taking into account the feeding from the higher isomer by applying a ``double-perturbation formalism"~\cite{vyve02b}. The results from this measurement and other experiments related to the magnetic and quadrupole moments of the blades of the shears in the $^{191-197}$Pb nuclei are presented in Table~\ref{tab::moments_blades}. A measurement of the $g$ factors of the $12^+$ $\nu i_{13/2}^{-2}$ and the $11^-$ $\pi h_{9/2}i_{13/2}$ isomers in $^{188}$Pb was carried out with the TDPAD technique~\cite{ione10}. The $g$ factor of the $12^+$ state, $g = -0.179(6)$ follows the observed slight down-sloping evolution of the $g$ factors of the $i_{13/2}^{-2}$ neutron states with decreasing $N$. The $g$ factor of the $11^-$ state, $g = +1.03(3)$, takes a similar value to the $g$ factors of the corresponding states in the heavier $^{194,196}$Pb isotopes.

 The $^{194,196}$Pb data on the quadrupole moments provide a possibility to approach the deformations of the $I^\pi = 16^-$ shears states. This was done using the additivity formalism, assuming the additivity of the $\hat{E}^0_2$ quadrupole operator in the intrinsic frame, $\hat{E}^0_2(\mathrm{tot}) = \hat{E}^0_2(\pi) + \hat{E}^0_2(\nu)$. Thus, the quadrupole moment of the $16^+$ shears state in $^{196}$Pb was found to be $Q_s = -0.32(10)$~$e$b, a value which is four times smaller than theoretical predictions within the tilted cranking model~\cite{vyve02}, {\it i.e.}, quenching of additivity was observed in this case.


\begin{table}[h!]
\centering
\caption{
Magnetic dipole, $\mu$, and spectroscopic electric quadrupole, $Q_s$, moments of states identified with shears magnetic rotational bands in $^{191-197}$Pb. States identified with the proton and neutron blades are also included.
The assigned configurations are indicated along with spin-parity assignments.
Original references are provided in each case. References after shears-state configurations correspond to spectroscopy work where the bands were identified. Data are from Refs.~\cite{ston05a,kuma23}, unless explicitly indicated. The proton ($h^2_{9/2}$)$_{8^+}$ and ($h_{9/2}i_{13/2}$)$_{11^-}$ configurations, and the neutron ($i_{13/2}^{-2}$)$_{12^+}$ configuration, are denoted as $\pi^2 8^+$, $\pi^2 11^-$ and $\nu^2 12^+$, respectively.}
\begin{tabular}{|c|c|c|c|c|}
 \hline
 Nucleus & $I^\pi$ & $\mu$ ($\mu_N^2$) & $Q_s$ ($e$b) & configuration \\
 \hline
 $^{191}$Pb & $13/2^+$ & -1.172(7) \cite{dutt91} & & $\nu i_{13/2}^{-1}$ \\
  & $27/2^-$ & & & $\pi^2 8^+\otimes \nu i_{13/2}^{-1}$ \cite{foti98} \\
  & ($29/2^-$) & & & $\pi^2 11^-\otimes \nu i_{13/2}^{-1}$ \cite{foti98}\\
  \hline
 $^{192}$Pb & $12^+$ & 2.08(2) \cite{sten83a} &  0.32(4) \cite{ione07} & $\nu^2 12^+$ \\
 & & 2.00(24) \cite{kmie10} & &  \\
 & $11^-$ & & 2.9(3) \cite{ione07} & $\pi^2 11^-$ \\
 & $15^- $ & & & $\pi^2 11^- \otimes \nu^2 12^+$ \cite{plom93} \\
 & $18^- $ & & & $\pi^2 11^- \otimes \nu^2 12^+$ \cite{plom93} \\
 \hline
 $^{193}$Pb & $13/2+$ & -1.150(7) \cite{dutt91} & +0.195(10) \cite{dutt91} & $\nu i_{13/2}^{-1}$ \\
 & ($21/2^-$) & -0.62(12) \cite{ione04} & 0.22(2) \cite{bala03,bala11} & $\nu^2 12^+ \nu 3p_{3/2}$ \\
 & ($27/2^-$)\tnote{a} & +9.9(4) \cite{chme97} & 2.6(3) \cite{bala03,bala11} & $\pi^2 11^- \otimes \nu i_{13/2}^{-1}$ \cite{ducr96,bald96} \\
 & ($33/2^+$) & -2.82(15) \cite{ione04} & 0.45(4) \cite{bala03,bala11} & $\nu i_{13/2}^{3}$ \\
 & ($45/2^-$) & & & $\pi^2 11^- \otimes \nu i_{13/2}^{3}$ \cite{ducr96,bald96} \\
 \hline
 $^{194}$Pb & $12^+$ & -2.00(2) \cite{sten85} & 0.49(3) \cite{sten85} & $\nu^2 12^+$ \\
 & & -1,90(7) \cite{roul77} & & \\
 & $11^-$ & +11.3(2) \cite{vyve04} & (-)4.8(7)\tnote{b} \cite{vyve02b} & $\pi^2 11^-$ \\
 & & & 3.6(4) \cite{ione07} &  \\
 & $14^-$ & & & $\pi^2 8^+ \otimes \nu i_{13/2}^{-1}p_{3/2}$ \cite{kuts09} \\
 & ($15^+$) & & & $\pi^2 8^+ \otimes \nu^2 12^+$ \cite{kuts09} \\
 & $15^+$ & & & $\pi^2 11^- \otimes \nu i_{13/2}^{-1}p_{3/2}$ \cite{kuts09} \\
 & $17^-$ & & & $\pi^2 11^- \otimes \nu^2 12^+$ \cite{kuts09} \\
 \hline
 $^{195}$Pb & $13/2^+$ & -1.128(7) \cite{dutt91} & +0.306(15) \cite{dutt91} & $\nu i_{13/2}^{-1}$ \\
 & & -1.1318(13) \cite{ding87} & +0.29(10) \cite{ding87} & \\
 & $27/2^-$ & & & $\pi^2 11^- \otimes \nu i_{13/2}^{-1}$ \cite{kaci96} \\
 & ($39/2^-$) & & & $\pi^2 11^- \otimes \nu i_{13/2}^{3}$ \cite{kaci96} \\
 \hline
 $^{196}$Pb & $12^+$ & -1.92(2) \cite{sten83} & 0.65(5) \cite{zywi81} & $\nu^2 12^+$ \\
 & & -1.88(8) \cite{roul77} & & \\
 & $11^-$ & +11.4(6) \cite{vyve04} & (-)3.6(6)\tnote{b} \cite{vyve02} & $\pi^2 11^-$ \\
 & $14^-$ & & & $\pi^2 8^+ \otimes \nu i_{13/2}^{-1}(p_{3/2}f_{5/2})^{-1}$ \cite{sing02} \\
 & $14^+$ & & & $\pi^2 8^+ \otimes \nu^2 12^+$ \cite{sing02} \\
 & $17^+$ & & & $\pi^2 11^- \otimes \nu i_{13/2}^{-1}(p_{3/2}f_{5/2})^{-1}$ \cite{sing02} \\
 & $16^-$ & & -0.32(10)\tnote{c} \cite{vyve02} & $\pi^2 11^- \otimes \nu^2 12^+$ \cite{sing02} \\
 \hline
 $^{197}$Pb & $13/2^-$ & -1.098(11) \cite{dutt91} & +0.38(2) \cite{dutt91} & $\nu i_{13/2}^{-1}$ \\
 & & -1.105(3) \cite{anse86} & +0.5(3) \cite{anse86} & \\
 & $21/2^-$ & -0.531(6) \cite{sten85} & & \\
 & ($33/2^+$) & -2.51(10) \cite{sten85} & & \\
 & $27/2^-$ & & & $\pi^2 11^- \otimes \nu i_{13/2}^{-1}$ \cite{gorg01} \\
 & $37/2^+$ & & & $\pi^2 11^- \otimes \nu i_{13/2}^{-2}f_{5/2}^{-1}$ \cite{gorg01} \\
 & $39/2^{(+)}$ & & & $\pi^2 11^- \otimes \nu i_{13/2}^{-2}p_{3/2}^{-1}$ \cite{gorg01} \\
 \hline
\end{tabular}
\begin{tablenotes}
\item[($^a$)]~$I^\pi = 27/2^-$ was reassigned for this state, similar to the states observed in $^{191,195}$Pb~\cite{bala11}.
\item[($^b$)]~Updated value from Ref.~\cite{vyve04a}.
\item[($^c$)]~Value obtained from additivity of the quadrupole moments of the blades~\cite{vyve02}.
\end{tablenotes}
\label{tab::moments_blades}
\end{table}

\begin{figure}[htbp]
	\centering
	\includegraphics[width=0.95\linewidth]{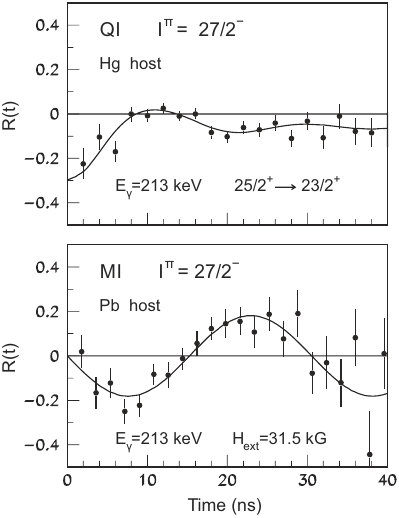}
	\caption{Experimental and theoretical TDPAD spectra for the $27/2^-$ $T_{1/2} = 9.4(5)$~ns isomer in $^{193}$Pb illustrating (top) the quadrupole interaction in solid Hg host and (bottom) the magnetic interaction in an external magnetic field. The figure is taken from Ref.~\cite{bala11}}
	\label{fig::31_193Pb}
 \end{figure}

The $T_{1/2} = 9.4(5)$~ns isomer in $^{193}$Pb, which is the bandhead of a magnetic rotational band, was used as a playground to study the configuration and the deformation of a nuclear shears state. The spin of this state was reassigned to $I^\pi = 27/2^-$ based on angular distribution measurements~\cite{bala11}. The magnetic dipole moment of the isomer was measured as $\mu = +9.9(4)$~$\mu_N$~\cite{chme97}, and its spectroscopic electric quadrupole moment as $\mid Q_s \mid = 2.6(3)$~$e$b~\cite{bala11}. The TDPAD technique was used in both experiments. The results from these experiments are displayed in Fig.~\ref{fig::31_193Pb}.

This state is understood to result from the coupling of a two-proton $K^\pi = 11^-$ excitation with an $i_{13/2}$ neutron hole. The $I^\pi = 13/2^+$ state, which is the neutron blade of the shears, is an isomeric state in $^{193}$Pb for which the magnetic and quadrupole moments were deduced as $\mu = -1.150(7)$~$\mu_N$ and $Q_s = +0.195(10)$~$e$b, respectively, from a collinear laser spectroscopy experiment~\cite{dutt91}. The magnetic dipole and electric quadrupole moments of the $11^-$ state, which is the proton blade of the shears, are known for the neighboring $^{192,194}$Pb nuclei; see Table~\ref{tab::moments_blades}. By applying the additivity formula, the $g$ factor of this state is obtained as $g_{calc} = 0.68(1)$~\cite{bala11} in perfect agreement with the measured value $g_{exp} = 0.68(3)$~\cite{chme97}.

Magnetic, antimagnetic and chiral rotation in atomic nuclei are described in the framework of the tilted-axis cranking model (TAC), which takes into consideration that the rotational axis can take any direction with respect to the deformation axes of a nucleus~\cite{frau93,frau01}. The concept considers excitations in nearly-spherical or in triaxial nuclei. In axially symmetric nuclei, the angle between the angular momentum and the nuclear symmetry axis is called the tilt angle. The experimental information about the shears bands in the $^{193-202}$ Pb nuclei, {\it e.g.}, spin-parity assignments, lifetime and electromagnetic moment measurements, was compared to systematic TAC calculations of the tilt angles, deformation parameters, angular momenta, and reduced magnetic dipole and electric quadrupole transition probabilities~\cite{chme07}. The tilt angles and deformation parameters were calculated self-consistently for all configurations and rotational frequencies. The self-consistently calculated quadrupole-deformation parameters for the various configurations take values around $\epsilon_2 = -0.1$, which correspond to small oblate deformations, and the triaxiality is small. The tilt angle remains almost constant within each band, showing only a small increase toward higher angular momenta. Calculated and experimental $M1$ transition rates are in good agreement. They decrease with increasing spin within the bands, as expected for the shears effect. The calculated $B(E2)$ values show only a weak spin dependence. This study provides a consistent picture of the shears excitations in the Pb region, but such systematic calculations, as well as magnetic dipole moment and electric quadrupole moment measurements, are missing for the other regions of the Segr\'e diagram where magnetic bands were observed.

There have been several attempts to measure high-spin $g$ factors in the mass $A=80$ region. The $g$ factors of the first four states in the magnetic rotational bands built on the $17/2^-$ state in $^{85}$Zr and of the $11^-$ band in $^{82}$Rb have been measured using the transient-field  method~\cite{yuan08,yuan10}. The $\pi g_{9/2}^2 \otimes \nu f_{7/2}$ configuration of the band was affirmed for the $17/2^-$ band in $^{85}$Zr. The measured $g$ factors in the $11^-$ band in $^{82}$Rb were found to correspond to the $\pi g_{9/2}^2 (p_{3/2},f_{5/2}) \otimes \nu g_{9/2}$ configuration.

Antimagnetic bands in $^{106,108}$Cd were reported at high spins for the positive parity yrast structures in  Ref.~\cite{simo03,datt05} and were suggested to originate from the $\pi g^{-2}_{9/2} \otimes \nu g_{7/2}^2h_{11/2}^2$ configuration. The evolution of the yrast structure in $^{106,108}$Cd is understood to exhibit a sharp alignment of a pair of $h_{11/2}$ neutrons, followed by a slow alignment of a pair of $g_{7/2}$ neutrons above $I^\pi = 10^+$. Finally, a double-shears structure is developed at high frequencies, $\hbar \omega \sim 0.5$~MeV. In this picture, both the $g_{7/2}$ and $h_{11/2}$ neutrons are fully aligned and their angular momentum vectors are in the direction of the rotational axis. The two $g_{9/2}$ proton holes are in time-reversed orbits, which form a symmetric double shears. When coupled to the neutron particle configuration, at spin $16^+$, their angular momenta are perpendicular to the direction of the rotational axis.

In $^{108}$Cd, lifetimes of levels in the antimagnetic band were measured, which correspond to a decreasing $B(E2)$ trend and large $\mathfrak{J}^{(2)}/B(E2)$ ratios. The $g$~factors of the $I^\pi = 10^+$ to $I^\pi = 16^+$ states in the yrast structure in $^{108}$Cd have been measured using the transient-field technique~\cite{fan15}. The $10^+$ state has the two-quasineutron $h_{11/2}^2$ configuration in which the two neutrons are fully aligned. The measured $g$ factor of this state is $g = -0.25(8)$~\cite{fan15}, which is consistent with the measured $g$ factors of the $10^+$ isomeric states in $^{116,118}$Sn with pure $h_{11/2}^2$ configuration, and $g = -0.2447(7)$ and $g = -0.2326(15)$, respectively~\cite{ston05a}. Above the $10^+$ state, the $g_{7/2}$ neutrons align and the $g$~factors increase gradually as the neutrons reach  full alignment at the $16^+$ state, for which the $\pi (g^{-2}_{9/2})_{J=0} \otimes \nu (g_{7/2}^2)_{J=6}(h_{11/2}^2)_{J=10}$ configuration was assigned~\cite{datt05}. It is also the bandhead of the antimagnetic band. The measured $g$ factor of this state, $g = -0.09(3)$ is in agreement with single-particle model calculations~\cite{fan15}.

 {Magnetic and anti-magnetic bands have been observed in several mass regions of the Segr\'e diagram. Nuclear moment measurements were employed in studies of the structure and deformations of the shears bands. The $g$ factors and electric quadrupole moments of both the 'blades' of the shears and the shears bandheads in the Pb nuclei were studied. The results for the quadrupole moments are summarized in Table~\ref{tab::moments_blades}. In general, the mechanism for the generation of angular momentum and the configurations of the magnetic bands are rather well understood. For reviews of the experimental findings, see \cite{hube05,pali25,teng24}. The structure of the bandhead of the anti-magnetic band in $^{108}$Cd was studied through a $g$-factor measurement, in agreement with single-particle model calculations. Certainly, more experimental results related to anti-magnetic rotation are needed for achieving a detailed description of these excitations. }

\subsubsection{Chiral rotation }
\label{par::chiral_bands}
In 2001, Starosta et al. observed pairs of strongly-coupled bands with almost identical energies in doubly-odd nuclei~\cite{star01}. Following the suggested existence of chirality in atomic nuclei~\cite{frau97}, these excitations were understood as due to the spontaneous breaking of the chiral symmetry. Since then, a number of such bands have been identified in atomic nuclei. For a compilation of these excitations, see Ref.~\cite{xion19}.

In doubly-odd nuclei, the three angular momentum vectors of the proton, the neutron, and the even-even core can have two distinct orientations in a three-dimensional space. The corresponding bands are characterized by near-degeneracy of the levels and characteristic patterns of the $E2$ and $M1$ electromagnetic transition probabilities~\cite{koik04}. Spontaneous chiral symmetry breaking in $^{128}$Cs was reported based on the observation of two nearly degenerate rotational bands and characteristic patterns of the transition probabilities~\cite{koik03,grod06,grod11}. These features were observed for rotational states with spins $I > 13 \hbar$.

\begin{figure}[htbp]
	\centering
	\includegraphics[width=0.95\linewidth]{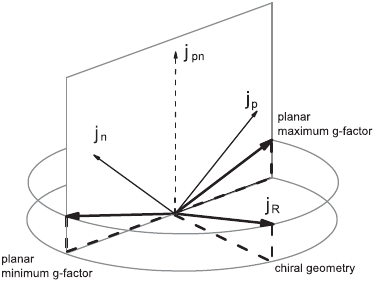}
	\caption{Schematic presentation of the coupling of the angular momentum vectors for the $(9^+)$ state in $^{128}$Cs. The core angular momentum $j_R$ may be coupled at different precession angles about the resultant $j_{pn}$ of proton and neutron angular momenta to form the specified spin of the isomeric state. The figure is taken from Ref.~\cite{grod18}.}
	\label{fig::31_chiral_vectors}
 \end{figure}

The $g$ factor of the $I^\pi = (9^+)$, $T_{1/2} = 50(6)$~ns isomer in $^{128}$Cs was measured using the TDPAD technique, yielding a value of $g = +0.59(1)$~\cite{grod18,grod22}. This state is the band head of the suggested yrast positive-parity chiral rotational band. The $\pi h_{11/2} \otimes \nu h_{11/2}^{-1}$ configuration is assigned to this band. A schematic presentation of the coupling of the angular momentum vectors for the $(9^+)$ state is presented in Fig.~\ref{fig::31_chiral_vectors}.

The additivity of the $g$ factors was used for understanding of the result. Assuming that the core rotation does not contribute to the total spin of the $\pi h_{11/2} \otimes \nu h^{-1}_{11/2}$ band head, the additivity formula gives the expected $g$-factor value, provided that the $g$ factors of the proton, $g_p^{\rm{conf}}$, and neutron, $g_n^{\rm{conf}}$, configurations are known. Considering that $g_p^{\rm{conf}} = g(\pi h_{11/2})$ and $g_n^{\rm{conf}} = g(\nu^{-1} h_{11/2})$ in the neighboring odd-$A$ nuclei, a value of $g = +0.519(2)$ was obtained, which is $\sim 15$\% smaller than the measured value.

For the band head of a chiral band, the three angular momentum vectors $\bm{j_p}$, $\bm{j_n}$ and $\bm{j_R}$ couple to a total angular momentum $\bm{I}$. The generalized additivity formula for a three-component system, considering that the core $g$ factor, $g_R$, provides a contribution, is

\begin{eqnarray}
& g & =\frac{1}{2} (g_p^{\rm{conf}} + g_n^{\rm{conf}} + g_R) \nonumber \\
&   & + \frac{\langle \bm{j_p^2} \rangle} {2\langle\bm{I^2}\rangle}(g_p^{\rm{conf}} - g_n^{\rm{conf}} - g_R) + \frac{\langle \bm{j_n}^2 \rangle}{2\langle\bm{I^2}\rangle}(g_n^{\rm{conf}} - g_p^{\rm{conf}} - g_R) \\
&    & + \frac{\langle \bm{j_R^2} \rangle}{2\langle\bm{I^2}\rangle}(g_R - g_p^{\rm{conf}} - g_n^{\rm{conf}}) - \frac{1}{\langle\bm{I^2}\rangle}(g_p^{\rm{conf}} \langle\bm{j_n}\cdot\bm{j_R}\rangle + g_n^{\rm{conf}}\langle\bm{j_p}\cdot\bm{j_R}\rangle + g_R\langle\bm{j_p}\cdot\bm{j_n}\rangle) \nonumber,
\label{eq:g_chiral}
\end{eqnarray}

\noindent where the last term contains the expectation values of the scalar products of the angular momentum vectors. Thus, it is sensitive to their mutual orientation; it is zero for an ideal chiral configuration and can take on positive or negative values for other cases. Therefore, the first four terms of Eq.~(\ref{eq:g_chiral}) correspond to the $g$ factor value with maximum chirality $g_{ch}$. Another possible scenario is the planar geometry of Fig.~\ref{fig::31_chiral_vectors}, where $j_R$ tends toward $j_p$, which gives the highest possible value of the $g$ factor, $g \approx 0.6$ for $j_p = j_n = 11/2$ and $j_R = 2$. A second planar-geometry case, where $j_R$ tends toward $j_n$, gives the lowest possible value of the $g$ factor,  $g \approx 0.4$ for $j_p = j_n = 11/2$ and $j_R = 2$. These two planar cases determine the limits of possible $g$-factor values for a given $j_p$ and $j_n$ coupling. Aplanar geometries corresponding to chiral configurations give $g$-factor values in between.

Grodner {\it et al.}~\cite{grod18,grod22} discuss the experimental results in the frame of quantum angular momentum algebra and semiclassical calculations, and in the framework of the particle-plus-rotor model. Based on the quantitative theoretical analysis, the conclusion is that the three angular momentum vectors lie almost in one plane, which suggests that the chiral configuration in $^{128}$Cs, demonstrated in previous works by characteristic patterns of electromagnetic transitions, appears to apply only above some critical rotational frequency.

 {Recently, chiral bands built on intruder orbitals were reported in $^{118}$I~\cite{bai25}. The bandhead of one of them is the $I^\pi = (7^-), T_{1/2} = 8.5(5)~\textrm{min}$ isomer which has the $\pi g_{7/2}^{-1} \otimes \nu h_{11/2}$ configuration. The $g$ factor of this isomer was known to be $g = 0.67(1)$~\cite{shaw85}. The additivity analysis demonstrates that, also in this case, the three angular momentum vectors are coupled to a planar geometry. It is impossible to draw conclusions based on only two measurements, but the results obtained so far indicate that the chiral configuration is realized above a certain rotational frequency. Certainly, this topic is far from been understood and more experimental data are needed. It will be interesting to know also the deformations of these states, which will constrain further the theoretical calculations.}

\section{Key results on exotic nuclei}
\label{sec:results}

\subsection{Anomalous isoscalar spin expectation values }
\label{sec:9Li9C}

The isoscalar, $\mu^{(0)}$, and isovector, $\mu^{(1)}$, magnetic dipole moments can be extracted from the measured magnetic moments of a mirror pair of nuclei with isospin $T=\pm T_{\textrm{z}}$, assuming that the isospin symmetry is preserved. They are defined  as~\cite{brow87}:
\begin{eqnarray}
  \begin{aligned}
    \mu^{(0)} &= \frac{1}{2}\left(
    \mu_- + \mu_+ \right)\ \textrm{and}\\
    \mu^{(1)} &= \frac{1}{2}\left(
    \mu_- - \mu_+ \right),
\label{Eq:MEC_muISIV}
  \end{aligned}
\end{eqnarray}
where $\mu_+$ is the experimental magnetic moment for
$T_\textrm{z} = +T$ and $\mu_-$ for $T_\textrm{z} = -T$.  Provided the single-particle estimate holds, {\it i.e.}, the spin of the valence nucleon forms the nuclear angular momentum $I$, for a mirror pair of odd-mass nuclei, the isoscalar
spin expectation value $\langle{\sigma_\textrm{z}}\rangle$ can be
deduced as~\cite{sugi69,brow87}
\begin{eqnarray}
  \langle \sigma_\textrm{z} \rangle
  = \frac{ 2\mu^{(0)} - I }
  {(g_\textrm{s}^{p,\textrm{free}}+g_\textrm{s}^{n,\textrm{free}}-1)/2},
\label{Eq:MEC_SpinMirror}
\end{eqnarray}
where $(g_\textrm{s}^{p,\textrm{free}}+g_\textrm{s}^{n,\textrm{free}}-1)/2$
$=$ 0.3796 using the $g$~factors of the free nucleons. For isospin symmetry and pure states, the expectation is that $ \langle \sigma_\textrm{z} \rangle=1$.

\begin{figure}[ht]
    \centering
    \includegraphics[width=0.95\columnwidth]{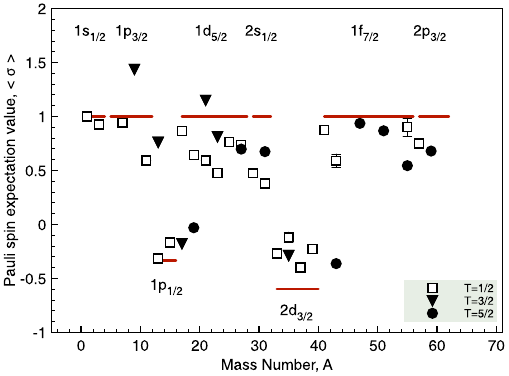}
    \caption{Spin expectation values as a function of $A$ for mirror nuclei pairs. $T=1/2$ nuclei are denoted with open squares, T$=3/2$ with triangles, and $T=5/2$ with circles. The latter have been estimated using the linear relation established for $T=1/2$ mirror pairs using the Buck-Perez approach. Values corresponding to the Schmidt limits for protons and neutrons in low-lying shells (marked accordingly) are shown as solid lines. The figure is taken from Ref.~\cite{mert16}.}
    \label{fig::spin_expectation}
\end{figure}

In light nuclei, the magnetic moments of most $T = 1/2$ mirror nuclei are known. The single-particle estimate, which was obtained from the measured magnetic moments of $T=1/2$  mirror nuclei, shows quenching from $\langle{\sigma_\textrm{z}}\rangle$~$=1$, as demonstrated in Fig.~\ref{fig::spin_expectation}. However, a significant stretching of the spin expectation value $\langle{\sigma_\textrm{z}}\rangle$ $=$ $1.44$ was observed in the $^{9}$Li and $^{9}$C mirror pair~\cite{mats95}. The spin expectation value was extracted from the measured magnetic dipole moments $\vert \mu(^9\textrm{Li}) \vert = 3.4391(6)$~\cite{corr83} and $\vert \mu(^9\textrm{C}) \vert = 1.3914(5)$~\cite{mats95}. The recent precise measurement of the magnetic dipole moment of $^9$Li, $\mu(^9\textrm{Li}) = 3.43678(6)$~\cite{neug08} will not change this result. The observed anomalous spin expectation value is due to the fact that the measured magnetic dipole moment for the $^9$C ground state is smaller than the model predictions.

The $^9\mathrm{Li}$$-$$^9\mathrm{C}$ $\langle{\sigma_\textrm{z}}\rangle$ anomaly was studied in the framework of the shell model~\cite{utsu04}, where the calculations were performed in a large model space to fully incorporate the important correlations, and adopting the $g$ factors of the free nucleons.  The $N=8$ shell gap for $Z=3$, which was determined empirically considering the data for $Z$ $=4$--$6$, turned out to be narrow. The $^{9}$Li nucleus has a usual nuclear structure with a dominant $p$-shell configuration, such that the experimental magnetic moment of $^{9}$C cannot be explained without a large amplitude of intruder configurations owing to the isospin symmetry breaking. It is caused by the Thomas--Ehrman shift (TES)~\cite{ehrm51,thom52}. The TES is due to the isobaric symmetry violation. For example, in {\it sd}-nuclei, the $l=0$ and $l=2$ orbits are quite close to each other. However, they have different radial extent, and their relative positions in isobaric partner states are strongly affected by the presence or absence of the Coulomb interaction. The admixture of the intruder configurations was calculated to be as high as 40\%.

The $\langle{\sigma_\textrm{z}}\rangle$ $^9$Li--$^9$C anomaly was further investigated in the framework of the Greens Function Monte Carlo (GFMC) model~\cite{past13} assuming mirror symmetry and mirror symmetry breaking. The former calculation yielded a value close to $\langle{\sigma_\textrm{z}}\rangle$ $=1$, which was consistent with the estimated value for a single particle, even including the contribution of the meson exchange currents (MEC).  In the latter case, the calculated $\langle{\sigma_\textrm{z}}\rangle$ value, including the effect of the MEC, increased such that it was closer to the experimental value, suggesting that the mirror symmetry of $^{9}$Li and $^{9}$C is disturbed, similar to that reported in a study by Utsuno~et~al.~\cite{utsu04}. However, the calculation uncertainty is large, and it is difficult to draw a definite conclusion.

\begin{figure}[ht]
    \centering
    \includegraphics[width=0.95\columnwidth]{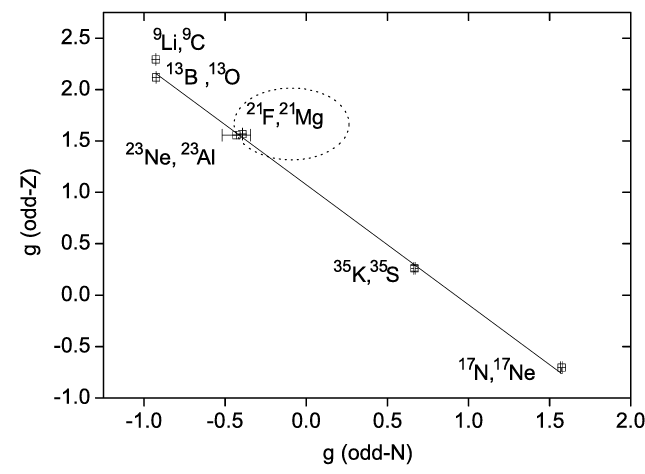}
    \caption{$g$-factors of odd-$Z$ {\it vs} odd-$N$ $T = 3/2$ mirror nuclei. The solid line shows a linear fit of the data with the parameters $\alpha = -1.167(32)$ and $\beta = 1.074(30)$. The figure is taken from Ref.~\cite{kram09}.}
    \label{fig::buck_perez_plot}
\end{figure}

An alternative approach for checking the isospin symmetry of mirror nuclei from magnetic moments was applied. A linear relationship between the $g$~factors of the odd-neutron and odd-proton nuclei of a mirror pair, $g_n^{\rm{conf}}$ and $g_p^{\rm{conf}}$, was suggested ~\cite{buck83} under the assumption that the contribution from the core is negligible
\begin{eqnarray}
  g_p^{\rm{conf}} \approx \alpha g_n^{\rm{conf}}+{\beta},
\label{Eq:MEC_OddEvenAna}
\end{eqnarray}
where the coefficients $\alpha$ and $\beta$ are related to the free nucleon $g$ factors,  $\alpha$ $=$ $(g_{s}^{p,\textrm{free}}-g_{\ell}^{p,\textrm{free}})/(g_{s}^{n,\textrm{free}}-g_{\ell}^{n,\textrm{free}})$ and $\beta$ $=$ $g_{\ell}^{p,\textrm{free}}-{\alpha}g_{\ell}^{n,\textrm{free}}$. This analysis, called the Buck-Perez analysis, using the magnetic moments of mirror pairs with mass number $A=$3--41, clearly demonstrates linearity with a correlation coefficient $r$ $=-0.999$.  A subsequent analysis of this linear correlation based on the updated dataset was then reported~\cite{buck01,pere08,mert16}.  The data point obtained for the $^{9}$Li--$^{9}$C mirror pair, however, shows a deviation from the straight line~\cite{mats95} (see Fig.~\ref{fig::buck_perez_plot}), supporting a disturbance in the mirror symmetry.

As described above, based on the study of nuclear moments, both experimental and theoretical studies have strongly indicated that a mirror symmetry breaking is observed for the $^{9}$Li--$^{9}$C mirror pair.  For a better understanding of this phenomenon, the measurement of other related observables, for example, the electric quadrupole~moment of $^{9}$C and further systematic studies on larger-$T_\mathrm{z}$ mirror pairs, are desired.

In the $sd$-shell, the spin expectation value $^{21}$Mg--$^{21}$F $A=21$ mirror pair also takes an anomalous value, $\langle{\sigma_\textrm{z}}\rangle = 1.15(2)$~\cite{kram09}, although not as large as for the $A=9$ pair. It is extracted from the measured magnetic dipole moments of the $5/2^+$ ground states in $^{21}$Mg and $^{21}$F, which take values $\mu(^{21}\textrm{Mg}) = -0.983(7) \mu_{\textrm{N}}$~\cite{kram09} and $\vert \mu(^{21}\textrm{F})\vert = 3.9194(12) \mu_{\textrm{N}}$~\cite{mats99}. The shell-model ground-state configuration of the $N=9$ isotones is governed by an unpaired neutron in the $0d_{5/2}$ orbit. For $^{21}$Mg, USD shell-model calculations~\cite{wild84} demonstrate that admixtures due to intruder proton configurations (i.e., from outside the USD model space) result in a much smaller value of the magnetic dipole moment compared to the single-particle Schmidt value~\cite{kram09}. The deduced spin expectation value for the $^{21}$Mg--$^{21}$F mirror pair is well reproduced by shell-model calculations if isospin-breaking interactions are taken into account.

The isospin symmetry in these nuclei was also studied based on the measured $B(E2; 5/2^+ \rightarrow 1/2^+)$ values.  The $B(E2; 5/2^+ \rightarrow 1/2^+)$ in $^{21}$Mg was found to be more than two times larger than the corresponding value in its mirror nucleus $^{21}$F~\cite{ruot19}. The measured values were compared to shell-model and {\it ab initio} calculations. Related to shell-model calculations, the authors discuss that the $B(E2)$ values are largely insensitive to the phenomenological isospin symmetry-breaking modifications of the USD interaction introduced in Ref.~\cite{door07}. The conclusion is that the associated $B(E2)$ values do not signal significant isospin symmetry breaking~\cite{ruot19}. Therefore, $g$-factor measurements provide a more sensitive isospin symmetry test.


\subsection{The archipelago of islands of inversion}\label{sub::islands_of_inversion}
The nuclear shell model has guided our understanding of the basic nuclear properties, which are benchmarked by the magic numbers, {\it i.e.}, certain isotopic or isotonic chains for which extra stability and other evidence of large shell gaps were observed. As nuclear structure studies expand towards more-and-more neutron-rich nuclei, such traditional benchmarks tend to disappear. For example, breaking of magicity has been reported for the $N=8$ nucleus $^{12}$Be, for which the ground state has been determined to include a significant occupancy of the intruder $d$-wave ~\cite{pain06}. Other examples are related to the $N \approx 20$ nuclei around $^{32}$Mg for which the shell gap closure is weakened. In these nuclei, the ground state and low-lying excitations originate from deformation-related intruder configurations. For example, the neutrons occupy the $s_{1/2}d_{3/2}$ orbitals, as expected from the shell model for the ground state of $^{34}$Si, while for $^{32}$Mg the ground state is a two particle, two hole (2p-2h) state, with two neutrons occupying the $f_{7/2}p_{3/2}$ orbitals. This is understood as due to the interplay between an enhanced proton-neutron interaction and a reduction of the $N=20$ shell gap~\cite{otsu01}. The inversion of the states is sudden, hence the term island of inversion (IoI). However, there are other regions at the Segr\'e diagram where quenching of magicity happens. In neutron-rich nuclei, such structural effects were observed also for $N=28$, $N=40$, and $N=50$, thus forming an ``archipelago of islands of inversion"~\cite{brow10}.

A distinct feature of atomic nuclei is the appearance of shape coexistence, which has been observed widely across the nuclear landscape. For a recent review, see Ref.~\cite{heyd11}. The simplified understanding of nuclear structure anticipates that nuclei take spherical shapes near closed shells and are deformed away from them. However, it has been demonstrated that shape co-existence occurs throughout the nuclear chart. To date, the term IoI has been exclusively applied to neutron-rich nuclei lying close to magic numbers whose ground states are deformed. For a recent review related to theory and experiment, see Ref.~\cite{nowa21}. Here we focus on nuclear moment measurements and how they help us understand nuclei in and near these IoIs. This aspect is often missing from discussions of the nuclear structure of IoI.

\subsubsection{The {\it \textbf{N}}~=~20 island of inversion}
\label{sec:N20}
After the discovery of anomalous features in $N \approx 20$ nuclei~\cite{thib75,detr79,guil84,hube78,warb90}, the $N=20$ IoI attracted considerable experimental and theoretical interest. Measurements include masses and separation energies, nuclear moments, excitation energies, transition strengths, and spectroscopy of the nuclear wave function through direct reactions, as summarized in Ref.~\cite{otsu20}. In all cases, the results challenged state-of-the-art theoretical calculations and pushed the development of nuclear models. For reviews see Ref.~\cite{caur05,otsu20}. Here we highlight the physics case and what we learn from nuclear moment measurements, following an earlier discussion in Ref.~\cite{neug06}. The $g$ factors and quadrupole moments of the nuclei around $^{32}$Mg were studied systematically to map the $N=20$ IoI. This program was carried out by the Leuven, RIKEN, and Tokyo Tech groups at GANIL, RIKEN, and ISOLDE. These measurements were discussed in Sect.~\ref{beta-NMR method}.

Single-particle properties underlie the basic aspects of nuclear structure. Therefore, it is of utmost importance to understand the composition of the nuclear wave functions. Magnetic dipole moment measurements were carried out throughout the region for this purpose. Since the ground states of the IoI nuclei originate from deformation-related intruder configurations, the electric quadrupole moments need to be studied to reveal the deviation from sphericity. These measurements, which mapped the north-western border of the $N = 20$ IoI, are summarized in Table~\ref{tab::IoI_moments}. Evaluated values for the quadrupole moments of $^{31}$Al, $Q_s = 134.0(16)$~mb, and $^{32}$Al, $Q_s = 25.0(21)$~mb were reported in Ref.~\cite{rydt13}. The nuclear moments and charge radii for the
$^{27-32}$Al isotopes were also measured at ISOLDE with the method of collinear laser spectroscopy~\cite{heyl21}.

\begin{table}[ht!]
\centering
\caption{Nuclear magnetic dipole and electric quadrupole moments of the $N=20$ IoI. In those cases where the sign of $g$ or $Q_s$ is not explicitly indicated, it has not been determined from the experiment.}
\begin{tabular}{|c|c|r|r|}
\hline
& N & $g$ & $Q_s [mb]$ \\
\hline
Na ($Z=11$) & 18 & 1.638(1)~\cite{keim00} & +86(3)~\cite{keim00} \\
            & 19 & 1.035(1)~\cite{keim00} & $\approx +145(10)$~\tnote{a}~\cite{geit00} \\
            & 20 & 1.532(1)~\cite{keim00} & $\approx +100(20)$~\tnote{a}~\cite{geit00}\\
 \hline
Mg ($Z = 12$) & 19 & - 1.7671(3)~\cite{neye05,kowa08} & \\
              & 21 & - 0.4971(3)~\cite{yord07} & \\
\hline
 Al ($Z = 13$) & 18 & 1.532(2)~\cite{himp06} &  136.5(23)~\cite{heyl16}   \\
               &    & 1.517(20)~\cite{borr02} & 134.0(16)~\cite{rydt09a} \\
               &    & +1.533(2)~\cite{heyl21} & 112(32)~\cite{naga09}  \\
               &    &  & +156(14)~\cite{heyl21}  \\
               & 19 & 1.9516(22)~\cite{himp06} & 25(5)~\cite{xu19}  \\
               &    & 1.959(9)~\cite{ueno05} &  24(2)~\cite{kame07}  \\
               &    & +1.92(4)~\cite{heyl21} & +10(50)~\cite{heyl21}   \\
               & 20 & 1.635(2)~\cite{himp06} & 141(3)~\cite{heyl16}  \\
               &    &  & 132(16)~\cite{shim12}  \\
               &    &  & $\approx 130$~\cite{naga09a}  \\
               & 21 & 0.539(2)~\cite{himp08}  & 38(5)~\tnote{b}~\cite{xu19}  \\
               &    & 1.757(14)~\tnote{b}~\cite{xu19} &   \\
\hline
\end{tabular}
\begin{tablenotes}
\item[$^a$]~The quadrupole moment values were plotted in a figure in Ref.~\cite{geit00} and reported as preliminary, {\it i.e.}, the values listed here were extracted from the figure.
\item[$^b$]~The listed $g$ factor and the quadrupole moment refer to the first-excited $1^+$ isomer in $^{34}$Al.
\end{tablenotes}
\label{tab::IoI_moments}
\end{table}

Historically, the chain of Na isotopes was measured first, and the charge radii from isotope shift measurements indicated an increase in radii towards $^{31}$Na, while the electric quadrupole moments were reported with large uncertainties~\cite{touc82}. In a follow-up experiment, the Mainz group utilized the $\beta$-NMR/NQR technique at ISOLDE~\cite{keim00,geit00} and pointed that up to $^{29}$Na the comparison of the experimental values of the $g$ factors with $sd$ shell model calculations is good, but becomes significantly worse for $^{30}$Na and $^{31}$Na~\cite{keim00}. The theoretical values for the quadrupole moments demonstrate that the $fp$-shell admixture has little effect for $^{29}$Na, but improves the theoretical value for $^{30}$Na~\cite {geit00} decisively.

The two $g$-factor measurements of the ground states in $^{31}$Mg and $^{33}$Mg were crucial for the understanding of the intruder states in the $N=20$ nuclei~\cite{neye05,yord07}. The experiments were carried out at the COLLAPS beamline at ISOLDE, where the method of collinear laser spectroscopy is utilized~\cite{neug78,anto78}. In the first experiment, the spin and magnetic moment of the $^{31}$Mg ground state were obtained by combining the results of a hyperfine-structure and a $\beta$-NMR measurement~\cite{neye05}.  The spin and parity of $I^\pi = 1/2^+$ were unambiguously assigned to the ground state, which corresponds to a prolate deformed intruder state. The comparison of the value of the magnetic moment with shell-model calculations suggests that the wave function of the observed ground state is close to a pure $2p-2h$ intruder configuration. The second experiment explored the ground-state spin and $g$~factor of $^{33}$Mg~\cite{yord07}. A ground-state spin and parity were tentatively assigned as $I^\pi = (3/2)^+$ from a $\beta$-decay measurement~\cite{numm01} and $I^\pi = (5/2)^+$ from Coulomb excitation~\cite{prit02} and inelastic proton scattering~\cite{elek06} experiments, which correspond to $1p-1h$ configurations. The measured spin and $g$-factor values are inconsistent with the suggested $1p-1h$ configurations and provide evidence for a $2p-2h$ intruder ground state with negative parity~\cite{yord07}. A critical evaluation of the experimental observables for $^{31}$Mg and $^{33}$Mg was carried out by Neyens~\cite{neye11}, which provided a consistent picture of the $2p-2h$ excitations in these nuclei.

A huge effort has been devoted to measurements of the nuclear moments of the Al isotopes (see Table~\ref{tab::IoI_moments}). The $g$ factors of $^{30-34}$Al were measured in projectile-fragmentation reactions in GANIL and RIKEN~\cite{borr02,ueno05,himp06,himp08}. The results are compared with large-scale shell model calculations. The $g$ factors of $^{30}$Al and $^{31}$Al are in very good agreement with the theoretical predictions in the $sd$-shell model space~\cite{himp06}. No evidence for the presence of intruder configurations in the ground state of $^{32}$Al was found either, but the experimental level scheme and the ground state $g$ factor might be better reproduced with an enhanced $Z = 16$ shell gap~\cite{himp06}. For $^{33}$Al, a non-negligible, $\approx$~25\%  contribution of intruder configurations into the ground state wave function was reported~\cite{himp06}. In the case of $^{34}$Al, a dominant amount of intruder components is needed in the ground-state wave function to account for the observed large $g$ factor~\cite{himp08}.

\begin{figure}[!ht]
	\centering
	\includegraphics[width=0.95\linewidth]{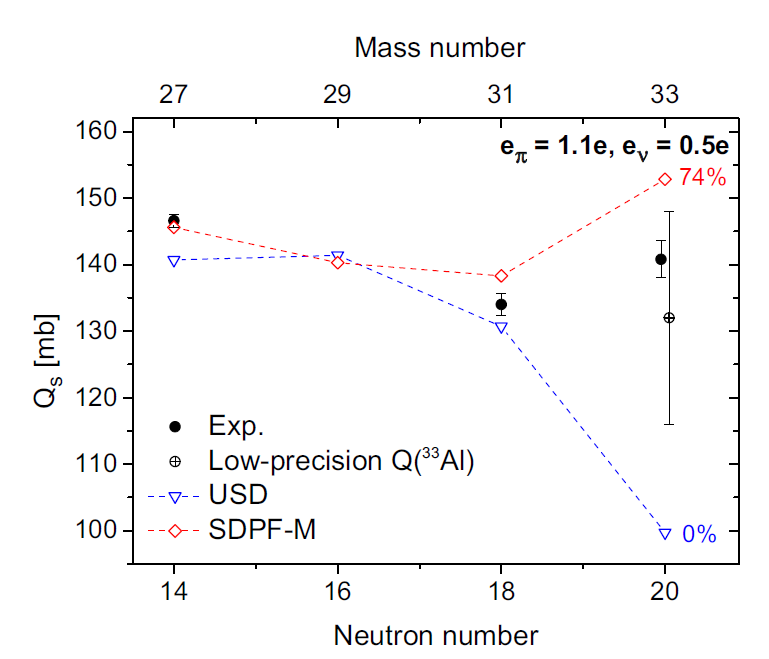}
	\caption{Experimental quadrupole moments of $^{27}$Al, $^{31}$Al, and $^{33}$Al. For $^{33}$Al both the high-precision value obtained in Ref.~\cite{heyl16}, denoted as 'this work', and a previous measurement~\cite{shim12}, denoted as 'low-precision $Q$($^{33}$Al)' are shown. These are compared to shell model calculations with the USD and SDPF-M effective interactions using constant effective charges $e_\pi = 1.1e$, $e_\nu = 0.5e$. The figure is taken from Ref.~\cite{heyl16}}
	\label{fig::33Al}
 \end{figure}

The measured quadrupole moments complete the understanding of the importance of intruder configurations in the Al isotopes. For almost a decade these studies were carried out at GANIL, improving the precision by an order of magnitude, {\it e.g.}, the uncertainties on the measured ground-state quadrupole moments of $^{31}$Al~\cite{naga09,heyl16} and $^{33}$Al~\cite{shim12,heyl16} were reduced by an order of magnitude from typically 20\% to about 2\%, as demonstrated in Fig.~\ref{fig::33Al} (see also Table~\ref{tab::IoI_moments}).

In these experiments, the deviation from sphericity has been studied. The nuclei at the edge of the $N=20$ IoI have a transitional character with a varying amount of normal and intruder configurations. Therefore, their study is of importance for the understanding of structural changes that are taking place. Prior to the high-precision ground-state quadrupole moment measurement in $^{33}$Al~\cite{heyl16}, there was conflicting information related to the existence of the intruder component. Mass measurements~\cite{kwia15} and $^{33}$Al $\beta$-decay studies~\cite{mort02,trip08} placed it outside the IoI, while the $g$-factor measurement suggested a sizable intruder component in the wave function~\cite{himp06}. There has been a debate on the reliability of structural interpretation based on spins and parities deduced from logft values~\cite{trip08} {\it vs} magnetic moment values~\cite{yord10}. Finally, the precise quadrupole moment measurement~\cite{heyl16} unambiguously demonstrated the existence of a sizable intruder component, as demonstrated in Fig.~\ref{fig::33Al}, but it also demonstrated the sensitivity of quadrupole moment measurements to intruder admixtures in the wave function.

The nuclear moment measurements in the doubly-odd $^{32}$Al and $^{34}$Al shed additional light on the excitations at the northern border of the $N=20$ IoI. The measured $g$ factor of $^{34\textrm{m}}$Al~\cite{xu19} takes a value close to that of $^{32\textrm{g}}$Al~\cite{himp06}, where `m' and `g' denote isomer or ground state, respectively. This similarity can be attributed to the very similar orbital occupations of the odd proton and neutron. A $1p-1h$ $\pi(d_{5/2})^{-1} \otimes \nu(d_{3/2})^{-1}(f_{7/2})^2$ configuration was assigned to the $I^\pi = 1^+$ ground state in $^{32}$Al and to the $I^\pi = 1^+$ isomer in $^{34\textrm{m}}$Al. The additivity of the empirical $g$ factors calculated from a simple proton–neutron coupling scheme supports the configuration assignment. The quadrupole moment of $^{34\textrm{m}}$Al increases about 50\% in amplitude with respect to that of $^{32}$Al, which is due to enhanced deformation originating from $2p-2h$ excitations across $N=20$~\cite{xu19}.

The nuclear charge radii of $^{27-32}$Al were measured at ISOLDE~\cite{heyl21}. These charge radii show a normal odd-even staggering on top of a generally increasing trend between $N = 14$ and $N = 18$. At $N = 19$, the observed decrease in charge radius appears larger than expected from the odd-even staggering alone. This has been interpreted as due to a potential shell effect at $N = 20$, although firm conclusions cannot be made due to its relatively large uncertainty. The measured charge radii for the isotopes of elements below Al, {\it e.g.} for Mg~\cite{yord12} and Na~\cite{hube78,touc82}, $Z=11,12$, demonstrate a sudden increase which is explained by the onset of deformation in the IoI around $N = 20$. On the other hand, the charge radii for the isotopes of elements above Al, {\it e.g.}, Ar, K, and Ca~\cite{ange13,blau08,ross15,kosz21a}, smoothly increase across $N=20$ which is interpreted as due to a balance between the monopole and the quadrupole proton-core polarization effects when neutrons fill the $sd$ shell below $N = 20$ and the $f_{7/2}$ orbital above it~\cite{blau08,ross15,kosz21a}. Further, more precise measurements for the Al isotopes across $N=20$ are required for the understanding of the subtle interplay between shell effects and deformation.

 {
There are very few microsecond isomeric states known in the neutron-rich $N=20$ region. The $g$ factor of only one of them, the $I^{\pi}=4^+$, $T_{1/2} = 200$ ns isomer in $^{32}\mathrm{Al}$ has been measured as $g = 1.32(1)$ \cite{ichi12}, though no information on the interpretation of its structure from nuclear moment point of view has been reported so far.
}
 {
An attempt has been made to approach the $N=20$ IoI by studies of short-lived excited states with a measurement of the $2^+$ excited state in $^{28}\mathrm{Mg}$ applying the TDRIV technique on a post-accelerated radioactive beam at HIE-ISOLDE. The challenges of the measurement and the details of the present status of the data analysis, related to a derivation of a final number for the $g(2^+, ^{28}\mathrm{Mg})$, are presented in Sect. \ref{sect:RIV-TDRIV}.
}

\subsubsection{The {\it \textbf{N}}~=~28 island of inversion}
\label{sec:N28}
The conventional magic number $N = 28$ breaks down in the region of neutron-rich nuclei centered around $^{42}_{14}$Si and $^{44}_{16}$S~\cite{glas97,bast07}, known as the $N = 28$ IoI~\cite{caur14}. The experimental discovery of multiple coexisting shapes or configurations in $^{44}$S~\cite{park17,forc10} set priority on understanding the structure of the odd-$A$ neighbor $^{43}$S, especially in confirming the pictures put forth by theory.

\begin{figure}[!ht]
	\centering
	\includegraphics[width=0.95\linewidth]{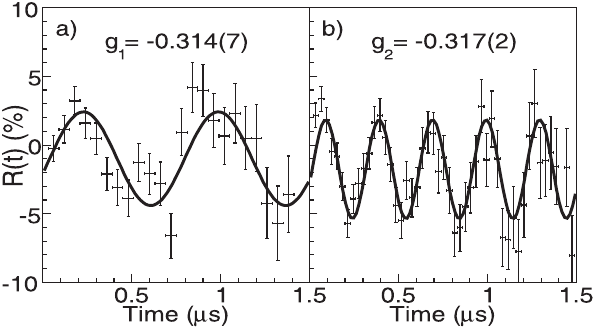}
	\caption{$R(t)$ functions associated with the 320.5 keV $\gamma$-ray in $^{43}$Sc for (a) magnetic field $B_1 = - 0.275(2)$~T and (b) $B_2 = 0.688(4)$~T. Solid curves display the result of the fit using the $R(t)$ function. The figure is taken from Ref.~\cite{gaud09}.}
	\label{fig::43S}
 \end{figure}

The existence of a low-lying 320-keV isomeric state in $^{43}$S was reported~\cite{sara00}. A $g$-factor measurement of the isomer was carried out at GANIL using the TDPAD technique, yielding a value $g = - 0.317(4)$~\cite{gaud09}. The results from the $g$-factor measurements are presented in Fig.~\ref{fig::43S}. This $g$~factor can only be reproduced by a single neutron in the $f_{7/2}$ orbit, which suggests a spin-parity assignment $I^\pi = 7/2^-$ for the state. The half-life unambiguously points towards a multipolarity $k = 2$ for the 320.5(5) keV transition, favoring $E2$ character. Thus, $I^\pi = 3/2^-$ was deduced for the ground state. Such an $I^\pi$ value is accounted for by neutron excitation across the $N = 28$ shell gap. Shell-model calculations predict that this $f_{7/2}$ state coexists with the intruder prolate deformed $3/2^-$ ground state. The experiment firmly establishes the erosion of the $N = 28$ gap, placing $^{43}$S inside the $N = 28$ IoI.

Next, the electric quadrupole moment of the isomer was measured at RIKEN, yielding a value $\mid Q_s\mid$ = 23(3)~$e$fm$^2$~\cite{chev12}, larger than that expected for a single-particle state. Shell model calculations show that intruder neutron configurations arising from proton-neutron correlations have a non-negligible impact on the structure of this state. The results suggest that the correlations drive the isomeric state away from a purely spherical shape.

\subsubsection{The {\it \textbf{N}}~=~40 island of inversion }
\label{sect::N40}
For the neutron-rich $N=40$ nuclei, the harmonic oscillator shell gap closure between the neutron $pf$-shell and the $\nu g_{9/2}$, $\nu d_{5/2}$ orbitals is reduced, resulting in collectivity and thus forming an IoI. When protons are removed from the spherical nucleus $^{68}$Ni ($N=40$), the energy of the first excited $2^+_1$ states, $E(2^+_1)$, of the $N=40$ isotones shifts from 2033 keV in $^{68}$Ni to 573 and 429 keV in $^{66}$Fe and $^{64}$Cr, respectively, which implies enhanced collectivity in these nuclei~\cite{sorl2008,gade10,cort20}. These nuclei are very difficult to produce. So far, spectroscopy in the Fe, Cr, and Ti chains has been reported; for a recent review, see Ref.~\cite{gade21}.

A measurement of the $g$ factors of isomeric states around $^{68}$Ni was carried out at GANIL and RIKEN. The $I^\pi = 9/2^+$ isomer in $^{67}$Ni and the $I^\pi = 13/2^+$ isomer in $^{69}$Cu were populated in projectile fragmentation and measured with the TDPAD technique~\cite{geor02,kuso16}.  {A recent work has revisited the isomeric state in $^{67}\mathrm{Ni}$, providing a $g$-factor value in better agreement with the expectations for a $\nu g_{9/2}$ state, and reported new values for the $17/2^-$ isomer in $^{69}$Ni ($E_x=2700$
keV, $T_{1/2}=439(3)$ ns), and the $8^+$ isomer in $^{70}$Ni ($E_x=2861$
keV, $T_{1/2}=232(1)$) \cite{stoy26}. The comparison of the experimental results with two different shell-model interactions, namely JUN45~\cite{honm09} and LNPS~\cite{lenz10}, elucidates the importance of proton excitations across the $Z=28$ shell gap. For example, considering an inert core of $^{56}\mathrm{Ni}$ requires a significant reduction of the single-nucleon spin $g$ factors to reproduce the data ($g_s^{\mathrm{eff}}\leq0.75~g_{s}^{\mathrm{free}}$). While in the LNPS interaction, in which $^{48}\mathrm{Ca}$ is considered as core, the experimental $g$ factors in the region can be reproduced using a smaller reduction of 0.9~$g_{s}^{\mathrm{free}}$.}

The $g$ factor of the $9/2^+$ ($T_{1/2} = 22$~ns, $E_{x} = 1017$~keV) isomeric state in $^{65}\mathrm{Ni}$ has been measured in a $(d,p)$ reaction, using the static hyperfine field of Ni(\underline{Ni}) \cite{geor06}.
The ground-state magnetic moments of $^{67}$Ni ($1/2^-$) and $^{69}$Cu ($3/2^-$) were measured at ISOLDE with the $\beta$-NMR/ON technique~\cite{riko00} and collinear laser spectroscopy~\cite{ving10}. So far, these are the only data related to nuclear moments of neutron-rich nuclei in the vicinity of $N=40$. For the Fe isotopes, the electromagnetic moments for the $I^\pi = (9/2^+)$ isomer in $^{61}$Fe were measured~\cite{mate04,verm07}. This is the heaviest Fe isotope for which such a measurement has been performed, but it is still a few mass units away from the border of the $N=40$ IoI.

 {
An attempt to approach the $N=40$ IoI has been made in the chromium isotopes as well, with a laser spectroscopy measurement at the CRIS setup \cite{coco16,atha23} at ISOLDE. The ground-state spin and magnetic moment of $^{61}\mathrm{Cr}$ have been determined \cite{lala25}, indicating a transition between the $2p-2h$ and the $4p-4h$ regime. The evolution of the wave function in $^{61}\mathrm{Cr}$ is understood as a second-order quantum phase transition at the border of the $N=40$ IoI.
}

\subsection{Nuclear moments in the vicinity of \texorpdfstring{$^{78}\mathrm{Ni}$}{} }
\label{sec:78Ni}
There are three doubly magic nuclei in the Ni isotopes, the harmonic oscillator doubly closed shells at $N=Z=28$, i.e., $^{56}$Ni, a shell closure at $N=40$, namely at $^{68}$Ni, and $N=50$ at $^{78}$Ni. This triplet of shell closures makes the Ni isotopes unique for testing nuclear models all the way from the $N=Z$ $^{56}$Ni to the extremely neutron-rich $^{78}$Ni. Nuclear moments of the odd-$A$ $Z=28$ Ni nuclei were studied systematically from $^{55}$Ni up to $^{67}$Ni, and the neighboring odd-proton $Z=29$ Cu isotopes have been measured between $N=28$ and  {$N=49$ \cite{degr17}}, aiming at understanding the single-particle states along the $Z=28$ chain of Ni isotopes.

Nuclear moments in the Ni nuclei were measured in a series of experiments~\cite{drai64,blec70,gori71,wend74,kran76,mull89,ohts96,riko00,geor02,geor05a,berr09}. On the neutron-deficient side, protons and neutrons occupy the negative-parity $pf$ orbits. It has been shown experimentally~\cite{kenn00,coco10, somm22} that the $^{56}$Ni core is rather soft. The $g$ factors for the $2^+_1$ states in the even-even $^{58-64}$Ni isotopes have been measured employing the technique of projectile Coulomb excitation and transient magnetic fields~\cite{kenn00} (see sections \ref{sect:TF-method} and \ref{sect:TF-gfactors-Z.le.50}). The theoretical understanding is that excitations of protons and neutrons across the $N = Z = 28$ shell closure from the $f_{7/2}$ orbital into the higher $pf$ orbits are needed to reproduce these $g$ factors~\cite{honm02,honm04}. A similar conclusion has been drawn as well from the $g$-factor measurement of the $9/2^+$ isomeric state in $^{65}\mathrm{Ni}$ \cite{geor05a}. On the neutron-rich side, the strong interaction between the $f_{5/2}$ protons and the $g_{9/2}$ neutrons plays a crucial role~\cite{otsu05,otsu10}. The $g$ factors around $N=40$ were discussed in Sect.~\ref{sect::N40}.

The ground-state magnetic dipole moments in the odd-mass nickel isotopes $^{59,63,65,67}$Ni, as well as first measurements of the spectroscopic quadrupole moments of $^{59,65}$Ni, were measured at the COLLAPS beamline at ISOLDE~\cite{neug17}, using the method of collinear laser spectroscopy~\cite{mull24}. The values are obtained from the determined hyperfine structure constants using the known $^{61}$Ni moments, which were critically reevaluated. These measurements, together with the measured nuclear charge radii in the $^{58-68,70}$Ni isotopes~\cite{kauf20,malb22} were used as a testing ground for comparison of different theoretical models. The nuclear moments of the odd-mass isotopes $^{59-67}$Ni were compared to phenomenological shell-model and {\it ab initio} valence-space in-medium similarity renormalization group (VS-IMSRG) calculations~\cite{herg16,stro19}. For the latter, two-body current contributions were included~\cite{past09} and improved the predictive power of the calculation in terms of the mean-square deviation over these isotopes. The best overall agreement with the experimental data was found for the VS-IMSRG calculations using the $\Delta$N$^2$LO$_{\mathrm{GO}}$~\cite{jian20} interaction, while the best shell-model calculations were obtained with the \texttt{GXPF1A}~\cite{honm05} interaction. Note that shell model calculations with the same phenomenological interactions were used to describe the moments in the Cu chain~\cite{ving10,ving11}.

 {
On the neutron-deficient side, studies at the BECOLA facility \cite{mina13} at NSCL have extended the experimental results on charge radii down to $^{54}\mathrm{Ni}$, providing as well information on the magnetic moment of $^{55}\mathrm{Ni}$ \cite{somm22}. The considerable reduction of the magnetic dipole moment (only 55\% of the single-particle value) indicates an important contribution of $M1$ excitations between the spin-orbit partners across the $N,Z=28$ shell gap. A kink in the charge radii of the Ni isotopes across $N=28$, very similar to the one observed in the Ca isotopes, was established as well. Comparing those observations with the very different $B(E2)$ values for the Ni and Ca isotopes \cite{prit16}, it was concluded that a kink in the charge radii cannot be considered as a direct indication of the strength of a shell closure.
}

\begin{figure}[htbp]
	\centering
	\includegraphics[width=0.95\linewidth]{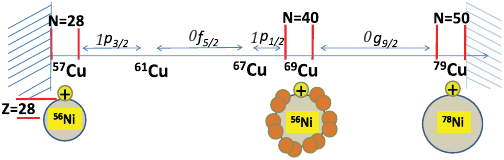}
	\caption{Illustration of the model space of large-scale shell model calculations for the Cu isotopes. It consists of a $^{56}$Ni core, with copper isotopes having one proton outside the magic $Z = 28$ shell. Beyond $^{57}$Cu$_{28}$ up to $^{79}$Cu$_{50}$, the negative parity neutron orbits $1p_{3/2}$, $0f_{5/2}$, and $1p_{1/2}$ and the positive parity $0g_{9/2}$ are filled, from the $N = 28$ to the $N = 50$ shell gap across the $N = 40$ harmonic oscillator sub-shell gap. The figure is taken from Ref.~\cite{ving10}.}
	\label{fig::Cu_isotopes}
 \end{figure}

The Cu nuclei were extensively studied starting from $^{55}$Cu up to  {$^{77}$Cu~\cite{dods66,phil68,blec72,blec73,lutz78,kawa83,lohm93,golo04,ghey04,mina06,ston08,ston08a,coco10,miha10,ving10,ving11,loze11,flan09,flan10,koes11,kuso16,ichi19,degr17}}. Of special interest in the region are the ground-state properties of the copper isotopes, which are dominated by a single proton coupling to the underlying Ni core. A large data set of ground-state magnetic dipole and electric quadrupole moments for 18 Cu isotopes, $^{58-75}$Cu, was obtained at ISOLDE, using the collinear laser spectroscopy method~\cite{ving10,ving11}. These measured moments were compared to large-scale shell model calculations. The model space for these calculations is schematically illustrated in Fig.~\ref{fig::Cu_isotopes}.

\begin{figure}[ht!]
	\centering
	\includegraphics[width=0.95\linewidth]{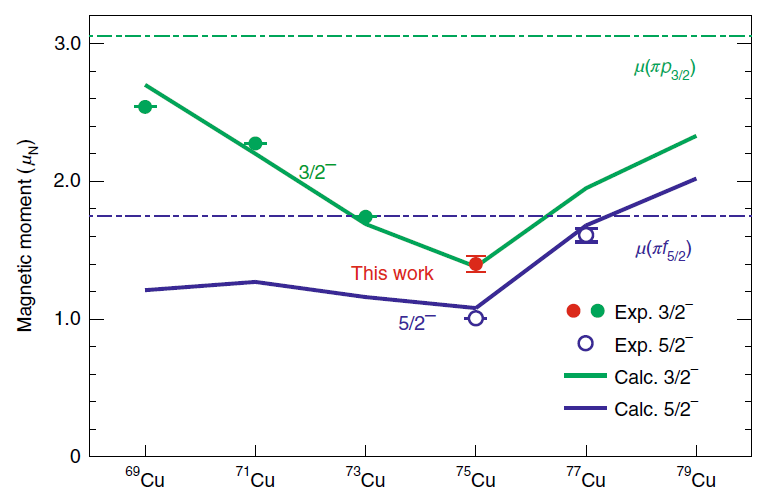}
	\caption{Systematics of the magnetic moments for odd-$A$ Cu isotopes. Filled (open) circles represent experimental data for the $3/2^-$ ($5/2^-$) states. The filled red circle represents the result obtained in Ref.~\cite{ichi19}. The solid green (blue) lines indicate MCSM calculations for the $3/2^-$ ($5/2^-$) states. $\mu$($\pi p_{3/2}$) and $\mu$($\pi f_{5/2}$) denote the proton Schmidt values for $2p_{3/2}$ and $1f_{5/2}$, respectively. The figure is taken from Ref.~\cite{ichi19}.}
	\label{fig::75Cu}
 \end{figure}

An inversion of the ground-state structure from $\pi 1p_{3/2}$ dominated to $\pi 0f_{5/2}$ dominated was established by the measured ground-state spins of $^{73,75}$Cu~\cite{flan09}. In a follow-up experiment, Ichikawa et al. measured the magnetic dipole moment of the $\pi 1p_{3/2}$ isomer in $^{75}$Cu~\cite{ichi19} using the TDPAD technique in two-step projectile fragmentation at RIKEN. This experiment revealed the subtle interplay between core excitations and shell evolution. The results for the measured magnetic dipole moments are presented in Fig.~\ref{fig::75Cu}, where they are compared with calculations within the Monte Carlo shell model~\cite{otsu01a,shim12a}, using effective spin and orbital $g$ factors ($g_s^{\rm eff} = 0.7 g_s^{\rm free}$, $g_{\ell}^p = 1.1$ and $g_{\ell}^n = -0.1$).

The measured nuclear moments both for the ground ($I^{\pi}=3/2^-$) \cite{riko00,ving10} and the isomeric ($I^{\pi}=13/2^+$) states in $^{69}$Cu~\cite{geor02,kuso16} are close to the Schmidt value, suggesting that the picture of a single proton orbiting an inert $^{68}\mathrm{Ni}$ core is reasonable. However, the values increasingly deviate with the increase of $N$. This deviation can have two different origins, namely shell evolution and core excitation. The MCSM calculations of \cite{ichi19} show that both the ground ($I^{\pi}=3/2^-$) and the isomeric ($I^{\pi}=5/2^-$) states of $^{75}\mathrm{Cu}$ appear to have the same characteristic features of a proton, orbiting around a $^{74}\mathrm{Ni}$ core with certain shape oscillations and deviations from a perfect sphere. The conclusion of the authors is that the shell evolution arises even in the presence of core excitations, while the core excitation is not a major driving force in this case.

\begin{figure}[ht!]
	\centering
	\includegraphics[width=0.95\linewidth]{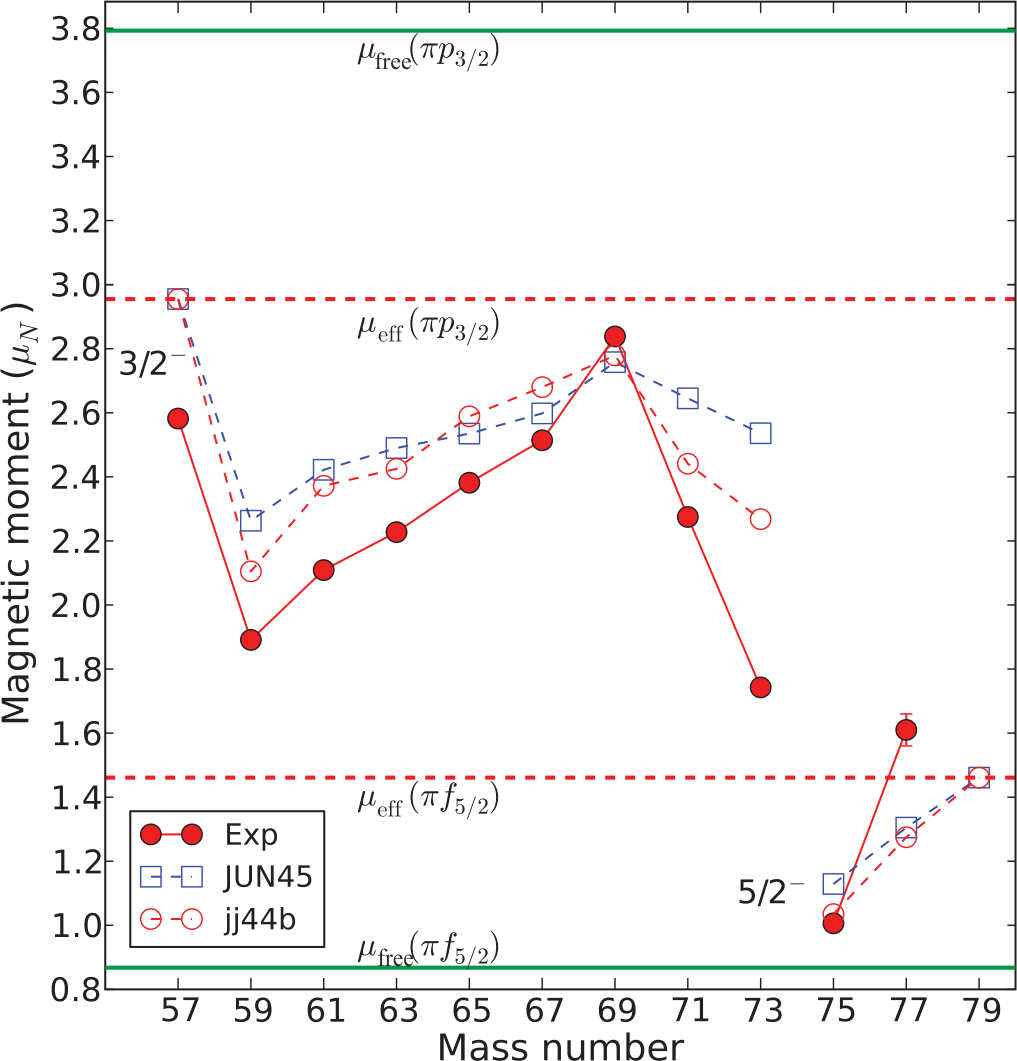}
	\caption{Comparison of experimental magnetic moments with shell-model calculations. The effective magnetic moments are shown as dashed lines and are calculated from the effective Schmidt estimates using $g_s = 0.7g_s^{\rm free}$. The free-nucleon Schmidt values are shown as full lines. The figure is taken and modified from Ref.~\cite{koes11}.}
	\label{fig::77Cu}
 \end{figure}

The decrease of the magnetic moment for the $I^{\pi}=3/2^-$ states is well reproduced both in MCSM calculations \cite{ichi19} and using the  \texttt{JUN45} and the  \texttt{jj44b} interactions in a relatively restricted model space ($^{56}\mathrm{Ni}$ core)\cite{ving10,koes11}, and the conclusion is that the Ni core is becoming softer towards the mid-shell, although not all authors agree on the importance of this effect on the observed magnetic moments trend and, especially, on the ground-state value of $^{75}\mathrm{Cu}$. The claim is that $\mu(^{75}\mathrm{Cu})$ is further deviating from the (effective) Schmidt value \cite{flan09,ving10,koes11}. Looking in more detail, it can be inferred that the free-nucleon Schmidt value for the $\pi p_{3/2}$ orbital is higher than its effective value, $\mu_{\rm free}(\pi p_{3/2}) = 3.793$ compared to $\mu_{\rm eff}(\pi p_{3/2}) = 3.0$, while for the $\pi f_{5/2}$ orbital the free-nucleon Schmidt value is smaller, $\mu_{\rm free}(\pi f_{5/2}) = 0.862$ compared to $\mu_{\rm eff}(\pi f_{5/2}) = 1.46$. Therefore, although the experimental magnetic moment of the ground state of $^{75}\mathrm{Cu}$ deviates more from the effective Schmidt value, it gets closer to the free-nucleon value, see Fig.~\ref{fig::77Cu}. A question arises: Is this a signature for a rather pure $\nu f_{5/2}$ configuration, as this is claimed in \cite{ichi19}?

Another question that can be posed is: Are the correlations in $^{75}\mathrm{Cu}$/$^{74}\mathrm{Ni}$ increasing significantly with respect to $^{73}\mathrm{Cu}$/$^{72}\mathrm{Ni}$? The comparison of the quadrupole moments of $^{73}\mathrm{Cu}$ and $^{75}\mathrm{Cu}$~\cite{ving10}, reveals that $Q(^{75}\mathrm{Cu})$ is $\sim 30\%$ higher than $Q(^{73}\mathrm{Cu})$. This can be due to two different contributions, namely, an increase of $Q_{core}$ or of $Q_{sp}$. The analysis done in Ref.~\cite{ving10} indicates that they can be of similar importance. Indeed, using Eq.~(\ref{eq:q_sp}) and the measured Cu charge radii~\cite{biss16} it can be estimated that $Q_{sp}(3/2^-, ^{73}\mathrm{Cu})\sim 1.4\cdot Q_{sp}(5/2^-, ^{75}\mathrm{Cu})$. On the other hand, the results of the JUN45 and the jj44b calculations related to $Q_{core}$ are not conclusive, one of them giving a stronger core contribution, while the other a stronger single-particle contribution to the quadrupole moment change.

The picture becomes even more complex when considering the magnetic moment of $^{77}\mathrm{Cu}_{48}$~\cite{koes11}; see Fig. \ref{fig::77Cu}. Its value goes back towards the effective Schmidt line, contrary to what one would expect when going closer to the $N=50$ shell closure. The authors of Ref.~\cite{koes11} state that the larger magnetic moment of $^{77}\mathrm{Cu}$ comes as a surprise. They looked for an explanation in terms of possible proton excitations across the $Z=28$ shell gap, an effect which is not included in the JUN45 and the jj44b interaction (using a $^{56}\mathrm{Ni}$ core). Such excitations are taken into account in Ref.~\cite{siej10}, suggesting a reduction of the $Z=28$ shell closure when approaching  $N=50$. A Coulomb excitation study of $^{75}\mathrm{Cu} $ should confirm or refute the hypothesis of significant core-excitation contributions in this nucleus. Another stringent test of the theory could come from the magnetic moment of $^{79}\mathrm{Cu}$ right at $N=50$. These studies are complemented by CFBLS measurements of charge radii, which were carried out at ISOLDE, CERN~\cite{flan09,biss16,degr20}. The results reveal a possible sub-shell closure at $N=40$~\cite{biss16} and a structural change as $N=50$ is approached~\cite{flan09,degr20}.

The next frontier to be explored concerns studies of ${N} \approx 50$ neutron-rich nuclei north of $^{78}$Ni. A first $g$-factor study of $^{79}$Zn~\cite{yang16} was carried out using collinear laser spectroscopy at the COLLAPS setup~\cite{muel83,nort10} at ISOLDE, CERN. The magnetic moments of the $9/2^+$ ground state and the $1/2^+, T_{1/2} \geq 200$~ms isomer were measured. The experiment confirms the assigned spin and parity of these states, which correspond to the $\nu g_{9/2}^{-1}$ configuration for the ground state and a $\nu s_{1/2}$ configuration that includes a $1p-2h$ excitation across the $N=50$ shell, for the isomer. In addition, a large isomer shift, $\delta \langle r_c^2\rangle^{gs,m} = +0.204(6)$ was measured. The increase of the mean square radius of the isomer compared to the ground state is the first evidence for shape coexistence due to excitations across the $N=50$ shell. This interpretation is supported by the observed $2p-2h$~intruder $0_2^+$ state in $^{80}$Ge~\cite{gott16}. Clearly, a further measurement of the half-life and the excitation energy of the $1/2^+$ isomer in $^{79}$Zn, a measurement of the properties of the $1/2^+$ isomer in $^{81}$Ge~\cite{hoff81}, and a study of the low-lying states in $^{77}$Ni are required for a more comprehensive understanding of $1p-2h$~intruder neutron excitations across the $N = 50$ shell.

\subsection{Onset of deformation in the \texorpdfstring{$A \sim 100$}{} region}
\label{sec:A100}

Nuclear electromagnetic moments can play an important role in the understanding of nuclear structure not only around shell closures but also in regions of sudden structural changes. In this respect, the neutron-rich nuclei around $A\sim 100$ can be cited as a textbook example.
This region attracted attention more than half a century ago when significant deformation was predicted theoretically \cite{arse69} and observed experimentally soon after \cite{chei70}. The accessibility of these nuclei in fission allowed for a large number of experimental investigations, including $\gamma$-ray spectroscopy, mass measurements, nuclear moments and charge radii studies, etc. It was initially suggested \cite{fede77} that the sudden onset of deformation across $N=60$ is due to the proton-neutron interaction between orbits with large spatial overlap. On the experimental side, it has been identified that the two-neutron separation energies ($\textrm{S}_{2n}$) are actually increasing for the more neutron-rich nuclei when crossing $N=20$ for $37<Z<42$ \cite{huan21,wang21}. It was suggested (see, {\it e.g.}, \cite{urba04}) that spherical structures are dominating the ground states of those nuclei up to $N=59$, while well-deformed configurations are observed at low excitation energies. From $N=60$ onward, the relative positions of the spherical and deformed configurations are exchanged, with the well-deformed configurations becoming the ground states while the spherical states are observed at higher excitation energies. This is a typical case of shape coexistence. The peculiarity of the region is the suddenness of the transformation. An interesting point to mention is that this region of sudden onset of deformation is well defined with the Kr isotopes (at its southern border) showing a gradual development of deformation as observed both in mass measurements \cite{naim10} and in Coulomb excitation \cite{albe12}. Thus, the Rb nuclei ($Z=37$) can be defined as the last isotopic chain in which this sudden structural change is observed. On the north side of the region, the clear coexistence between well-deformed and spherical configurations is gradually washed out with triaxiality emerging \cite{luo05}. No abrupt change in the deformation is observed beyond the Mo isotopes ($Z=42$). Studies in even-even nuclei in the region using Coulomb excitation techniques, in which both transition probabilities and spectroscopic quadrupole moments of the low-lying excited states could be derived, have been performed. They have demonstrated a clear signature of shape coexistence between highly deformed prolate and spherical configurations in $^{96,96}\mathrm{Sr}$ \cite{clem16}. Hyperfine structure studies, using low-energy beams at ISOLDE back in the 1980s, obtained the spins, nuclear moments, and mean-square charge radii differences in a long chain of Rb isotopes ($^{76-98}\mathrm{Rb}$) \cite{thib81}. These allowed the determination of $N=50$ as a robust shell closure, with nuclei in the $50 < N < 60$ region exhibiting predominantly spherical features. A sudden onset of deformation, evidenced in an abrupt increase of both the charge radii and the electric quadrupole moments, has been identified at $N=60$. The magnetic moment measurement of $^{97}\mathrm{Rb}$, the first nucleus in which the strongly-deformed configuration is observed in its ground state, could not distinguish between two possible Nilsson configurations, namely $3/2^-[301]$ and $3/2^+[431]$ \cite{thib81}. A complementary approach, using Coulomb-excitation of post-accelerated radioactive ion beams at ISOLDE \cite{sott15}, sheds light on the cornerstone of the region of deformation around $A \sim 100$. The experimentally observed $B(M1)/B(E2)$ ratios within the ground-state band of $^{97}\mathrm{Rb}$ were compared to particle-rotor calculations. This confirmed that the ground-state band originated from the $3/2^-[431]$ Nilsson configuration. This study is a clear example in which complementary investigations are necessary to understand nuclear structure in specific regions.

 {
The ground-state and some long-lived isomeric state moments of the neutron-rich yttrium, zirconium, and niobium isotopes have been studied extensively at the IGISOL facility in Jyv\"askyl\"a \cite{camp02,thay03,chea07,chea09} using collinear laser spectroscopy. These measurements, covering magnetic-dipole and electric quadrupole moments and charge radii, provide a continuous picture of the evolution of the onset of deformation across $N=60$ as the proton number increases.
}

The shape coexistence across $N=60$ in the Zr nuclei has been investigated as well by a $g$-factor measurement of a microsecond isomeric state in $^{99}\mathrm{Zr}$ \cite{boul20}. The 336 ns, $J^{\pi} = 7/2^+$ state was populated and spin-aligned in a single-step abrasion-fission reaction using a $^{238}\mathrm{U}$ beam at RIKEN RIBF. The experimentally determined $g$ factor of $\lvert g\rvert = 0.66(4)$ differs from the $g$ factor of a previously observed $7/2^+$ isomeric state in $^{97}\mathrm{Zr}$ ($g = + 0.39(4)$) \cite{bera85}, which gives a value consistent with a simple shell model prediction for a single-particle $\nu g_{7/2}$ state. This observation clearly indicates that the structure of the two isomers is considerably different. The experimental result of $^{99}\mathrm{Zr}$ is compared to $\mathrm{IBFM-1}$ calculations, which consistently reproduce the $g$ factors, $\mathrm{M1}$ and $\mathrm{E2}$ reduced matrix elements of the low-lying states, using theoretical single-particle energies at $\mathrm{N=59}$ in an abrupt deviation from those of the less neutron-rich $^{91-97}\mathrm{Zr}$ isotopes. The obtained wave function indicates a predominant $\nu d_{5/2}$ composition and the presence of collective modes. This is in line with the predictions of a quantum phase transition in the neutron-rich $N=60$ region \cite{toga16}, as a result of a type-II shell evolution, which was defined as the coexistence of different shapes within the same nucleus as a result of the balance of the central and the tensor components of the effective nucleon-nucleon interaction  \cite{otsu16}.

The magnetic dipole moment of the $11/2^-$ ($T_{1/2}= 742$ ns; $E_x = 684$ keV) isomeric state in $^{99}\mathrm{Mo}$ has been studied recently in a $(d,p)$ reaction \cite{daug21} and the results were compared to $\mathrm{IBMF-1}$ calculations. A strong quenching of the spin $g$ factor $g_s = 0.45\:g_s^{free}$ appeared necessary in order to reproduce the experimental result. This indicates that the wave function of the supposed $h_{11/2}$ orbital could be strongly mixed and that the deformation, observed in the ground states of the $N \ge 60$ nuclei in the region, might extend to low-excited states down to this $N=57$ isomer.

The quadrupole moment of a nuclear state is the key experimental observable that can provide information about the deformation of a state of interest. Therefore, nuclear quadrupole moment studies across ${N=60}$ are of utmost importance. Steps in this direction have begun with a proposal to measure $Q_s$ of the $7/2^+$ isomer in $^{99}\mathrm{Zr}$ at RIKEN RIBF. Other isomeric states that can shed light on the shape transition in the region are the microsecond isomers in $^{96}\mathrm{Rb}$ \cite{gene99,pins05} ($J^{\pi}=10^-, T_{1/2}=2.0(1) \mu s$) and $^{97}\mathrm{Rb}$ ($J^{\pi}=(1/2,3/2)^-, T_{1/2}=5.7(6) \mu s$) \cite{kame12,rudi13}. The former one is suggested to have a $[\pi g_{9/2}\otimes\nu h_{11/2}]_{10^-}$ configuration and is situated above a $\pi 3/2[431] \otimes \nu 3/2[541], K^{\pi}= 3^-$ rotational band. The latter one is suggested to have either a prolate ($\pi 3/2[312]$) or an oblate ($\pi 1/2[321]$, $\pi 3/2[321]$) configuration. Those isomers are already reachable at projectile-fragmentation facilities such as RIBF at RIKEN or FRIB at MSU.

Independent of the TDPAD studies in the $A\sim100$ region, there have been some developments for time-integral measurements. A new Time Integral Perturbed Angular Correlations technique had been proposed \cite{pate02}. It was used with a multi-detector array and a $^{252}\mathrm{Cf}$ fission source. It is based on the implantation of the fission fragments in ferromagnetic host (Fe foil in this case), subjected to the static hyperfine fields, and the $\gamma - \gamma$ angular correlations. A number of experimental results, obtained using this approach, have been reported. The first one has confirmed and improved the precision of the $g$ factor of the first $2^+$ state in $^{104}\mathrm{Mo}$, thus demonstrating the feasibility of the approach. Other studies aimed at $g$-factor studies of the $2^+$ states in the Pd, Ru, Mo, and Zr isotopes \cite{smit04,smit05} and of a number of low-lying excited states in the odd-mass $^{101}\mathrm{Zr}$ and $^{103 - 105}\mathrm{Mo}$ isotopes \cite{orla06}.
 {
See also section \ref{sect:TF-gfactors-Z.le.50} for a discussion of trends in 2$^+_1$-state $g$~factors in this region.
}

\subsection{Nuclear moments in the \texorpdfstring{$^{132}\mathrm{Sn}$}{} region}
\label{sec:132Sn}
The nuclear shell model is often viewed as one of the most fundamental approaches we have at present for the understanding of nuclear structure. It is based on a number of ``pillars" - doubly magic nuclei in which both the proton and neutron numbers are equal to one of the magic numbers (8, 20, 28, 50, 82, 126, etc.) among which could be enlisted those that are stable (e.g. $^{16}\mathrm{O}$, $^{40}\mathrm{Ca}$, $^{48}\mathrm{Ca}$, $^{208}\mathrm{Pb}$) and have been extensively studied, and some radioactive ones (e.g. $^{48}\mathrm{Ni}$, $^{56}\mathrm{Ni}$, $^{78}\mathrm{Ni}$, $^{132}\mathrm{Sn}$), that have been a subject of extensive studies in the last decades with the advances of the radioactive beam facilities. The last one on this list, $^{132}\mathrm{Sn}$, is often considered a strong analog to $^{208}\mathrm{Pb}$, but is placed a single nuclear shell lower, both on the proton and the neutron side. Often, the question is asked whether $^{132}\mathrm{Sn}$ is a better doubly magic nucleus than $^{208}\mathrm{Pb}$. The experimental and theoretical studies in the region around $^{132}\mathrm{Sn}$ over the last decades are too numerous to be covered completely within the scope of the present review. Here, we limit ourselves to nuclear structure from a nuclear moments perspective.
This includes recent studies in the $\mathrm{Cd}$, $\mathrm{In}$, $\mathrm{Sn}$, $\mathrm{Sb}$ and $\mathrm{Te}$ isotopic chains that have been studied by laser spectroscopy, covering the ground states and long-lived isomeric states, TDPAD measurements of $\mu s$ isomeric states, and RIV investigations of short-lived excited states.

\subsubsection{Key results from laser spectroscopy studies}
\label{subsubsec:132Sn_lasers}
The recent laser spectroscopy studies in the $^{132}\mathrm{Sn}$ region started with the investigation of the $\mathrm{Cd}$ isotopes \cite{yord13} (two proton holes away from the semimagic $\mathrm{Sn}$ isotopes) at the COLLAPS setup at ISOLDE. The experimental challenge of that measurement was related to the use of a deep-UV excitation (214.5 nm).  {In addition, the bunching of the radioactive ion beam allowed for an increase in} the experimental sensitivity and provided high-precision results for a chain of ten $11/2^-$ states from $N=63$ up to $N=81$, just below the neutron shell closure. In addition, the nuclear moments of a number of positive-parity isomeric states ($1/2^+$ and $3/2^+$) were measured as well. A  striking linear increase of the quadrupole moments of the $11/2^-$ states, associated with the neutron $h_{11/2}$ orbital, is in agreement with the expectations of the extreme single-particle shell model, although the valence space extends beyond a single $h_{11/2}$ shell. However, the relatively large effective charge ($e_n = 2.5 e$) as well as the strong deviation of the magnetic moments from the single-particle (Schmidt) limit call for a cautious interpretation of the structure of the $\mathrm{Cd}$ isotopes as a very good example of spherical nuclei, adjacent to a proton shell closure.

\begin{figure}
    \centering
    \includegraphics[width=0.95\linewidth]{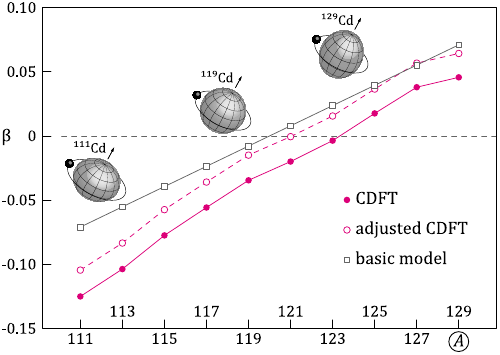}
    \caption{Quadrupole deformation of the proton distribution of the $11/2^-$ states in the Cd isotopes. The calculations of the basic model (open squares) are compared to the covariant density functional theory, CDFT (full circles). The adjusted CDFT (open circles) results are in good agreement with the basic model. The figure is taken from Ref.~\cite{yord16}.}
    \label{fig:Cd_beta}
\end{figure}

The charge radii differences between the ground states (either $1/2^+$ or $3/2^+$) and the $11/2^-$ isomeric states in the cadmium isotopes have been examined in Ref. \cite{yord16}. A distinct parabolic behavior, as a function of the mass number, has been observed in the entire chain between $^{111}\mathrm{Cd}$ and $^{129}\mathrm{Cd}$. The authors propose a qualitative description of the observed pattern, based on the decomposition of the nuclear charge radii into a spherical and a deformed component. The spherical part is canceled in the radii differences, while the ground-state deformations are assumed to be quasiconstant and relatively small. The quadratic trend in the isomer shifts \cite{yord16} therefore implies a linear mass dependence of the quadrupole deformation in the $11/2^-$ states, which is in line with the linear increase in the measured quadrupole moments \cite{yord13}. The results were analyzed as well in the frame of a covariant density functional theory (CDFT). The calculated deformations within the CDFT are associated with the collective part of the quadrupole moment. The $\beta$ (vertical) axis can be corrected for contributions from the non-collective parts (dashed line in Fig. \ref{fig:Cd_beta}), which brings the CDFT results into very good agreement with the simple model. Altogether, this is a very good example of the simple concept that the single-particle quadrupole moment is expected to be negative (oblate deformation) at the beginning of the shell, increasing with the addition of extra particles, and going to a positive (prolate deformation) towards the end of the shell.

The ground-state charge radii of the Cd isotopes from $A=100$ to $A=130$ are reported in Ref. \cite{hamm18}. A smooth parabolic behavior is observed on top of a linear trend and a regular odd-even staggering, covering the entire shell, although it falls short of confirming the theoretically predicted kinks at the shell closures. Very good consistency of the results with theory based on a Fayans parametrization \cite{faya98, faya00} of the density functional is observed.

\begin{figure}
    \centering
    \includegraphics[width=0.95\linewidth]{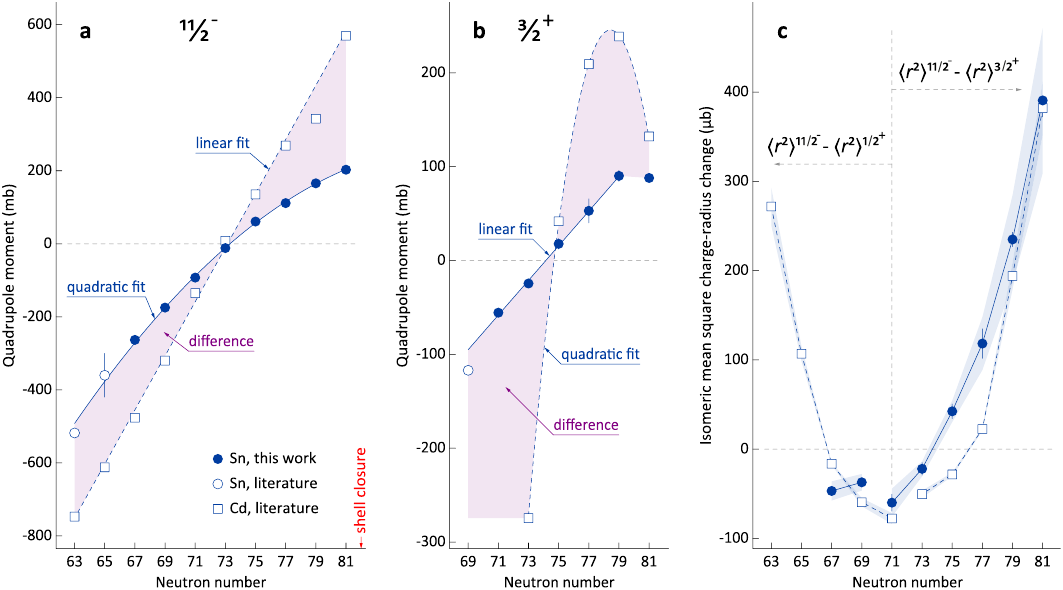}
    \caption{Comparison of the electric quadrupole moments of the tin and the cadmium isotopes for the $11/2^-$ (a) and the $3/2^+$ (b) states. The charge radii differences between the $11/2^-$ and the $3/2^+$ states (c) are very much similar to the trend in the Cd isotopes. The figure is taken from Ref.~\cite{yord20}.}
    \label{fig:Sn_q_mu_r}
\end{figure}

Further studies in the neutron-rich Sn isotopes ($^{117-131}\mathrm{Sn}$), covering electromagnetic moments and charge-radii differences between the lowest $1/2^+$, $3/2^+$ and the $11/2^-$ states \cite{yord20} have shown a considerable attenuation of the quadrupole moment of the Sn isotopes with respect to the Cd isotopes (see Fig. \ref{fig:Sn_q_mu_r}). The zero crossing of the quadrupole moments of the $11/2^-$ states appears at ${N=73}$, and at ${N=75}$ for the $3/2^+$ states, and this is both for the Sn and Cd isotopes. The quadrupole moments of the $11/2^-$ states are a factor of 2 larger than those of the $3/2^+$ states, which can be explained by a stronger polarization of the core by the unique-parity $h_{11/2}$ neutrons. The mass dependence of the quadrupole moments of the $11/2^-$ states in the Sn isotopes shows a quadratic trend,  whereas the trend in the corresponding Cd cases is linear \cite{wibo25,wibo25a}. The opposite is true for the $3/2^+$ states, where the linear dependence is observed for the Sn isotopes and the quadratic one for the Cd isotopes. The experimentally observed charge radii differences between the $11/2^-$ and the $3/2^+$ states are very similar to those previously observed in the Cd isotopes. Density functional theory calculations explain the global behavior of the electromagnetic moments and the charge-radii differences.

One step beyond the $N=82$ line has been taken in the Sn isotopes with a measurement of the electromagnetic moments of $^{133}\mathrm{Sn}$ \cite{rodr20}. Both the magnetic dipole moment and the electric quadrupole moments were well reproduced in large-scale shell model calculations using $^{132}\mathrm{Sn}$ as an inert core and effective $\mathrm{M1}$ and $\mathrm{E2}$ operators, confirming the doubly-magic character of $^{132}\mathrm{Sn}$. Another indirect confirmation of the doubly magic character of $^{132}\mathrm{Sn}$ came from charge-radii measurements of the even-even $\mathrm{Sn}$ isotopes. The characteristic kink signalling shell closure has been observed at ${N=82}$ \cite{gorg19}, similar to previous observations in other isotopic chains, including the ${N=126}$ shell closure in doubly magic $^{208}\mathrm{Pb}$. Once again, DFT calculations, using a recently developed Fayans functional, reproduce well the experimental observations.

\begin{figure}
    \centering
    \includegraphics[width=0.95\linewidth]{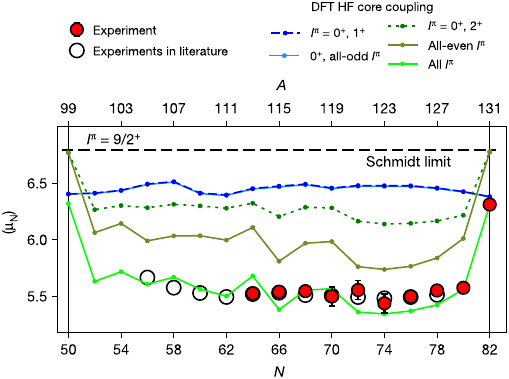}
    \caption{Comparison of the experimental magnetic moments of the indium isotopes to DFT calculations. A breakdown of the different contributions to the core polarization is given. The figure is taken from Ref.~\cite{vern22}.}
    \label{fig:in_nature_mu}
\end{figure}

The indium isotopes, with a single proton hole in the ${Z=50}$ shell closure, are expected to have their low-energy structure dominated by the single-hole configuration in the $\pi g_{9/2}$ orbit. Indeed, the previously observed magnetic moments of the $9/2^+$ ground states have shown remarkably little variation in a mass range spanning from $A = 105$  to $A=127$. These were considered a textbook example of single-particle behavior near a proton shell closure. However, a recent work has reported an abrupt change in the magnetic dipole moment of $^{131}\mathrm{In}$, right at $N=82$ (see Fig. \ref{fig:in_nature_mu}) \cite{vern22}.

The experimental results were compared with two different theoretical approaches, namely, ab initio valence-space
in-medium similarity renormalization group (VS-IMSRG) calculations and symmetry-breaking nuclear density functional theory (DFT). Both reproduce the general trend of the quadrupole moments, however, the VS-IMSRG underestimates their magnitudes, whereas they are closely reproduced by the DFT. A more stringent test of the theoretical models is provided by the experimental magnetic dipole moments and, more precisely, by the abrupt increase of $\mu (^{131}\mathrm{In})$. This magnetic moment is much closer to the extreme single-particle ``Schmidt limit" and reaches $93 \%$ of the \emph{free-nucleon value}. Looking in more detail into the DFT calculations, it appears that the addition of time-odd fields is essential for the reproduction of both the quadrupole and the magnetic moments of the studied cases. The effect of the odd-spin and even-spin states of the polarized core has been studied independently in the DFT theoretical approach. It is clear (see Fig. \ref{fig:in_nature_mu}) that the addition of $2^+$ ($\mathrm{E2}$) core polarization (dashed green line in Fig. \ref{fig:in_nature_mu}) can account for about half of the deviation of the experimental magnetic moments from the Schmidt limit for all cases of ${N} < 82$. Adding higher-order even-spin core contributions, e.g., $4^+, 6^+ ...$ (full dark-green line), brings the theoretical results closer to the experimental values but is still substantially off, especially for the $N=82$ case. The addition of the odd-spin core polarization, which is not vanishing only if time-odd terms are considered, is the key ingredient that brings the theory into excellent agreement with the experiment. Contrary to the even-spin core polarization, where the higher-order terms contribute considerably, the $1^+$ ($\mathrm{M1}$) component is sufficient, and higher-order terms $3^+, 5^+ ...$ are negligible. Due to the doubly magic character of $^{131}\mathrm{In}$, the even-spin core polarization terms have no influence on its magnetic moment - any $2^+, 4^+ ...$ contributions would require excitations across the shell closures. The experimental value of $^{131}\mathrm{In}$ is reproduced only by the $1^+$ ($\mathrm{M1}$) first-order core excitation, as it was suggested for nuclei in the extreme single-particle limit \cite{arim54}. It should be noted that in the present calculations only \emph{free-nucleon} $g$-factor values have been used.

The result on $^{131}\mathrm{In}$ and the neighboring indium isotopes shows that the general concept of using the same \emph{effective} $g$ factor throughout the nuclear chart (usually $g_s^{\rm eff} = 0.7 g_s^{\rm free}$) may need to be reconsidered.

The results of a laser spectroscopy study of the electromagnetic moments of $^{133,134}\mathrm{Sb}$ \cite{lech21}  are compared to large-scale shell model calculations using two different model spaces, namely with $^{132}\mathrm{Sn}$ and $^{88}\mathrm{Sr}$ cores. The calculated magnetic moments of the $N=82$ isotones are compared to the experimental results for the odd-proton isotopes in the range $Z=51 - 57$. The deviations between the theoretical calculations and the experimental result are considerable (a factor of about 1.5) for the $^{132}\mathrm{Sn}$-core calculations, using effective $g$ factors ($g_s^{\rm eff} = 0.7 g_s^{\rm free}$ both for protons and neutrons and $g_{\ell}^{\rm eff}= 1.18$ for protons). It is suggested that this can result from the exclusion of the spin-orbit partner from the model space (e.g., while the $\pi g_{7/2}$ is active, the $\pi g_{9/2}$ is blocked in the inert core). The experimentally observed decrease of the magnetic moments with an increase of $Z$ is explained as due to the Pauli blocking of the core polarization by adding extra protons in the $g_{7/2}$ orbital. This interpretation is supported by a calculation using a $^{88}\mathrm{Sr}$-core (using effective $g$ factors), which allows for cross-shell particle-hole excitations. In the later calculation, the theoretical magnetic moments follow the experimental trend closely. The quadrupole moments are well reproduced in the $^{132}\mathrm{Sn}$-core calculation for the states under study using effective charges ($e^{\rm eff}_p = 1.7$ and $e^{\rm eff}_n = 0.7$), showing that, contrary to the M1 operator, the E2 operator is insensitive to the presence of the $g_{9/2}$ orbital in the model space. On the contrary, the quadrupole moments obtained by the $^{88}\mathrm{Sr}$-core calculations are considerably off the experimental values, especially for the heavier isotones. However, it is worth mentioning that the effective charges used in the $^{88}\mathrm{Sr}$-core calculations are much smaller compared to those used in the $^{132}\mathrm{Sn}$-core calculations.

$^{134}\mathrm{Sb}$ is another nucleus considered in the Ref. \cite{lech21} study. It can be regarded as a single valence proton above $Z = 50$ and a single valence neutron above $N = 82$. For such a case, the additivity rule for nuclear magnetic moments can be used, making use of the moment values of the neighboring odd-mass nuclei $^{133}\mathrm{Sb}$ and $^{133}\mathrm{Sn}$. Indeed, the validity of the additivity relation is confirmed with the magnetic moment of $^{134}\mathrm{Sb}$ well reproduced using empirical (experimental) magnetic moments of the odd-mass neighboring isotopes. This supports the assumption of pure shell-model configurations of a single proton and a single neutron above $^{132}\mathrm{Sn}$.

Extension of the electromagnetic moments studies of the antimony isotopes towards the neutron-deficient side of the nuclear chart has been reported in Ref. \cite{lech23}, where the experimental results are compared both to shell-model and ab-initio valence-space in-medium similarity renormalization group (VS-IMSRG) calculations. The shell-model calculations are in excellent agreement with the experimental results when using effective $g$ factors ($g_s^{\rm eff} = 0.7 g_s^{\rm free}$). The \textit{ab initio} calculations do a very good job, similar to the shell-model calculations, on the neutron-deficient side. However, some deviations from the experiment are observed on the neutron-rich side while using the same effective $g$ factors as in the shell model calculations. It is suggested that the M1 (spin-flip) $g_{9/2}$ - $g_{7/2}$ excitations play an essential role, although these are insufficient for explaining the deviations from the Schmidt limits.

\subsubsection{Magnetic moments of microsecond isomeric states}
\label{subsubsec:132Sn_isomers}

\begin{figure}
    \centering
    \includegraphics[width=0.95\linewidth]{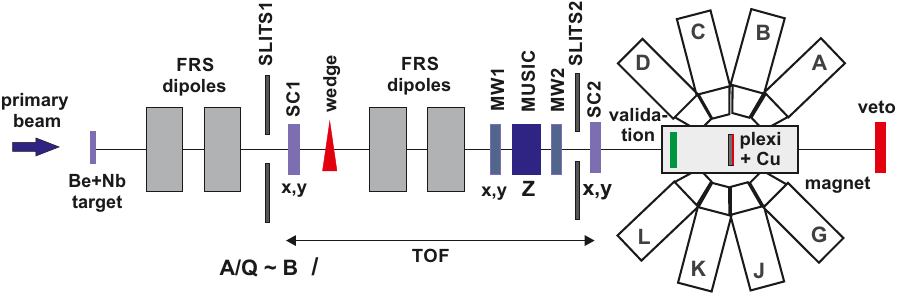}
    \caption{Schematic drawing of the beam-line detectors and the TDPAD setup of the $g$-RISING campaign. The figure is taken from Ref.~\cite{neye07a}.}
    \label{fig:g-RISING_setup}
\end{figure}

Studies of microsecond isomeric states in the neutron-rich $^{132}\mathrm{Sn}$ region were initiated during the $g$-RISING campaign at GSI \cite{neye07a} in 2005. The first proof of principle measurement of Schmidt-Ott and collaborators \cite{schm94}, proving the feasibility of nuclear moment studies in projectile fragmentation, was done at GSI  in 1994, by measuring the spin alignment of $^{43m}$Sc ($I=19/2^-$, $T_{1/2} = 473$ ns) produced by fragmentation of a 500 MeV/u $^{46}$Ti beam.
However, no further nuclear studies were performed there until the $g$-RISING campaign. This campaign can be considered the first $g$-factor measurements of microsecond isomeric states with relativistic radioactive ion beams at GSI. It had to meet a number of experimental challenges. For example, the beam transport included detection systems that could not be moved out of the beam (see Fig. \ref{fig:g-RISING_setup}). Therefore, to select only fully stripped nuclei, the beam energy had to be kept relatively high. For example, this required a beam energy of $\sim$ 300 MeV/u for the neutron-rich Sn region \cite{neye07a}. A similar beam energy at the implantation point was also used for a study in the $A \sim 200$ mass region \cite{kmie10} in which the authors claim $90 \%$ fully stripped ions, although different charge-state models give values in the $30 \% - 40 \%$ range. This high beam energy leads to a strong prompt $\gamma$-ray flash, created by the stopping of the heavy ions in the implantation host,  from which the $\gamma$ray detectors take some significant time to recover. Although a total of 8 Ge Cluster detectors were used in the setup, which increases the total $\gamma$-ray efficiency, only 4 of them proved useful for the $g$-factor studies. In addition, the nuclear spin orientation in projectile fragmentation was still not fully understood or well controlled. These experimental compromises altogether led to a quite poor signal-to-noise ratio for the results obtained, requiring some level of additional caution when interpreting them.

One of the first studies of the campaign aimed at the measurement of the magnetic moment of the $7^-$ ($T_{1/2} = 5.9 \mu s$) isomeric state in $^{126}\mathrm{Sn}$ \cite{ilie10}. This was the first measurement in which nuclear spin alignment was observed in relativistic fission. The authors report a total momentum selection of 2 \% without being able to identify quantitatively how far off the center of the momentum distribution their selection was. From an experimental point of view, there was an additional difficulty encountered in this measurement: Namely, the $7^-$ isomeric state of interest was populated both directly in the reaction and also through a decay from a higher-lying $10^+$ ($T_{1/2}= 7.5 \mu s$) isomer. The authors estimated that both population channels had identical intensities, within the experimental uncertainty. This, inevitably, induces a reduction of the amplitude of the observed $R(t)$ function since any fraction of the $7^-$ isomeric decay populated through the $10^+$ isomer will have its spin ensemble perturbed by the $g$~factor of the higher-lying isomeric state.

The lowest $7^-$ states in the Sn isotopes are expected to have a rather pure $\nu h_{11/2}^{-1} \otimes \nu d_{3/2}^{-1}$ configuration with the component $g$ factors having opposite signs. This leads to a partial canceling of the contributions; thus, small variations of the wave function could cause significantly different $g$ factors for the $7^-$ state. It is interesting to note that the derived $g$ factor, $g(7^-, ^{126}\mathrm{Sn}) = - 0.098(9)$, is closer to the experimental $g$ factors of the corresponding states in the more neutron-deficient isotopes, e.g. $^{118}\mathrm{Sn}$ and $^{114}\mathrm{Sn}$, and a factor of two different from the previously reported value in $^{130}\mathrm{Sn}$. Detailed shell model studies revealed the necessity to consider core polarization across the $N=82$ and $Z=50$ shell gaps in order to reproduce the experimental value. It was observed, as well, that the additivity relation does not hold in this case, contrary to the previously discussed case of $^{134}\mathrm{Sb}$.
An important outcome of this work is the observed spin alignment in relativistic fission of $A = 18 (8) \%$. This result still has to be taken with caution, however, considering that the observed value has a statistical significance considerably less than $2\sigma$.

Atanasova {\it et al.}~\cite{atan10} reported another result from the $g$-RISING campaign on the $19/2^+$, $T_{1/2} = 4.5$~$\mu \: \mathrm{s}$ isomer in $^{127}\mathrm{Sn}$ and the $10^+$, $T_{1/2} = 2.69$~$\mu \mathrm{s}$ in $^{128}\mathrm{Sn}$. The beam settings in this work were aiming at the former isomer. Therefore, the momentum selection was tuned at the wing of the momentum distribution of $^{127}\mathrm{Sn}$;  $^{128}\mathrm{Sn}$ came as a byproduct. Its momentum selection was identified as being in the center of the momentum distribution. In the $^{127}\mathrm{Sn}$ case, clear and consistent oscillations were observed for the $R(t)$ functions of both the 715 keV (supposed mixed E1/M2) transition and the 1095 keV (pure E2) transition. They are in anti-phase, indicating that the angular distribution coefficients for the two transitions should have opposite signs. In order that the $15/2 \rightarrow 15/2$ transition has an asymmetry parameter with opposite sign compared to the $15/2 \rightarrow 11/2$ (pure E2) transition, the former one should either have a strong multipolarity 2 admixture or the initial and/or the final spins need to be different. This second option cannot be excluded, considering that the authors clearly indicate that the spin assignments of the isomer are based on the similarity with the level schemes of the lighter Sn isotopes and theoretical calculations.

For the $10^+$ isomer in $^{128}\mathrm{Sn}$, an oscillation pattern in the $R(t)$ function was observed only for the 321 keV (supposed E1/M2) transition, with the other isomeric transition having an energy (79 keV) well below the detection threshold.

The obtained $g$-factor values of $g(19/2^+, ^{127}\mathrm{Sn}) = - 0.17(2)$ and $g(10^+, ^{128}\mathrm{Sn}) = - 0.20(4)$ have relatively large uncertainties, about 10\% and 20\% respectively, which is attributed to the low number of detected $\gamma$-rays of about $10^4$. An important result from this work, from an experimental point of view, is the level of spin alignment, obtained as $- 19(5)\%$ at the wing of the momentum distribution and $+12(4)\%$ in the center, respectively,  for the $^{127}\mathrm{Sn}$ and the $^{128}\mathrm{Sn}$ cases.

The comparison of the experimental results with shell model calculations was done using two different model spaces, namely a $^{132}\mathrm{Sn}$ core and a $^{88}\mathrm{Sr}$ core, similar to the previously discussed $^{134,134}\mathrm{Sb}$ cases. The $^{88}\mathrm{Sr}$-core calculations reproduce very well the experimental result for the $^{127}\mathrm{Sb}$ isomer with free nucleon $g$ factors. This might be attributed to the fact that the M1 (spin-flip) core polarization between the spin-orbit partner orbitals is included in this model space. In contrast, both theoretical approaches overestimate the $g$ factor of the $10^+$ isomer in $^{128}\mathrm{Sn}$ when using $g_s^{\rm free}$, and a reasonable agreement between the experimental and the theoretical values is reached only by use of an effective $g_s^{\rm eff}$.

It is worth mentioning that the data from the $^{127,128}\mathrm{Sn}$ experiment were reanalyzed as part of the PhD thesis of T.J. Gray~\cite{gray21}. The results for the ${19/2^+}$ isomer in $^{127}\mathrm{Sn}$ were confirmed, while those for the $10^+$ isomer in $^{128}\mathrm{Sn}$ were not reproduced, so comparisons between theory and experiment for this case may not be meaningful. This case also highlights the need for caution when interpreting results based on very low statistics.

\subsubsection{Magnetic moments by the recoil in vacuum (RIV) and transient-field methods}
\label{subsubsec:Te_RIV}

The previous two subsections have reviewed ground-state moments and moments of isomeric states, both near the ground state and at high spin. These measurements are most sensitive to the single-particle structure of the wavefunction. At low excitation energy and specifically for the ground state, the measurements are on odd-$A$ isotopes, and the moment is primarily attributed to that of a single shell-model orbit. The high-spin isomers reviewed above include both even-even and odd-$A$ nuclei, but again the structure is dominated by a single few-nucleon configuration. Due to the high spin of the isomer, the unique-parity high-$j$ intruder orbit in the major shell is invariably present. Thus, while these measurements give important insights into nuclear structure and the effective $\hat{M1}$ operator, there is little sensitivity to the structure of the core and the emergence of nuclear collectivity as the isotopes move further from doubly magic $^{132}$Sn. To track emerging collectivity, it is important to study the excited states of the even isotopes, especially the $g$~factor, quadrupole moment, and $B(E2)\uparrow$ of the $2^+_1$ state.

This section largely concerns measurements in the $^{132}$Sn region performed at the Holifield Radioactive Beam Facility (HRIBF) at Oak Ridge National Laboratory
\cite{ston05,stuc07a,danc11,stuc13,allm13,allm15,allm17,stuc17},
which demonstrated that it is possible to measure $Q(2^+_1)$, $B(E2; 0^+_1 \rightarrow 2^+_1)$ and $g(2^+_1)$,  simultaneously in one measurement by taking advantage of the recoil in vacuum (RIV) technique introduced in section~\ref{sect:RIV-method}. Comparisons are made with transient-field measurements \cite{benc08,kumb12a}, which were also performed for radioactive $^{132}$Te and $^{126}$Sn, and with related transient-field measurements on stable isotopes of Sn, Te, and Xe.

\paragraph{The pioneering case of \texorpdfstring{$^{132}\mathrm{Te}$}{}}

The first RIV $g$-factor measurement at HRIBF was performed on $^{132}$Te and reported in 2005 \cite{ston05}. The experimental $g$~factor was subsequently revised as a result of a more detailed analysis and calibration of the RIV interaction \cite{stuc07a}. This revision included new $g$-factor measurements on the stable Te isotopes \cite{stuc07} on which the empirical RIV calibration was based. A further revision was required when it was discovered that the incorrect target thickness had been used in the evaluation of the $^{132}$Te $B(E2; 0_1^+ \rightarrow 2_1^+)$, from which the lifetime of the $2^+_1$ state was determined. The RIV method determined the product $ \vert g \vert \tau$, so a revision of the lifetime required a revision of the $g$~factor.

The reevaluated lifetime of the $2^+_1$ state [$\tau=2.2(5)$~ps] together with the RIV calibration from \cite{stuc07a} gives the current adopted $g$ factor of the $2^+_1$-level in $^{132}$Te as $g=(+)0.46(5)$. This value agrees very well with shell model calculations \cite{brow05,stuc17}. The sign of the $g$~factor is taken from the transient-field measurement \cite{benc08}, which gives a less precise magnitude of the $g$~factor [$g(2^+_1) = + 0.28(15)$], but affirms a positive sign.

\paragraph{The puzzling case of \texorpdfstring{$^{136}\mathrm{Te}$}{}   }

\begin{figure}
    \centering
    \includegraphics[width=0.95\linewidth]{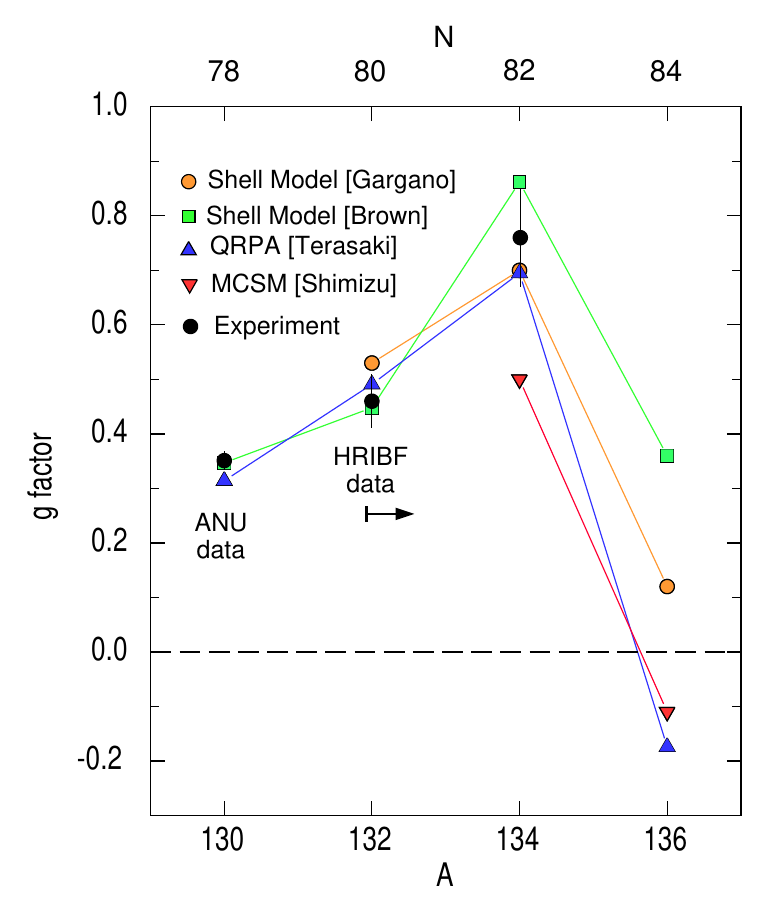}
    \caption{Various model calculations of the first-excited state $g$~factors in the Te isotopes. The predictions for $^{136}$Te vary considerably in magnitude and sign. Shell model calculations with alternative Hamiltonians are from   Brown~\cite{brow05} and Gargano \cite{allm17}. Quasiparticle random phase approximation (QRPA ) calculations are from Terasaki \cite{tera02}, and the Monte Carlo shell model (MCSM) is from Shimizu \cite{shim04}.}
    \label{fig:Te130-136g-predictions}
\end{figure}

From the theoretical perspective, the positive sign of $g(2^+_1)$ in $^{132}$Te is not questioned. In contrast, the theoretical $g(2^+_1)$ in $^{136}$Te varies greatly depending on the model used, as shown in Fig.~\ref{fig:Te130-136g-predictions}. The theories include the possibility of either a positive or a negative sign.

The RIV measurement on $^{136}$Te yielded $ \vert g(2^+_1) \vert = 0.34^{+0.08}_{-0.06}$ \cite{allm17,stuc17}, which, among the calculations shown in Fig.~\ref{fig:Te130-136g-predictions}, agrees best with that of Brown et al. \cite{brow05}. Given the magnitude of the $g$~factor, a positive sign is strongly favored. Nevertheless, it would be valuable to measure the sign and magnitude of $g(2^+_1)$ in $^{136}$Te by an independent technique. A realistic possibility is to use perturbed $\gamma\gamma$ angular correlations after the implantation and beta decay of $^{136}$Sb into an iron host at a facility such as nuCARIBU at Argonne National Laboratory. A pilot experiment on $^{138}$Xe used beams of $^{138}$I from CARIBU implanted into a polarized iron foil centered in Gammasphere. Perturbed $\gamma\gamma$ correlations were measured. A wealth of spectroscopic data was obtained, but the $g$-factor measurement did not succeed because the effective static hyperfine field for Xe in iron proved to be too small \cite{gray21}.


\paragraph{\texorpdfstring{$^{134}\mathrm{Te}$}{} and the quality of the \texorpdfstring{$^{132}\mathrm{Sn}$}{} double shell closure }


One of the last experiments performed at the HRIBF was a combined measurement of the $g$~factor and $B(E2)\uparrow$ of the first excited 2$^+$ state of the neutron-rich semimagic nuclide $^{134}$Te, which has two protons added to $^{132}$Sn. The isotope $^{134}$Te was produced as a radioactive beam and Coulomb-excited on a carbon target. The precision achieved for the $g$-factor measurement matched that of related measurements on stable beams. There was therefore sufficient precision to distinguish between alternative models.

The $B(E2)$ measurement exposed quadrupole strength in the $2^+_1$ state beyond that predicted by large-basis shell-model calculations. This additional quadrupole strength could be attributed to coupling between
the two valence protons and excitations of the $^{132}$Sn core. However, it was concluded that the wave functions of the low-excitation positive-parity states in $^{134}$Te up to $6^+_1$ remain dominated by the  $\pi (g_{7/2})^2$ configuration.

The quality of $^{132}$Sn as a doubly magic core can be assessed in comparison to other doubly magic nuclei by considering the electromagnetic properties of isotopes with two added protons, notionally in a single $j$ orbit. In this approximation, there are simple relationships for
the $B(E2)$ values between, and $g$-factors of, the states comprising the  $j^2$ multiplet.

It has been noted above that the $g$~factors of the states belonging to the $j^2$ configuration are identical.
For transitions between the states of the pure $j^2$ configuration, the $B(E2)$ values are related to the single-particle matrix element $\langle j \| T(E2)\| j \rangle$, by
\begin{equation}\label{eq:BE2-ratio}
B(E2; I_i \rightarrow I_i-2) = 4 (2I_i-3)
 \left \{ \begin{array}{ccc} j & I_i-2 & j \\
I_i & j & 2\\ \end{array} \right \}^2
\langle j \| T(E2) \| j \rangle ^2,
\end{equation}
where $I_i$ is the spin of the initial state.

\begin{table}[ht!]
\caption{\label{tab:g-fact-ratio}
Nominal structure and $g$-factor ratios for selected nuclei that can be approximated as a closed shell plus two protons (or proton holes) in an orbit with $j=7/2$.}
\centering
\begin{tabular}{|l|l|l|l|}
\hline
Nuclide & Nominal structure & $g(2^+_1)/g(6^+_1)$ & Reference \\
\hline
$^{50}$Ti    & $^{48} {\rm Ca} + \pi (f_{7/2})^2$       &   0.93(11) & \cite{spei00},\cite{boze76}  \\
$^{54}$Fe    & $^{56} {\rm Ni} + \pi (f_{7/2})^{-2}$    &   0.69(8)  & \cite{east09a},\cite{hens71} \\
$^{134}$Te   & $^{132} {\rm Sn} + \pi (g_{7/2})^2$      &   0.90(11) & \cite{stuc13},\cite{wolf76} \\
\hline
\end{tabular}
\end{table}

In Table~\ref{tab:g-fact-ratio} are shown the  $g(2^+_1)/g(6^+_1)$ ratios for $^{50}$Ti, $^{54}$Fe and $^{134}$Te, all of which nominally represent the coupling of a pair of $j=7/2$ protons to a doubly magic core. If the simple model were correct, this ratio would be unity. It is seen that $g(2^+_1)/g(6^+_1) \approx 0.9$ for $^{50}$Ti and $^{134}$Te, which can be taken to indicate that the doubly magic nature of  $^{132}$Sn is comparable to that of $^{48}$Ca. In contrast $g(2^+_1)/g(6^+_1) \approx 0.7$ for $^{54}$Fe, which suggests additional collectivity in the 2$^+_1$ state ($Z/A \approx 0.48$) and indicates that $^{56}$Ni is not a good doubly magic core. Indeed, for this reason, shell model calculations for nuclei near $^{56}$Ni require a diagonalization in the full $fp$ model space.

\begin{figure}
    \centering
    \includegraphics[width=0.95\linewidth]{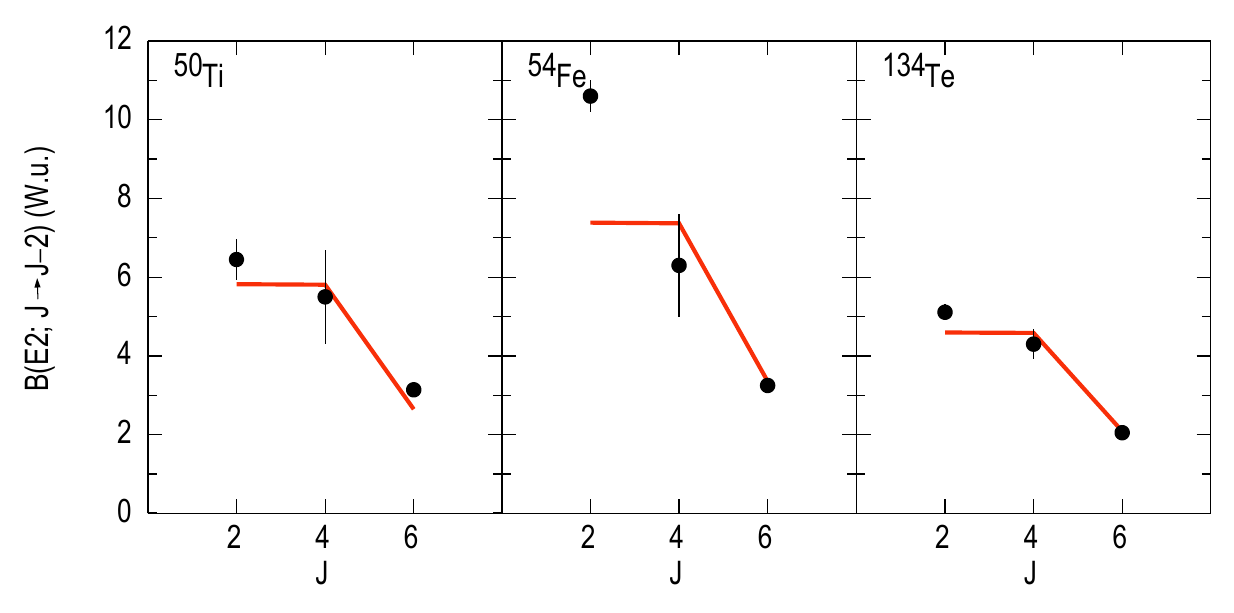}
    \caption{Data points indicate $E2$ transition strengths in nuclei with two valence protons, nominally in a $j=7/2$ orbit, adjacent to a doubly magic core. The lines show the predictions of the $j^2$ model. Data are from ENSDF, as well as Refs. \cite{stuc13,gray24}.}
    \label{fig:Be2-in-j-squared-model}
\end{figure}

These observations are confirmed by the $B(E2)$ decay behavior shown in Fig.~\ref{fig:Be2-in-j-squared-model}, again comparing data with the simple $j^2$ model. Additional $E2$ strength, implying increased collectivity, is observed in the $2^+_1 \rightarrow 0^+_1$ transition, markedly so for $^{54}$Fe. The $B(E2)$ data displayed in Fig.~\ref{fig:Be2-in-j-squared-model} are also listed in Table~\ref{tab:BE2}, along with theoretical $B(E2)$ ratios and two examples where $J=9/2$. The $B(E2; 2^+_1 \rightarrow 0^+_1)$ in $^{210}$Po, the sister of $^{134}$Te adjacent to $^{208}$Pb, does not fit the expected pattern. This case should be given priority for a future radioactive beam $B(E2)$ measurement by Coulomb excitation and considered for a $g$-factor measurement either by RIV or by the transient-field technique.

\begin{table}[ht!]
\begin{centering}
\caption{\label{tab:BE2}$B(E2)$ ratios for the $j^2$ configuration and experimental $B(E2)$ values in W.u. from Nuclear Data Sheets unless otherwise indicated.}
\begin{tabular}{llllllll}
\hline
 & & &
\multicolumn{5}{c}{Experimental $B(E2; J_i \rightarrow J_i-2)$ in W.u.}\\ \cline{4-8}
\multicolumn{1}{l}{$j$} &
\multicolumn{1}{l}{$J_i$} &
\multicolumn{1}{l}{B(E2) ratio \tnote{a}} &
\multicolumn{1}{l}{$^{50}$Ti} &
\multicolumn{1}{l}{$^{54}$Fe} &
\multicolumn{1}{l}{$^{134}$Te} &
\multicolumn{1}{l}{$^{92}$Mo} &
\multicolumn{1}{l}{$^{210}$Po} \\
\hline
${7}/{2}$  &     6  & $5/11 \simeq 0.455$     & 3.14(13) & 3.25(5) & $2.05(4)$ \\
     &     4  & $440/441 \simeq 0.998$  & 5.5(12)  & 6.3(13) & $4.3(4)$ \\
     &     2  & 1        & 6.4(5)\tnote{b}   & 10.6(4) & $5.1(2)$ \tnote{c}\\
${9}/{2}$  &     8  & $91/286 \simeq 0.318$  &          &      &   & 1.31(2)& 1.10(5)\\
     &     6  & $1250/1573 \simeq 0.795$  &          &      &   & 3.26(11) & 3.00(12)\\
     &     4  & $455/396 \simeq 1.149$    &          &     &    & $<24$    & 4.53(15)\\
     &     2  & 1          &          &       &  & 8.4(5)   & 1.83(8) \tnote{d}\\
\hline
\end{tabular}
\begin{tablenotes}
    \item[$^a$]~Ratio of $B(E2; J_i \rightarrow J_i-2)/B(E2; 2 \rightarrow 0)$ in the $j^2$ model; See Eq.~(\ref{eq:BE2-ratio}).
    \item[$^b$]~From \cite{gray24}.
    \item[$^c$]~From \cite{stuc13}.
    \item[$^d$]~From \cite{koch17}.
\end{tablenotes}
\end{centering}
\end{table}

A more comprehensive analysis of $g$~factors and $B(E2)$ transition strengths in doubly magic plus or minus two like nucleons has been presented in Ref.~\cite{stuc22}.

Returning to the case of $^{134}$Te, it was shown in Ref.~\cite{stuc13} that the simplified $j^2$ model can be brought into agreement with experiment by coupling the two valence protons to a $^{132}$Sn core vibration as determined from the properties of its 2$^+_1$ state at an excitation energy of 4.041~MeV.
The level of mixing was small, so it was concluded that $^{132}$Sn can be considered as a reliable closed core for shell model calculations.

\paragraph{Emerging collectivity in the Te and Xe isotopes below \texorpdfstring{$^{132}\mathrm{Sn}$}{}}

The Te and Xe isotopes near $^{132}$Sn provide a fertile testing ground to map the path of emerging nuclear collectivity from seniority structures in the semimagic $N=82$  isotopes  $^{134}$Te and $^{136}$Xe towards the isotopes near the neutron mid-shell at $N=66$, which show vibrational level structures. These sequences of isotopes have the advantage that their stable isotopes ($^{120}$Te - $^{130}$Te and $^{124}$Xe - $^{136}$Xe) span this sequence of level structures - i.e., from few-nucleon excitations near $N=82$ to vibrational-like structures nearer to midshell. Beginning from the semimagic cases, it is possible to track the emerging collectivity both theoretically through large-basis shell model calculations and experimentally by several techniques such as Coulomb excitation to measure transition rates and $g$~factors \cite{jako02,stuc07,coom20,coom22} and neutron scattering $(n,n^{\prime} \gamma)$ to determine rather complete spectroscopy of the low-spin ($I \lesssim 6$) and low-excitation spectrum \cite{pete18,pete19,hick22}.

\begin{figure}[hbt]
    \centering
    \includegraphics[width=0.95\linewidth]{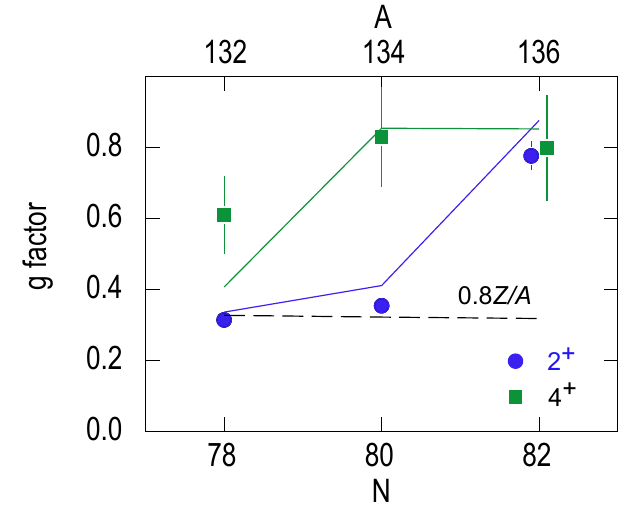}
    \caption{
    Experimental and shell model $g$~factors of $^{132}$Xe, $^{134}$Xe, and $^{136}$Xe.
    The data are from \cite{jako02}. The figure is adapted from \cite{pete19}, where details of the shell model calculations may be found.}
    \label{fig:xegfact}
\end{figure}

As noted in the previous subsection, the excitation spectrum of semimagic $^{134}$Te up to the 6$^+_1$ state corresponds closely to that associated with the two-proton $\pi (g_{7/2})^2$ configuration. Likewise,  the excitation spectrum of semimagic $^{136}$Xe up to the 6$^+_1$ state corresponds closely to the seniority-2 states of the  $\pi (0g_{7/2})^4$ configuration. As successive pairs of neutrons are removed, the nucleus becomes more collective, eventually showing a spectrum that is associated with surface vibrations. However, the evidence from these studies is that collectivity does not emerge suddenly, with the nucleus becoming collective as a whole, as might be inferred by examining energy patterns such as $R_{4/2} = E_x(4^+_1)/E_x(2^+_1)$ ratios alone. Instead, the $E2$ transition strengths and $g$~factors show that collectivity develops in subsets of nuclear excitation: the 2$^+_1$ states become collective first while the 4$^+_1$ and 6$^+_1$ states retain a significant  $\pi (0g_{7/2})^2$ component. To illustrate this point, Fig.~\ref{fig:xegfact} shows the $g$~factors of the 2$^+_1$ and 4$^+_1$ states in $^{132}$Xe, $^{134}$Xe and $^{136}$Xe. It is seen that the $g$~factor of the 2$^+_1$ state has reached the collective value of $g \approx 0.8Z/A$ at $N=78$, whereas $g(4^+_1)$ is about twice that of the $2^+_1$ state.

In a study of the $g(4^+_1)/g(2^+_1)$ ratios in the sequence of even Te isotopes from $^{124}$Te to $^{130}$Te similar behavior is observed \cite{coom22}: The 2$^+_1$ state $g$~factor reaches the collective value first at $^{130}$Te, followed by the 4$^+_1$ state at $^{124}$Te, whereas the seniority structure apparently persists in the 6$^+_1$ states even at $^{124}$Te. This last inference is based on a $B(E2)$ measurement and shell model calculations \cite{reec25} as measurements of $g(6^+_1)$ have not yet been performed for $^{124-130}$Te.

The behavior of these Te and Xe isotopes close to $^{132}$Te is not unique. There is evidence that the Nd isotopes above $N=82$ show similar trends, but in this case, it is the $1f_{7/2}$ neutrons that retain seniority components in the first 4$^+$ and 6$^+$ states \cite{hold01}. Here, $^{146}$Gd is serving as an approximate doubly magic nucleus. The transient-field $g$-factor measurements on the Nd isotopes predated the studies on Te and Xe isotopes and were discussed above in section~\ref{sect:TF-gfactors-Z.gt.50}.

It is evident that $g$-factor ratios such as $g(4^+_1)/g(2^+_1)$ and $g(6^+_1)/g(2^+_1)$ are important indicators of emerging nuclear collectivity versus the persistence of seniority structure.

\section{Summary and outlook}

It has become customary to classify present-day nuclear physics experiments as aiming at one of two frontiers, namely the discovery frontier or the precision frontier. A particular moment measurement might be assigned to one or the other of these categories, although there is a subtle interplay between them.  For example, laser spectroscopy is clearly a precision technique, however, a particular measurement might also be classified as being at the discovery frontier if it is performed on a very exotic nuclide and/or it reveals a new aspect of nuclear structure. Likewise with spin-precession measurements. Here, the precision is generally much lower, especially for measurements on excited states, however, such measurements can discover new physics, and they can be classed as precision measurements as they require specialized apparatus and techniques and a significant amount of beam time.

In the push to study nuclei at the extremes toward the drip lines, it is important to emphasize that precision measurements of electromagnetic observables closer to stability are far from complete, and that these measurements give more reliable insights into nuclear structure than do energy patterns alone. Understanding these less exotic nuclei is a prerequisite for understanding more exotic cases. Moreover, measurements of electromagnetic properties of excited states in nuclei away from stability often rely on normalization to a measurement on a stable nuclide, and these historical ``calibration values" have sometimes been open to question.

As a final remark on frontiers, whether discovery or precision, it has to be noted that there is a continual improvement in experimental techniques which moves both frontiers as time goes by.


The experimental focus of this review has been on spin-precession methods for nuclear moment measurements. A detailed discussion at the level of a practitioner's guide has been given to describe the experimental observables, the methodology, and the specifics of the experiments. An attempt is also made to identify directions for future development.

The TDPAD method as presented in section \ref{sec:isomers} has been used successfully at intermediate beam-energy facilities such as GANIL, RIKEN, and NSCL, MSU, for measurements of nuclear moments of short-lived, hundreds of ns to $\mu$s, isomers. The 2005 GSI $g$-RISING TDPAD campaign revealed several problems that need to be addressed when using beams at relativistic energies. These were discussed in Sect.~\ref{subsubsec:132Sn_isomers}. The main issue that was discussed is that the beam-transport detection system could not be removed from the beam line, which requires sufficiently high beam energies to ensure that only fully stripped ions are implanted in the host lattice. This results in an intense prompt $\gamma$-ray flush, and typically, the first 500 ns after the implantation could not be used in the experiments. The passage of the beam through the detection system of the spectrometer also limits its total intensity to a few times $10^4$ ions/s. In addition, the nuclear spin orientation in relativistic projectile fragmentation and fission is not fully understood and is not well controlled. On the other hand, RIKEN experiments using the two-step fragmentation method~\cite{ichi12} have demonstrated that the spin orientation can be controlled and maximal values can be achieved. This opens the possibility for obtaining high-confidence results in low-statistics experiments at the discovery frontier~\cite{ichi19}. The applicability of this technique needs to be explored at future relativistic beam facilities such as FAIR and China's High Intensity Accelerator Facility HIAF.

The application of the TDPAC technique for $\gamma\gamma$-correlation experiments was discussed in section {\ref{sec:isomers}}. An interesting possibility for future application of the method is in $\beta$-decay experiments, {\it e.g.}, at the ISOLDE decay station~\cite{fynb17} or any other ISOL facility. Extending this technique to projectile fragmentation facilities could open the possibility of studies of isomers, populated in the decay of microsecond nuclear states, contrary to the ISOL approach, where mother states of millisecond lifetimes or longer are accessible. An interesting possibility has been suggested recently, namely, to explore the low half-life limit in nuclear-moment experiments and study $g$ factors of $\sim 1$~ns isomers~\cite{geor22}. We believe that this technique has great potential to be applied at both ISOL and projectile-fragmentation facilities, an option that is still to be explored.

The $\beta$-NMR/NQR technique, discussed in section \ref{sec:ground}, has been used for measurements of ground-state nuclear moments in light nuclei and in a few cases of ms isomers at intermediate beam energies at GANIL and RIKEN. Extensive nuclear-moment studies on the $N=20$ IoI nuclei were performed using this method, as discussed in Sect.~\ref{sec:N20}. Extending this technique to higher-energy projectile fragmentation reactions, {\it e.g.}, above 100 MeV/u, encounters some difficulties, related to the mechanism of generation of nuclear spin polarization. At present, some experimental groups are working on different ideas on how those difficulties could be tackled at higher-energy projectile fragmentation. Another approach forward to nuclear moment studies using the $\beta$-NMR/NQR technique would be to apply different methods for spin polarization, {\it e.g.}, using laser techniques, at ISOL facilities. Either of those options would allow exploring the applicability of the technique to heavier neutron-rich nuclei.

The transient magnetic field method has been applied for studies of nuclear magnetic moments of picosecond states in a large number of nuclei, especially stable nuclei, as discussed in Sect.~\ref{sect:TF-method}.
 {
and chapter~\ref{sect:short-lived}.
}
An immediate pressing need is for an independent, precise, and accurate calibration of the transient-field strength for ions with $Z$ between about 12 and 40. There are a large number of precise relative measurements of $g$~factors (particularly for 2$^+_1$ states) in this atomic number range for which the uncertainties on the absolute values have been inflated in the most recent table of recommended values \cite{ston20}. The time-dependent recoil in vacuum method with Li-Like and Na-like ions (section \ref{sect:RIV-TDRIV}) may provide sufficiently precise and accurate calibration $g$~factors for resolving this problem.

The transient-field method has also been employed to measure the $g$~factors of the first-excited states of a few even-even radioactive beams in two variations called high-velocity transient field (HVTF) and low-velocity transient field (LVTF) techniques applicable at fragmentation-type and with post-accelerated beams at ISOL-type facilities, respectively. As discussed in \cite{fior12}, the transient field may decrease in strength to a negligible level for very high velocity heavy ions. Measurements on $^{72}$Zn by both the HVTF and LVTF techniques suggest that the two methods give rise to similar precession angles, {\it i.e.}, to similar average field strengths, at about $Z=30$. Thus, for $Z>30$ the LVTF method is expected to become more favorable than the HVTF. In the evaluation of the possibilities for future research in section
\ref{sect:TF_prospects} it is suggested that the LVTF method might be applicable for moment measurements on heavy nuclei near and beyond $^{208}$Pb. Such studies are essential if a measurement of the sign of the $g$~factor of a short-lived state is critical.

The hyperfine interaction of free ions recoiling in vacuum is exploited in the RIV method, and its time-dependent plunger variation, TDRIV, shows considerable promise. The RIV method has proven to give more precise $g$ factors for nuclei near $^{132}$Sn than the transient-field method, see sections \ref{sect:RIV-complex-ions} and \ref{sect:RIV-future}. A shortcoming is that only the magnitude of the $g$~factor is determined. An advantage of the method is that it can run as part of a precise $B(E2)$ measurement, provided that the attenuation factors are appropriately calibrated.

In recent years, much effort has been made to develop post-accelerated RIBs. As a result, facilities such as HIE-ISOLDE at CERN and CARIBU plus ATLAS at ANL became operational. It will be interesting to investigate the possibilities of employing these beams in nuclear-moment experiments. So far, only two studies have been performed using post-accelerated short-lived RIB's at HIE-ISOLDE utilizing the transient field technique on $\mathrm{^{72}Zn}$~\cite{illa14} and $^{28}$Mg beam for a TDRIV measurement~\cite{stoy23}. A number of transient field \cite{benc08,kumb12a} and RIV measurements in the neutron-rich Sn-Te region were performed at the Holifield Radioactive Beam Facility (HRIBF) at the Oak Ridge National Laboratory \cite{allm13,allm15,stuc07a,stuc13,allm17,stuc17,danc11}. The RIV measurements demonstrated the possibility of measuring transition probabilities, quadrupole moments, and $g$ factors of $2^+$ states in a single experiment simultaneously.

There are a number of open questions related to the methodology of spin-precession experiments. As already discussed throughout the text, they are primarily related to the calibration of the fields.  {The current understanding of the problems and the approaches accepted for their solution were commented in each particular case.} In nuclear moment measurements, the experimental observable is the interaction frequency. Except for applying external magnetic fields in TDPAD and $\beta$-NMR experiments, which can be mapped with high precision, all other spin-precession methods require an independent measurement of the utilized hyperfine fields.  {It is clear that further effort is needed for obtaining more precise values of these fields.}

In the case of transient field experiments, several calibrations have been applied, as discussed in Sect.~\ref{sect:TF-param}, {\it i.e.}, the Bonn~\cite{spei91a}, Rutgers~\cite{shu80}, Chalk River~\cite{andr82,haus83}, and the more recent model-based parametrization for high-velocity ions~\cite{stuc04}. The accumulated data demonstrate that there is no universal parametrization of the transient field in terms of ion velocity and atomic number, and that the field in alternative ferromagnetic hosts does not always scale simply with the magnetization.

In contrast to the transient fields, the RIV hyperfine fields are well defined, in principle, particularly for simple systems such as $1s$ (H-like), $2s$ (Li-like), and $3s$ (Na-like) ions. The data so far have shown that for H-like ions, the fields can be calculated from first principles and directly used in the experimental study. It is still an open question what the influence of excited atomic states could be in the case of Li-like and Na-like charge states on the observed hyperfine fields. An important point to investigate is the effect of excited electronic configurations of these ions, which may have lifetimes that are comparable to the lifetime of the nuclear state under investigation.

Things are getting more complex for many-electron systems. In practice, several - up to many - different atomic configurations can be involved. For any given atomic configuration, the RIV hyperfine interactions can be calculated with sufficient accuracy for applications to $g$-factor measurements using atomic structure codes. The limitation for complex many-electron ions is that the profile of atomic excitation energies applicable for ions entering vacuum is not well known, see Sect.~\ref{sect:RIV-abinitio}. However, there is a sufficient body of experimental data on RIV to suggest that a renewed atomic computational effort might resolve this matter. For simple ions, not only H-like but also Li-like and Na-like, the time-dependent RIV method could produce precise absolute $g$~factor measurements, including those needed to calibrate the transient field. In any case, the RIV method based on empirical calibration of the hyperfine interactions is solidly established; it can be rated as the preferred method for magnetic moment measurements on short-lived excited states of RIBs.

In quadrupole moment measurements, the electric field gradients needed to interpret the observed frequency have to be calculated within density functional theory (DFT). A long-lasting attempt to provide a `golden standard' EFG for Cd implanted in Cd (Cd\underline{Cd}) has been carried out recently~\cite{haas21}. As a result, the $5/2^+$ isomer in $^{111}$Cd is suggested to be used for calibration of other EFGs. The conclusion of this recent study is that measurements with simple free molecules, such as Cd or Hg halides, in a gas phase, are a reliable tool for defining EFGs because the molecular geometry enables accurate DFT calculations. In the future, such studies may need to be extended to the Sn\underline{Sn} and Pb\underline{Pb} systems.

A compilation of the nuclear moment data is provided under the auspices of the Nuclear Data Section of the International Atomic Energy Agency (IAEA)~\cite{mert16a}. It is based on several earlier compilations~\cite{ragh89,ston05a}. Over the last several years, N.J. Stone put considerable effort into keeping the compilation updated and determining recommended values~\cite{ston05a,ston13a,ston14}. As noted in this review, field calibrations are a key factor in determining nuclear moments, and the raw data in many cases require the application of particular corrections.   Some guiding materials on how the different corrections have been applied to the compiled data were published~\cite{ston15a,ston15b}, however, in some cases, it is difficult to figure out how the corrections were done.

The situation with the choice of recommended values is also complicated. Several sets of recommended nuclear magnetic dipole and electric quadrupole moments have been published~\cite{ston19,ston20,ston21}, and in some cases, the recommended values appear to have been derived based on subjective judgements. The guiding principle of his work has been that a recommended value must be based on a reliable measurement and the provided uncertainty should be realistic and by no means underestimated. Moreover, this database needs to be further maintained and enlarged, and occasional misprints should be corrected.

To understand nuclear structure physics, it is required to obtain information from many observables. As demonstrated in this review, nuclear moments should be considered together with other observables, such as nuclear charge radii and $B(E2)$ reduced transition probabilities. Concerning charge radii, the last compilation is somewhat outdated~\cite{ange13}, because a large number of new measurements have been completed in the last few years. On the other hand, the database for the $B(E2; 0^+ \rightarrow 2^+)$ reduced transition probabilities has been regularly updated and can be accessed at the NNDC~\cite{nndc} webpage; the last printed copy was published in 2016~\cite{prit16}.

The primary importance of most nuclear moments studies is found in the comparison of experimental results with the theoretical models. Often, the experimental results are taken at face value. However, some data need to be taken with caution, especially for the cases with large experimental uncertainties (either statistical or systematic). When pushing the discovery frontier, experimentalists are always trying to extract a result from low-statistics data. This is a consequence of the extremely competitive allocation of expensive beam time at radioactive beam facilities, where most such experiments are performed. In our view, results obtained with low statistical validity might be better considered as a ``good indication'' rather than a firm result, especially when they point at new physics; more precise studies may need to be performed when the experimental conditions are available.

Within this review, we have addressed a number of physics cases that were at the central focus of nuclear physics research over the last few decades. At the turn of the century, high-spin physics was still dominating nuclear physics research, but the interest of the community has steadily been shifting towards the structure of exotic neutron-rich nuclei.
We have covered the basic achievements of high-spin nuclear physics research and the contribution of nuclear moment measurements to the understanding of different structural effects.
Nowadays, the research focus is largely on the evolution of nuclear structure when moving away from the valley of $\beta$ stability. We have discussed a number of situations where nuclear moments play a key role, including nuclear moment measurements related to the so-called islands of inversion, where deformed ground state configurations occur for semi-magic nuclei.
We have also discussed studies in the $A \approx 100$ deformed mass region, as well as around the $Z = 50$ closed shell and the $^{132}$Sn region. It has been demonstrated that for isotopes with at least one type of nucleon close to a magic number, the nuclear moments are sensitive to different types of configuration mixing and core polarization effects. At this point in time, core excitations in the Sn region need to be understood in more detail. Certainly, nuclear-moment studies will continue, pushing the limits towards $N=50$ in the Ni region. In the Sn region, by exploring the precision frontier, simple trends were observed for both the quadrupole moments and nuclear charge radii~\cite{yord13,yord16,yord18,hamm18}. Such studies need to be expanded to other parts of the Segr\'e chart.

An improvement in the precision of moment measurements will reveal subtle nuclear structure effects, similarly to what was observed for precise mass measurements. Studies of higher-order nuclear moments are certainly to be addressed through improvements at the precision frontier. The improvement of the methodology of the spin-precession experiments, which were discussed here, is going to push the discovery frontier to new mass regions further from the valley of $\beta$ stability.

\backmatter



\bmhead{Acknowledgments}
DLB acknowledges partial financial support through grants from the Romanian Ministry of Research, Innovation and Digitalization under contracts No. PN 23 21 01 06 and ELI-RO-RDI-2024-007 (ELITE). AES was supported in part by Australian Research Council Discovery Grants nos.
DP170101673, DP210101201, and DP250100400. HU was partially  supported by JSPS KAKENHI 24H00229






\bigskip

\begin{appendices}
\section{Abbreviations}\label{abbreviations}
\begin{longtable}{|l |p{0.7\linewidth}|}
\hline
 \centering
AFP & adiabatic fast passage technique \\
AFR & adiabatic field rotation\\
ANU & Australian National University, Canberra, Australia \\
ATLAS & Argonne Tandem Linac Accelerator System, USA\\
BareBall &  CsI(Tl) scintillator particle detector at HRIBF\\
CDFT & covariant density functional theory \\
CERN & Conseil Européen pour la Recherche Nucléaire (European Organization for Nuclear Research in Geneva, Switzerland) \\
CFBLS & collinear fast-beam laser spectroscopy \\
CLARION & CLover Array for Radioactive ION Beams\\
COLLAPS & collinear laser spectroscopy (setup at ISOLDE, CERN) \\
CSM & cranked shell model \\
DBLS & Doppler-broadened line shape \\
DFT & density functional theory \\
DIPM & deformed independent particle model \\
DSAM & Doppler-shift attenuation method \\
EFG & electric field gradient \\
EXOGAM & high efficiency, gamma-ray spectrometer at GANIL, France\\
FLAPW & full-potential linearized augmented-plane-wave (method) \\
FOM & figure of merit \\
FRIB & Facility for rare isotope beams (at MSU, East Lansing, MI, USA) \\
GANIL & Grand Accélérateur National d’Ions Lourds (in Caen, France) \\
GFMC & Green's function Monte Carlo (method) \\
GRASP & general relativistic atomic structure package \\
gs & ground state \\
GSI & Gesellschaft für Schwerionenforschung (in Darmstadt, Germany) \\
\texttt{GXPF1A} & shell-model effective interaction for the $pf$ shell\\ 
HF & Hartree-Fock (model) \\
HIAF & Heavy-ion accelerator facility (at ANU, Canberra, Australia) \\
HIMAC & Heavy Ion Medical Accelerator in Chiba (at the QST hospital in Chiba, Japan) \\
HRIBF & Holifield radioactive ion-beam facility (at Oak Ridge, TN, USA) \\
HVTF & high-velocity transient field (technique) \\
IBM & interacting boson model \\
IBFM & interacting boson-fermion (model) \\
IoI & island of inversion \\
ISOL & isotope separation on line (technique) \\
ISOLDE & isotope mass separator on line (experiment at CERN) \\
\texttt{jj44b} & shell-model effective interaction proton and neutrons in the $0f_{5/2}$, $1p_{3/2}$, $1p_{1/2}$, and $0g_{9/2}$ orbits \\
\texttt{JUN45} & shell-model effective interaction  for proton and neutrons in the $0f_{5/2}$, $1p_{3/2}$, $1p_{1/2}$, and $0g_{9/2}$ orbits\\
LEMS & level-mixing spectroscopy (technique) \\
LS & spin-orbit (interaction) \\
LVTF & low-velocity transient field (technique) \\
MCSM & Monte Carlo shell model \\
MEC & meson-exchange current \\
MRI & magnetic resonance imaging \\
MSU & Michigan State University (at East Lansing, MI, USA) \\
MCHF & multi-configuration HF (atomic structure package) \\
NMR & nuclear magnetic resonance \\
NNQR & newly-developed NQR \\
NQCC & nuclear quadrupole coupling constant \\
NQR & nuclear quadrupole resonance \\
NSCL & National superconducting cyclotron laboratory (at MSU, East Lansing, MI, USA) \\
ORGAM & Orsay Gamma Array \\
ORNL & Oak Ridge National Laboratory (at Oak Ridge, TN, USA) \\
OPSA & Orsay particle scintilator array\\
OUPS & Orsay universal plunger device \\
PAC & perturbed angular correlations \\
PAW & projection augmented waves (method) \\
RF & radio frequency \\
RI & radioactive isotope \\
RIB & radioactive ion beam \\
RIKEN & Rikagaku Kenkyūjo (National Institute of Physical and Chemical Research in Japan) \\
RIV & recoil-in-vacuum (technique) \\
QI & quadrupole interaction \\
qp & quasi-particle (1qp, 2qp, 3qp, ...) \\
SOPAD & stroboscopic observation perturbed angular distribution (technique) \\
TAC &  {tilted}-axis cranking (model) \\
TDPAD & time-dependent perturbed angular distribution (technique) \\
TDPAC & time-dependent perturbed angular correlations (technique) \\
TDRIV & time-dependent recoil-in-vacuum (technique) \\
TF & transient field (technique) \\
TFT & tilted-foil technique \\
TRIUMF & Three universities meson facility (in Vancouver, Canada) \\
$U(6/4)$ & boson-fermion single-$j$ super-symmetry scheme \\
$U(6/12)$, $U(6/20)$ & boson-fermion multi-$j$ super-symmetry schemes \\
USD & universal {\it sd} (interaction) \\
VASP & Vienna {\it ab initio} simulation package \\
WIEN2k  & all-electron code implementing the FLAPW method\\
VS-IMSRG & valence-space in-medium similarity renormalization group ({\it ab initio} calculation) \\
$\beta$-NMR & $\beta$-nuclear magnetic resonance \\
$\beta$-NMR & $\beta$-nuclear quadrupole resonance \\
$\Delta \textrm{N}^2\textrm{LO}_{GO}$  & nuclear interaction potential\\
\hline
\end{longtable}

\newpage
\section{Index of mathematical symbols}\label{symbols}

a: \\
\begin{tabular}{l p{0.85\linewidth}}
$a$ & nuclear surface thickness, Eq.~(\ref{eq:q_moment_core}) \\
$A$ & nuclear mass number, Eqs.~(\ref{eq:g_collective}, \ref{eq:dev_angle}, \ref{eq:Rt_2_frag_pi}) \\
$\cal A$ & spin alignment for an axially-symmetric spin ensemble, Eq.~(\ref{eq:alignment}) \\
$A_J$ &  hyperfine interaction constant, Eqs.~(\ref{eq:omegaFF}, \ref{eq:AJ}, \ref{eq:BH}) \\
$A_\beta$ & $\beta$-decay asymmetry parameter, Eqs.~(\ref{Eq:bNMR_Wtheta}, \ref{Eq:bNMR_R}, \ref{eq:bNMR_on_off}, \ref{Eq:bNMR_FOM-T}, \ref{Eq:bNMR_Rdepol})  \\
$A_{k}$ & angular distribution coefficients, Eqs.~(\ref{eq:w_general}, \ref{eq:angular_distribution_coeff}, \ref{eq:w_distribution_axial}, \ref{eq:W_theta_t}, \ref{eq:w_correlation_gamma}, \ref{eq:Rt_gen}, \ref{eq:Rt_2}, \ref{eq:Rt_2pm},  \ref{eq:Rt_2_frag}, \ref{eq:Rt_2_frag_pi}, \ref{eq:QI_pol}, \ref{eq:pacTF}, \ref{eq:ac-sym}) \\
\end{tabular}
\ \\
b: \\
\begin{tabular}{l p{0.85\linewidth}}
 $b$ & temperature-dependence factor of the EFG, Eq.~(\ref{eq:T3/2}) \\
 $b_0$ & magnetic decoupling parameter, Eqs.~(\ref{eq:gnils}, \ref{eq::bm1_strong_coupling}) \\
 $\bm B$ & vector of an external magnetic field, Eqs.~(\ref{eq:Hamiltonian-magnetic}, \ref{Eq:bNMR_Bloch0}, \ref{Eq:bNMR_Bloch-omg}, \ref{Eq:bNMR_Beff}, \ref{Eq:bNMR_Bloch-T1T2}, \ref{Eq:bNQR_Hml}), Fig.~\ref{fig:TDPAD_scheme} \\
 $B$ & strength of an external magnetic field, Eqs.~(\ref{eq:gammaR}, \ref{eq:Larmor}, \ref{eq:E_zeeman}, \ref{eq:Heff}, \ref{Eq:bNMR_dThdT}, \ref{Eq:bNQR_Em}, \ref{Eq:bNMR_AFRcond2}, \ref{eq:Rt_2_frag_pi}) \\
 $B^\prime$ & static magnetic field in the rotating frame of reference, Eq.~(\ref{Eq:bNMR_FastCond}) Sect.~\ref{Sec:AFP} \\
 $B_{\textrm{eff}}$ & strength of an effective magnetic field, Eqs.~(\ref{Eq:bNMR_Beff}, \ref{eq:bNQR_comp}) \\
 ${\bm B}_\textrm{eff}$ & vector of the effective magnetic field, Eq.~(\ref{Eq:bNMR_Beff}), Fig.~(\ref{FIG:bNMR_B0B1}) \\
 $B_{\rm{H}}$ & hyperfine magnetic field at a nucleus, Eqs.~(\ref{eq:AJ}, \ref{eq:BH}) \\
 $B_1$ & amplitude of an oscillating magnetic field, Eqs.~(\ref{Eq:bNMR_B1-RotField}, \ref{Eq:bNMR_Beff}, \ref{Eq:bNMR_omgeff}, \ref{Eq:bNMR_dThdT},  \ref{Eq:bNMR_AdiabaticCond}, \ref{Eq:bNMR_Bloch-T1T2}, \ref{Eq:bNMR_FastCond}, \ref{Eq:bNMR_AFPCond}), Fig.~(\ref{FIG:bNMR_B0B1}) \\
 $\bm B_1$ & vector of the oscillating magnetic  {field}, Eq.~(\ref{Eq:bNMR_Beff}) \\
 $B_\textrm{1s}$ &hyperfine field at the nucleus due to the 1s electron, Eqs.~(\ref{eq:B1s}, \ref{eq:omegas}) \\
$B_{k}$ & orientation parameters for an axially symmetric spin ensemble, Eqs.~(\ref{eq:B_orientation}, \ref{eq:alignment}, \ref{eq:w_distribution_axial}, \ref{eq:W_theta_t}, \ref{eq:Rt_gen}, \ref{eq:Rt_2}, \ref{eq:Rt_2pm}, \ref{eq:Rt_2_frag}, \ref{eq:Rt_2_frag_pi}, \ref{eq:QI_pol}, \ref{eq:ac-sym})  \\
$B_{k}^q$ & spin orientation statistical tensor, Eqs.(\ref{eq:B_k^q}, \ref{eq:pacTF}, \ref{eq:pacRIV}), related to $\rho_q^k$ through Eq.~(\ref{eq:B_k^q}) \\
$B_\textrm{ns}$ & magnetic field 
at the nucleus from an electron in the n-th {\it s} orbit, Eqs.~(\ref{eq:B_k^q}, \ref{eq:BTF_model}, \ref{eq:B_ns}) \\
$B_{\textrm{TF}}$ & TF strength, Eq.~(\ref{eq:dtheta}) \\
$B(E2)$ & $E2$ reduced transition probability, Eqs.~(\ref{eq::bm1be2}, \ref{eq::BE2_rotational}, \ref{eq::BE2_rotational}, \ref{eq:BE2-ratio}) \\
$B(M1)$ & $M1$ reduced transition probability, Eqs.~(\ref{eq::bm1_strong_coupling}, \ref{eq::bm1be2} \ref{eq::BM1_CSM}) \\
$B(v,Z)$ & empirical TF parametrization function, Eqs.~(\ref{eq:lin-param}, \ref{eq:RUparam}, \ref{eq:CRparam}) \\
\end{tabular}
\ \\
c: \\
\begin{tabular}{l p{0.85\linewidth}}
$C_k$ & de-orientation parameter for the  strength of the RIV interaction, Eq.~(\ref{eq:Ck-defn}, \ref{eq:Gkemp})\\
\end{tabular}
\ \\
d: \\
\begin{tabular}{l p{0.85\linewidth}}
$d_\textrm{f}$ & demagnetization factor, Eq.~(\ref{eq:df}) \\
$D_{m^\prime m}^j(\alpha,\beta,\gamma)$ & Wigner rotational D-matrix, Eqs.~(\ref{eq:DYP}, \ref{eq:pacTF}, \ref{eq:pacRIV}) \\
\end{tabular}
\ \\
\newpage
\noindent e: \\
\begin{tabular}{l p{0.85\linewidth}}
$e$ & elementary electric charge \\
$e_{j}$ & charge of a nucleon in $j$-th orbit, Eqs.~(\ref{eq:q_moment_Cart}, \ref{eq:q_moment_spherical}, \ref{eq:q_sp}) \\
$e_\omega(Z,N,\beta$) & Routhian at rotational frequency $\omega$ and deformation $\beta$, Sect.~\ref{subsec:axially_def} \\
$E_0$ & bandhead energy in a rotational band, Eq.~(\ref{eq::K_rotational_energies}, \ref{eq::rotational_energies})\\
$E_m$ & energy of the $m$-th magnetic substate, Eqs.~(\ref{eq:E_zeeman}, \ref{Eq:bNQR_Em}) \\
$E(I)$ & energy of a state in a rotational band with spin $I$, Eqs.~(\ref{eq::K_rotational_energies}, \ref{eq::rotational_energies}) \\
$E_{\gamma}$ & energy of a $\gamma$-ray transition, Eqs.~(\ref{eq::pureK-delta}, \ref{eq:gKgRQ0}, \ref{eq::bm1be2}) \\
\end{tabular}
\ \\
\noindent f: \\
\begin{tabular}{l p{0.85\linewidth}}
$F$ & total angular momentum of an atom or molecule, Eqs.~(\ref{eq:omegaFF}, \ref{eq:Gk}, \ref{eq:alphak}) \\
$\bm F$ & total angular momentum vector of an atom or a molecule, Fig.~\ref{fig:free-ion} \\
$F_{k}$ & ordinary $F$ coefficients, Eqs.~(\ref{eq:angular_distribution_coeff}, \ref{eq:f_ordinary}, \ref{eq:pacRIV}) \\
$F_{k}^{k_1,k_0}$ & generalized $F$ coefficients Eq.~(\ref{eq:f_generalized})\\
$F_{ns}^1$ & fraction of ions with a single electron in the n-th orbit, Eq:~(\ref{eq:BTF_model}) \\
\end{tabular}
\ \\

\noindent g: \\
\begin{tabular}{l p{0.85\linewidth}}
$g$ & gyromagnetic factor for a nuclear state, Eqs.~(\ref{eqn:mu_nuclear}, \ref{eq:gammaR}, \ref{eq:mu_substate}, \ref{eq:mu_WE}, \ref{eq:additivity}, \ref{eq:sm_gl_gs}, \ref{Eq:MEC_muSchmidt}, \ref{eq:mucollectivsL},  \ref{eq:gcollectiveL}, \ref{eq:g_collective}, \ref{eq:gIBM2a}, \ref{eq:gIBM2}, \ref{eq:gcollectiveMoI}, \ref{eq:gnils}, \ref{eq::bm1_strong_coupling}, \ref{eq:Larmor}, \ref{eq:E_zeeman}, \ref{eq:dev_angle}, \ref{eq:Rt_2_frag_pi}, \ref{eq:dtheta}, \ref{eq:AJ}, \ref{eq:BH}, \ref{eq:omegas}, \ref{eq:Ck-defn}, \ref{eq:Gkemp}, \ref{eq::gK_additivity}) \\
$\langle g \rangle$ & average quasi-continuum $g$ factor, Sect.~\ref{par::superdef} \\
$g^{(0)}$ & isoscalar part of the $g$ factor, Eq.~(\ref{Eq:MEC_gISIV}) \\
$g^{(1)}$ & isovector part of the $g$ factor, Eq.~(\ref{Eq:MEC_gISIV}) \\
$g_i$ & $g$ factor of an aligned nucleon, Eq.~(\ref{eq::g-average}) \\
$g_K$ & single-particle $g$ factor in rotational band, Eqs.~(\ref{eq:gnils}, \ref{eq::bm1_strong_coupling}, \ref{eq:gKgRQ0}, \ref{eq::BM1_CSM}, \ref{eq::mu_rotational_bandhead}, \ref{eq::mu_rotational}, \ref{eq::gK_additivity}) \\
$g_{\ell}$ & orbital $g$ factor of a valence nucleon, Eqs.~(\ref{eq:sm_gl_gs}, \ref{Eq:MEC_muSchmidt}) \\
$g_{\ell}^\textrm{free}$ & orbital $g$ factor of a free nucleon, Eqs.~(\ref{eq:mu_eff}, \ref{Eq:MEC_mu}) \\
$g_{\ell}^{n,\textrm{free}}$ & orbital $g$ factor of a free neutron, Sect.~\ref{sec:collective} \\
$g_{\ell}^{p,\textrm{free}}$ & orbital $g$ factor of a free proton, Sect.~\ref{sec:collective} \\
$g_n^{\rm{conf}}$ & $g$ factor of a neutron configuration, Eqs.~(\ref{eq:g_chiral}, \ref{Eq:MEC_OddEvenAna}) \\
$g_\textrm{R}$ & $g$ factor of the rotational core of the nucleus, Eqs.~(\ref{eq:gnils},
\ref{eq::bm1_strong_coupling}, \ref{eq:gKgRQ0}, \ref{eq::BM1_CSM}, \ref{eq::g-average}, \ref{eq::mu_rotational_bandhead}, \ref{eq::mu_rotational}, \ref{eq:g_chiral}) \\
$g_\textrm{p}$ & tensor term of a $g$ factor, Eqs.~(\ref{eq:mu_eff}, \ref{Eq:MEC_mu}) \\
$g_p^{\rm{conf}}$ & $g$ factor of a proton configuration, Eqs.~(\ref{eq:g_chiral}, \ref{Eq:MEC_muSchmidt}) \\
$g_s$ & intrinsic (spin) $g$ factor of a valence nucleon, Eqs.~(\ref{eq:sm_gl_gs}, \ref{Eq:MEC_muSchmidt}) \\
$g_s^\textrm{free}$ & intrinsic (spin) $g$ factor of a free nucleon, Eqs.~(\ref{eq:mu_eff}, \ref{Eq:MEC_mu}) \\
\end{tabular}
\begin{tabular}{l p{0.85\linewidth}}
$g_{s}^{n,\textrm{free}}$ & intrinsic (spin) $g$ factor of a free neutron, Eq.~(\ref{Eq:MEC_SpinMirror}), Sect.~\ref{intro_nucleon} \\
$g_{s}^{p,\textrm{free}}$ & intrinsic (spin) $g$ factor of a free proton, Eq.~(\ref{Eq:MEC_SpinMirror}), Sect.~\ref{intro_nucleon} \\
$g_{\textrm{S}}$ & $g$ factor of S-band states, Eq.~(\ref{eq::g-average}) \\
$\langle g_{\rm{SD}}\rangle$ & average $g$ factor of a superdeformed band, Sect.~\ref{par::superdef} \\
$g_{\pi}$ & $g$ factor of proton bosons, Eq.~(\ref{eq:gIBM2}) \\
$g_{\nu}$ & $g$ factor of neutron bosons, Eq.~(\ref{eq:gIBM2}) \\
$G_{\textrm{beam}}$ & TF multiplicative correction factor, Eq.~(\ref{eq:Gbeam}) \\
$G_{k}(t)$ & time-differential attenuation coefficient, Eqs.~(\ref{eq:rhot}, \ref{eq:Gk}, \ref{eq:Gkinf}, \ref{eq:GK}, \ref{eq:flight}, \ref{eq:lorentz-Gkt}, \ref{eq:Gkemp}), Fig.~\ref{fig:Gkexamples} \\
$G_{k}(t,B)$ & interaction coefficient of the nuclear spin ensemble and external magnetic field, Eqs.~(\ref{eq:f_ordinary}, \ref{eq:W_theta_t}) \\
$G_{k}^\textrm{flight}(T)$ & RIV average de-orientation coefficient for ions in flight, Eq.~(\ref{eq:flight}) \\
$G_k^{\rm{slow}}$ & RIV de-orientation coefficient for ions after the reset foil, Eq.~(\ref{eq:Gkslow}) \\
$G_{k}^\textrm{stopped}(t_\textrm{f})$ & RIV perturbation factor for stopped ions, Eq.~(\ref{eq:stopped}) \\
$G_{k}(\infty)$ & RIV integral perturbation factor, Eqs.~(\ref{eq:Ginfty}, \ref{eq:Gkslow}) \\
$G_{k}^\infty(\tau)$ & time-integral attenuation factor, Eqs.~(\ref{eq:Gkinf}, \ref{eq:lorentz-Gktau}, \ref{eq:Gkemp}), Fig.~\ref{fig:Gkexamples} \\
$G_{\textrm{k}_1\textrm{k}_2}^{\textrm{N}_1\textrm{N}_2}$ & perturbation factors, Eqs.~(\ref{eq:QI_pol}, \ref{eq:G_k_correlations}) \\
\end{tabular}
\ \\
\noindent h: \\
\begin{tabular}{l p{0.85\linewidth}}
$h$ & Plank constant \\
$\hbar$ & reduced Plank constant \\
$\hbar \omega$ & rotational frequency with respect to the rotational frame, Eq.~(\ref{eq::BM1_CSM}) \\
$\hbar\omega_c$ & backbenging (band-crossing) frequency, Sect.~\ref{par::high-spin} \\
$H_\textrm{demag}$ & demagnitization field strength, Eq.~(\ref{eq:Heff}) \\
$H_\textrm{eff}$ & effective hyperfine field strength at a nucleus, Eq.~(\ref{eq:Heff}) \\
$H_\textrm{static}$ & hyperfine field strength, Eq.~(\ref{eq:Heff}) \\
$\hat{\cal H}^\textrm{(B)}$ & Hamiltonian of the nuclear boson core, Eq.~(\ref{eq:HIBFM}) \\
$\hat{\cal H}^\textrm{(F)}$ & Hamiltonian of the uncoupled fermion, Eq.~(\ref{eq:HIBFM}) \\
$\hat{\cal H}_\textrm{IBFM}$ & IBFM Hamiltonian, Eq.~(\ref{eq:HIBFM}) \\
$\hat{\cal H}_\textrm{R}$ & rotational Hamiltonian, Sect.~\ref{subsub:nilsson} \\
$\hat {\cal H}_{\textrm{Q}}$ & Hamiltonian of the quadrupole interaction, Eq.~(\ref{eq:Q_hamilt_general}, \ref{eq:Hamiltonian-Q+EFQ}, \ref{eq:Hamiltonian-etazeero}) \\
$\hat {\cal H}_{\mu}$ & Hamiltonian of the combined Zeeman and quadrupole interaction, Eq.~(\ref{eq:Hamiltonian-magnetic}) \\
$\hat {\cal H}_{\mu \textrm{Q}}$ & Hamiltonian of the combined Zeeman and quadrupole interaction, Eq.~(\ref{Eq:bNQR_Hml}) \\
\end{tabular}
\ \\

\noindent i: \\
\begin{tabular}{l p{0.85\linewidth}}
$i$ & aligned angular momentum in the rotating reference frame, Eq.~(\ref{eq::BM1_CSM}) \\
$i_\textrm{S}$ & aligned angular momentum of the $S$-band in the rotating reference frame, Eq.~(\ref{eq::g-average}) \\
$I$ & nuclear total angular momentum (nuclear spin) value, Eqs.~(\ref{eq:q_moment_spec}, \ref{eq:q_s-q_0}, \ref{eq:mu_substate}, \ref{eq:mu_WE}, \ref{eq:additivity}, \ref{eq:wf_sp_conf}, \ref{eq:wf_miced_conf}, \ref{eq:mucollectivsL}, \ref{eq:gnils}, \ref{eq::bm1_strong_coupling}, \ref{eq::pureK-delta}, \ref{eq:gKgRQ0}, \ref{eq::bm1be2}, \ref{eq:Hamiltonian-magnetic},  \ref{eq:Q_tensor}, \ref{eq:Q_tensor_comp}, \ref{eq:Hamiltonian-Q+EFQ}, \ref{eq:Hamiltonian-etazeero}, \ref{eq:quadrupole_frequency}, \ref{eq:rho_axial}, \ref{eq:B_orientation}, \ref{eq:B_k^q}, \ref{eq:alignment}, \ref{eq:alpha2max}, \ref{eq:w_general}, \ref{eq:f_ordinary}, \ref{eq:f_generalized}, \ref{eq:de-orientation}, \ref{eq:de-orientation_mixed}, \ref{eq:de-orientation_total}, \ref{Eq:bNQR_Hml}, \ref{Eq:bNQR_Em}, \ref{Eq:bNQR_NuQ}, \ref{eq:GK},  \ref{eq::K_rotational_energies}, \ref{eq::rotational_energies}, \ref{eq::BM1_CSM}, \ref{eq::mu_rotational_bandhead}, \ref{eq::mu_rotational}, \ref{Eq:MEC_SpinMirror}, \ref{eq:BE2-ratio})
\\
$\bm I$ & nuclear total angular momentum (nuclear spin) vector, Eqs.~(\ref{eqn:mu_nuclear}, \ref{eq:g_chiral}), Fig.~\ref{fig:free-ion} \\
$\hat{I}$ & nuclear total angular momentum (spin) operator, Eqs.~(\ref{eq:Q_tensor}, \ref{eq:Q_tensor_comp}, \ref{eq:Hamiltonian-Q+EFQ}, \ref{eq:Hamiltonian-etazeero}, \ref{Eq:bNQR_Hml}, \ref{Eq:bNQR_Em}) \\
$I_\textrm{c}$ & angular momentum of the nuclear core, Eqs.~(\ref{eq:wf_sp_conf}, \ref{eq:wf_miced_conf}) \\
$\hat I_i$ & components of the spin operator on the principal axes, Eqs.~(\ref{eq:Q_tensor}, \ref{eq:Q_tensor_comp}, \ref{eq:Hamiltonian-Q+EFQ}, \ref{eq:Hamiltonian-etazeero}, \ref{Eq:bNQR_Hml})  \\
$I_\textrm{S}$ & total angular momentum of the $S$-band, Eq.~(\ref{eq::g-average}) \\
$\hat I_{\pm}$ & spin ladder operators, Eqs.~(\ref{eq:Q_tensor}, \ref{eq:Q_tensor_comp}, \ref{eq:Hamiltonian-Q+EFQ}, \ref{Eq:bNQR_Hml}) \\
$I_\gamma$ & intensity of a $\gamma$-ray transition, Eqs.~(\ref{eq::pureK-delta}, \ref{eq::bm1be2}, \ref{eq:int_theta_t}, \ref{eq:Rt_90}, \ref{eq:Rt_B})  \\
$I_{\beta}$ & average $\beta$-decay counting rate, Eq.~(\ref{Eq:bNMR_FOM-T}) \\
\end{tabular}
\ \\

\noindent j: \\
\begin{tabular}{l p{0.85\linewidth}}
$j$ & total angular momentum of the nucleon, Eqs.~(\ref{eq:q_sp}, \ref{eq:sm_gl_gs}, \ref{Eq:MEC_muSchmidt}, \ref{eq:wf_sp_conf}, \ref{eq:wf_miced_conf}) \\
$\bm j$ & total angular momentum vector of the nucleon \\
$\bm j_n$ & total angular momentum vector of a neutron configuration, Eq.~(\ref{eq:g_chiral}), Fig.~\ref{fig::31_magnetic_vectors}, \ref{fig::31_chiral_vectors} \\
$\bm j_p$ & total angular momentum vector of a neutron configuration, Eq.~(\ref{eq:g_chiral}), Fig.~\ref{fig::31_magnetic_vectors}, \ref{fig::31_chiral_vectors} \\
$\bm j_R$ & core angular momentum vector of a nucleus, Fig.~\ref{fig::31_chiral_vectors} \\
$\hat j_{\pm}$ & single nucleon total angular momentum ladder operators, Sect.~\ref{subsub:nilsson} \\
$J$ & total angular momentum vector of atomic of molecular configuration, Eqs.~(\ref{eq:AJ}, \ref{eq:BH}, \ref{eq:Gk}) \\
$\bm J$ & rotational angular momentum vector of an ion, atom or molecule, Sect.~\ref{sec:EFG}, Fig.~\ref{fig:free-ion} \\
$\cal J$ & moment of inertia of the nucleus, Eqs.~(\ref{eq:gcollectiveMoI}, \ref{eq::K_rotational_energies}, \ref{eq::rotational_energies}) \\
$\cal J_{\textrm{n}}$ & moment of inertia of the neutrons, Eq.~(\ref{eq:gcollectiveMoI}) \\
$\cal J_{\textrm{p}}$ & moment of inertia of the protons, Eq.~(\ref{eq:gcollectiveMoI}) \\
${\cal J}^{(2)}$ & dynamical moment of inertia, Sect.~\ref{par::K_isomers} \\
\end{tabular}
\ \\

\noindent k: \\
\begin{tabular}{l p{0.85\linewidth}}
$\bm k$ & emission direction vector of a particle, Sect.~\ref{sec:orientation} \\
$K$ & projection of the total angular momentum on the symmetry axis of the nucleus, Eqs.~(\ref{eq:q_s-q_0}, \ref{eq:gnils}, \ref{eq::bm1_strong_coupling}, \ref{eq::pureK-delta}, \ref{eq::K_rotational_energies}, \ref{eq::rotational_energies}, \ref{eq::BM1_CSM}, \ref{eq::mu_rotational_bandhead}, \ref{eq::gK_additivity}), Sect.~\ref{subsub:nilsson} \\
$K_\textrm{S}$ & Knight shift, Eq.~(\ref{eq:Heff}) \\
\end{tabular}
\ \\

\noindent l: \\
\begin{tabular}{l p{0.85\linewidth}}
$\ell$ & orbital angular momentum of a nucleon, Eq.~(\ref{eq:sm_gl_gs}, \ref{Eq:MEC_muSchmidt}) \\
$\bm \ell$ & orbital angular momentum vector of a nucleon, Eqs.~(\ref{mu_classical}, \ref{mu_qm}) \\
$\hat \ell$ & orbital angular momentum operator of a nucleon, Eqs.~(\ref{eq:M1}, \ref{eq:mu_eff}, \ref{Eq:MEC_mu}) \\
$L$ & total orbital angular momentum of the nucleons in a nucleus, Eqs.~(\ref{eq:mucollectivsL},\ref{eq:gcollectiveL}) \\
$L_n$ & orbital angular momentum of the neutrons in a nucleus, Eqs.~(\ref{eq:mucollectivsL}, \ref{eq:gcollectiveL}) \\
$L_p$ & orbital angular momentum of the protons in a nucleus, Eqs.~(\ref{eq:mucollectivsL}, \ref{eq:gcollectiveL}) \\
\end{tabular}
\ \\

\noindent m: \\
\begin{tabular}{l p{0.85\linewidth}}
$m_e$ & mass of the electron, Sect.~\ref{intro_basic_concepts} \\
$m_p$ & mass of the proton, Eq.~(\ref{eq:muN}) \\
$M$ & foil magnetization, Sect.~\ref{sect:SF} \\
$\bm M$ & magnetization vector, Eqs.~(\ref{Eq:bNMR_Bloch0}, \ref{Eq:bNMR_Bloch-omg}, \ref{Eq:bNMR_Bloch-T1T2}), Sect.~\ref{Sec:AFP} \\
$M_0$ & thermal equilibrium magnetization, Eq.~(\ref{Eq:bNMR_Bloch-T1T2}) \\
${\hat{M1}}$ & magnetic dipole moment operator, Sect.~\ref{sec:eff_M1} \\
$\hat{M}_3^0$ & magnetic octupole moment operator, Sect.~\ref{intro_basic_concepts} \\
\end{tabular}
\ \\

\noindent n: \\
\begin{tabular}{l p{0.85\linewidth}}
$N$ & nuclear neutron number \\
$N_\gamma^{\uparrow(\downarrow)}(\theta_\gamma)$ & $\gamma$-ray rate with field up(down), Eq.~(\ref{eq:rho}) \\
$N_{\pi}$ & number of proton bosons, Eqs.~(\ref{eq:gIBM2a}, \ref{eq:gIBM2}) \\
$N_{\nu}$ & number of neutron bosons, Eqs.~(\ref{eq:gIBM2a}, \ref{eq:gIBM2}) \\
\end{tabular}
\ \\

\noindent p: \\
\begin{tabular}{l p{0.85\linewidth}}
$p(m)$ & population probability of the $m$-th substate, Eqs.~(\ref{eq:rho_axial}, \ref{eq:pm_alignment}, \ref{eq:pm_polarization}, \ref{eq:B_orientation}) \\
$P$ & spin polarization, Eqs.~(\ref{Eq:bNMR_Wtheta}, \ref{Eq:bNMR_R}, \ref{eq:bNMR_on_off}, \ref{Eq:bNMR_dRR}, \ref{Eq:bNMR_FOM-T}, \ref{Eq:bNMR_Rdepol}) \\
$\bm P$ & polarization vector of a particle \\
$P_k(\cos\theta$) & Legendre polynomials, Eqs.~(\ref{eq:w_distribution_axial}, \ref{eq:W_theta_t}, \ref{eq:w_correlation_gamma}, \ref{Eq:bNMR_Wtheta}, \ref{eq:ac-sym}) \\
$P_k^q(\cos\theta$) & associated Legendre polynomials, Eq.~(\ref{eq:DYP}) \\
\end{tabular}
\ \\

\noindent q: \\
\begin{tabular}{l p{0.85\linewidth}}
$q$ & ion charge, Eqs.~(\ref{eq:dev_angle},  \ref{eq:Rt_2_frag_pi})  \\
$\hat Q$ & nuclear electric quadrupole moment operator, Eqs.~(\ref{eq:q_moment_Cart}, \ref{eq:q_moment_spherical}, \ref{eq:q_moment_spec})  \\
$Q_0$ & intrinsic quadrupole moment, Eqs.~(\ref{eq:q_s-q_0}, \ref{eq:gKgRQ0}, \ref{eq::BE2_rotational}) \\
$Q_2^{0,\pm 1,\pm 2}$ & components of $Q_{ij}$ on the principal axes, Eq.~(\ref{eq:Q_tensor_comp}) \\
$Q_c$ & quadrupole moment of the nuclear core, Eqs.~(\ref{eq:q_moment_core_sharp}, \ref{eq:q_moment_core}) \\
$Q_{ij}$ & tensor of the nuclear electric quadrupole moment, Eq.~(\ref{eq:Q_tensor}) \\
$\hat{Q}_{ij}$ & nuclear electric quadrupole moment tensor operator , Eqs.~(\ref{eq:Q_hamilt_general}, \ref{eq:Q_tensor}, \ref{eq:Q_tensor_comp}) \\
$Q_{\textrm{s}}$ & spectroscopic quadrupole moment, Eqs.~(\ref{eq:q_moment_spec}, \ref{eq:q_s-q_0}, \ref{eq:Q_tensor}, \ref{eq:Q_tensor_comp}, \ref{eq:Hamiltonian-Q+EFQ}, \ref{eq:Hamiltonian-etazeero}, \ref{eq:quadrupole constatnt}, \ref{eq:quadrupole_frequency}, \ref{Eq:bNQR_Hml}, \ref{Eq:bNQR_Em}) \\
$Q_{sp}$ & single-particle quadrupole moment, Eq.~(\ref{eq:q_sp}) \\
\end{tabular}
\ \\

\noindent r: \\
\begin{tabular}{l p{0.85\linewidth}}
$\langle r^2 \rangle$ & mean-square radius \\
$R$ & nuclear radius, Eqs.~(\ref{eq:q_moment_core_sharp}, \ref{eq:q_moment_core}) \\
$\langle r_\textrm{j}^2 \rangle$ & mean-square radius of a nucleon in orbit $j$, Eq.~(\ref{eq:q_sp}) \\
$R$ & nuclear radius, Eqs.~(\ref{eq:q_moment_core_sharp}, \ref{eq:q_moment_core}) \\
$\bm R$ & core rotation vector, Sect.~\ref{subsub:nilsson} \\
$R(t)$ & TDPAD/TDPAC ratio function, Eqs.~(\ref{eq:Rt_90}, \ref{eq:Rt_B}, \ref{eq:Rt_gen}, \ref{eq:Rt_2}, \ref{eq:Rt_2pm}, \ref{eq:Rt_2_frag}, \ref{eq:Rt_2_frag_pi}) \\
$R(Z)$ & TF (RIV) relativistic correction factor, Eqs.~(\ref{eq:lin-param}, \ref{eq:B_ns}, \ref{eq:B1s}) \\
$R_\textrm{on}$ & up/down ratio of $I_\beta$ with applied $B_1$ field, Eqs.~(\ref{Eq:bNMR_R}, \ref{eq:bNMR_on_off}, \ref{Eq:bNMR_dRR}, \ref{Eq:bNMR_Rdepol}) \\
$R_\textrm{off}$ & up/down ratio of $I_\beta$ without $B_1$ field applied , Eqs.~(\ref{eq:bNMR_on_off},  \ref{Eq:bNMR_dRR}, \ref{Eq:bNMR_Rdepol}) \\
\end{tabular}
\ \\ \\
s: \\
\begin{tabular}{l p{0.85\linewidth}}
$s$ & intrinsic spin of the nucleon \\
$\hat s$ & spin operator of a nucleon, Eqs.~(\ref{eq:M1}, \ref{eq:mu_eff}, \ref{Eq:MEC_mu}) \\
$S_{\textrm{N}_1\textrm{N}_2}^{\textrm{k}_1\textrm{k}_2}$ & perturbation coefficients, Eqs.~(\ref{eq:G_k_correlations}) \\
$S(\theta_\gamma)$ & slope (logarithmic derivative) of the TF angular correlation function, Eq.~(\ref{eq:S}) \\
\end{tabular}
\ \\
\noindent t: \\
\begin{tabular}{l p{0.85\linewidth}}
$T_{\rm f}$ & flight time, Eqs.~(\ref{eq:stopped},\ref{eq:flight}, \ref{eq:FT},\ref{eq:Gkslow}) \\
$t_\textrm{hr}$ & time of half rotation of $B$, Eqs.~(\ref{eq:bNQR_comp}, \ref{Eq:bNMR_AFRcond1}, \ref{Eq:bNMR_AFRcond2}), Fig.~\ref{FIG:bNMR_AFR} \\
$t_\textrm{m}$ & measurement time, Eq.~(\ref{Eq:bNMR_FOM-T}) \\
$t_\textrm{RF}$ & RF application time, Sect.~\ref{subsub:bNQR} \\
$t_\textrm{sl}$ & spin-lattice relaxation time, Eq.~(\ref{Eq:bNMR_Bloch-T1T2}) \\
$t_\textrm{ss}$ & spin-spin relaxation time, Eqs.~(\ref{Eq:bNMR_Bloch-T1T2}, \ref{Eq:bNMR_FastCond}, \ref{Eq:bNMR_AFPCond}) \\
$T_{1/2}$ & half-life of a nuclear state or radioisotope \\
$T, T_z$ & isospin quantum number \\
$T_\gamma(L)$ & partial $\gamma$-ray intensity for a given multipolarity, Sect.~\ref{sec:orientation} \\
\end{tabular}
\ \\
u: \\
\begin{tabular}{l p{0.85\linewidth}}
$U$ & Coulomb potential, Eq.~(\ref{eq:V_tensor}), Sect. \ref{sec:EFG} \\
$U_{k}(I_1I_2L)$ & de-orientation coefficients for a pure $\gamma$-ray transition, Eq.~(\ref{eq:de-orientation})\\
$U_{k}(I_1I_2LL^\prime)$ & de-orientation coefficients for a mixed $\gamma$-ray transition, Eq.~(\ref{eq:de-orientation_mixed})\\
\end{tabular}
\ \\
v: \\
\begin{tabular}{l p{0.85\linewidth}}
$v$ & particle velocity Eqs.~(\ref{mu_classical}, \ref{eq:lin-param}, \ref{eq:RUparam}, \ref{eq:CRparam}, \ref{eq:aes-param}, \ref{eq:BTF_model}) \\
$v_0$ & Bohr velocity of the electron, Eq.~(\ref{eq:lin-param}) \\
$\hat{V}^\textrm{(BF)}$ & boson-fermion interaction operator, Eq.~(\ref{eq:HIBFM}) \\
$V_\textrm{ij}$ & EFG tensor, Eqs.~(\ref{eq:V_tensor}, \ref{eq:V_tensor_diagonal}, \ref{eq:Q_hamilt_general}) \\
$V_{ii}$ & values of $V_\textrm{ij}$ on the principal axes, Eq.~(\ref{eq:V_tensor_diagonal}), Sect.~\ref{sec:EFG} \\
$V_\textrm{zz}$ & EFG value, Eqs.~(\ref{eq:V_tensor_diagonal}, \ref{eq::eta}, \ref{eq:T3/2}), Sect.~\ref{sec:EFG} \\
\end{tabular}
\ \\
\newpage
\noindent w: \\
\begin{tabular}{l p{0.85\linewidth}}
$W({\bm k})$ & directional angular distribution function,  Sect.~\ref{sec:orientation} \\
$W({\bm k},{\bm P})$ & polarization angular distribution function, Sect.~\ref{sec:orientation} \\
$W(t)$ & time-dependence of the $\gamma$-ray anisotropic distribution, Eq.~(\ref{eq:QI_pol}) \\
$W(\theta)$ & axially symmetric angular distribution function, Eqs.~(\ref{eq:w_distribution_axial}, \ref{eq:w_correlation_gamma}, \ref{Eq:bNMR_Wtheta}) \\
$W(\theta,\phi)$ & angular distribution of the emitted radiation,  Eq.~(\ref{eq:w_general}) \\
$W(\theta,\phi,B)$ & angular distribution of the emitted radiation with external magnetic field,  Eqs.~(\ref{eq:W_theta_t}, \ref{eq:int_theta_t}) \\
$W(\theta_\gamma,\theta_\textrm{p},\Delta\phi)$ & particle-$\gamma$ angular correlation,  Eq.~(\ref{eq:pacTF}) \\
$W(\theta_\gamma,\theta_\textrm{p},\Delta\phi,t)$ & time-dependent particle-$\gamma$ angular correlation,  Eq.~(\ref{eq:pacRIV}) \\
$W^{\uparrow(\downarrow)}$ & TF up/down perturbed angular correlations, Eqs.~(\ref{eq:W_up_down}, \ref{eq:epsilon1}) \\
\end{tabular}
\ \\
y: \\
\begin{tabular}{l p{0.85\linewidth}}
$Y_k^q(\theta,\phi)$ & Laplace spherical harmonics,  Eqs.~(\ref{eq:q_moment_spherical}, \ref{eq:mu_eff}, \ref{Eq:MEC_mu}, \ref{eq:w_general}, \ref{eq:w_distribution_axial}, \ref{eq:DYP}, \ref{eq:QI_pol})  \\
\end{tabular}
\ \\
z: \\
\begin{tabular}{l p{0.85\linewidth}}
$Z$ & atomic number, Eqs.~(\ref{eq:q_moment_core_sharp}, \ref{eq:q_moment_core}, \ref{eq:g_collective}) \\
\end{tabular}
\ \\ \\
$\alpha$: \\
\begin{tabular}{l p{0.85\linewidth}}
$\alpha$ & angle between the beam axis and the ion spin orientation axis, Eqs.~(\ref{eq:dev_angle}, \ref{eq:Rt_2_frag}), Fig.~\ref{fig:phase_alpha} \\
$\alpha_k$ & hard-core value of a time-differential attenuation coefficient, Eqs.~(\ref{eq:alphak}, \ref{eq:lorentz-Gkt}, \ref{eq:lorentz-Gktau}) \\
$\alpha_k(m)$ & spin-orientation attenuation coefficient, Eqs.~(\ref{eq:B_orientation}, \ref{eq:alignment}, \ref{eq:alpha2max}) \\
\end{tabular}
\ \\ \\
$\beta$: \\
\begin{tabular}{l p{0.85\linewidth}}
$\beta$ & nuclear deformation parameter, Eqs.~(\ref{eq:q_moment_core_sharp}, \ref{eq:q_moment_core}) \\
$\beta_2$ & nuclear quadrupole deformation parameter \\
\end{tabular}
\ \\ \\
$\gamma$: \\
\begin{tabular}{l p{0.85\linewidth}}
$\gamma_{\textrm{R}}$ & gyromagnetic ratio, Eqs.~(\ref{eq:gammaR}, \ref{Eq:bNMR_Bloch0}, \ref{Eq:bNMR_Bloch-omg}, \ref{Eq:bNMR_Beff}, \ref{Eq:bNMR_omgeff}, \ref{Eq:bNMR_dThdT}, \ref{Eq:bNMR_AdiabaticCond}, \ref{Eq:bNMR_Bloch-T1T2}, \ref{Eq:bNMR_FastCond}, \ref{Eq:bNMR_AFPCond}, \ref{Eq:bNMR_AFRcond1}, \ref{Eq:bNMR_AFRcond2}) \\
$\Gamma_k$ & RIV relaxation constant, Eqs.~(\ref{eq:lorentz-Gkt}, \ref{eq:lorentz-Gktau}, \ref{eq:Ck-defn}) \\
\end{tabular}
\ \\ \\
$\delta$: \\
\begin{tabular}{l p{0.85\linewidth}}
$\delta$ & multipolarity mixing ratio for a $\gamma$-ray transition,  Eqs.~(\ref{eq::pureK-delta}, \ref{eq:gKgRQ0}, \ref{eq::bm1be2}, \ref{eq:angular_distribution_coeff}) \\
$\delta g_{\ell}$ & correction factor to $g_{\ell}$, Eqs.~(\ref{eq:mu_eff}, \ref{Eq:MEC_mu}) \\
$\delta g_{s}$ & correction factor to $g_{s}$, Eqs.~(\ref{eq:mu_eff}, \ref{Eq:MEC_mu}) \\
$\Delta e^\prime$ & signature splitting in the rotating reference frame,  Eqs.~(\ref{eq::bm1be2}, \ref{eq::BM1_CSM}) \\
$\Delta p/p$ & momentum selection (in fragmentation reaction), Sect.~\ref{subsubsec:spin_orientations} \\
$\Delta\theta$ & transient-field precession angle, Eqs.~(\ref{eq:W_up_down}, \ref{eq:delta-theta}) \\
\end{tabular}
\ \\ \\
$\epsilon$: \\
\begin{tabular}{l p{0.85\linewidth}}
$\epsilon_\textrm{TF}$ & transient-field measured ``effect", Eqs.~(\ref{eq:epsilon}, \ref{eq:epsilon1}) \\
\end{tabular}
\ \\ \\
$\eta$: \\
\begin{tabular}{l p{0.85\linewidth}}
$\eta$ & asymmetry parameter of the EFG, Eqs.~(\ref{eq::eta}, \ref{eq:Hamiltonian-Q+EFQ}, \ref{Eq:bNQR_Hml}, \ref{Eq:bNQR_Em}) \\
\end{tabular}
\ \\ \\
$\xi$: \\
\begin{tabular}{l p{0.85\linewidth}}
$\xi_\textrm{ns}(v,Z)$ & TF degree of polarization caused by a single electron in the n-th orbit, Eqs.~(\ref{eq:BTF_model}, \ref{eq:p1s-param}) \\
$\xi_\textrm{1s}$ & $K$-shell polarization caused by a single electron, Eqs.~(\ref{eq:p1s-param}), Fig.~\ref{fig:TF-p1s} \\
\end{tabular}
\ \\

\noindent $\lambda:$ \\
\begin{tabular}{l p{0.85\linewidth}}
$\lambda$ & decay constant of a nuclear state, Eqs.~(\ref{eq:flight}, \ref{eq:FT}) \ \\
\end{tabular}
\ \\ \\
\noindent $\mu$: \\
\begin{tabular}{l p{0.85\linewidth}}
$\mu$ & expectation value of the nuclear magnetic dipole moment, Eqs.~(\ref{eq:mu_substate}, \ref{eq:mu_WE}, \ref{Eq:MEC_muSchmidt}, \ref{eq:mucollectivsL}, \ref{Eq:bNQR_Em}, \ref{eq::mu_rotational_bandhead}, \ref{eq::mu_rotational}) \\
$\bm \mu$ & vector of the nuclear magnetic dipole moment, Eqs.~(\ref{mu_classical}, \ref{mu_qm}, \ref{eqn:mu_nucleon}, \ref{eqn:mu_nuclear}, \ref{eq:Hamiltonian-magnetic}) \\
${\hat \mu}$ & magnetic moment operator, Eqs.~(\ref{eq:M1_sum}, \ref{eq:mu_eff}, \ref{Eq:MEC_mu}, \ref{Eq:MEC_muISIV}) \\
$\mu^{(0)}$ & isoscalar magnetic moment, Eq.~(\ref{Eq:MEC_muISIV}) \\
$\mu^{(1)}$ & isovector magnetic moment, Eq.~(\ref{Eq:MEC_muISIV}) \\
$\hat \mu^{(i)}$ & magnetic moment operator of a nucleon, Eqs.~(\ref{eq:M1}, \ref{eq:M1_sum}) \\
$\mu_{\textrm{N}}$ & nuclear magneton, Eqs.~(\ref{eq:muN}, \ref{eqn:mu_nuclear},  \ref{eq::bm1_strong_coupling}, \ref{eq:Larmor}, \ref{eq:E_zeeman}, \ref{eq:Rt_2_frag_pi}) \\
$\mu_\pm$ & magnetic moment for a $T_{\rm{z}}=\pm T$ state, Eq.~(\ref{Eq:MEC_muISIV}) \\
\ \\
\end{tabular}
\ \\
\noindent $\nu$: \\
\begin{tabular}{l p{0.85\linewidth}}
$\nu$ & oscillation frequency \\
$\nu_\textrm{L}$ & Larmor frequency, Eq.~(\ref{Eq:bNQR_NuQ}) \\
$\nu_{\textrm{Q}}$ & nuclear quadrupole coupling constant, Eqs.~(\ref{eq:quadrupole constatnt}, \ref{Eq:bNQR_NuQ}) \\
$\nu_{m,m+1}$ & resonance frequency between $m$-th and ($m$+1)-th magnetic sub-states, Eq.~(\ref{Eq:bNQR_NuQ}) \\
\end{tabular}
\ \\
$\rho$: \\
\begin{tabular}{l p{0.85\linewidth}}
$\rho(I)$ & spin-distribution matrix,  Sect.~(\ref{sec:orientation}) \\
$\rho^\textrm{k}_\textrm{q}$ & spin orientation statistical tensor, Eqs.~(\ref{eq:rho_axial}, \ref{eq:B_orientation}, \ref{eq:B_k^q}, \ref{eq:w_general}, \ref{eq:rhoFFt}, \ref{eq:rhot}), related to $B^q_k$ through Eq.~(\ref{eq:B_k^q}) \\
$\rho_\textrm{TF}$ & double-ratio of $\gamma$-ray detection rates in TF experiment, Eqs.~(\ref{eq:rho}, \ref{eq:epsilon}) \\
\end{tabular}
\ \\ \\
$\theta$: \\
\begin{tabular}{l p{0.85\linewidth}}
$\theta_{\textrm{c}}$ & beam deviation angle,  Eqs.~(\ref{eq:dev_angle}, \ref{eq:Rt_2_frag_pi}), Fig.~\ref{fig:phase_alpha} \\
$\theta_{\textrm{L}}$ & rotation angle of the spin ensemble in dipole magnetic field,  Eq.~(\ref{eq:dev_angle}), Fig.~\ref{fig:phase_alpha} \\
$\theta_\gamma$ & position angles of TF detectors, Eq.~(\ref{eq:W_up_down}) \\
\end{tabular}
\ \\ \\
$\sigma$: \\
\begin{tabular}{l p{0.85\linewidth}}
$\langle\sigma_\textrm{z}\rangle$ & isoscalar spin expectation value, Eq.~(\ref{Eq:MEC_SpinMirror}) \\
$\bm \sigma$ & Pauli spin matrix, Eq.~(\ref{eqn:mu_nucleon}) \\
\end{tabular}
\ \\ \\
$\tau$: \\
\begin{tabular}{l p{0.85\linewidth}}
$\tau$ & mean lifetime of a nuclear state, Eqs.~(\ref{eq:int_theta_t}, \ref{eq:FT}, \ref{eq:Gkinf}, \ref{eq:lorentz-Gktau}, \ref{eq:Gkemp}) \\
$\tau_\textrm{p}$ & resonance passage time, Eq.~(\ref{Eq:bNMR_FastCond}) \\
\end{tabular}
\ \\

$\omega$: \\
\begin{tabular}{l p{0.85\linewidth}}
$\omega$ & angular velocity, Eqs. (\ref{Eq:bNMR_B1-RotField}, \ref{Eq:bNMR_omgeff}, \ref{Eq:bNMR_dThdT}) \\
$\bm \omega$ & angular velocity vector, Eqs.~(\ref{Eq:bNMR_Bloch-omg}, \ref{Eq:bNMR_Beff}) \\
$\omega_\textrm{c}$ & cyclotron frequency, Sect.~\ref{sec:isomers} \\
$\omega_\textrm{eff}$ & effective angular velocity, Eqs.~(\ref{Eq:bNMR_omgeff}, \ref{Eq:bNMR_AFP0}, \ref{Eq:bNMR_dThdT}, \ref{Eq:bNMR_AFPomg}), Fig.~\ref{FIG:bNMR_B0B1} \\
$\omega_{\rm FF^\prime}$ & hyperfine precession frequency, Eqs.~(\ref{eq:omegaFF}, \ref{eq:Gk}) \\
$\omega_\textrm{L}$ & Larmor precession, Eqs.~(\ref{eq:gammaR}, \ref{eq:Hamiltonian-magnetic}, \ref{eq:Larmor}, \ref{Eq:bNMR_omgeff}, \ref{Eq:bNMR_AFP0}, \ref{Eq:bNMR_dThdT}, \ref{Eq:bNMR_AFPomg}, \ref{Eq:bNMR_AdiabaticCond}, \ref{eq:Rt_gen}, \ref{eq:Rt_2}, \ref{eq:Rt_2pm}, \ref{eq:Rt_2_frag}) \\
$\omega_\textrm{Q}$ & nuclear quadrupole frequency, Eqs.~(\ref{eq:quadrupole_frequency}, \ref{eq:G_k_correlations}, \ref{eq:omega0}) \\
$\omega_\textrm{s}$ & RIV precession frequency for a H-like ion, Eqs.~(\ref{eq:omegas}, \ref{eq:GK}, \ref{eq:FT}, \ref{eq:Gkinf})\\
$\Omega$ & projection of the angular momentum of a nucleon on the symmetry axis (Nilsson model quantum number), Sect.~\ref{subsub:nilsson} \\
$\Omega_3$ & magnetic  {octupole} moment, Sect.~\ref{intro_basic_concepts} \\
\end{tabular}





\end{appendices}



\bibliography{EM_review}


\end{document}